\documentclass[acmtog,nonacm,screen]{acmart}
\usepackage{algorithm,algpseudocode,tabularx,array,booktabs,multirow}
\usepackage{graphicx,subcaption,placeins,xcolor,needspace}
\algrenewcommand\textproc{}
\acmJournal{TOG}
\setcopyright{none}
\acmDOI{}
\allowdisplaybreaks[1]
\newif\ifrevisionnotes
\revisionnotestrue
\ifdefined\DiffGIPCReading\revisionnotesfalse\fi

\newcommand{\newtablemark}{}
\newcommand{\R}{\mathbb{R}}
\newcommand{\vx}{\mathbf{x}}
\newcommand{\vq}{\mathbf{q}}
\newcommand{\vy}{\mathbf{y}}
\newcommand{\vg}{\mathbf{g}}
\newcommand{\vtheta}{\boldsymbol{\theta}}
\newcommand{\vlambda}{\boldsymbol{\lambda}}
\newcommand{\vzero}{\mathbf{0}}

\newcommand{\T}{\mathsf{T}}

\newcolumntype{Y}{>{\raggedright\arraybackslash}X}

\begin{document}
% Keep references and citations black in this clean copy.
\hypersetup{hidelinks}
% Avoid stretching inter-paragraph space to justify float-heavy columns.
\raggedbottom
\title{DiffGIPC: Differentiable Mixed-Coordinate Contact Mechanics on GPU}
\author{Kemeng Huang}
\orcid{0000-0001-9147-2289}
\email{kmhuang@connect.hku.hk}
\email{kmhuang819@gmail.com}
\affiliation{%
  \institution{The University of Hong Kong}
  \city{Hong Kong}
  \country{Hong Kong SAR, China}
}

\author{Bowen Ren}
\email{boweyna@gmail.com}
\affiliation{%
  \institution{The University of Hong Kong}
  \city{Hong Kong}
  \country{Hong Kong SAR, China}
}

\author{Xinyu Lu}
\email{lxy819469559@gmail.com}
\affiliation{%
  \institution{TransGP}
  \city{Hong Kong}
  \country{Hong Kong SAR, China}
}

\author{Minchen Li}
\orcid{0000-0001-9868-7311}
\email{minchernl@gmail.com}
\affiliation{%
  \institution{Carnegie Mellon University}
  \city{Pittsburgh}
  \country{USA}
}

\author{Taku Komura}
\orcid{0000-0002-2729-5860}
\email{taku@cs.hku.hk}
\affiliation{%
  \institution{The University of Hong Kong}
  \city{Hong Kong}
  \country{Hong Kong SAR, China}
}

\renewcommand{\shortauthors}{Huang et al.}

\begin{abstract}

Contact-rich inverse problems require efficient gradients and reliable enforcement of non-penetration. Existing differentiable methods provide fast simulation and rigid-deformable coupling, but contact treatments based on penalties or velocity-level constraints can permit penetration or positional drift. Incremental Potential Contact (IPC) maintains non-penetration through geometric barriers and {continuous collision detection (CCD)}. However, repeated forward and backward evaluations remain costly in full-space finite-element (FEM) implementations, particularly when nearly rigid components retain many vertex {degrees of freedom (DoFs)}. We present DiffGIPC, a GPU-accelerated differentiable IPC framework for optimizing interacting deformable and nearly rigid bodies. Our affine-body derivatives enable {rest-shape} design, initial-state optimization, and articulated control while retaining only twelve motion DoFs per affine body. These gradients account for how changes in geometry, placement, and actuation affect other objects through collision and friction. We preserve these interactions in one coupled FEM-affine adjoint system per frame, reducing the backward-system size without discarding the dependencies that drive the inverse problem. GPU-optimized backward computation further reduces the cost of repeated gradient evaluations. A reusable PyTorch interface connects supported physical parameters to user-defined parameter maps and losses across design, identification, motion, placement, and articulated control. Experiments demonstrate peak speedups over DiffIPC of approximately $150.0\times$ with full-space FEM and {$4133.3\times$} with affine bodies for combined forward simulation and backward differentiation.
\end{abstract}


\begin{teaserfigure}
  \centering
  \begin{subfigure}[t]{\dimexpr0.25\textwidth-0.5pt\relax}
    \centering
    \includegraphics[width=\linewidth, trim=0 0 0 0, clip]{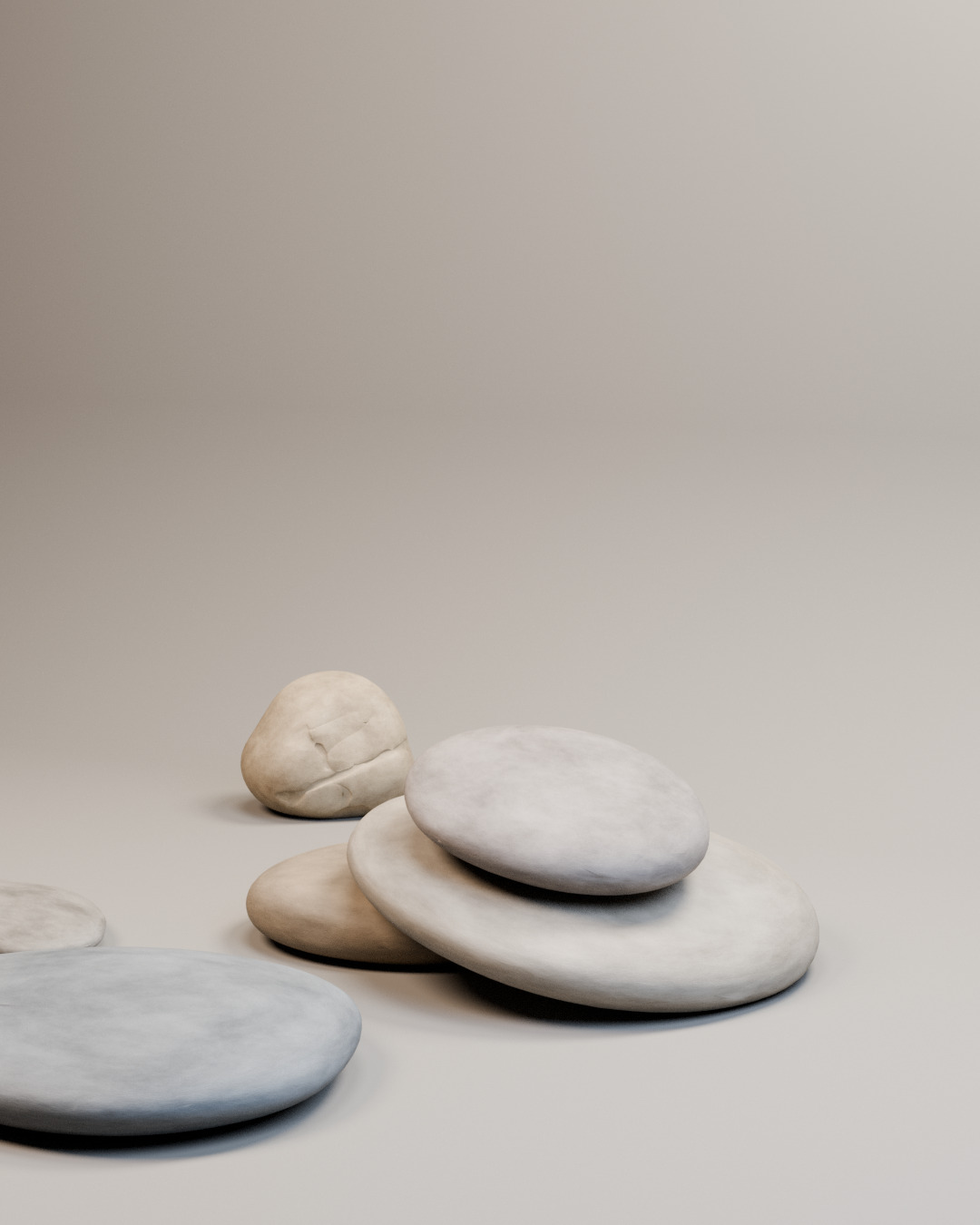}
    \caption{Initial placement}
  \end{subfigure}\hspace{0.5pt}%
  \begin{subfigure}[t]{\dimexpr0.25\textwidth-0.5pt\relax}
    \centering
    \includegraphics[width=\linewidth, trim=0 0 0 0, clip]{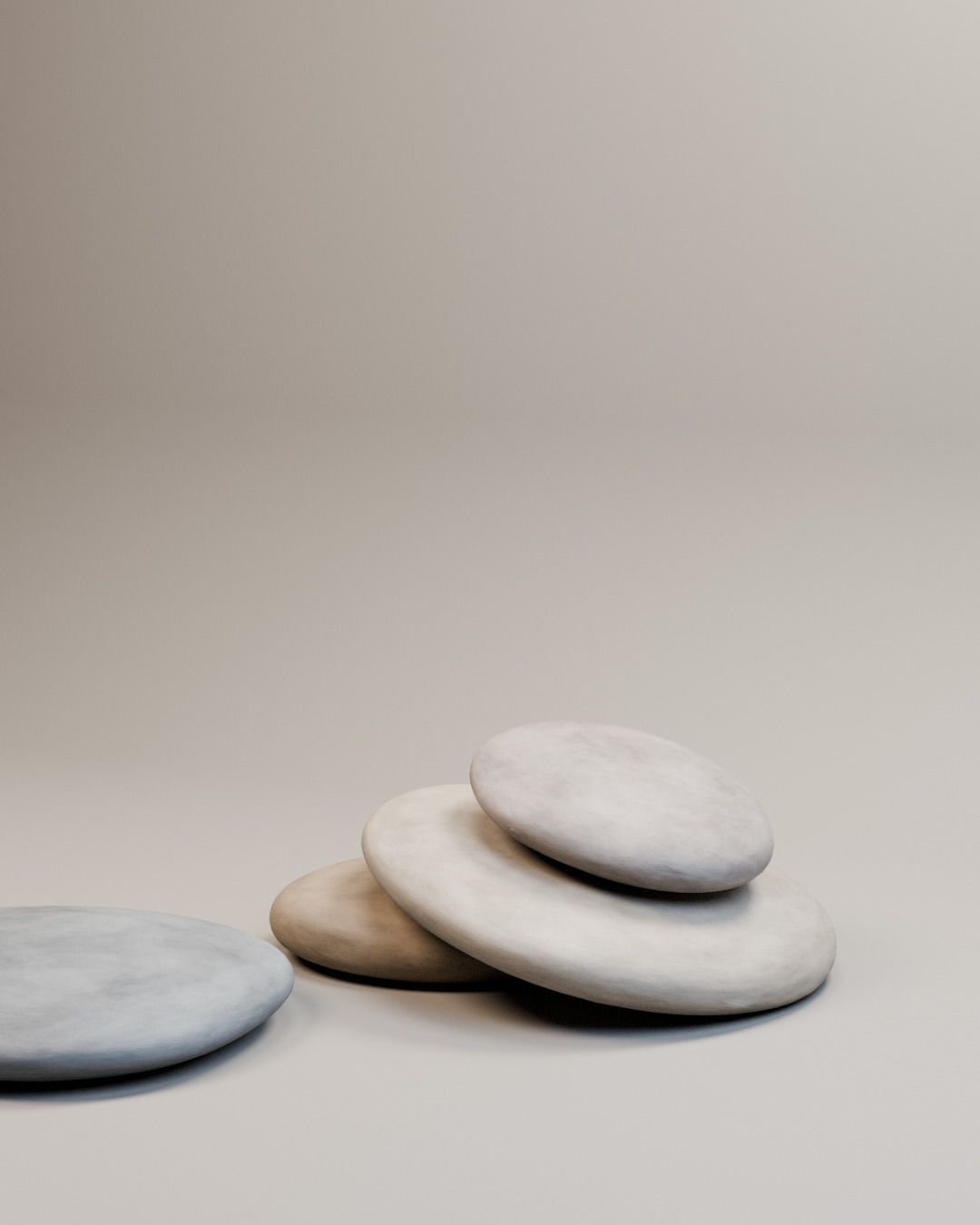}
    \caption{Iteration 10}
  \end{subfigure}\hspace{0.5pt}%
  \begin{subfigure}[t]{\dimexpr0.25\textwidth-0.5pt\relax}
    \centering
    \includegraphics[width=\linewidth, trim=0 0 0 0, clip]{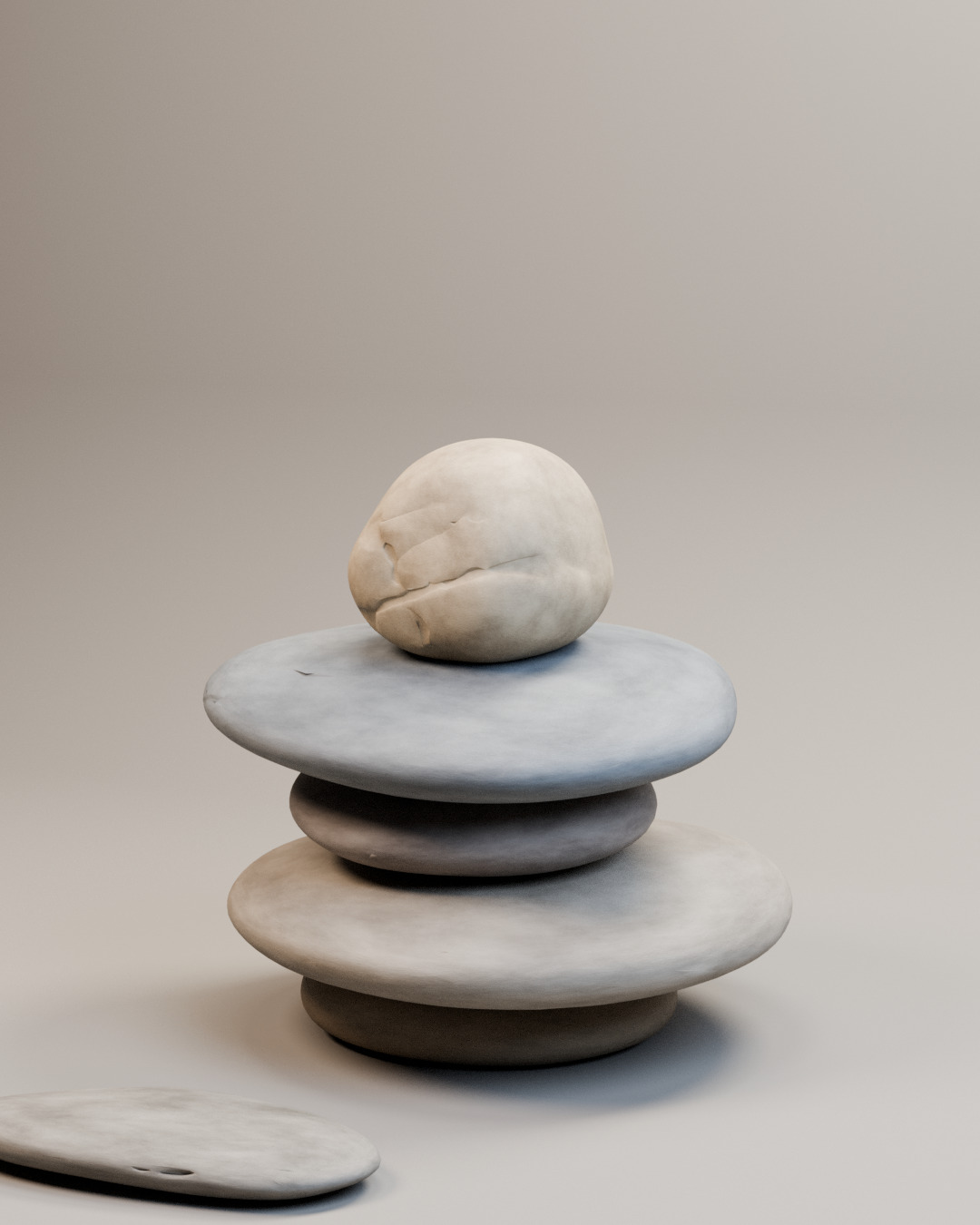}
    \caption{Iteration 50}
  \end{subfigure}\hspace{0.5pt}%
  \begin{subfigure}[t]{\dimexpr0.25\textwidth-0.5pt\relax}
    \centering
    \includegraphics[width=\linewidth, trim=0 0 0 0, clip]{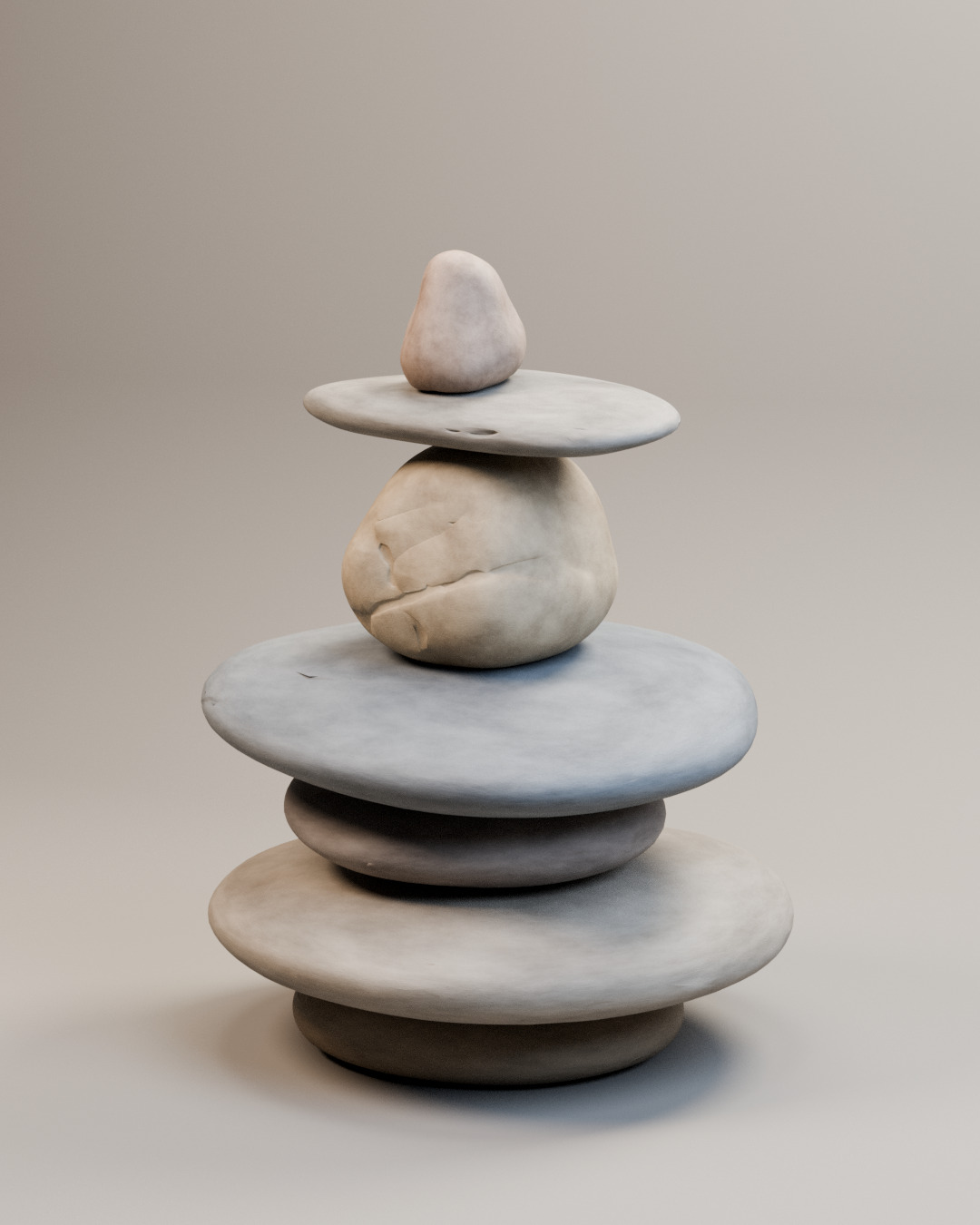}
    \caption{Iteration 78}
  \end{subfigure}
  \caption{Optimizing the placement of seven affine bodies. Keeping
  their shapes fixed, DiffGIPC adjusts only the horizontal initial
  positions to form a stable stack. The panels show the unoptimized
  configuration and the simulated results after 10, 50, and 78
  iterations.}
  \Description{Four panels show the initial stone placement and
  optimization updates 10, 50, and 78 at selected rendered states.}
  \label{fig:stone-placement}
\end{teaserfigure}
\maketitle

% Compact caption spacing in the body, leaving the title and teaser unchanged.
\captionsetup[figure]{skip=4pt}
\captionsetup[subfigure]{skip=2pt}
\captionsetup[table]{skip=4pt}

\section{Introduction}
\label{sec:introduction}

\begin{figure*}[tp]
  \centering
  \includegraphics[width=0.95\textwidth]{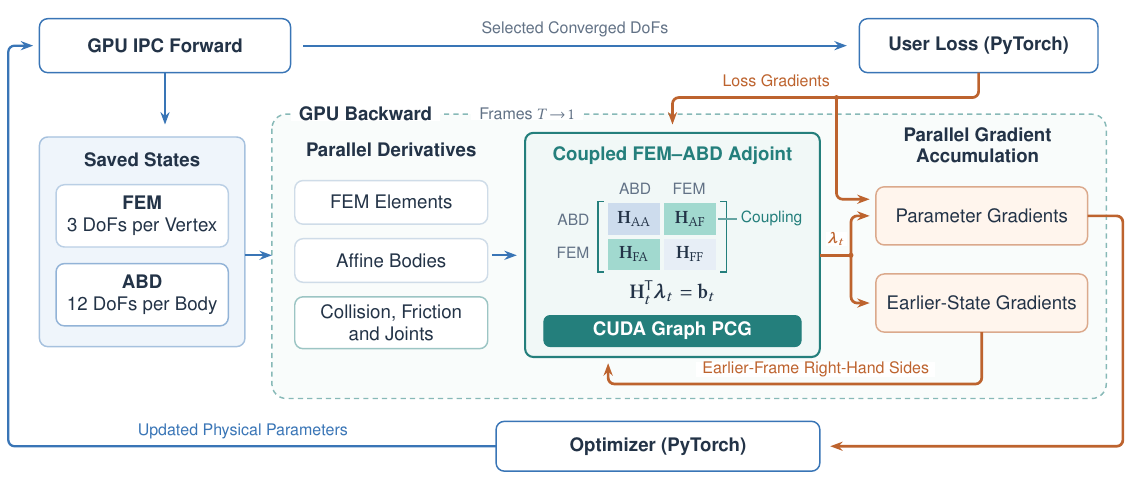}
  \caption{DiffGIPC's backward pipeline. Each reverse frame assembles one
  coupled FEM-ABD adjoint system, retaining collision, friction, and joint
  interactions with twelve DoFs per affine body. The GPU solve and local
  accumulation propagate gradients to earlier states and physical
  parameters for optimization through PyTorch.}
  \Description{Saved FEM and affine states feed local derivative evaluation, matrix assembly, one coupled adjoint solve, and gradient accumulation. The PyTorch loss supplies gradients, and the parameter map returns physical-parameter gradients to the optimizer.}
\label{fig:schematic-overview}
\end{figure*}

Gradient-based design, system identification, and control often optimize one object's parameters for their effect on another object. {A hanger's shape determines whether a garment remains supported. An initial velocity determines where another body moves after collision, and a robot command changes the motion of a grasped deformable object.} In each case, the gradient must follow forces transmitted through contact and the subsequent motion. Computing these derivatives efficiently is particularly challenging when a scene combines deformable objects with detailed, nearly rigid components: the representation must resolve the interactions without making every stiff component expensive to differentiate.

Existing methods address different parts of this problem. Differentiable projective-dynamics methods, including DiffPD and DiffCloth~\cite{du2021_diffpd,Li2022diffcloth}, accelerate repeated solves through local-global iterations and matrix prefactorization. Representing nearly rigid components as stiff deformable meshes nevertheless retains vertex-level motion DoFs, increasing simulation and differentiation costs even when only near-rigid motion is required~\cite{lai2026diffphd}. ADD~\cite{add2020} provides fully implicit, two-way coupled rigid and deformable dynamics with analytic adjoints, but uses penalty-based normal contact, which does not strictly prevent interpenetration. DiffIPC~\cite{diffipc} differentiates fully implicit FEM dynamics with friction, using IPC barriers and continuous collision detection to maintain non-penetration. Its full-space representation, however, carries the motion DoFs of every stiff component into the backward solve, making repeated evaluations costly in its CPU implementation. The resulting opportunity is to combine contact-coupled differentiation with compact motion coordinates and efficient GPU execution.

We present DiffGIPC, a GPU-accelerated framework for differentiating
coupled FEM-affine IPC simulations
(Figure~\ref{fig:schematic-overview}). Building on the forward
formulation of StiffGIPC~\cite{stiffgipc2025}, we derive frame-wise
adjoints directly in the mixed motion coordinates. Full-space FEM
retains the freedom needed for local deformation, while each affine
body contributes only twelve motion DoFs regardless of its surface
resolution. Contact and joint derivatives couple interacting bodies
within each frame. Inertia and friction make the time integration equations depend
on earlier states.
The backward systems therefore remain compact while preserving cross-object coupling and the temporal dependencies required for coupled inverse problems.

Compact motion coordinates alone do not provide shape or control gradients. Changing an affine rest surface changes its mass, the positions at which contact is evaluated, and the maps that convert point-space forces to affine coordinates. Under friction, the same rest change also moves the previous-frame points used to define sliding and the lagged contact quantities. We differentiate these dependencies explicitly, together with initial states and revolute or prismatic joint controls, and accumulate them in
parallel without enlarging the adjoint system. To retain the advantage of compact systems, conditional CUDA Graphs are employed to keep {preconditioned conjugate gradient (PCG)} iterations and convergence checks on the GPU, thereby reducing the overhead of repeated kernel launches and CPU–GPU synchronization. A PyTorch interface connects these
physical gradients to user-defined parameter maps and losses.

In summary, our contributions are:
% \begin{itemize}
%     \item \textbf{Coupled FEM-affine adjoints for differentiable IPC.}
%     We derive a trajectory adjoint directly in mixed FEM and affine coordinates, retaining contact, friction, and joint coupling in one backward system per frame. This formulation captures interactions between deformable and nearly rigid bodies while keeping only twelve motion DoFs per affine body.  

%     \item \textbf{Affine-body derivatives for shape design and articulated control.}
%     We derive gradients for affine rest shapes, initial states, and joint controls, including the effects of shape-dependent inertia and frictional contact. These derivatives enable detailed rest shape design and articulated control without introducing vertex-level motion DoFs for affine bodies.

%     \item \textbf{A reusable GPU optimization pipeline.}
%     We introduce a GPU-parallel pipeline for adjoint computation and parameter-gradient accumulation, using conditional CUDA Graphs to reduce iteration overhead for compact backward systems. A reusable PyTorch interface connects these gradients to user-defined parameter maps and losses, allowing the same backward computation to support design, identification, and control.

%     \item \textbf{Fully GPU-parallel differentiable IPC.}
%     We present the first fully GPU-parallel differentiable IPC framework for coupled nearly rigid and deformable bodies. Extensive experiments demonstrate its computational efficiency and effectiveness across design, identification, and control tasks.
% \end{itemize}
\begin{itemize}
\item \textbf{Coupled FEM-affine adjoints for differentiable IPC.}
We derive trajectory adjoints in mixed FEM and affine coordinates, preserving contact, friction, and joint coupling while using only twelve motion DoFs per affine body.

\item \textbf{Affine-body derivatives for shape design and articulated control.}
We derive gradients for affine rest shapes, initial states, and joint controls, including shape-dependent inertia and frictional contact.

\item \textbf{A reusable GPU optimization pipeline.}
We develop a GPU-parallel adjoint and gradient-accumulation pipeline with conditional CUDA Graphs and a PyTorch interface for design, system identification, and control.

\item \textbf{Fully GPU-parallel differentiable IPC.}
We present the first fully GPU-parallel differentiable IPC framework for coupled nearly rigid and deformable bodies, with substantial speedups demonstrated across contact-rich inverse problems.
\end{itemize}

Our evaluation examines the benefits of GPU execution and compact motion representations across contact-rich inverse problems. For comparison with DiffIPC, we retain full-space FEM in both systems and use relatively low-resolution benchmarks to keep evaluation times manageable for the CPU-based baseline. Even at these resolutions, DiffGIPC achieves {$12.3$--$34.1\times$} speedups in complete backward evaluation and {$10.6$--$60.9\times$} speedups in combined forward and backward evaluation, covering initial motion, rest shape, material, and friction optimization. A separate resolution study examines how performance scales with geometric detail and the additional benefit of affine coordinates. Further experiments demonstrate cloth design, affine shape optimization, stable placement, and articulated control. {The source code of our framework will be released.}

\section{Related Work}
\label{sec:related}

% Related-work reconstruction: problem-driven comparisons, 2026-09-30.
% All 40 works cited in the uploaded Related Work are retained in the text.

\paragraph{Efficient differentiable simulation and contact}
Inverse problems such as parameter identification from measured motion~\cite{hahn2019real2sim,jatavallabhula2021gradsim} require repeated simulation and gradient evaluations, making the efficiency of both computations important. DiffTaichi automates the differentiation of parallel simulation
programs through source-code transformations
\cite{hu2019difftaichi}. Brax integrates differentiable rigid-body
simulation and learning on accelerators, reducing data transfer
between simulation and optimization \cite{freeman2021brax}.
ChainQueen~\cite{hu2019chainqueen} uses explicit MLS-MPM with parallel local updates, but stiff materials require small time steps.
Implicit deformable simulation methods~\cite{rojas2021implicit,liang2019cloth,li2026physics} offer greater time-step stability, at the cost of solving linear systems during both simulation and differentiation.
{Projective dynamics (PD) reduces this cost through local projections and a global system whose matrix can be prefactorized for fixed discretization and coefficients~\cite{bouaziz2014pd}. DiffPD reuses this factorization in its backward iterations~\cite{du2021_diffpd}.}
Quasi-Newton methods support more general hyperelastic energies through efficient approximate systems~\cite{liu2017quasinewton}, and DiffQN further reduces differentiation costs by reusing factorizations from the forward simulation~\cite{cai2025diffqn}.
More recently, DiffPhD combines GPU factor reuse with techniques for heterogeneous and hyperelastic materials~\cite{lai2026diffphd}.

For contact-rich inverse problems, these computational benefits must also be considered alongside the contact response being differentiated.
DiffPD's penalty formulation supports frictional sliding but permits interpenetration, whereas its complementarity formulation enforces non-penetration under a sticky-contact assumption~\cite{du2021_diffpd}.
DiffCloth differentiates the dry-friction PD formulation of \citet{ly2020pdfriction}, supporting sliding cloth contact~\cite{Li2022diffcloth}.
However, its self-collision treatment considers only vertex-vertex pairs and does not handle vertex-face or edge-edge self-collisions.
DiffPhD supports Signorini-Coulomb contact on the GPU, but its velocity-level enforcement can produce position drift under quasi-static support~\cite{lai2026diffphd}.
These limitations are particularly relevant when the optimized behavior depends on sustained support, sliding, or contact between thin surfaces.
They motivate efficient differentiation of contact models that preserve geometric non-penetration throughout the simulated motion.

\paragraph{Differentiable contact and coupled dynamics}
\citet{qiao2020scalable} enable differentiable two-way coupling between rigid bodies and cloth, using local impact zones to reduce the cost of collision handling and differentiation.
ADD provides fully implicit rigid-deformable coupling with analytic adjoints~\cite{add2020}.
Its normal penalty makes contact differentiable while allowing finite
{overlap. Increasing the penalty stiffness reduces overlap at the cost
of poorer conditioning.} IPC instead combines a geometric
barrier with continuous collision detection to maintain
intersection-free motion from feasible initial configurations
\cite{Li2020IPC}. DiffIPC brings this contact model to differentiable
FEM simulation and supports shape, material, friction, and initial-state
optimization \cite{diffipc}.
It therefore provides a strong basis for inverse problems in which
penetration would change the intended interaction.
However, representing nearly rigid components with highly stiff,
full-space FEM retains vertex-level motion DoFs even when their local
deformation is negligible. In its CPU-based implementation, repeated
simulation and differentiation of these large systems remain costly.
DiffGIPC addresses this cost with coupled FEM-ABD adjoints and GPU
backward computation while retaining IPC's geometric contact safeguards.

\paragraph{Compact coordinates for coupled design and control}
Reducing motion DoFs addresses a cost that faster linear algebra alone
does not remove. Modal and adaptive subspaces reduce deformable motion
to a smaller set of directions
{\cite{barbic2005subspace,trusty2024tradingspaces}. Their accuracy depends
on whether those directions capture the deformation induced by loads
and contact.} Hyper-reduced PD also approximates local constraint
projections \cite{brandt2018hyperreduced}, while subspace mixed FEM
combines reduced positions with stretch variables to handle
heterogeneous materials \cite{trusty2023subspace}.
These approaches retain selected deformation modes and reduce their
evaluation cost. For components whose internal deformation is negligible,
rigid-body coordinates avoid resolving these modes altogether.
Rigid IPC preserves non-penetration in this representation
by accounting for curved rotational trajectories in collision detection
\cite{ferguson2021rigidipc}.
ABD relaxes exact rigidity through a stiff orthogonality energy and
uses twelve affine DoFs per body \cite{ABD}.
Its surface positions are linear in these DoFs, allowing piecewise-linear
collision trajectories and a pose-independent mass matrix at fixed
rest geometry. This offers a compact representation for stiff parts,
while full-space FEM retains local deformation where it is needed.
DiffGIPC carries both representations into the same backward system,
retaining contact, friction, and joint coupling between bodies.

Compact rigid-body states have supported differentiable learning and
control \cite{belbuteperes2018,degrave2019physics}, including
articulated dynamics in Nimble \cite{werling2021nimble}.
Shape optimization requires an additional distinction between the
motion representation and the geometry being designed.
Contact-based assemblies and shock-protecting microstructures
\cite{tozoni2021assemblies,huang2024shock} use geometric changes to
{alter load transfer and contact-dependent mechanical response.
Pneumatic actuator design and garment reconstruction
\cite{gjoka2024pneumatic,li2025dress123} likewise optimize rest geometry
through its effect on deformation.}
These applications motivate retaining a rich design space even when
the motion state can be compact.

SDRS uses convex polyhedra and auxiliary
separating planes to maintain differentiability under large shape
changes \cite{ye2025sdrs}.
Our focus is mesh-based design and control of affine bodies interacting
with deformable objects under IPC.
Affine {rest-shape} changes affect both the mass matrix and the maps
{from affine DoFs to surface points. Differentiating these dependencies
allows a body's shape to be optimized for its frictional interactions
with other objects.} Derivatives of joint targets and forces
likewise account for the coupled response of articulated and deformable bodies.
For a fixed motion representation, additional design parameters do not enlarge the per-frame backward system. Each affine body therefore contributes only twelve motion DoFs, even when its rest shape is described by many design parameters. Contact processing still depends on surface resolution, while parameter-gradient evaluation requires additional work as the number of design parameters grows.

\paragraph{GPU computation for trajectory adjoints}
Compact coordinates reduce adjoint system size, while efficient
linear solvers further reduce the cost of trajectory differentiation.
PD-based local-global methods exploit a fixed global matrix
to reuse precomputed factors. \citet{zeng2025fba} use a sparse
representation of its inverse to accelerate GPU forward simulation.
DiffPhD adopts this strategy for differentiable simulation, reusing
an inverse factor across forward, backward, and frictional-contact
computations \cite{lai2026diffphd}. In IPC, however, the frame Hessian varies with deformation and contact, so a fixed factorization cannot be reused across all adjoint systems along a trajectory. Preconditioned
iterative solvers provide an alternative, with effective
preconditioning reducing the number of iterations
\cite{chen2021multiscale,wu2022mas}. GIPC parallelizes IPC computation
\cite{gipc2024}, while StiffGIPC provides efficient FEM-ABD coupling
and GPU solvers across a wide range of stiffnesses
\cite{stiffgipc2025}, and AGIPC further reduces GPU linear-solve costs
through adaptive in-solve algebraic coarsening \cite{wang2026agipc}.
These methods advance GPU IPC efficiency, with StiffGIPC supplying the
forward simulation foundation for our work.

DiffGIPC combines coupled frame-wise adjoint solves with parallel
gradient accumulation on the GPU. Compact affine coordinates reduce
PCG's numerical workload, making repeated CPU kernel launches and
convergence synchronization more prominent. Conditional CUDA Graphs
reduce this overhead by keeping iteration control and convergence checks
on the GPU, with compatible graphs reused across frames as the Hessian
changes. After each solve, local derivative products are accumulated in
parallel to compute physical-parameter gradients and update earlier-frame
adjoint right-hand sides.

\section{A Unified Differentiable IPC Framework}

\label{sec:unified-framework}

The framework separates the coordinates needed to describe contact geometry from those needed to solve motion. Collision and friction derivatives are evaluated at world-space surface points and mapped into a coupled system of FEM vertex positions and affine body coordinates. This preserves cross-object derivatives while assigning twelve motion DoFs to each affine body.

The forward formulation builds on StiffGIPC~\cite{stiffgipc2025}. Simulation DoFs describe solved motion, whereas physical parameters are inputs changed by the optimizer. One frame is an accepted time step of length $h$, and mesh connectivity is fixed within each differentiated trajectory. The state and incremental potential below define the single-step adjoint. Section~\ref{sec:adjoint} extends it through time, Section~\ref{sec:fem-abd} supplies affine design and control derivatives, and Section~\ref{sec:pcg} develops their GPU computation and task interface.

\subsection{Simulation degrees of freedom and contact points}
\label{sec:affine-details}

A free FEM vertex $i$ is represented by its world-space position
$\vx_{t,i}\in\R^3$ at frame $t$. An affine body $b$ is represented by
its translation $\mathbf p_{t,b}\in\R^3$ and affine transformation
$\mathbf{A}_{t,b}\in\R^{3\times3}$. Let $\boldsymbol{\alpha}_{t,b,1}^{\T}$,
$\boldsymbol{\alpha}_{t,b,2}^{\T}$, and $\boldsymbol{\alpha}_{t,b,3}^{\T}$ be the rows of
$\mathbf{A}_{t,b}$. Stacking these vectors with the translation defines the
body's twelve DoFs:
\begin{equation}
 \begin{aligned}
 \mathbf{A}_{t,b}&=\begin{bmatrix}
 \boldsymbol{\alpha}_{t,b,1}^{\T}\\
 \boldsymbol{\alpha}_{t,b,2}^{\T}\\
 \boldsymbol{\alpha}_{t,b,3}^{\T}
 \end{bmatrix},\\
 \vq_{t,b}&=[\mathbf p_{t,b}^{\T},\boldsymbol{\alpha}_{t,b,1}^{\T},
 \boldsymbol{\alpha}_{t,b,2}^{\T},\boldsymbol{\alpha}_{t,b,3}^{\T}]^{\T}\in\R^{12}.
 \end{aligned}
 \label{seq:row-order}
\end{equation}
For a surface vertex with rest position $\bar\vx_i$, the world-space
position is
\begin{equation}
 \vx_{t,i}=\mathbf p_{t,b}+\mathbf{A}_{t,b}\bar\vx_i.
 \label{eq:abd-map}
\end{equation}
Let $\mathbf{I}_3$ denote the $3\times3$ identity matrix and $\mathbf e_1$,
$\mathbf e_2$, and $\mathbf e_3$ the Cartesian unit vectors. With
rest position fixed, the Jacobian of this position with respect to
$\vq_{t,b}$ is the $3\times12$ matrix
\begin{equation}
 \begin{aligned}
 \mathbf{J}_i=\mathbf{J}(\bar\vx_i)
   &=[\mathbf{I}_3\ \mathbf e_1\bar\vx_i^{\T}\
      \mathbf e_2\bar\vx_i^{\T}\ \mathbf e_3\bar\vx_i^{\T}],\\
 \vx_{t,i}&=\mathbf{J}_i\vq_{t,b}.
 \end{aligned}
 \label{seq:point-map}
\end{equation}
All surface vertices on body $b$ share the same $\vq_{t,b}$, but each
has its own rest-dependent map $\mathbf{J}_i$.

For $N_{\mathrm A}$ free affine bodies and $N_{\mathrm F}$ free FEM vertices, we collect the
simulation DoFs in
\begin{equation}
 \vy_t=[\vq_{t,1}^{\T},\ldots,\vq_{t,N_{\mathrm A}}^{\T},
        \vx_{t,1}^{\T},\ldots,\vx_{t,N_{\mathrm F}}^{\T}]^{\T}.
 \label{eq:mixed-frame-coordinates}
\end{equation}
The total number of simulation DoFs is
$n=12N_{\mathrm A}+3N_{\mathrm F}$.
The set and ordering of free DoFs do not change
{throughout each trajectory}. Adding surface vertices to an affine body can increase contact work,
but does not enlarge its block in the linear system. Rest coordinates
are design parameters, not additional simulation DoFs.

For contact $c$, we collect the world-space positions of its $n_c$
surface points into $\vx_t^c$. Let $\vy_t\in\R^n$ contain the free
simulation DoFs. The Jacobian $\mathbf{K}_{t,c}$ describes how
changes in these DoFs affect the contact-point positions:
\begin{equation}
    \begin{aligned}
        \vx_t^c
        &= [\vx_{t,i_1}^{\T},\ldots,
            \vx_{t,i_{n_c}}^{\T}]^{\T},\\
        \mathbf{K}_{t,c}
        &= \frac{\partial\vx_t^c}{\partial\vy_t}
        \in\R^{3n_c\times n}.
    \end{aligned}
    \label{eq:contact-point-jacobian}
\end{equation}
{The three-row map for vertex $i$ is denoted by
$\mathbf K_t^{\,i}=\partial\vx_{t,i}/\partial\vy_t\in\R^{3\times n}$.
The contact map $\mathbf K_{t,c}$ stacks the maps of its participating
vertices.}
For a free FEM vertex, its three positional DoFs appear directly
in $\vy_t$, so these rows contain an identity block $\mathbf{I}_3$
in the corresponding columns.
For a point on a free affine body, its position depends on the body's
twelve affine DoFs through $\mathbf{J}_i$, so these rows contain $\mathbf{J}_i$
in that body's columns.
For each free point, columns outside its associated vertex or body block are zero. A fixed point contributes three zero rows.
{For fixed rest geometry and DoF ordering,
$\mathbf K_t^{\,i}$ is unchanged across frames.}
When differentiating with respect to the free motion DoFs, we hold the rest
geometry and prescribed motion fixed.
When differentiating with respect to the rest shape, we also
differentiate the dependence of the affine point maps $\mathbf{J}_i$,
and hence $\mathbf{K}_{t,c}$, on the rest positions
(Section~\ref{sec:rest-gradient}).

\subsection{Incremental potential and its energy gradient}
\label{sec:step-equation}

Let $\vtheta\in\R^{n_\theta}$ collect the physical parameters selected for optimization,
such as material and friction coefficients, rest coordinates, initial
states, and joint controls. Here $n_\theta$ is the number of
selected scalar parameters. These parameters are held fixed during one
forward simulation. A joint-control schedule can still contain
different prescribed values for different frames. For a given $\vtheta$, each time step of length $h$
computes the simulation DoFs $\vy_t$ by minimizing the incremental
potential
\begin{equation}
 \begin{aligned}
 \Phi_t(\vy_t)
 &=\tfrac12\|\vy_t-\widehat\vy_t\|_{\mathbf{M}}^2\\
 &\quad+\sum_{e\in\mathcal E}w_e(h,\vtheta)
       \Psi_{t,e}(\vy_t,\vy_{t-1};\vtheta).
 \end{aligned}
 \label{eq:incremental-potential}
\end{equation}
The first term accounts for inertia, with
$\|\mathbf v\|_{\mathbf{M}}^2=\mathbf v^{\T}\mathbf{M}\mathbf v$.
The mass matrix $\mathbf{M}$ is expressed in the simulation coordinates:
it contains vertex-mass blocks for FEM and a $12\times12$ block
for each affine body. Section~\ref{sec:fem-abd} gives the affine
mass matrix and its {rest-shape} derivatives.
The predictor $\widehat\vy_t$ is computed from earlier states and
prescribed accelerations $\mathbf a_t$. For dynamic bodies, backward Euler (BDF1)
gives
\begin{equation}
 \widehat\vy_t=2\vy_{t-1}-\vy_{t-2}+h^2\mathbf a_t,
 \qquad t\geq2{.}
 \label{eq:bdf1-predictor}
\end{equation}
At the first step, the predictor is $\widehat\vy_1=\vy_0+h\dot\vy_0+h^2\mathbf a_1$,
where $\vy_0$ specifies the initial configuration and $\dot\vy_0$
contains the initial FEM vertex velocities and the initial
velocities of the affine coordinates.

The second term in Equation~\eqref{eq:incremental-potential} sums
the contributions from FEM elastic energy, affine shape energy,
collision, friction, and joints. Each contribution is indexed by $e\in\mathcal E$ and written as $w_e(h,\vtheta)\Psi_{t,e}$, with energy function $\Psi_{t,e}$ and scalar weight $w_e(h,\vtheta)$.
The collision energy is the non-penetration barrier, distinct
from the friction energy for tangential resistance.
IPC's friction energy depends on the relative tangential displacement
of the two configurations, determined by $\vy_t$ and $\vy_{t-1}$.
Its tangent bases and normal forces are held fixed from the
previous-frame positions when solving for $\vy_t$, and their
dependence on those positions is included in the derivative with
respect to $\vy_{t-1}$.

At a minimum of the incremental potential, its gradient $\vg_t$ with respect to the free simulation DoFs satisfies the stationarity equation
\begin{equation}
 \begin{aligned}
 \vg_t(\vy_t,\vy_{t-1},\vy_{t-2};\vtheta)
 &:=\nabla_{\vy_t}\Phi_t\\
 &=\mathbf{M}(\vy_t-\widehat\vy_t)\\
 &\quad+\sum_{e\in\mathcal E}
       w_e(h,\vtheta)\nabla_{\vy_t}\Psi_{t,e}\\
 &=\vzero.
 \end{aligned}
 \label{eq:residual}
\end{equation}
The energy gradient $\vg_t$ is also the nonlinear residual of this
equation. For each parameter choice, the forward solver finds DoFs
that satisfy this stationarity condition. Differentiating the condition determines how a parameter change affects the solved DoFs and the loss evaluated on them.

\subsection{Parameter gradients from the step equation}
\label{sec:single-step-adjoint}

Consider first a single time step from a prescribed initial state,
omitting the frame index. The solved DoFs satisfy
$\vg(\vy,\vtheta)=\vzero$. A parameter change affects this equation
both directly and through the resulting change in $\vy$.
Differentiating the stationarity equation gives
\begin{equation}
 \underbrace{\frac{\partial\vg}{\partial\vy}}_{\mathbf{H}}
 \frac{\mathrm{d}\vy}{\mathrm{d}\vtheta}
 +\underbrace{\frac{\partial\vg}{\partial\vtheta}}_{\mathbf{G}}
 =\vzero.
 \label{eq:static-sensitivity}
\end{equation}
Here, $\mathbf{G}$ is the partial derivative of the energy gradient with
respect to the physical parameters. The derivative $\mathbf{H}$ with
respect to the DoFs is the Hessian of the incremental potential:
\begin{equation}
 \mathbf{H}=\frac{\partial\vg}{\partial\vy}
  =\nabla_{\vy}^{2}\Phi.
 \label{eq:frame-energy-hessian}
\end{equation}
Both matrices are evaluated at the converged DoFs satisfying
$\vg(\vy,\vtheta)=\vzero$. Rather than forming
$\mathrm{d}\vy/\mathrm{d}\vtheta$ explicitly, we use the adjoint below.
Restoring the frame index, we denote these matrices by
$\mathbf{H}_t\in\R^{n\times n}$ and $\mathbf{G}_t\in\R^{n\times n_\theta}$.

For a scalar loss $L(\vy,\vtheta)$, let
$\boldsymbol\ell_y=\nabla_{\vy}L$ and
$\boldsymbol\ell_\theta=\nabla_{\vtheta}L$
denote its partial gradients with respect to the DoFs and
physical parameters, respectively. The chain rule gives
\begin{equation}
 \nabla_{\vtheta}L=\boldsymbol\ell_\theta+
 \left(\frac{\mathrm{d}\vy}{\mathrm{d}\vtheta}\right)^{\T}\boldsymbol\ell_y.
 \label{eq:static-loss-chain}
\end{equation}
Explicitly computing $\mathrm{d}\vy/\mathrm{d}\vtheta$ becomes
costly as the number of optimized parameters grows. We avoid
forming this matrix and compute the loss gradient by solving
the adjoint system:
\begin{equation}
 \mathbf{H}^{\T}\vlambda=\boldsymbol\ell_y.
 \label{eq:static-adjoint}
\end{equation}
The solution $\vlambda$ is the adjoint vector. Combining this solve
with Equation~\eqref{eq:static-sensitivity} gives the parameter gradient
\begin{equation}
 \nabla_{\vtheta}L=\boldsymbol\ell_\theta-\mathbf{G}^{\T}\vlambda.
 \label{eq:static-parameter-gradient}
\end{equation}
For a single time step, all parameter gradients share the same coupled
FEM-affine adjoint. Section~\ref{sec:adjoint} extends this formulation
through time.

\subsection{Collision and friction derivatives across representations}
\label{sec:coupled-contact}

Collision and friction energies are evaluated at world-space surface points, whereas the adjoint in Equation~\eqref{eq:static-adjoint} acts on simulation DoFs. The map $\mathbf{K}_{t,c}$ from Equation~\eqref{eq:contact-point-jacobian} converts the point-space derivatives into FEM and affine coordinates. The collision energy uses GIPC's squared-log barrier on squared distance~\cite{gipc2024}
\begin{equation}
 b_{\mathrm{bar}}(d)=
 \begin{cases}
 \kappa(d-\widehat d)^2\log^2(d/\widehat d),
     &0<d<\widehat d,\\
 0,  &d\geq\widehat d,
 \end{cases}
 \label{eq:rank2-barrier}
\end{equation}
where $d$ is the squared distance, $\widehat d$ is its activation
threshold, and $\kappa$ is the barrier coefficient. This squared-log
form is the GIPC barrier, rather than the original IPC single-log form. At each Newton iteration, we truncate the step
along the search direction using continuous collision detection to
prevent intersections, followed by line search to ensure a decrease
in the incremental potential. Backward differentiation uses the converged frame state and its associated collision and friction data.

For a contact $c$ at frame $t$, let $\vg_{t,c}^x$ and $\mathbf{H}_{t,c}^x$
denote the gradient and Hessian of the collision energy
with respect to the stacked world-space point positions
$\vx_t^c$. The superscript $x$ denotes point coordinates, whereas
$y$ denotes simulation DoFs. With rest geometry fixed, the point
map $\mathbf{K}_{t,c}$ from Equation~\eqref{eq:contact-point-jacobian} is constant
with respect to $\vy_t$. The chain rule therefore gives the
contributions to the frame energy gradient and Hessian:
\begin{equation}
 \vg_{t,c}^y=\mathbf{K}_{t,c}^{\T}\vg_{t,c}^x,\qquad
 \mathbf{H}_{t,c}^y=\mathbf{K}_{t,c}^{\T}\mathbf{H}_{t,c}^x\mathbf{K}_{t,c}.
 \label{eq:common-contact-map}
\end{equation}
The same transformation applies to friction derivatives with
respect to the current-frame DoFs. Friction derivatives with
respect to earlier states are included in the trajectory
calculation in Section~\ref{sec:adjoint}.

Each point contributes through the map associated with its object.
For point $i$ on affine body $b$ and point $j$ on affine body $b'$,
let $\mathbf{H}_{ij}^x$ be their $3\times3$ point-space Hessian block.
Mapping this block gives a $12\times12$ contribution to $\mathbf{H}_{bb'}^{\mathrm{AA}}$.
If the second point is instead a FEM vertex $j$, its identity map
gives a $12\times3$ contribution to $\mathbf{H}_{b j}^{\mathrm{AF}}$.
{For the point pair $(i,j)$, the block
contributions are}
{
\begin{equation}
 \begin{aligned}
 \Delta\mathbf{H}_{bb'}^{\mathrm{AA},(i,j)}
 &=\mathbf{J}_i^{\T}\mathbf{H}_{ij}^x\mathbf{J}_j,\\
 \Delta\mathbf{H}_{b j}^{\mathrm{AF},(i,j)}
 &=\mathbf{J}_i^{\T}\mathbf{H}_{ij}^x.
 \end{aligned}
 \label{eq:contact-body-blocks}
\end{equation}
}
All surface vertices of an affine body share its twelve DoFs.
{The contributions of all relevant point
pairs are summed into the corresponding body blocks, without
introducing independent vertex DoFs.} For distinct points on the same body, both ordered
point-block contributions must be included when they map to the
same diagonal body block.

After assembly, grouping affine ($\mathrm A$) and finite-element ($\mathrm F$)
DoFs gives the frame Hessian
\begin{equation}
 \mathbf{H}_t=
 \begin{bmatrix}
 \mathbf{H}_{\mathrm{AA}} & \mathbf{H}_{\mathrm{AF}}\\
 \mathbf{H}_{\mathrm{FA}} & \mathbf{H}_{\mathrm{FF}}
 \end{bmatrix},
 \qquad \mathbf{H}_{\mathrm{FA}}=\mathbf{H}_{\mathrm{AF}}^{\T}.
 \label{eq:coupled-frame-hessian}
\end{equation}
The diagonal blocks $\mathbf{H}_{\mathrm{AA}}$ and $\mathbf{H}_{\mathrm{FF}}$ collect derivatives within the affine and FEM coordinates. Collision and friction contribute $\mathbf{H}_{\mathrm{AF}}$ and $\mathbf{H}_{\mathrm{FA}}$, while joints couple affine bodies within $\mathbf{H}_{\mathrm{AA}}$. These cross blocks let a loss on one object influence parameters of another. Compact affine coordinates reduce system size while the cross terms preserve the interacting response.

The same point maps form the loss-gradient vector
$\boldsymbol\ell_y$ on the right-hand side of
Equation~\eqref{eq:static-adjoint}. For a loss evaluated on world-space
surface points, its gradient with respect to the affine coordinates
of body $b$, whose surface-vertex index set is $\mathcal V_b$, is
\begin{equation}
 \nabla_{\vq_b}L
 =\sum_{i\in\mathcal V_b}\mathbf{J}_i^{\T}\nabla_{\vx_i}L.
 \label{eq:abd-loss-map}
\end{equation}
For a FEM vertex, the loss gradient with respect to its DoFs equals
that with respect to its world-space position. {These mappings hold with rest geometry fixed.
Section~\ref{sec:rest-gradient} includes the additional dependence
of the point maps and surface positions on rest shape.}
Section~\ref{sec:adjoint} extends this coupled adjoint to trajectories,
where earlier states affect both later motion and the loss.

\section{Frame-Wise Adjoint}
\label{sec:adjoint}

% Chapter 4 overview: the frame-wise backward pass only.

The single-step adjoint captures interactions at one frame. A trajectory also depends on how earlier motion changes later inertia and friction. We account for these dependencies by backward substitution through the causal sequence of frame equations. Each reverse frame solves one coupled FEM-affine system, then contributes to the parameters and earlier states that influenced it. Algorithm~\ref{alg:adjoint} summarizes this recurrence. We derive its temporal structure and then specify the inertia and friction terms, using the converged states and recorded data that define the forward equations.

\begin{algorithm}[!t]
\caption{Frame-wise adjoint for a converged trajectory}
\label{alg:adjoint}
{
\small
\algrenewcommand\algorithmicrequire{\textbf{Input:}}
\algrenewcommand\algorithmicensure{\textbf{Output:}}
\begin{algorithmic}[1]
\Require Converged states $\vy_1,\ldots,\vy_T$ and their frame
  {data, and loss partial gradients} $\boldsymbol\ell_1,\ldots,
  \boldsymbol\ell_T$ and $\boldsymbol\ell_\theta$
  (\S\ref{sec:reverse-order})
\Ensure Parameter gradient $\bar\vtheta=\nabla_{\vtheta}L$
\State $\mathbf b_t\gets\boldsymbol\ell_t$ for $t=1,\ldots,T$;
  $\bar\vtheta\gets\boldsymbol\ell_\theta$
  \hfill(\S\ref{sec:frame-computation})
\For{$t=T,T-1,\ldots,1$}
  \hfill(\S\ref{sec:reverse-order})
  \State Restore frame $t$'s converged state and earlier-frame data
  \State Assemble the coupled Hessian $\mathbf{H}_t$
  \State Solve $\mathbf{H}_t^{\T}\vlambda_t=\mathbf b_t$
    \hfill(\S\ref{sec:frame-computation})
  \State $\bar\vtheta\gets\bar\vtheta-\mathbf{G}_t^{\T}\vlambda_t$
    \hfill(\S\ref{sec:frame-computation})
  \Statex \hspace{\algorithmicindent}%
    Include optimized initial-state derivatives in $\mathbf{G}_t$
    (\S\ref{sec:adjoint-history})
  \For{each earlier solved frame $s$ affecting $\vg_t$}
    \hfill(\S\ref{sec:trajectory-equations})
    \State $\mathbf b_s\gets\mathbf b_s-
      (\partial\vg_t/\partial\vy_s)^{\T}\vlambda_t$
    \Statex \hspace{\algorithmicindent}\hspace{\algorithmicindent}%
      {Inertia: \S\ref{sec:adjoint-history}, and
      friction: \S\ref{sec:friction}}
  \EndFor
\EndFor
\State \Return $\bar\vtheta$
\end{algorithmic}
}
\end{algorithm}

\subsection{Equations for the whole trajectory}
\label{sec:trajectory-equations}

Let $\vy_t\in\R^n$ contain the free simulation DoFs in
the order defined in Section~\ref{sec:affine-details}.
The equation at frame $t$ depends on its current state, a subset of
earlier solved states, and the physical parameters:
\begin{equation}
 \vg_t(\vy_t,\{\vy_s:s\in\mathcal D(t)\},\vtheta)=\vzero.
 \label{eq:dynamic-residual}
\end{equation}
The set $\mathcal D(t)\subseteq\{1,\ldots,t-1\}$ identifies those
earlier frames. The initial state is an input rather than a solved
frame: its optimized components contribute to $\mathbf{G}_t$, while unselected
components remain fixed. Prescribed joint commands likewise enter
$\mathbf{G}_t$ through the applied controls.
The matrices $\mathbf{H}_t$ and $\mathbf{G}_t$ retain their state and parameter
definitions from Section~\ref{sec:single-step-adjoint}.
Temporal dependence introduces the additional derivative
\begin{equation}
 \mathbf{C}_{t,s}=\frac{\partial\vg_t}{\partial\vy_s}
 \in\R^{n\times n},
 \qquad s\in\mathcal D(t).
 \label{eq:derivative-blocks}
\end{equation}
Thus $\mathbf{H}_t$ describes the coupled response within
frame $t$, whereas $\mathbf{C}_{t,s}$ describes its dependence on an earlier
state and vanishes for $s\notin\mathcal D(t)$.
Both current and earlier solved states vary with $\vtheta$.
Differentiating Equation~\eqref{eq:dynamic-residual} includes these
changes together with the direct parameter derivative:
\begin{equation}
 \mathbf{H}_t\frac{\mathrm{d}\vy_t}{\mathrm{d}\vtheta}
 +\sum_{s\in\mathcal D(t)}\mathbf{C}_{t,s}\frac{\mathrm{d}\vy_s}{\mathrm{d}\vtheta}
 +\mathbf{G}_t=0.
 \label{eq:step-sensitivity}
\end{equation}

With
$\mathbf{S}_t=\mathrm{d}\vy_t/\mathrm{d}\vtheta\in\R^{n\times n_\theta}$,
the equations for all frames form the block system
\begin{equation}
 \underbrace{\begin{bmatrix}
 \mathbf{H}_1&0&\cdots&0\\
 \mathbf{C}_{2,1}&\mathbf{H}_2&\ddots&\vdots\\
 \vdots&\ddots&\ddots&0\\
 \mathbf{C}_{T,1}&\cdots&\mathbf{C}_{T,T-1}&\mathbf{H}_T
 \end{bmatrix}}_{\boldsymbol{\mathcal J}}
 \underbrace{\begin{bmatrix}\mathbf{S}_1\\\mathbf{S}_2\\\vdots\\\mathbf{S}_T\end{bmatrix}}_{\boldsymbol{\mathcal S}}
 =-\underbrace{\begin{bmatrix}\mathbf{G}_1\\\mathbf{G}_2\\\vdots\\\mathbf{G}_T\end{bmatrix}}_{\boldsymbol{\mathcal G}}.
 \label{eq:stacked-sensitivity}
\end{equation}
Causality makes $\boldsymbol{\mathcal J}$ block lower triangular, with the
frame Hessians $\mathbf{H}_t$ on the diagonal and nonzero off-diagonal
blocks only for dependencies in $\mathcal D(t)$.

The predictor in Equation~\eqref{eq:bdf1-predictor} depends on
$\vy_{t-1}$ and $\vy_{t-2}$. Friction adds to the first of these
dependencies, through $\mathbf{C}_{t,t-1}$.
For the first three solved frames, the system is
\begin{equation}
 \begin{bmatrix}
 \mathbf{H}_1 & 0 & 0\\
 \mathbf{C}_{2,1} & \mathbf{H}_2 & 0\\
 \mathbf{C}_{3,1} & \mathbf{C}_{3,2} & \mathbf{H}_3
 \end{bmatrix}
 \begin{bmatrix}\mathbf{S}_1\\\mathbf{S}_2\\\mathbf{S}_3\end{bmatrix}
 =-\begin{bmatrix}\mathbf{G}_1\\\mathbf{G}_2\\\mathbf{G}_3\end{bmatrix}.
 \label{eq:three-frame-sensitivity}
\end{equation}
This three-frame system illustrates the reverse
calculation below. The sensitivities $\mathbf{S}_t$ and the assembled
trajectory matrix are used for the derivation, but are not formed
by the backward algorithm.

\subsection{Reverse-time adjoint equations}
\label{sec:reverse-order}

We extend the single-step adjoint in
Equation~\eqref{eq:static-adjoint} to the trajectory loss
$L(\vy_1,\ldots,\vy_T;\vtheta)$.
Let $\boldsymbol\ell_t=\nabla_{\vy_t}L$ denote its partial
gradient with respect to the simulation DoFs at frame $t$,
and let $\boldsymbol\ell_\theta$ denote its partial parameter
gradient with all simulated states fixed.
For a loss evaluated only on the final state,
$\boldsymbol\ell_t=\vzero$ for $t<T$, although earlier states
can still affect the loss through the subsequent motion.
Using $\mathbf{S}_t=\mathrm{d}\vy_t/\mathrm{d}\vtheta$ from
Section~\ref{sec:trajectory-equations}, the chain rule extends
Equation~\eqref{eq:static-loss-chain} to
\begin{equation}
 \begin{aligned}
 \nabla_{\vtheta}L
 &=\boldsymbol\ell_\theta+
   \sum_{t=1}^{T}
   \left(\frac{\mathrm{d}\vy_t}{\mathrm{d}\vtheta}\right)^{\T}
   \boldsymbol\ell_t\\
 &=\boldsymbol\ell_\theta+
   \sum_{t=1}^{T}\mathbf{S}_t^{\T}\boldsymbol\ell_t\\
 &=\boldsymbol\ell_\theta+
   \boldsymbol{\mathcal S}^{\T}\boldsymbol\ell_{\mathrm{all}},
 \end{aligned}
 \label{eq:trajectory-loss-chain}
\end{equation}
where $\boldsymbol\ell_{\mathrm{all}}$ stacks the per-frame
loss gradients in the same order as the blocks of $\boldsymbol{\mathcal S}$.
The first term accounts for the explicit parameter dependence
{of the loss. The second accounts for changes in the simulated
trajectory.}

To evaluate the trajectory contribution without computing
$\boldsymbol{\mathcal S}$, we introduce an adjoint vector $\vlambda_t$
for each frame and solve
\begin{equation}
 \boldsymbol{\mathcal J}^{\T}
 \underbrace{
 \begin{bmatrix}
 \vlambda_1\\
 \vdots\\
 \vlambda_T
 \end{bmatrix}}_{\boldsymbol\Lambda}
 =
 \underbrace{
 \begin{bmatrix}
 \boldsymbol\ell_1\\
 \vdots\\
 \boldsymbol\ell_T
 \end{bmatrix}}_{\boldsymbol\ell_{\mathrm{all}}}.
 \label{eq:stacked-adjoint}
\end{equation}
Combining this equation with
$\boldsymbol{\mathcal J}\boldsymbol{\mathcal S}=-\boldsymbol{\mathcal G}$ from
Equation~\eqref{eq:stacked-sensitivity} gives
\begin{equation}
 \boldsymbol{\mathcal S}^{\T}\boldsymbol\ell_{\mathrm{all}}
 =
 \boldsymbol{\mathcal S}^{\T}\boldsymbol{\mathcal J}^{\T}\boldsymbol\Lambda
 =
 -\boldsymbol{\mathcal G}^{\T}\boldsymbol\Lambda.
 \label{eq:adjoint-chain-elimination}
\end{equation}
The transpose reverses the causal order: $\boldsymbol{\mathcal J}^{\T}$ is
block upper triangular, so the trajectory adjoint is solved from the last
frame backward.
For the three-frame example,
\begin{equation}
 \begin{bmatrix}
 \mathbf{H}_1^{\T} & \mathbf{C}_{2,1}^{\T} & \mathbf{C}_{3,1}^{\T}\\
 0 & \mathbf{H}_2^{\T} & \mathbf{C}_{3,2}^{\T}\\
 0 & 0 & \mathbf{H}_3^{\T}
 \end{bmatrix}
 \begin{bmatrix}
 \vlambda_1\\
 \vlambda_2\\
 \vlambda_3
 \end{bmatrix}
 =
 \begin{bmatrix}
 \boldsymbol\ell_1\\
 \boldsymbol\ell_2\\
 \boldsymbol\ell_3
 \end{bmatrix}.
 \label{eq:three-frame-adjoint}
\end{equation}
Backward substitution first determines $\vlambda_3$,
then includes its effect when solving for the earlier adjoints:
\begin{align}
 \mathbf{H}_3^{\T}\vlambda_3
 &=\boldsymbol\ell_3,\notag\\
 \mathbf{H}_2^{\T}\vlambda_2
 &=\boldsymbol\ell_2-\mathbf{C}_{3,2}^{\T}\vlambda_3,\notag\\
 \mathbf{H}_1^{\T}\vlambda_1
 &=\boldsymbol\ell_1-\mathbf{C}_{2,1}^{\T}\vlambda_2
   -\mathbf{C}_{3,1}^{\T}\vlambda_3.
 \label{eq:three-frame-backward-steps}
\end{align}
Applying the same backward substitution to all $T$ frames gives
\begin{equation}
 \mathbf{H}_t^{\T}\vlambda_t
 =\boldsymbol\ell_t
 -\sum_{\substack{\tau>t\\t\in\mathcal D(\tau)}}
   \mathbf{C}_{\tau,t}^{\T}\vlambda_\tau,
 \qquad t=T,\ldots,1.
 \label{eq:exact-frame-adjoint}
\end{equation}
Only later equations that depend on $\vy_t$ contribute
to its right-hand side. The resulting adjoints account for these
temporal effects in the parameter gradient:
\begin{equation}
 \begin{aligned}
 \nabla_{\vtheta}L
 &=\boldsymbol\ell_\theta-\boldsymbol{\mathcal G}^{\T}\boldsymbol\Lambda\\
 &=\boldsymbol\ell_\theta-\sum_{t=1}^{T}\mathbf{G}_t^{\T}\vlambda_t.
 \end{aligned}
 \label{eq:dynamic-parameter-gradient}
\end{equation}

\subsection{Computing one coupled system per frame}
\label{sec:frame-computation}

{For each frame $t$, $\mathbf b_t$ denotes
the right-hand side of its adjoint equation. The vector
$\bar\vtheta$ accumulates contributions to the loss gradient with
respect to the physical parameters $\vtheta$. 
The reverse pass starts from the direct loss derivatives,
$\mathbf b_t=\boldsymbol\ell_t$ for $t=1,\ldots,T$ and
$\bar\vtheta=\boldsymbol\ell_\theta$.
Before frame $t$ is solved, $\mathbf b_t$ combines the direct loss
gradient with respect to its motion DoFs $\vy_t$ with all
contributions propagated from later frames that depend on $\vy_t$.}
{The backward pass processes frames in
reverse order. For each frame, $\mathbf{H}_t$ is first assembled
from that frame's saved converged state. The adjoint is then
obtained by solving}
\begin{equation}
 \mathbf{H}_t^{\T}\vlambda_t=\mathbf b_t.
 \label{eq:frame-adjoint}
\end{equation}
{Let $\Delta\mathbf b_s^{(t)}$ and
$\Delta\bar\vtheta^{(t)}$ denote frame $t$'s contributions to an
earlier-frame right-hand side and the parameter gradient, respectively.
They are given by}
{
\begin{align}
 \Delta\mathbf b_s^{(t)}&=-\mathbf{C}_{t,s}^{\T}\vlambda_t,
 &&s\in\mathcal D(t),\label{eq:history-update}\\
 \Delta\bar\vtheta^{(t)}&=-\mathbf{G}_t^{\T}\vlambda_t.
 &&\label{eq:parameter-gradient}
\end{align}
}
{The history contributions are summed into
the corresponding $\mathbf b_s$, giving the right-hand side in
Equation~\eqref{eq:exact-frame-adjoint}. The parameter contributions
are summed into $\bar\vtheta$. After all frames have been processed,
$\bar\vtheta=\nabla_{\vtheta}L$, as given by
Equation~\eqref{eq:dynamic-parameter-gradient}.}

\begin{figure}[!tbp]
  \centering
  % Fixed physical text size; content-sized frames determine the figure footprint.
  \includegraphics[width=0.9\linewidth]{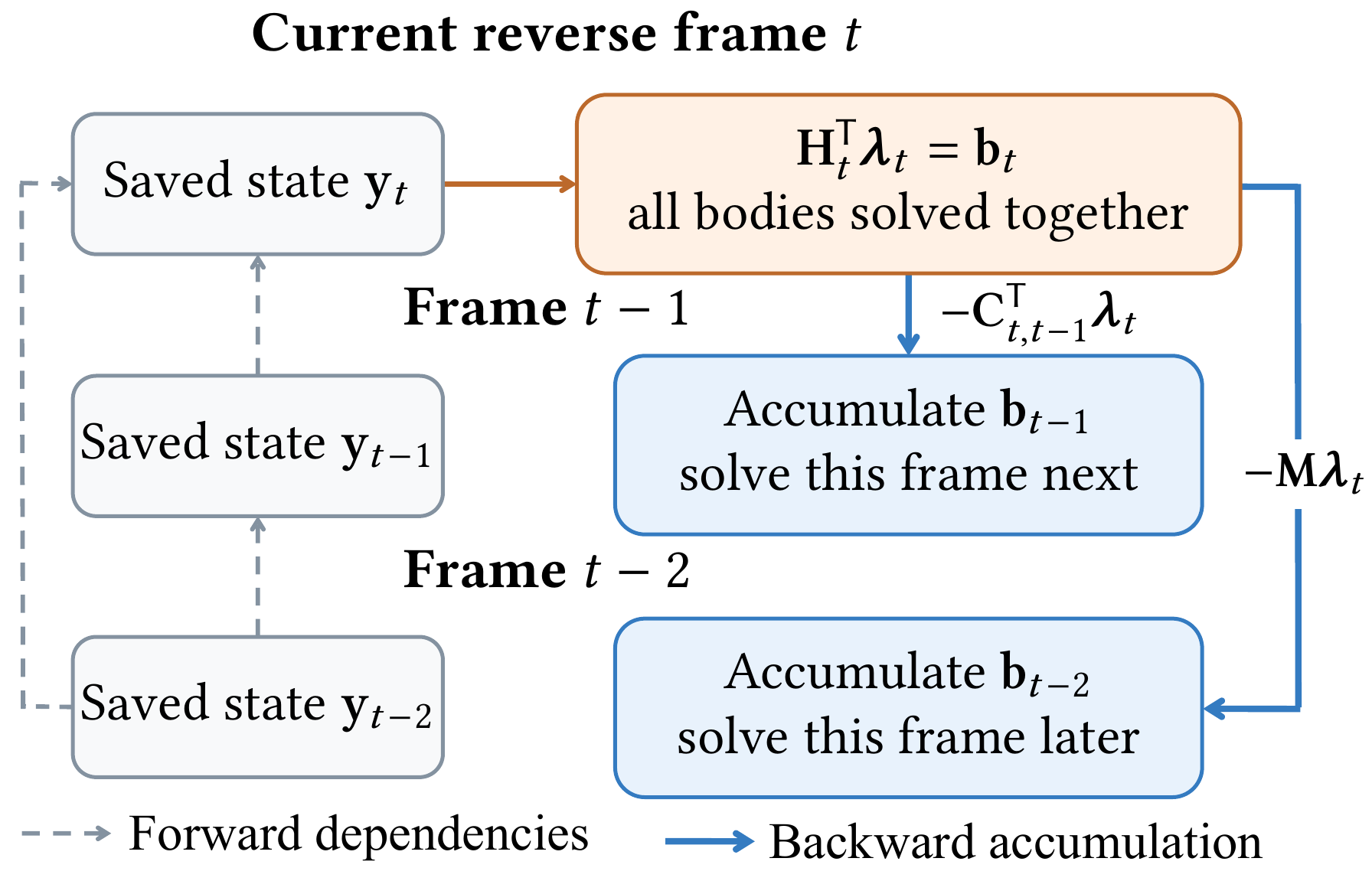}
  \caption{Reverse-time accumulation in the frame-wise adjoint.
  {Dashed arrows indicate forward state dependencies. Blue arrows carry
  contributions from the solved $\boldsymbol\lambda_t$ to the earlier
  right-hand sides $\mathbf b_{t-1}$ and $\mathbf b_{t-2}$.}
  For the dynamic predictor shown, inertia and friction contribute
  to $\mathbf b_{t-1}$, while inertia contributes
  $-\mathbf{M}\boldsymbol\lambda_t$ to $\mathbf b_{t-2}$. Each right-hand side
  also includes its direct loss gradient.}
  \Description{Three time levels are stacked vertically in a single-column figure. Saved states occupy the left lane, and the coupled solve and earlier adjoint right-hand sides occupy the right lane. Dashed arrows point toward newer states. Blue arrows carry the current adjoint's contributions to the two earlier frames.}
  \label{fig:schematic-framewise}
\end{figure}

Each history contribution reaches frame $s<t$ before that frame is
solved (Figure~\ref{fig:schematic-framewise}).
{The products in
Equations~\eqref{eq:history-update} and \eqref{eq:parameter-gradient}
are evaluated locally without assembling full
$\mathbf{C}_{t,s}$ or $\mathbf{G}_t$.}
{Only the current frame's Hessian is
retained during this reverse traversal, together with the saved
trajectory and accumulated right-hand sides.}

Each retained $\mathbf{H}_t$ preserves cross-object coupling, while backward substitution carries that response through time. The trajectory length determines the number of frame solves. Additional parameters contribute local derivative products and share these solves, making the same recurrence suitable for low-dimensional controls and detailed shape design.

% Revision of Sections 4.3-7 in main_update(3).tex.

\subsection{Inertia and initial-state dependencies}
\label{sec:adjoint-history}

{The history contribution in
Equation~\eqref{eq:history-update} requires}
the derivative of $\vg_t$ with respect to each earlier state.
For inertia, $\vg_t^{\mathrm{in}}=\mathbf{M}(\vy_t-\widehat\vy_t)$.
Substituting the BDF1 predictor from
Equation~\eqref{eq:bdf1-predictor} and differentiating with respect
to $\vy_{t-1}$ and $\vy_{t-2}$ gives
\begin{equation}
 \mathbf{C}_{t,t-1}^{\mathrm{in}}=-2{\mathbf M},\qquad
 \mathbf{C}_{t,t-2}^{\mathrm{in}}={\mathbf M}.
 \label{eq:inertial-history-blocks}
\end{equation}
{Substituting these blocks into
Equation~\eqref{eq:history-update} gives the inertial parts of
$\Delta\mathbf b_s^{(t)}$, denoted by the superscript $\mathrm{in}$:}
{
\begin{equation}
 \begin{aligned}
 \Delta\mathbf b_{t-1}^{(t),\mathrm{in}}&=2\mathbf{M}\vlambda_t,
 &&t\geq2,\\
 \Delta\mathbf b_{t-2}^{(t),\mathrm{in}}&=-\mathbf{M}\vlambda_t,
 &&t\geq3.
 \end{aligned}
 \label{eq:bdf1-adjoint}
\end{equation}
}
The same mass products apply to both
representations: $\mathbf{M}$ contains the FEM vertex-mass blocks and the
$12\times12$ mass block of each affine body.

The initial configuration and velocity are inputs to the simulation,
not additional solved frames. Their contributions therefore enter
{the parameter-gradient contribution in
Equation~\eqref{eq:parameter-gradient}},
rather than a right-hand-side vector $\mathbf b_0$.
The first-step predictor is
$\widehat\vy_1=\vy_0+h\dot\vy_0+h^2\mathbf a_1$,
so the inertial derivatives with respect to $\vy_0$ and
$\dot\vy_0$ are $-\mathbf{M}$ and $-h\mathbf{M}$, respectively.
For each frame $t$, the partial derivatives of $\vg_t$ with
respect to the optimized initial configuration and velocity
form the corresponding $\mathbf{G}_t$.
We evaluate these derivatives at the converged state of frame $t$
and accumulate their contributions to the initial-state gradients
using Equation~\eqref{eq:parameter-gradient}.

For the velocity-reset approximation of quasi-static behavior, we set the predictor velocity to zero at every step, giving $\widehat\vy_t=\vy_{t-1}+h^2\mathbf a_t$. The mass-weighted quadratic term in Equation~\eqref{eq:incremental-potential} remains, so each update still depends on the previous configuration.
Differentiating its energy gradient with respect to $\vy_{t-1}$
contributes $-\mathbf{M}$ to $\mathbf{C}_{t,t-1}$.
{Substituting this contribution into
Equation~\eqref{eq:history-update} gives the inertial contribution}
{
\begin{equation}
 \Delta\mathbf b_{t-1}^{(t),\mathrm{in}}=\mathbf{M}\vlambda_t,
 \qquad t\geq2.
 \label{eq:quasistatic-abd-adjoint}
\end{equation}
}

\subsection{Friction between frames}
\label{sec:friction}

Friction adds a second source of temporal dependence: its energy uses both the current displacement and contact quantities computed from the previous configuration. DiffIPC differentiates this dependence for finite elements~\cite{diffipc}. Here the point maps carry these history derivatives into the coupled FEM-affine recurrence, including contacts whose points use different motion representations.
For contact $c$, let $\vg_{t,c}^{\mathrm{fr},x}$ be the friction
energy gradient with respect to the current world-space point
positions $\vx_t^c$. The maps $\mathbf{K}_{t,c}$ and $\mathbf{K}_{t-1,c}$ relate
the same contact points to the simulation DoFs at the two frames.
With rest geometry fixed, these maps do not depend on the solved
DoFs. Differentiating
$\vg_{t,c}^{\mathrm{fr},y}=\mathbf{K}_{t,c}^{\T}\vg_{t,c}^{\mathrm{fr},x}$
with respect to $\vy_{t-1}$ therefore gives
\begin{equation}
 \frac{\partial\vg_{t,c}^{\mathrm{fr},y}}{\partial\vy_{t-1}}
 =\mathbf{K}_{t,c}^{\T}
   \frac{\partial\vg_{t,c}^{\mathrm{fr},x}}{\partial\vx_{t-1}^c}
   \mathbf{K}_{t-1,c}.
 \label{eq:friction-history-map}
\end{equation}
This is the contact's contribution to $\mathbf{C}_{t,t-1}$.
{For $t\geq2$, contact $c$ contributes
$\Delta\mathbf b_{t-1}^{(t,c),\mathrm{fr}}$ to the earlier-frame
right-hand side. Equation~\eqref{eq:history-update} gives}
{
\begin{equation}
 \Delta\mathbf b_{t-1}^{(t,c),\mathrm{fr}}=-
 \mathbf{K}_{t-1,c}^{\T}
 \left(\frac{\partial\vg_{t,c}^{\mathrm{fr},x}}
              {\partial\vx_{t-1}^c}\right)^{\T}
 \mathbf{K}_{t,c}\vlambda_t.
 \label{eq:friction-history-gradient}
\end{equation}
}
{Summing these per-contact vectors gives
the friction part of $\Delta\mathbf b_{t-1}^{(t)}$.}
Each contact point retains its FEM or affine map, so the same expression covers contact between the two representations.

The relative tangential displacement is computed from the
corresponding vertex displacements using closest-point
interpolation weights and a tangent basis.
These weights, the tangent basis, and the normal force are
computed from the converged configuration of the previous frame.
They remain fixed throughout the current frame's solve.
When differentiating with respect to the previous-frame positions,
we also include the derivatives of these quantities.
Appendix~\ref{sec:friction-details} gives the friction law
and its derivatives.

\section{Affine-Body Derivatives for Shape, Initial States, and Control}
\label{sec:fem-abd}

The coupled adjoint describes how a loss responds to motion. To optimize an affine body's shape or controls, we must also differentiate how those parameters enter the frame equations. In particular, an affine body can expose many {rest-shape} coordinates for design, while adding only twelve motion DoFs per solve. A {rest-shape} change affects mass and inertia, world-space contact geometry, and the maps from point-space forces to affine coordinates. Friction adds the dependence on that geometry at the previous frame.

We derive these parameter derivatives, followed by initial-state and joint-control derivatives. For a scalar parameter $\theta_k$, the partial derivative $\partial\vg_t/\partial\theta_k$ forms column $k$ of $\mathbf{G}_t$, evaluated at fixed solved DoFs. The adjoint accounts for the resulting motion response, and Equation~\eqref{eq:dynamic-parameter-gradient} combines $-\mathbf{G}_t^{\T}\vlambda_t$ with the explicit loss derivative $\boldsymbol\ell_\theta$. The following derivatives preserve this separation, allowing surface design, placement, and control to share the twelve-DoF affine representation.

\subsection{Rest-dependent mass and inertia}

For affine body $b$, let $\Omega_b$ be its rest domain and
$\rho_b$ its constant density. For a material point with rest position $\bar\vx$,
the map $\mathbf{J}(\bar\vx)$ in Equation~\eqref{seq:point-map}
gives its world-space position as
$\vx_t=\mathbf{J}(\bar\vx)\vq_{t,b}$.
With rest geometry fixed, its velocity is
$\dot\vx_t=\mathbf{J}(\bar\vx)\dot\vq_{t,b}$. The corresponding mass
block is
\begin{equation}
 \mathbf{M}_b=\int_{\Omega_b}\rho_b \mathbf{J}(\bar\vx)^{\T}\mathbf{J}(\bar\vx)\,
       \mathrm{d}V.
 \label{eq:abd-generalized-mass}
\end{equation}
Both $\mathbf{M}_b$ and its shape derivative can be computed from the closed surface mesh. We express the volume integral as signed triangle contributions, each depending explicitly on the triangle's rest vertices. Differentiating these expressions gives the mass-matrix derivatives directly from the design surface, without an interior tetrahedral discretization.

Expanding $\mathbf{J}(\bar\vx)^{\T}\mathbf{J}(\bar\vx)$ in
Equation~\eqref{eq:abd-generalized-mass} gives terms that are
constant, linear, or quadratic in the rest position $\bar\vx$.
The mass matrix $\mathbf{M}_b$ is determined by three integrals over the body's rest volume:
\begin{equation}
 \begin{aligned}
 m_b
 &=\int_{\Omega_b}\rho_b\,\mathrm{d}V,\\
 \mathbf m_b^{(1)}
 &=\int_{\Omega_b}\rho_b\bar\vx\,\mathrm{d}V,\\
 \mathbf{Q}_b^{(2)}
 &=\int_{\Omega_b}\rho_b\bar\vx\bar\vx^{\T}\,\mathrm{d}V.
 \end{aligned}
 \label{eq:abd-mass-integrals}
\end{equation}
Here $m_b$ is the total mass.
The vector $\mathbf m_b^{(1)}$ is the first mass moment and
the matrix $\mathbf{Q}_b^{(2)}$ is the second mass moment.
With density fixed, $m_b$, $\mathbf m_b^{(1)}$, and
$\mathbf{Q}_b^{(2)}$ depend only on the body's rest shape.

The closed surface mesh provides these volume integrals. For an outward-oriented triangle $f$ with rest vertices
$\bar\vx_{f,1}$, $\bar\vx_{f,2}$, and $\bar\vx_{f,3}$,
the three vertices and the origin define a signed tetrahedron.
Its volume and mass-weighted integrals can be calculated
directly from the three vertex positions.
With
$\mathbf s_f=\bar\vx_{f,1}+\bar\vx_{f,2}+\bar\vx_{f,3}$,
the signed contributions are
\begin{align}
 V_f
 &=\tfrac16\bar\vx_{f,1}^{\T}
       (\bar\vx_{f,2}\times\bar\vx_{f,3}),\notag\\
 m_f
 &=\rho_b V_f,\qquad
 \mathbf m_f^{(1)}=\tfrac{m_f}{4}\mathbf s_f,\notag\\
 \mathbf{Q}_f^{(2)}
 &=\tfrac{m_f}{20}
   \left(
     \sum_{j=1}^{3}
       \bar\vx_{f,j}\bar\vx_{f,j}^{\T}
     +\mathbf s_f\mathbf s_f^{\T}
   \right).
 \label{seq:surface-moments}
\end{align}
Summing $m_f$, $\mathbf m_f^{(1)}$, and $\mathbf{Q}_f^{(2)}$ over
all surface triangles gives $m_b$, $\mathbf m_b^{(1)}$,
and $\mathbf{Q}_b^{(2)}$, respectively.
Retaining the signs in these sums recovers the volume moments
from the oriented surface triangles.

Using these quantities and the coordinate order in
Equation~\eqref{seq:row-order}, the affine-body mass matrix is
\begin{equation}
 \mathbf{M}_b=
 \begin{bmatrix}
 m_b\mathbf{I}_3
 &\mathbf e_1(\mathbf m_b^{(1)})^{\T}
 &\mathbf e_2(\mathbf m_b^{(1)})^{\T}
 &\mathbf e_3(\mathbf m_b^{(1)})^{\T}\\
 \mathbf m_b^{(1)}\mathbf e_1^{\T}
 &\mathbf{Q}_b^{(2)}&0&0\\
 \mathbf m_b^{(1)}\mathbf e_2^{\T}
 &0&\mathbf{Q}_b^{(2)}&0\\
 \mathbf m_b^{(1)}\mathbf e_3^{\T}
 &0&0&\mathbf{Q}_b^{(2)}
 \end{bmatrix}.
 \label{seq:abd-mass}
\end{equation}
The total mass determines the translation block.
The first mass moment determines the blocks between translation
and the affine transformation, while the second mass moment
determines the blocks for the affine transformation.

Each contribution in Equation~\eqref{seq:surface-moments}
is an explicit function of the triangle's rest vertex positions.
With density and mesh connectivity fixed, summing their derivatives over the surface triangles gives $\mathrm{d}m_b$, $\mathrm{d}\mathbf m_b^{(1)}$, and $\mathrm{d}\mathbf{Q}_b^{(2)}$.
Substituting them into Equation~\eqref{seq:abd-mass} gives
$\mathrm{d}\mathbf{M}_b$.
When the inverse mass matrix is needed, its derivative is
$\mathrm{d}(\mathbf{M}_b^{-1})
=-\mathbf{M}_b^{-1}(\mathrm{d}\mathbf{M}_b)\mathbf{M}_b^{-1}$.
Shape updates preserve the closed, consistently oriented,
{nondegenerate surface and the validity of its mass distribution.
Invalid candidates are rejected before updating the simulation
geometry.}

\subsection{Rest-shape gradients}
\label{sec:rest-gradient}

Rest-shape derivatives express how the frame equations change when the affine design surface changes. Solved FEM positions and affine states are held fixed when forming $\mathbf{G}_t$, while an affine rest variation still moves its surface point by $\mathrm{d}\vx_{t,i}=\mathbf{A}_{t,b}\mathrm{d}\bar\vx_i$. The resulting mass, contact-map, and energy-gradient variations are denoted by $\mathrm{d}\mathbf{M}_b$, $\mathrm{d}\mathbf{K}_{t,c}$, and $\mathrm{d}\vg_t$. They supply the corresponding parameter derivatives, with direct surface-loss dependence added separately to $\boldsymbol\ell_\theta$. 

\paragraph{Mass, inertia, and shape energy}
For body $b$ at frame $t$, the inertial energy gradient is
$\vg_{t,b}^{\mathrm{in}}
=\mathbf{M}_b(\vq_{t,b}-\widehat\vq_{t,b})$.
The product rule gives the corresponding first-order change:
\begin{equation}
 \mathrm{d}\vg_{t,b}^{\mathrm{in}}
 =(\mathrm{d}\mathbf{M}_b)(\vq_{t,b}-\widehat\vq_{t,b})
 -\mathbf{M}_b\,\mathrm{d}\widehat\vq_{t,b}.
 \label{eq:abd-inertia-rest}
\end{equation}
The first term uses the mass derivative derived above.
The second includes any {rest-shape} dependence in the inputs to
the predictor.
The affine shape incremental potential is
$h^2\kappa_{\mathrm A}V_b\psi_{\mathrm A}(\vq_{t,b})$,
where $\kappa_{\mathrm A}$ is the affine stiffness coefficient,
$V_b$ is the rest volume, and $\psi_{\mathrm A}$ is the
near-rigidity penalty. Let $\vg_{t,b}^{\mathrm{shape}}$ be its
gradient with respect to $\vq_{t,b}$.
With $\kappa_{\mathrm A}$ and $\vq_{t,b}$ fixed, the {rest-shape}
derivative of the combined inertial and shape contributions is
\begin{equation}
 \begin{aligned}
 \mathrm{d}(\vg_{t,b}^{\mathrm{in}}
            +\vg_{t,b}^{\mathrm{shape}})
 &=\mathrm{d}\vg_{t,b}^{\mathrm{in}}\\
 &\quad+h^2\kappa_{\mathrm A}(\mathrm{d}V_b)
                    \nabla_{\vq_{t,b}}\psi_{\mathrm A}.
 \end{aligned}
 \label{seq:affine-rest}
\end{equation}
The two contributions describe different physical effects of redesign:
redistributing material changes the inertial response through $\mathbf{M}_b$,
while changing the rest volume changes the scale of the near-rigidity
{energy. Both enter the parameter derivative at fixed affine coordinates.
The adjoint then accounts for the resulting change in motion.}

\paragraph{Collision-barrier derivatives and the point map}
For a surface vertex $i$ on body $b$, partial differentiation with respect to $\bar\vx_i$ holds the translation $\mathbf p_{t,b}$ and affine transformation $\mathbf{A}_{t,b}$ fixed.
Equation~\eqref{eq:abd-map} gives the resulting world-space
position change as
$\mathrm{d}\vx_{t,i}=\mathbf{A}_{t,b}\,\mathrm{d}\bar\vx_i$.
This variation also changes $\mathbf{J}_i$ and the corresponding entries
of the contact map $\mathbf{K}_{t,c}$.
For contact $c$, let $\mathrm{d}\vx_t^c$ collect the changes
in its world-space point positions.
Applying the product rule to the collision energy gradient
$\vg_{t,c}^{\mathrm{ct},y}
=\mathbf{K}_{t,c}^{\T}\vg_{t,c}^{\mathrm{ct},x}$
gives its first-order change:
\begin{equation}
 \begin{aligned}
 \mathrm{d}\vg_{t,c}^{\mathrm{ct},y}
 &=(\mathrm{d}\mathbf{K}_{t,c})^{\T}\vg_{t,c}^{\mathrm{ct},x}\\
 &\quad+\mathbf{K}_{t,c}^{\T}\mathbf{H}_{t,c}^{\mathrm{ct},x}
          \mathrm{d}\vx_t^c.
 \end{aligned}
 \label{eq:abd-rest-contact}
\end{equation}
Here $\mathrm{ct}$ denotes the collision-barrier term, and
$\mathbf{H}_{t,c}^{\mathrm{ct},x}$ is its Hessian with respect to the
world-space point positions $\vx_t^c$.
The first term accounts for the changed mapping from a point force
to the body's affine equations: even an unchanged point-space force
has a different contribution when its material point moves.
The second term accounts for the force change caused by the altered
world-space collision geometry. {Rest-shape} changes affect both the world-space collision gradient
and the mapping $\mathbf{K}_{t,c}$, so both terms must be included.
Varying only a selected scalar rest-coordinate
$\theta_k$, while holding all other parameters and solved DoFs fixed,
gives the corresponding partial derivative:
\begin{equation}
 \begin{aligned}
 \frac{\partial\vg_{t,c}^{\mathrm{ct},y}}{\partial\theta_k}
 &=
 \left(\frac{\partial \mathbf{K}_{t,c}}{\partial\theta_k}\right)^{\T}
 \vg_{t,c}^{\mathrm{ct},x}\\
 &\quad+
 \mathbf{K}_{t,c}^{\T}\mathbf{H}_{t,c}^{\mathrm{ct},x}
 \frac{\partial\vx_t^c}{\partial\theta_k}.
 \end{aligned}
 \label{eq:abd-rest-collision-parameter}
\end{equation}
$\partial\vg_{t,c}^{\mathrm{ct},y}/\partial\theta_k$ is contact $c$'s collision contribution to
column $k$ of $\mathbf{G}_t$.

\paragraph{Rest shape under frictional contact}
The friction energy at frame $t$ depends on the surface-point
positions at both frames $t$ and $t-1$.
For vertex $i$ on affine body $b$, these two positions are
computed from the same rest position $\bar\vx_i$ using
Equation~\eqref{eq:abd-map}.
We vary $\bar\vx_i$ while keeping each frame's translation
and affine transformation at their converged values.
The resulting world-space position changes are
\[
 \mathrm{d}\vx_{t,i}
 =\mathbf{A}_{t,b}\,\mathrm{d}\bar\vx_i,
 \qquad
 \mathrm{d}\vx_{t-1,i}
 =\mathbf{A}_{t-1,b}\,\mathrm{d}\bar\vx_i.
\]
A rest vertex identifies the same material point at both frames.
Moving it can therefore change the relative sliding displacement even
when the two affine transforms are held fixed. Its previous-frame
position also determines the interpolation weights, contact-plane
basis, and normal force used by the friction law. The shape derivative
must include these changes together with the current point displacement.

Let
$\mathbf{H}_{t,c}^{\mathrm{fr},x}
=\partial\vg_{t,c}^{\mathrm{fr},x}/\partial\vx_t^c$
and
$\mathbf{C}_{t,c}^{\mathrm{fr},x}
=\partial\vg_{t,c}^{\mathrm{fr},x}/\partial\vx_{t-1}^c$
be the derivatives with respect to the current and previous
point positions. The block $\mathbf{C}_{t,c}^{\mathrm{fr},x}$ includes the
changes in the interpolation weights, tangent basis, and normal force.
The {rest-shape} variation also changes $\mathbf{K}_{t,c}$. Applying the
product rule to
$\vg_{t,c}^{\mathrm{fr},y}
=\mathbf{K}_{t,c}^{\T}\vg_{t,c}^{\mathrm{fr},x}$ therefore gives all three
contributions:
\begin{equation}
 \begin{aligned}
 \mathrm{d}\vg_{t,c}^{\mathrm{fr},y}
 &=(\mathrm{d}\mathbf{K}_{t,c})^{\T}\vg_{t,c}^{\mathrm{fr},x}\\
 &\quad+\mathbf{K}_{t,c}^{\T}\mathbf{H}_{t,c}^{\mathrm{fr},x}\,\mathrm{d}\vx_t^c\\
 &\quad+\mathbf{K}_{t,c}^{\T}\mathbf{C}_{t,c}^{\mathrm{fr},x}\,\mathrm{d}\vx_{t-1}^c.
 \end{aligned}
 \label{eq:affine-rest-friction-map}
\end{equation}
These terms capture the changed force map and the current- and
previous-frame contact geometry at fixed simulated DoFs. Their contribution
to $\mathbf{G}_t$ complements the earlier-state dependence propagated by
the history update in Section~\ref{sec:friction}.
%The partial derivative with respect to a selected rest coordinate is obtained using Equation~\eqref{eq:abd-rest-collision-parameter}.
Appendix~\ref{sec:friction-details} provides the friction law in terms of contact-point positions and its derivatives.

\begin{figure}[!t]
  \centering
  \setlength{\tabcolsep}{0pt}
  \begin{tabular}{@{}cc@{}}
    \small Initial & \small Optimized\\[2pt]
    \includegraphics[width=0.47\columnwidth,trim=467bp 683bp 467bp 133bp,clip]{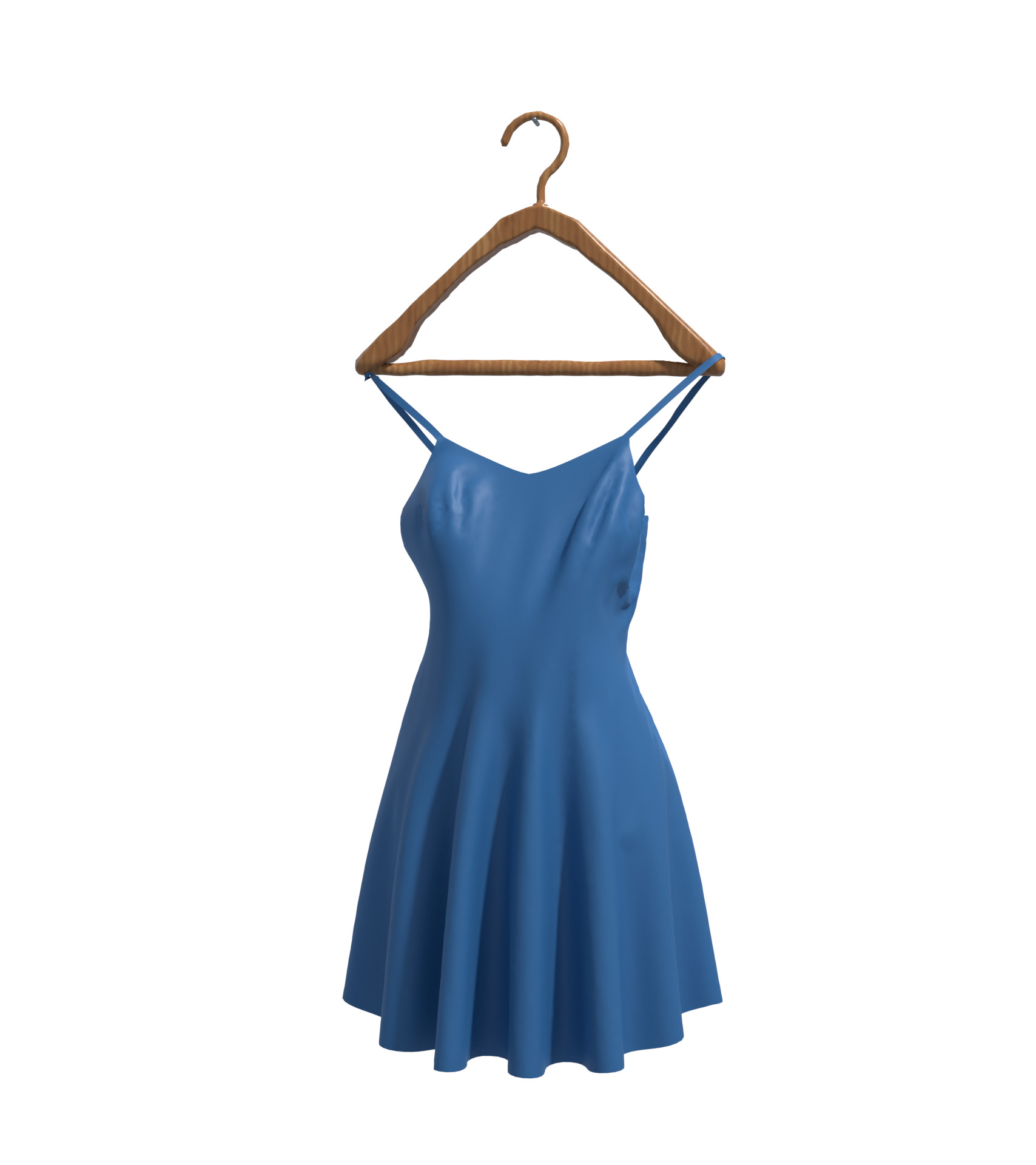} &
    \includegraphics[width=0.47\columnwidth,trim=467bp 683bp 467bp 133bp,clip]{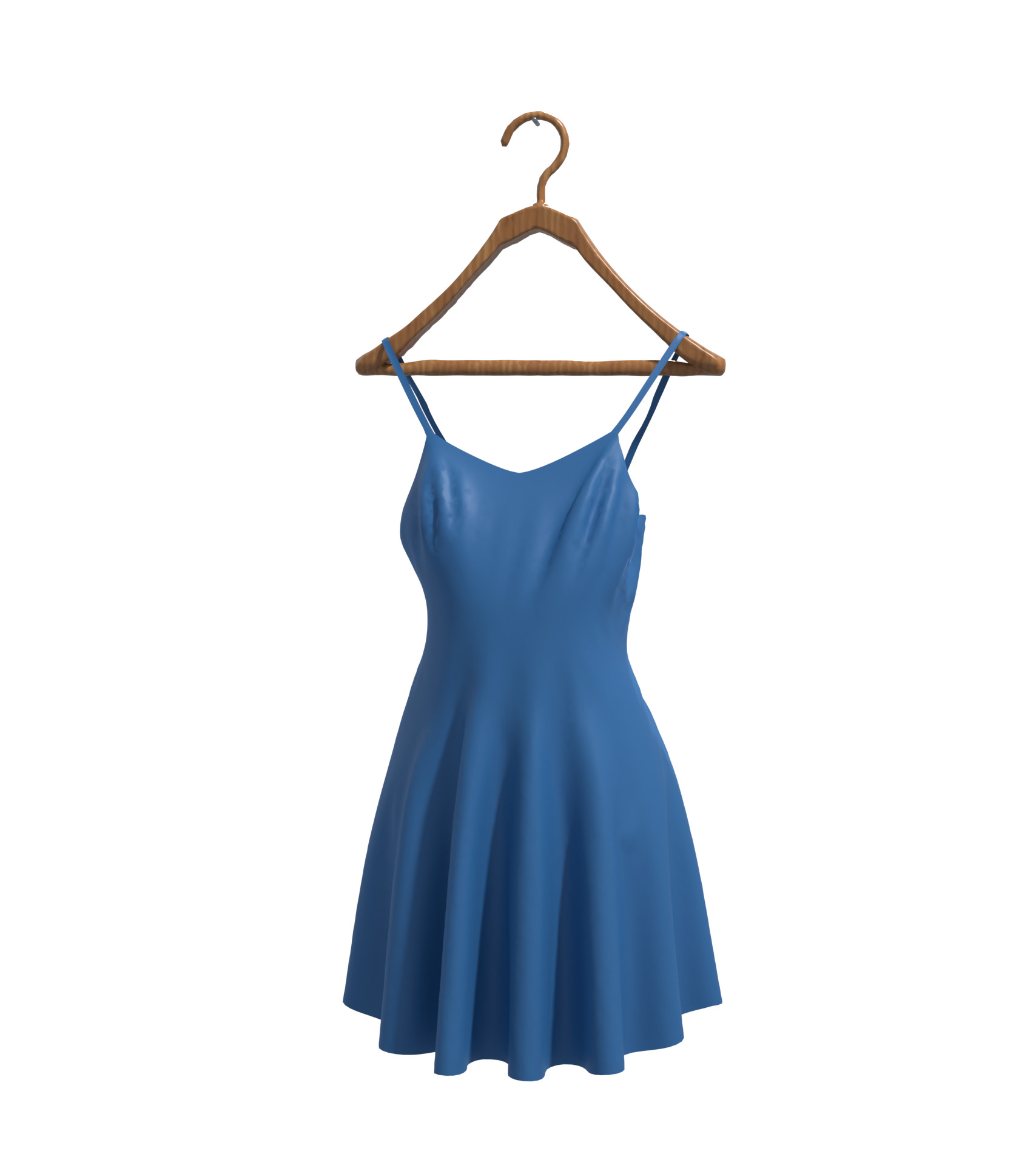}
  \end{tabular}
  \caption{Garment hanger shape design. {Two shared
  B\'ezier control scales modify both shoulders of the affine rest surface
  under frictional cloth contact.} The
  {initial and optimized configurations are shown at frame 33.
  The optimized hanger retains both garment straps.}}
  \Description{Initial and optimized garment hangers shown side by side
  with their supported cloth at frame 33.}
  \label{fig:abd-hanger-rest}
\end{figure}

\paragraph{Direct dependence of a surface loss}
A loss evaluated on an affine surface also depends on rest
geometry through $\vx_{t,i}=\mathbf{J}_i\vq_{t,b}$.
For a loss on surface positions at one frame $t$, holding the
simulation DoFs fixed gives the direct contribution
\begin{equation}
 \left.\nabla_{\bar\vx_i}L\right|_{\vy_t\ \mathrm{fixed}}
 =\mathbf{A}_{t,b}^{\T}\nabla_{\vx_{t,i}}L.
 \label{eq:abd-explicit-rest}
\end{equation}
For losses on several frames, these direct contributions are
summed over the corresponding frames and included in
$\boldsymbol\ell_\theta$. They are added to the contributions
through the frame equations in
Equation~\eqref{eq:dynamic-parameter-gradient}.
{Garment hanger
(Figure~\ref{fig:abd-hanger-rest}) illustrates both contributions. Its
support objective depends on the relative positions of the cloth straps
and hanger shoulders. Changing the rest surface therefore affects the
loss directly through the surface map and indirectly through the coupled
motion, including frictional contact. Both contributions enter the
rest-shape gradient.}

\subsection{Initial-state gradients}
\label{sec:affine-initial-state}

Following Section~\ref{sec:adjoint-history}, we include the
selected components of an affine body's initial configuration
$\vq_{0,b}$ and initial velocity $\dot\vq_{0,b}$ in $\vtheta$.
{Their parameter derivatives enter
$\mathbf{G}_t$ and determine the initial-state components of
$\Delta\bar\vtheta^{(t)}$ in
Equation~\eqref{eq:parameter-gradient}.}
For these partial derivatives, the rest geometry, prescribed
accelerations, and computed states $\vy_1,\ldots,\vy_T$
{remain unchanged. Only the selected initial values vary.}

The first-frame inertial derivatives in
Section~\ref{sec:adjoint-history} use the mass block $\mathbf{M}_b$
for affine body $b$:
\begin{equation}
 \frac{\partial\vg_{1,b}^{\mathrm{in}}}{\partial\vq_{0,b}}
 =-\mathbf{M}_b,
 \qquad
 \frac{\partial\vg_{1,b}^{\mathrm{in}}}{\partial\dot\vq_{0,b}}
 =-h\mathbf{M}_b.
 \label{eq:abd-initial-derivatives}
\end{equation}
The initial configuration also appears in the second-frame
BDF1 predictor,
$\widehat\vq_{2,b}
=2\vq_{1,b}-\vq_{0,b}+h^2\mathbf a_{2,b}$.
Differentiating
$\vg_{2,b}^{\mathrm{in}}
=\mathbf{M}_b(\vq_{2,b}-\widehat\vq_{2,b})$
with respect to $\vq_{0,b}$ gives
\begin{equation}
 \frac{\partial\vg_{2,b}^{\mathrm{in}}}{\partial\vq_{0,b}}
 =\mathbf{M}_b.
 \label{eq:abd-second-frame-initial-derivative}
\end{equation}
The first-frame derivatives contribute to $\mathbf{G}_1$, and the
second-frame derivative contributes to $\mathbf{G}_2$.
Let $\vlambda_{t,b}$ contain the twelve entries of the
coupled adjoint $\vlambda_t$ corresponding to body $b$.
{Substituting these derivatives into
Equation~\eqref{eq:parameter-gradient} gives the inertial
contributions to the initial-state gradients:}
{
\begin{equation}
 \begin{aligned}
 \Delta\bar\vq_{0,b}
 &=\mathbf{M}_b\vlambda_{1,b}-\mathbf{M}_b\vlambda_{2,b},\\
 \Delta\overline{\dot\vq}_{0,b}
 &=h\mathbf{M}_b\vlambda_{1,b}.
 \end{aligned}
 \label{eq:abd-initial-inertial-gradient}
\end{equation}
}
For a one-frame trajectory, the term containing
$\vlambda_{2,b}$ is omitted.

Friction provides an additional initial-configuration
derivative. At the first frame, its previous-frame positions
are the initial surface positions
$\vx_{0,i}=\mathbf{J}_i\vq_{0,b}$.
Changing $\vq_{0,b}$ changes these positions and the
interpolation weights, tangent basis, and normal force
computed from them.
We compute this derivative using
Equation~\eqref{eq:friction-history-map} with $t=1$.
Since $\vq_{0,b}$ is an input parameter, this derivative
contributes to $\mathbf{G}_1$ and the parameter-gradient update,
rather than an earlier-frame vector $\mathbf b_0$.

For the velocity-reset updates used to approximate
quasi-static behavior in Section~\ref{sec:adjoint-history},
the first-frame predictor is
$\widehat\vq_{1,b}=\vq_{0,b}+h^2\mathbf a_{1,b}$.
Its inertial derivatives are $-\mathbf{M}_b$ with respect to
$\vq_{0,b}$ and zero with respect to $\dot\vq_{0,b}$.
Subsequent predictors use the previous solved configuration
without an explicit $\vq_{0,b}$ term.
{In this mode, the inertial contributions are
${\Delta\bar\vq_{0,b}=\mathbf{M}_b\vlambda_{1,b}}$ and
${\Delta\overline{\dot\vq}_{0,b}=\vzero}$.
The first-frame friction contribution is included as described above.}

\subsection{Joint targets and \texorpdfstring{forces}{forces}}
\label{sec:joint-derivatives}

Revolute and prismatic joint targets and forces provide actuation parameters alongside rest shapes and initial states. Their reference geometry and driving stiffness remain fixed during control optimization.
Let $u_{j,t}$ denote the scalar target, force, or torque applied to
joint $j$ at frame $t$, and let $\vg_{t,j}^{\mathrm{joint}}$ denote
its contribution to the frame energy gradient.
{Equation~\eqref{eq:parameter-gradient}
gives this joint's contribution $\Delta\bar u_{j,t}$ to
the control gradient:}
{
\begin{equation}
 \Delta\bar u_{j,t}=-
 \left(\frac{\partial\vg_{t,j}^{\mathrm{joint}}}
              {\partial u_{j,t}}\right)^{\T}\vlambda_t.
 \label{eq:joint-control-gradient}
\end{equation}
}
Although the local control derivative involves only the two joint-connected bodies, its product with the coupled adjoint includes their response through contact with the rest of the scene. Joint targets and forces can therefore be optimized for objectives on interacting objects, using the same backward system as shape and initial-state parameters.

\paragraph{Actuation energies and parameter derivatives}
For the local formulas, we omit the joint and frame indices.
Let $a$ be the applied target, $f$ the joint force (a
linear force for a prismatic joint and a torque for a revolute joint),
and $k$ the effective driving stiffness. The affine DoFs of the two bodies connected by the joint are
denoted by $\vq_p$ and $\vq_c$. The vector $\mathbf q=[\vq_p^{\T},\vq_c^{\T}]^{\T}$ collects both bodies' DoFs, with $\nabla=\nabla_{\mathbf q}$.
For a prismatic joint, let $\xi$ denote the relative
displacement between the two joint attachment points, projected
onto the joint axis of body $c$ in world space.
Its energy and control derivatives are
\begin{align}
 \Phi^{\mathrm{pri}}&=\tfrac12 k(\xi-a)^2-f\xi,\notag\\
 \mathbf g^{\mathrm{pri}}
 &=\nabla\Phi^{\mathrm{pri}}=[k(\xi-a)-f]\nabla\xi,\notag\\
 \frac{\partial\mathbf g^{\mathrm{pri}}}{\partial a}
 &=-k\nabla\xi,\qquad
 \frac{\partial\mathbf g^{\mathrm{pri}}}{\partial f}
 =-\nabla\xi.
 \label{eq:prismatic-control-derivatives}
\end{align}
The gradient $\nabla\xi$ also accounts for how the world-space joint axis of body $c$ changes with $\vq_c$.
The quadratic term penalizes displacement away from
the target $a$, whereas $-f\xi$ specifies the applied
joint force along the joint coordinate. Targets and joint forces
therefore enter the same frame equation through different parameter
derivatives.

For a revolute joint, two scalar quantities $s_{\mathrm r}$ and
$c_{\mathrm r}$ computed from the bodies' transformed reference
directions define the relative angle
$\alpha=\operatorname{atan2}(s_{\mathrm r},c_{\mathrm r})$.
For target $a$, define
\begin{equation}
 S_{\mathrm r}=s_{\mathrm r}\cos a-c_{\mathrm r}\sin a,
 \qquad
 C_{\mathrm r}=c_{\mathrm r}\cos a+s_{\mathrm r}\sin a.
 \label{eq:revolute-target-signals}
\end{equation}
These quantities compare the relative orientation with the target.
For rigid rotations with
orthonormal reference directions and aligned joint axes, they reduce
to $S_{\mathrm r}=\sin(\alpha-a)$ and
$C_{\mathrm r}=\cos(\alpha-a)$.
{The squared sine vanishes both at the target and a half-turn away.
The cosine term below gives the half-turn a positive penalty.}
The periodic target energy is
\begin{align}
 \Phi^{\mathrm{rev}}
 &=\tfrac12 k\bigl[S_{\mathrm r}^2
                  +w_c(1-C_{\mathrm r})^2\bigr]-f\alpha,\notag\\
 \mathbf g^{\mathrm{rev}}
 &=\nabla\Phi^{\mathrm{rev}}\notag\\
 &=k\bigl[S_{\mathrm r}\nabla S_{\mathrm r}
       -w_c(1-C_{\mathrm r})\nabla C_{\mathrm r}\bigr]
       -f\nabla\alpha.
 \label{eq:revolute-joint-gradient}
\end{align}
Here $w_c=0.02$ is the fixed relative weight of the cosine penalty,
not an optimized control. The joint energy uses \(s_{\mathrm r}\) and \(c_{\mathrm r}\) computed directly from the affine-transformed joint reference directions. Their derivatives account for changes in both the lengths and orientations of these directions.
The angle derivatives used by the torque term $-f\alpha$ are given
in Appendix~\ref{sec:joint-actuation-details}.

The partial derivatives with respect to the applied target and
joint force are
\begin{equation}
 \begin{aligned}
 \frac{\partial\mathbf g^{\mathrm{rev}}}{\partial a}
 &=k\bigl[(-C_{\mathrm r}-w_c(1-C_{\mathrm r}))\nabla S_{\mathrm r}\\
 &\qquad +(w_c-1)S_{\mathrm r}\nabla C_{\mathrm r}\bigr],\\
 \frac{\partial\mathbf g^{\mathrm{rev}}}{\partial f}
 &=-\nabla\alpha.
 \end{aligned}
 \label{eq:revolute-control-derivatives}
\end{equation}
All control derivatives are evaluated at the converged frame
states. At fixed control values, differentiating the state gradients
gives the full joint Hessians $\nabla^2\Phi^{\mathrm{pri}}$ and
$\nabla^2\Phi^{\mathrm{rev}}$. These include the second derivatives
of the joint coordinates and of the potential terms
linear in the joint force.
Appendix~\ref{sec:joint-actuation-details} gives the
geometric definitions of $\xi$, $s_{\mathrm r}$, and
$c_{\mathrm r}$, their first and second coordinate derivatives,
and the full Hessians in Equation~\eqref{eq:joint-solve-blocks}.

% Reorganized GPU backward computation and task interface.

\begin{figure*}[t]
  \centering
  % Fixed physical text size; content-sized frames determine the figure footprint.
  \includegraphics[width=0.9\linewidth]{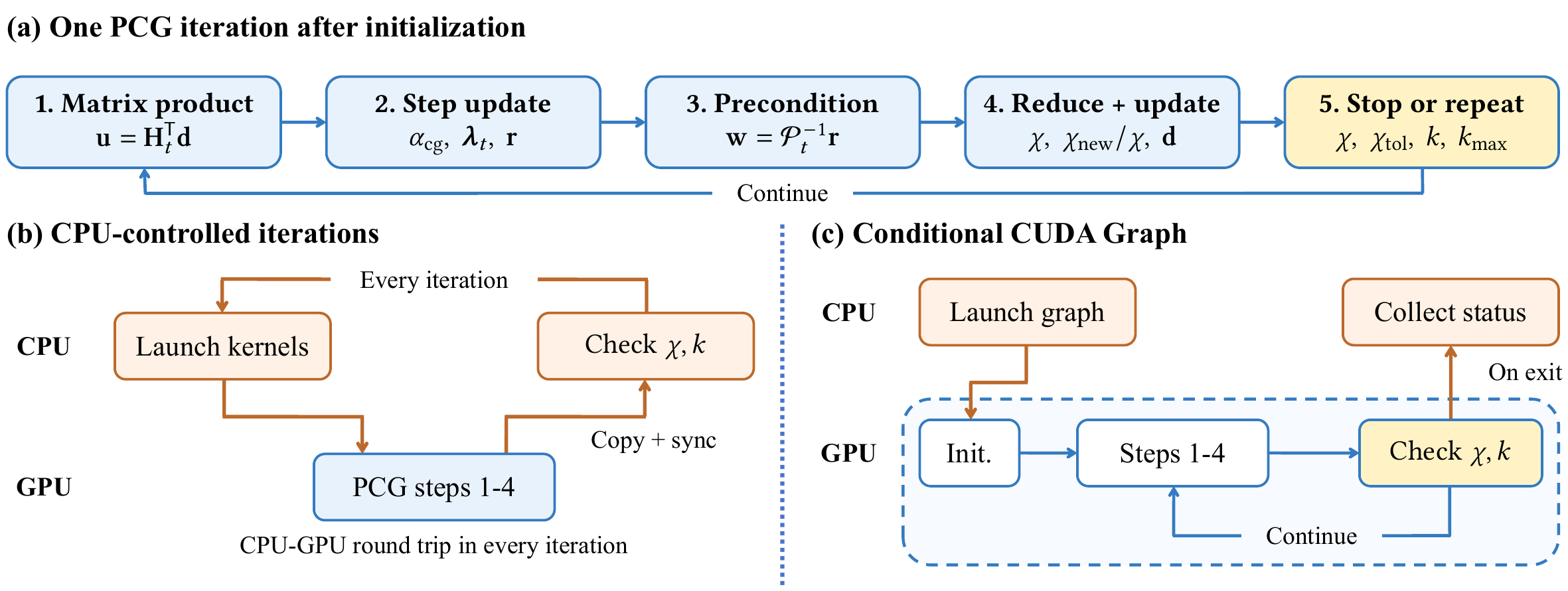}
  \caption{PCG computation and execution control.
  (a) Five stages of one iteration (Algorithm~\ref{alg:pcg-gpu}).
  (b) A host-controlled loop retrieves a convergence value after each
  GPU iteration. (c) A conditional CUDA Graph keeps iteration and
  stopping on the GPU. Frame restoration, Hessian assembly, and gradient
  {accumulation remain outside the graph. Both execution modes use the
  same PCG recurrence and stopping criterion.}}
  \Description{The top row shows matrix multiplication, solution and residual update, preconditioning, reduction and search-direction update, and the stopping test. Two CPU and GPU lane diagrams compare a host-controlled loop with a device-controlled conditional graph. Only the host-controlled loop returns a convergence scalar to the CPU at every iteration.}
  \label{fig:schematic-gpu-backward}
\end{figure*}
\section{GPU Backward and Reusable Interface}
\label{sec:pcg}

Compact coordinates and GPU parallelism reduce different parts of the backward cost. The coupled adjoint solve captures interactions across the scene, while local derivative products expose parallel work over elements, surface triangles, contacts, and joints. Conditional CUDA Graphs reduce repeated CPU coordination during the solve, and parallel accumulation returns gradients to earlier states and physical parameters (Figure~\ref{fig:schematic-gpu-backward}). A PyTorch interface makes this computation reusable across task-specific designs and objectives.

The adjoint system is assembled from each frame's converged FEM and affine states. Equation~\eqref{eq:abd-map} recovers the affine surface positions. The backward pass evaluates contact derivatives using the same contact pairs and distance formulas as the forward solve at this converged state. Previous-frame friction weights, contact-plane bases, and normal forces retain the history dependence derived in Section~\ref{sec:friction}. StiffGIPC's parallel global Hessian assembly~\cite{stiffgipc2025} combines element, body, collision, friction, and joint contributions, including FEM-affine cross terms. The resulting Hessian is assembled once per reverse frame and remains fixed throughout its PCG solve, avoiding a repeat of the nonlinear forward iterations.

\subsection{\texorpdfstring{{PCG with conditional CUDA Graphs}}{PCG with conditional CUDA Graphs}}
\label{sec:parallel-solve}

The adjoint in Equation~\eqref{eq:frame-adjoint} uses the same frame Hessian as the converged forward equations, allowing StiffGIPC's highly parallel GPU PCG solver~\cite{stiffgipc2025} to serve both forward and backward computations. Compact affine coordinates reduce PCG's numerical workload, making repeated CPU kernel launches and convergence synchronization a larger fraction of solve time. The stricter backward residual tolerance can amplify this cost over more {iterations. To address this}, we build a PCG pipeline managed by a conditional CUDA Graph, based on the PCG recurrence kernels~\cite{baraff1998largesteps}, their execution dependencies, and the convergence checks (Figure~\ref{fig:schematic-gpu-backward}). Once launched from the CPU, the graph initializes PCG and then repeats its iterations and convergence checks entirely on the GPU until convergence or the iteration limit is reached, eliminating per-iteration kernel launches and CPU–GPU synchronization.

\begin{algorithm}[!t]
\caption{PCG with Conditional CUDA Graphs}
\label{alg:pcg-gpu}
{
\small

\algrenewcommand\algorithmicrequire{\textbf{Input:}}
\algrenewcommand\algorithmicensure{\textbf{Output:}}
\begin{algorithmic}[1]
\Require Current $\mathbf{H}_t^\T$, $\mathbf b_t$, $\mathcal P_t$, relative target $\eta_{\mathrm{cg}}$, iteration limit $k_{\max}$
\Ensure Candidate $\vlambda_t$ and convergence status

\Statex \textbf{CPU: construct or reuse CUDA Graph}
\State Use $\mathbf{H}_t^\T$, $\mathbf b_t$, and $\mathcal P_t$ as inputs to the CUDA Graph
\If{no compatible CUDA Graph exists}
  \State Construct a CUDA Graph from the GPU kernels of PCG
  \State Connect dependent kernels
  \State Add conditional loop node that reads GPU-updated continue flag
\EndIf
\State Launch the CUDA Graph once from the CPU

\Statex \textbf{GPU: PCG initialization within CUDA Graph}
\State $\vlambda_t\gets\vzero$; $\mathbf r\gets\mathbf b_t-\mathbf{H}_t^{\T}\vlambda_t$; $\mathbf w\gets\mathcal P_t^{-1}\mathbf r$; $\mathbf d\gets\mathbf w$
\State $\chi\gets\mathbf r^{\T}\mathbf w$; $\chi_{\mathrm{tol}}\gets\eta_{\mathrm{cg}}|\chi|$; $k\gets0$
\State Set the continue flag from $|\chi|>\chi_{\mathrm{tol}}\ \land\ k<k_{\max}$ on the GPU

\Statex \textbf{GPU: PCG iterations within CUDA Graph}
\While{the graph's continue flag is true}
  \State $\mathbf u\gets \mathbf{H}_t^{{\T}}\mathbf d$; $\alpha_{\mathrm{cg}}\gets\chi/(\mathbf d^{\T}\mathbf u)$
  \State $\vlambda_t\gets\vlambda_t+\alpha_{\mathrm{cg}}\mathbf d$; $\mathbf r\gets\mathbf r-\alpha_{\mathrm{cg}}\mathbf u$
  \State $\mathbf w\gets\mathcal P_t^{-1}\mathbf r$; $\chi_{\mathrm{new}}\gets\mathbf r^{\T}\mathbf w$
  \State $\mathbf d\gets\mathbf w+(\chi_{\mathrm{new}}/\chi)\mathbf d$; $\chi\gets\chi_{\mathrm{new}}$; $k\gets k+1$
  \State A GPU kernel clears the continue flag if $|\chi|\leq\chi_{\mathrm{tol}}$ or $k\geq k_{\max}$
\EndWhile

\Statex \textbf{CPU: read final convergence status}
\State Read $\chi$ and $k$
\State \Return $\vlambda_t$
\end{algorithmic}
}
\end{algorithm}

For each reverse frame, we solve the assembled adjoint system using a reusable CUDA Graph (Algorithm~\ref{alg:pcg-gpu}). Its inputs $\mathbf{H}_t^\T$, $\mathbf b_t$, and preconditioner $\mathcal P_t$ are updated between frames and stay fixed during each solve. The graph preserves the dependencies between PCG kernels, while GPU convergence checks govern its conditional loop, so the same graph can run a different number of iterations. It can still be reused when contact sets change, as long as the kernel configuration and storage stay compatible. This reuse spreads graph construction across frames and removes per-iteration CPU coordination, so compact FEM-affine adjoint systems can benefit from their reduced numerical workload.
The GPU convergence check compares the preconditioned residual measure $|\chi^{(k)}|=|(\mathbf r^{(k)})^{\T}\mathcal P_t^{-1}\mathbf r^{(k)}|$ with the target $\chi_{\mathrm{tol}}=\eta_{\mathrm{cg}}|\chi^{(0)}|$, set at the start of each solve. The conditional loop stops when this target is reached or the iteration count reaches $k_{\max}$. The default backward settings are $\eta_{\mathrm{cg}}=10^{-14}$ and $k_{\max}=3\times10^{4}$. Only an adjoint whose final residual measure meets the target is accepted for the gradient accumulation described next.

\subsection{GPU-parallel gradient accumulation}

\label{sec:parallel-accumulation}

{The parameter-gradient and earlier-frame
contributions in Equations~\eqref{eq:parameter-gradient} and
\eqref{eq:history-update} share the coupled adjoint $\vlambda_t$.}
{Its entries contain the influence of interactions across the scene, but each derivative product accesses the DoFs of a local element, contact or joint. Surface triangles provide the local mass and volume derivatives for affine shape design. This locality exposes parallel work even when a body has a compact motion state.}

The parameter-gradient update is organized as parallel GPU tasks over FEM
elements, contact pairs, joints, and affine surface triangles. Each task
computes one local contribution to the gradient of a selected parameter.
For affine {rest-shape} gradients, tasks assigned to individual surface
triangles compute the mass-matrix and volume derivative contributions.
Each task reads only the adjoint entries for the DoFs involved and
evaluates their inner product with the corresponding parameter derivative.
With the frame adjoint $\vlambda_t$ solved, the local derivative products are independent and evaluated in parallel. Atomic additions combine contributions to the same parameter-gradient entry.
{Let $\vg_t^{[\iota]}$ denote the energy-gradient
contribution of one local term, expressed in the DoFs it involves.
The vector $\vlambda_t^{[\iota]}$ contains the matching entries of
the coupled adjoint $\vlambda_t$. The total loss gradient for
parameter $\theta_k$ is}
{
\begin{equation}
 \bar\theta_k=(\boldsymbol\ell_\theta)_k
 -\sum_{t=1}^{T}\sum_{\iota}
 \left(\frac{\partial\vg_t^{[\iota]}}{\partial\theta_k}\right)^{\T}
 \vlambda_t^{[\iota]}.
 \label{eq:sparse-parameter-dot-products}
\end{equation}
}
{Here $(\boldsymbol\ell_\theta)_k$ is the
direct loss derivative with all solved states fixed.}
{This decomposition sums the frame
contributions in Equation~\eqref{eq:parameter-gradient}}
without assembling the global parameter-derivative matrix
$\mathbf{G}_t$. The required derivatives come from Section~\ref{sec:fem-abd}, including
inertia in Equation~\eqref{eq:abd-inertia-rest}, initial states in
Equations~\eqref{eq:abd-initial-derivatives}--\eqref{eq:abd-second-frame-initial-derivative},
and joint controls in Equation~\eqref{eq:joint-control-gradient}.
Material derivatives are given in
Appendix~\ref{sec:fem-derivative-details}.

Friction follows the same decomposition. The mapped adjoint
${\mathbf K_t^{\,i}}\vlambda_t$ is computed once per vertex $i$ and reused by all
contacts involving that vertex. {For contact $c$, define}
\[
 \boldsymbol{\gamma}_{t,c}
 =
 \left(
   \frac{\partial\vg_{t,c}^{\mathrm{fr},x}}
        {\partial\vx_{t-1}^{c}}
 \right)^{\T}
 \mathbf{K}_{t,c}\vlambda_t.
\]
{Each component of $\boldsymbol\gamma_{t,c}$ is computed
independently by a GPU thread. Each thread evaluates the
inner product of the corresponding derivative column with
$\mathbf K_{t,c}\vlambda_t$, without assembling the full contact
derivative matrix.}
{For each vertex $i$, let
$\boldsymbol\gamma_{t,c}^{\,i}\in\R^3$ contain the three entries of
$\boldsymbol\gamma_{t,c}$ associated with that vertex. Summing over
all contacts containing $i$ gives}
{
\[
 \boldsymbol\gamma_t^i
 =\sum_{c\ni i}\boldsymbol\gamma_{t,c}^{\,i}\in\R^3.
\]
}
{This vertex sum is mapped to previous-frame DoFs}
(Equation~\eqref{eq:friction-history-gradient}). Contacts sharing a vertex
therefore apply ${(\mathbf K_{t-1}^{\,i})^{\T}}$ only once. For a free FEM vertex $i$, $-\boldsymbol{\gamma}_{t}^i$ is accumulated
directly into the corresponding entries of $\mathbf b_{t-1}$.
For affine bodies, the contributions
$-{(\mathbf K_{t-1}^{\,i})^{\T}}\boldsymbol{\gamma}_{t}^i$ are summed over
their vertices into the body's twelve entries in $\mathbf b_{t-1}$.
The per-vertex vectors $\boldsymbol{\gamma}_{t}^i$ are also reused
for the {rest-shape} gradient contribution arising from previous-frame
contact positions (Equation~\eqref{eq:affine-rest-friction-map}). Motion changes propagate
through the earlier-frame right-hand sides, and the remaining {rest-shape}
terms reuse the point gradients and Hessians assembled before affine
conversion, without adding motion DoFs.

The inertial history terms in
Equations~\eqref{eq:bdf1-adjoint}--\eqref{eq:quasistatic-abd-adjoint}
are likewise evaluated independently across vertices and bodies. All
contributions to an earlier right-hand side must be accumulated before its
adjoint solve. Parallelism therefore lies within each reverse frame, while
frames follow the dependency order in
Section~\ref{sec:frame-computation}. For a fixed trajectory and scalar
loss, all selected parameters share the same frame adjoint, so adding
parameters requires local derivative computation rather than additional
coupled solves.

\subsection{A reusable interface for inverse problems}

\label{sec:interface-backend}

The parameter map converts user-defined design variables into physical
parameters such as rest geometry, material properties, or joint controls.
The output map converts simulated DoFs into quantities used to evaluate
the loss, such as surface vertex positions. PyTorch differentiates both
maps and the loss, while DiffGIPC computes derivatives through the
coupled simulation. A loss evaluated on one object can therefore guide
the optimization of another object's parameters through their physical
interactions. %For affine bodies, the surface geometry remains available
%for shape optimization and loss evaluation, while the motion and adjoint
%solves retain only twelve DoFs per body.

Let $\mathbf z\in\R^{n_z}$ contain the optimizer
variables, $\phi$ map them to physical parameters
$\vtheta\in\R^{n_\theta}$, and $\mathcal O$ form the quantities used by
the scalar loss $\ell$ from the converged states:
\begin{equation}
 \begin{aligned}
 \vtheta&=\phi(\mathbf z),\\
 \mathbf o&=\mathcal O(\vy_1,\ldots,\vy_T,\vtheta),\\
 L&=\ell(\mathbf o,\vtheta).
 \end{aligned}
 \label{eq:task-composition}
\end{equation}
The parameter map imposes task-specific constraints or
shares design variables across physical parameters, so $n_z$ need not
equal $n_\theta$. The output map combines quantities from different
objects and frames without changing the coordinates of the coupled
solve.

PyTorch differentiates the loss and output map to obtain
\begin{equation}
 \boldsymbol\ell_o=\nabla_{\mathbf o}\ell,\qquad
 \boldsymbol\ell_t=
 \left(\frac{\partial\mathcal O}{\partial\vy_t}\right)^{\T}
 \boldsymbol\ell_o.
 \label{eq:interface-state-gradients}
\end{equation}
For affine surface points, $\mathbf{J}_i^{\T}$ maps
point gradients to body DoFs as in Equation~\eqref{eq:abd-loss-map},
whereas FEM positions use the identity map. Selecting outputs sets
the direct loss gradients, not the dependencies retained by the
adjoint. A frame absent from $\mathcal O$ has
$\boldsymbol\ell_t=\vzero$ but can still receive contributions from
later frames. The output map and loss may also depend explicitly on
physical parameters. With the simulated DoFs fixed, their contribution
is
\begin{equation}
 \boldsymbol\ell_\theta=
 \left.\nabla_{\vtheta}\ell\right|_{\mathbf o\ \mathrm{fixed}}
 +\left(\frac{\partial\mathcal O}{\partial\vtheta}\right)^{\T}
  \boldsymbol\ell_o.
 \label{eq:interface-direct-gradient}
\end{equation}
For example, changing an affine rest surface moves its
world-space points even at fixed motion DoFs. This explicit geometric
effect is distinct from the change in simulated motion accounted for
by the adjoint. The vectors $\boldsymbol\ell_t$ and
$\boldsymbol\ell_\theta$ initialize the right-hand sides and parameter
gradient in Section~\ref{sec:frame-computation}. The GPU backward pass
then supplies the physical gradient, which PyTorch maps to the
optimizer variables:
\begin{equation}
 \begin{aligned}
 \bar\vtheta&=\boldsymbol\ell_\theta
                -\sum_{t=1}^{T}\mathbf{G}_t^{\T}\vlambda_t,\\
 \nabla_{\mathbf z}L&=
 \left(\frac{\partial\phi}{\partial\mathbf z}\right)^{\T}
 \bar\vtheta.
 \end{aligned}
 \label{eq:task-chain-rule}
\end{equation}

Task parameterizations and objectives can therefore change without changing the coupled backward equations. {In garment hanger
(Figure~\ref{fig:abd-hanger-rest}), two shared shoulder-control scales
specify the rest surface, while the loss measures support using both
hanger and cloth positions. PyTorch differentiates the shape
parameterization and geometric loss. DiffGIPC combines the direct
surface dependence with the gradient through their coupled motion,
retaining twelve motion DoFs for the hanger.}
{The same composition applies to the physical parameter
families listed in Table~\ref{stab:registered-parameters}
(Appendix~\ref{sec:experimental-details}).}
Additional physical models enter through their local derivatives, preserving the distinction between task design and coupled simulation.

\begin{figure}[!tbp]

  \centering
  \begin{subfigure}[t]{0.48\linewidth}
    \centering
    \includegraphics[width=\linewidth,trim=110bp 2bp 212bp 320bp,clip]{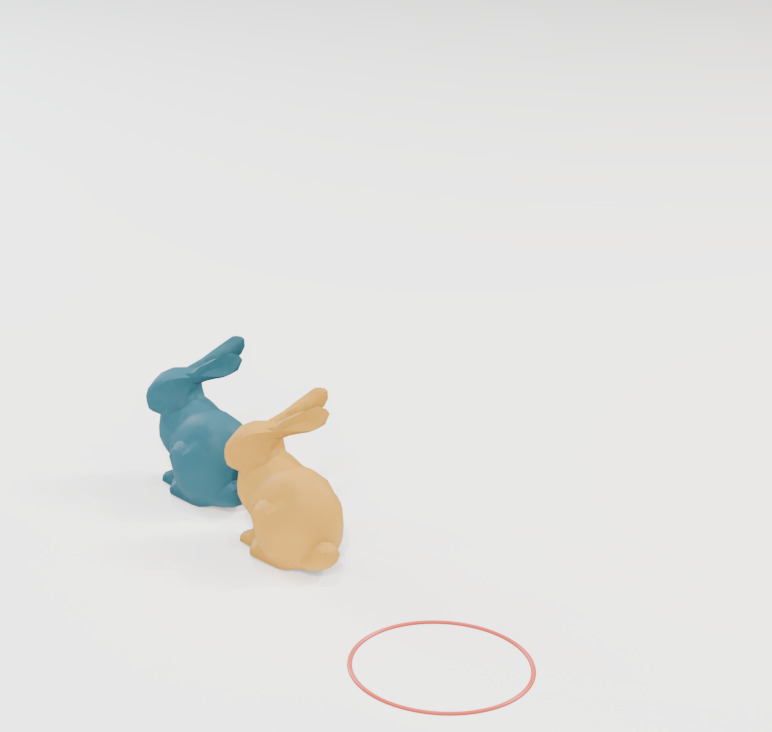}
    \caption{Iteration 0}
  \end{subfigure}\hfill
  \begin{subfigure}[t]{0.48\linewidth}
    \centering
    \includegraphics[width=\linewidth,trim=110bp 2bp 212bp 320bp,clip]{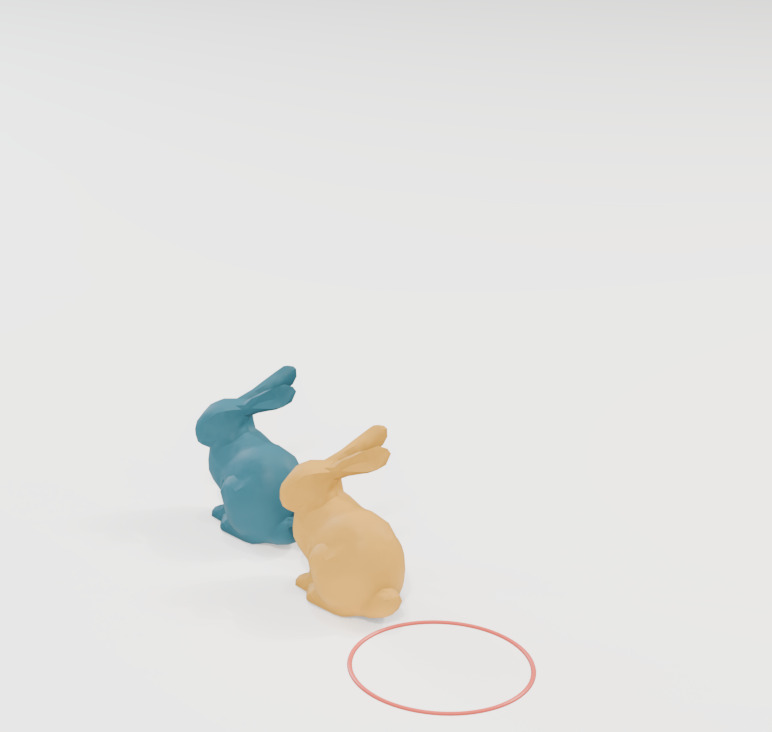}
    \caption{Iteration 1}
  \end{subfigure}\par\smallskip
  \begin{subfigure}[t]{0.48\linewidth}
    \centering
    \includegraphics[width=\linewidth,trim=110bp 2bp 212bp 320bp,clip]{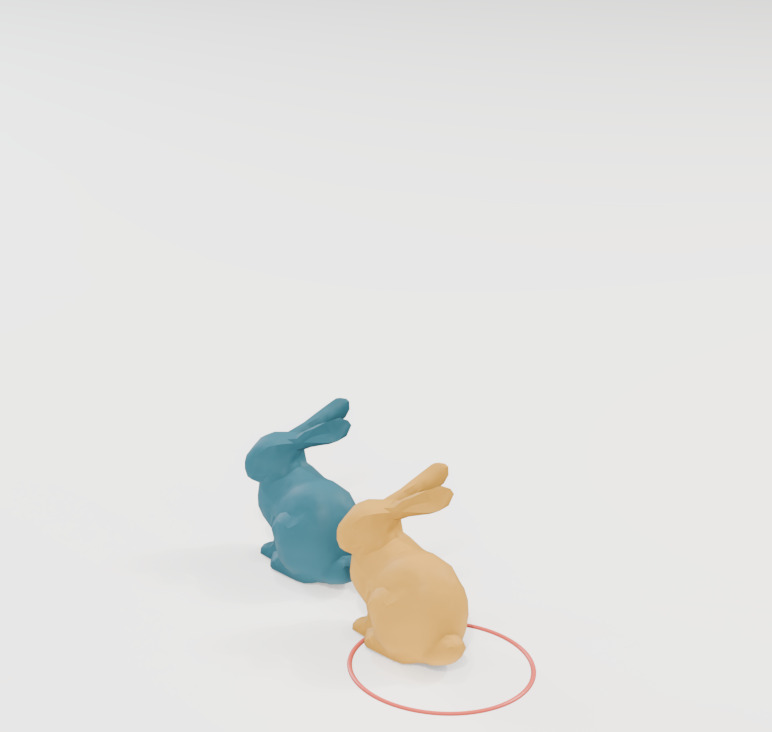}
    \caption{Iteration 3}
  \end{subfigure}\hfill
  \begin{subfigure}[t]{0.48\linewidth}
    \centering
    \includegraphics[width=\linewidth,trim=110bp 2bp 212bp 320bp,clip]{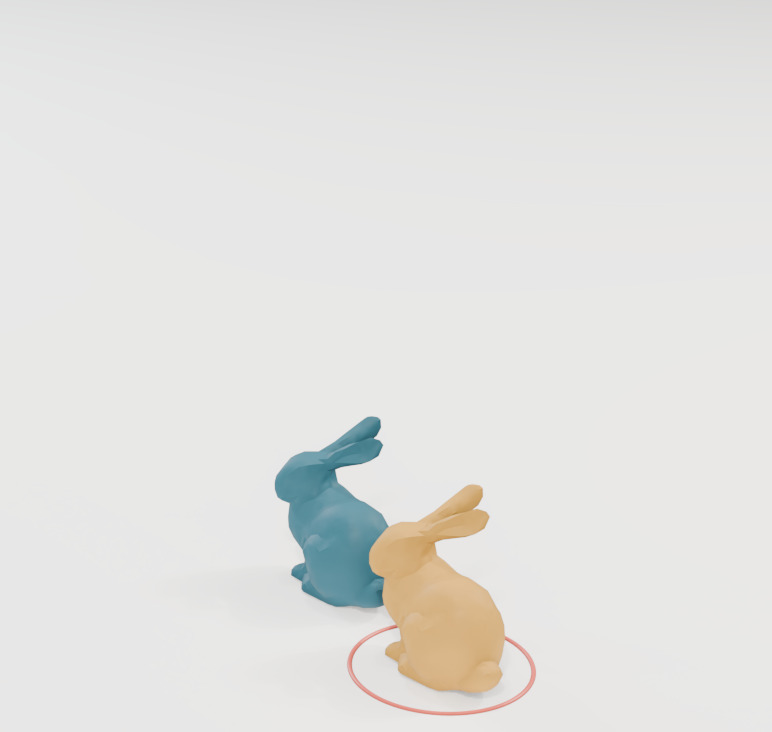}
    \caption{Iteration 10}
  \end{subfigure}
  \caption{Collision-mediated initial-velocity optimization. The first
  bunny's initial velocity is adjusted so that the second bunny's
  horizontal centroid reaches its target. The panels compare frame 40
  before optimization and after 1, 3, and 10 iterations.}
  \Description{Four panels show the initial placement and optimization updates
  1, 3, and 10 at simulation frame 40.}
  \label{fig:bunny-initial-velocity}
\end{figure}

\section{Evaluation}
\label{sec:evaluation}

Our evaluation examines the parameter-to-output
dependencies supported by DiffGIPC and the cost of forward and backward
evaluations across full-space and affine motion representations.
The shared full-space FEM benchmarks from DiffIPC~\cite{diffipc} cover
initial motion, rest design, and material or friction fitting at their
original low mesh resolutions. In these benchmarks, both methods use full-space FEM. Further experiments exercise
cloth design, affine shape and placement, and articulated control,
including mixed-coordinate inverse problems.
{We first compare optimization outcomes and
evaluation times on the shared FEM benchmarks, then demonstrate the
coupled design and control tasks. Gradient validation assesses agreement between analytic gradients and finite differences, while studies of motion representation,
resolution, and GPU execution explain the optimization cost.}

For the shared benchmarks, DiffGIPC runs
{on an NVIDIA GeForce RTX 4090 and DiffIPC on an Intel Core i9-13900K.
Both use double-precision arithmetic and the same convergence tolerances.}
{Appendix~\ref{sec:experimental-details} summarizes
the simulation settings in Table~\ref{tab:sim-settings} and the
differentiated parameter families in Table~\ref{stab:registered-parameters}.}
{A forward evaluation simulates a trajectory to the
task's observation horizon. A backward
evaluation computes its loss gradient over that trajectory.
$N_{\mathrm{fwd}}$ counts forward evaluations, including line-search
trials and verification runs, while $N_{\mathrm{bwd}}$ counts complete
gradient evaluations used for optimization.} Let $\bar t_f$ and
$\bar t_b$ denote their mean times. The combined mean time
{is $\bar t_{f+b}=\bar t_f+\bar t_b$. The accumulated evaluation time is}
$T_{\mathrm{eval}}={N_{\mathrm{fwd}}}\bar t_f+
{N_{\mathrm{bwd}}}\bar t_b$.
{In the loss definitions below, $\vx_{t,i}$ is the
world-space position of vertex $i$ at observation frame $t$, and a star
denotes a prescribed target held fixed during optimization. Each task
specifies its observation endpoint $T$ or selected frame set.
Unsubscripted norms are Euclidean norms, and $\|\cdot\|_{\mathrm F}$
denotes the Frobenius norm for matrices.}

\subsection{\texorpdfstring{{Comparison with DiffIPC}}{Comparison with DiffIPC}}
\label{sec:diffipc-comparison}

The shared FEM benchmarks establish the framework's coverage in full-space FEM before introducing affine design and control. They progress from contact-driven motion to {rest-shape} design and parameter fitting, changing both the optimized input and the observed response. {Figure~\ref{fig:loss-curves-shared} compares their optimization histories, and Table~\ref{tab:diffipc-phase-comparison} reports the corresponding evaluation times. We relate each task's objective to its physical outcome before examining the computational cost.}

\begin{figure}[!tbp]

  \centering
  \begin{subfigure}[t]{0.45\linewidth}
    \centering
    \includegraphics[width=\linewidth,trim=50bp 150bp 50bp 150bp,clip]{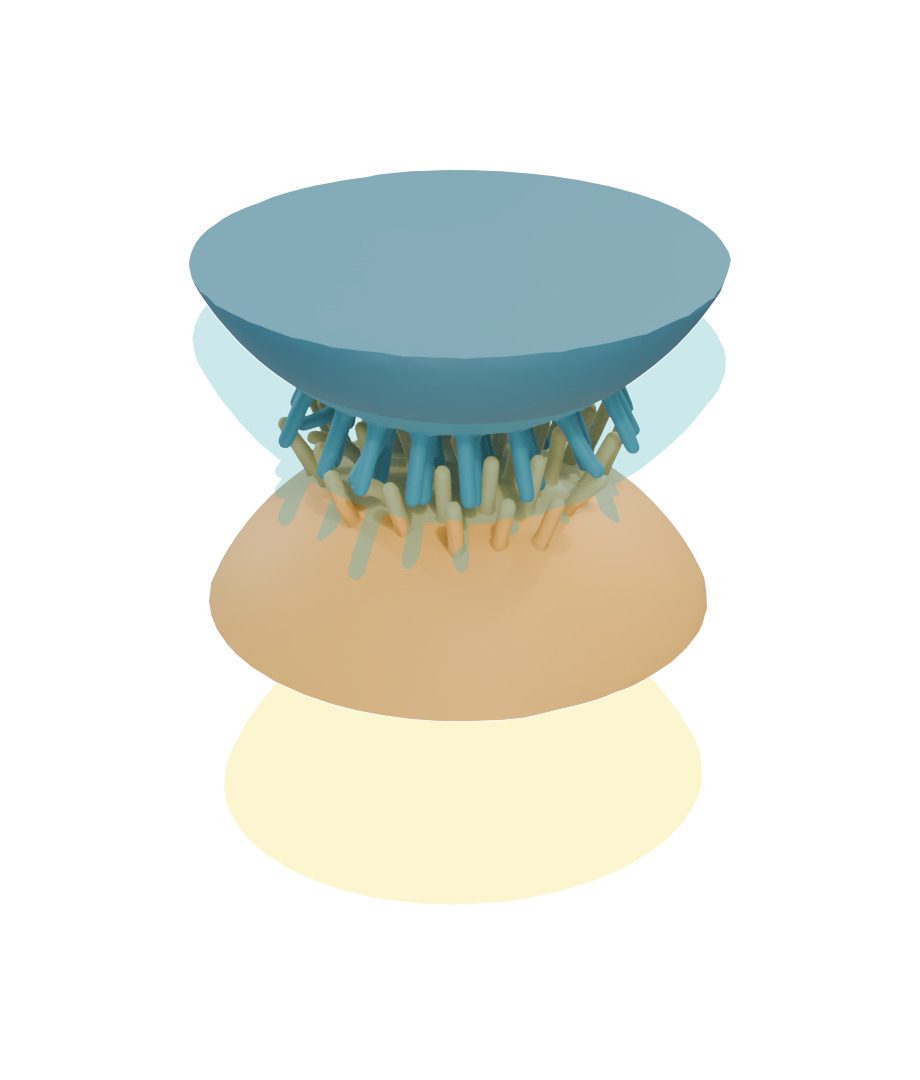}
    \caption{Iteration 0}
  \end{subfigure}\hfill
  \begin{subfigure}[t]{0.45\linewidth}
    \centering
    \includegraphics[width=\linewidth,trim=50bp 150bp 50bp 150bp,clip]{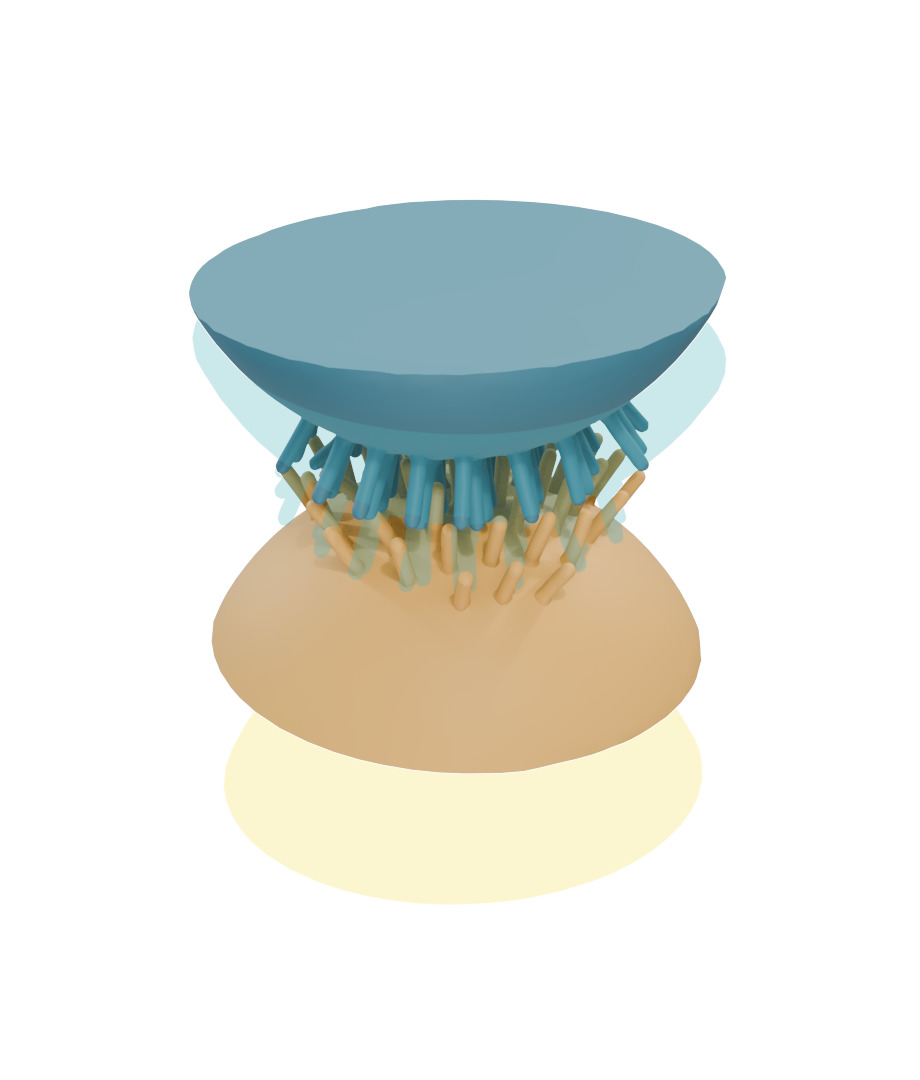}
    \caption{Iteration 1}
  \end{subfigure}\par\smallskip
  \begin{subfigure}[t]{0.45\linewidth}
    \centering
    \includegraphics[width=\linewidth,trim=50bp 150bp 50bp 150bp,clip]{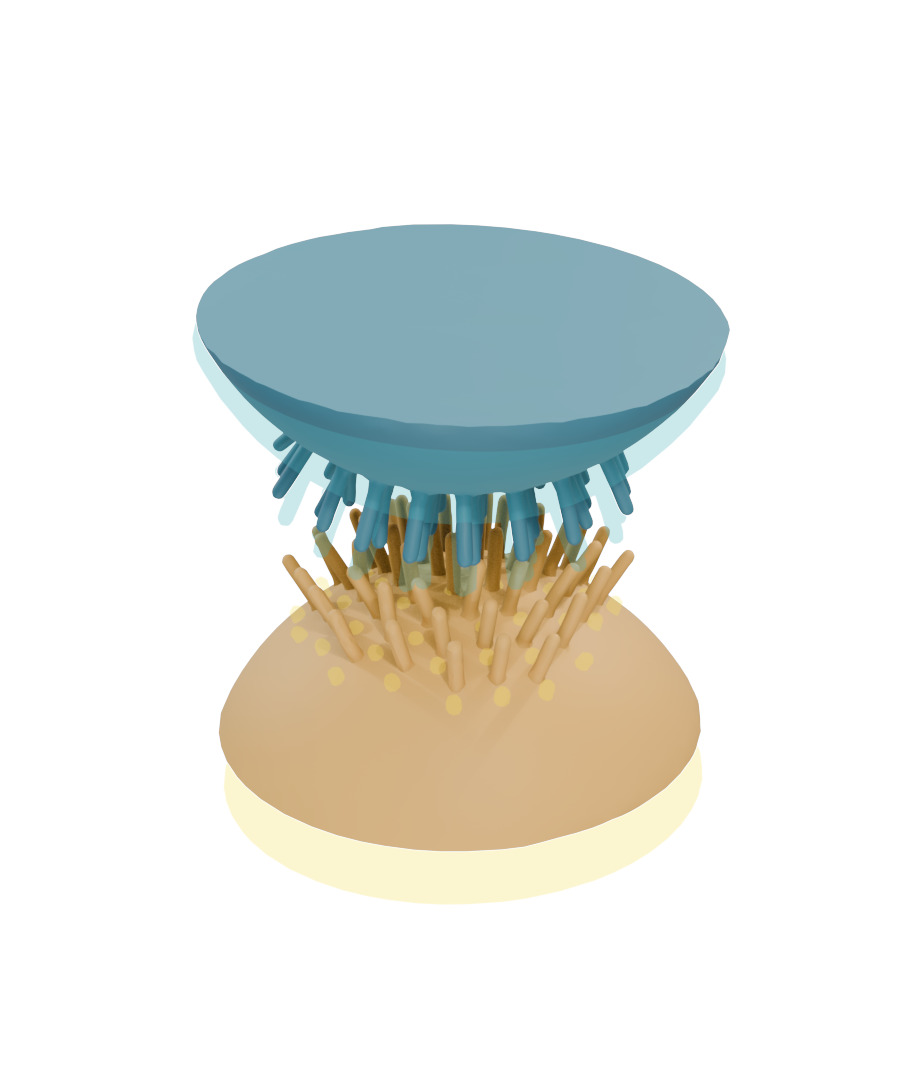}
    \caption{Iteration 3}
  \end{subfigure}\hfill
  \begin{subfigure}[t]{0.45\linewidth}
    \centering
    \includegraphics[width=\linewidth,trim=50bp 150bp 50bp 150bp,clip]{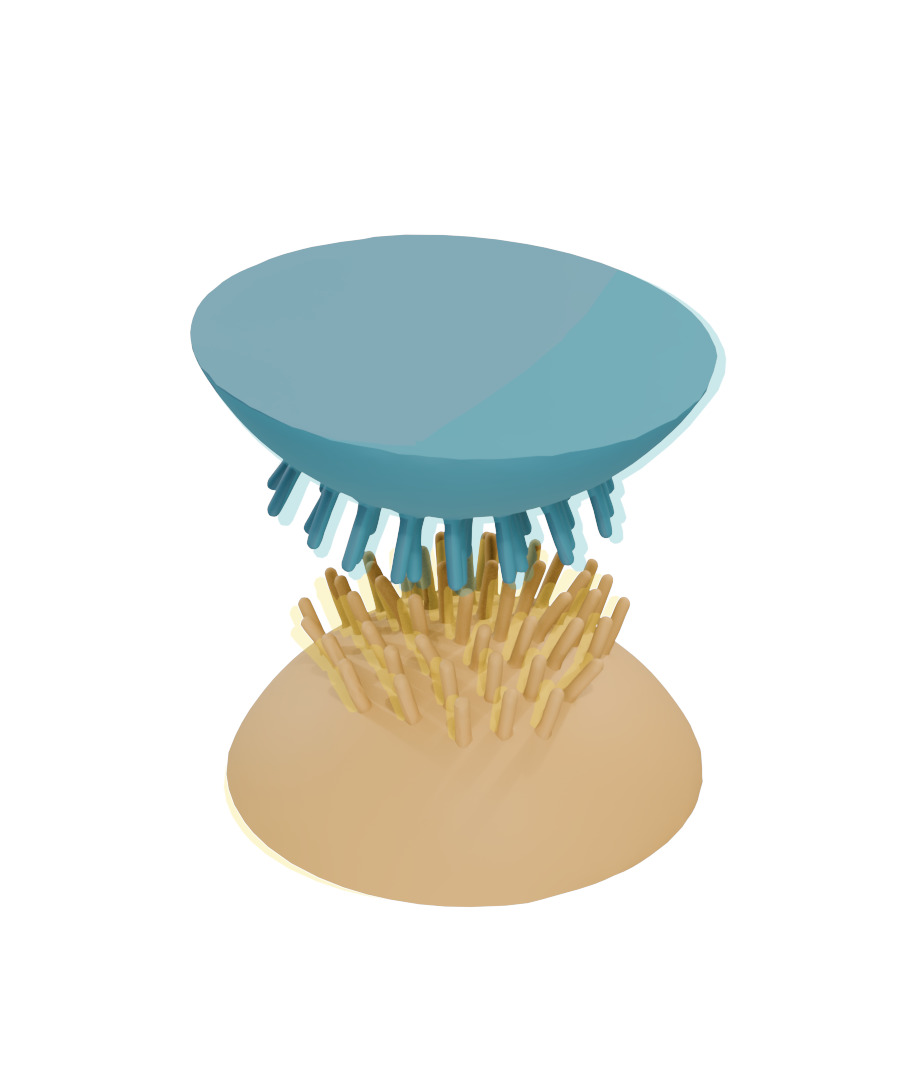}
    \caption{Iteration 10}
  \end{subfigure}
  \caption{Trajectory matching through contact between two deformable
  bodies. The first body's initial velocity is optimized to reproduce
  the target motion. Opaque surfaces show the simulation and translucent
  surfaces the target at frame 100, before optimization and after 1, 3,
  and 10 iterations.}
  \Description{Trajectory matching for two contacting deformable bodies. The shared initial velocity of the first body is optimized to reproduce a prescribed motion. Opaque surfaces show the simulated bodies and translucent surfaces show the target configuration. The panels show frame 100.}
  \label{fig:tentacle-trajectory}
\end{figure}

\paragraph{Bunny collision}
We optimize a shared initial velocity for the vertices of one bunny so
that collision moves the other bunny to a target. The objective uses
only the second bunny's volume-weighted horizontal center after 40 frames:
\begin{equation}
 L_{\mathrm{bunny}}=
 \|\mathbf c_{xz}(\vx_T^{(2)})-\mathbf c_{xz}^{*}\|^2.
 \label{eq:fig1-loss}
\end{equation}
Here $\mathbf c_{xz}$ is the volume-weighted horizontal $(x,z)$ center
of the second bunny and $T=40$. Because the loss observes only that body,
the optimized velocity affects it through collision. Figure~\ref{fig:bunny-initial-velocity}
shows the final position approaching the target. Section~\ref{sec:representation-study}
revisits this task across motion representations.

\paragraph{Tentacle trajectory matching}
The first deformable body's shared initial velocity is optimized while the second body's initial velocity remains fixed. Let $\mathbf c_t^{\mathrm{tent}}$ be the first body's volume-weighted position, using fixed rest-tetrahedron weights. Its trajectory-matching objective is
\begin{equation}
 L_{\mathrm{tentacle}}=200h\sum_{t=1}^{T}
       \|\mathbf c_t^{\mathrm{tent}}-\mathbf c_t^{\mathrm{tent},*}\|^2,
 \label{eq:fig15-loss}
\end{equation}
{where $T=100$ and $\mathbf c_t^{\mathrm{tent},*}$ is the prescribed target.} Thus observations throughout the
trajectory contribute to the initial-velocity gradient through motion and
friction history; Figure~\ref{fig:tentacle-trajectory} shows the endpoint.

% Preview DiffIPC comparison with TOG figure references.
\begin{table*}[!t]\centering\small
\newtablemark
\caption{Mean time per forward and backward evaluation (seconds)
on the shared FEM benchmarks. Forward + Backward gives
the sum of the two per-evaluation means.}
\label{tab:diffipc-phase-comparison}
\setlength{\tabcolsep}{2.4pt}
\renewcommand{\arraystretch}{1.15}
\setlength{\aboverulesep}{0pt}
\setlength{\belowrulesep}{0pt}
\setlength{\cmidrulesep}{0pt}
\begin{tabularx}{\textwidth}{@{}c|c|r|r|r|r|r|>{\centering\arraybackslash}X|r|r|r|>{\centering\arraybackslash}X|>{\centering\arraybackslash}X@{}}\toprule
\multirow{2}{*}{Example} & \multirow{2}{*}{System} & \multicolumn{6}{c|}{Forward} & \multicolumn{4}{c|}{Backward} & \multicolumn{1}{c}{\multirow{2}{*}{\shortstack{Forward +\\Backward}}} \\
\cmidrule{3-8}\cmidrule{9-12}
 &  & \multicolumn{1}{c|}{Linear} & \multicolumn{1}{c|}{CCD} & \multicolumn{1}{c|}{Hessian} & \multicolumn{1}{c|}{Search} & \multicolumn{1}{c|}{Other} & \multicolumn{1}{c|}{Total} & \multicolumn{1}{c|}{Hessian} & \multicolumn{1}{c|}{Linear} & \multicolumn{1}{c|}{Other} & \multicolumn{1}{c|}{Total} & \multicolumn{1}{c}{} \\
\midrule
\multirow{3}{*}{Bunny collision (Fig.~\ref{fig:bunny-initial-velocity})} & DiffIPC & 4.5 & 10 & 4.6 & 1.1 & 1.9 & 22 & 1.4 & 1.1 & 0.44 & 2.9 & 25 \\
 & DiffGIPC & 0.65 & 0.044 & 0.037 & 0.017 & 0.002 & 0.75 & 0.0069 & 0.22 & 0.0057 & 0.24 & 0.98 \\
 & Speedup & $6.9\times$ & $232.0\times$ & $123.9\times$ & $67.7\times$ & $963.8\times$ & $30.0\times$ & $204.7\times$ & $4.9\times$ & $76.3\times$ & $12.5\times$ & $25.8\times$ \\
\specialrule{\lightrulewidth}{0pt}{0pt}
\multirow{3}{*}{Tentacle matching (Fig.~\ref{fig:tentacle-trajectory})} & DiffIPC & 140 & 14 & 120 & 66 & 11 & 350 & 7.4 & 11 & 2.8 & 21 & 370 \\
 & DiffGIPC & 8.1 & 2.4 & 2.2 & 1.4 & 0.011 & 14 & 0.04 & 1 & 0.024 & 1.1 & 15 \\
 & Speedup & $16.9\times$ & $5.9\times$ & $55.3\times$ & $46.7\times$ & $1037.5\times$ & $24.6\times$ & $184.8\times$ & $10.9\times$ & $120.7\times$ & $19.7\times$ & $24.3\times$ \\
\specialrule{\lightrulewidth}{0pt}{0pt}
\multirow{3}{*}{Loaded support (Fig.~\ref{fig:fem-rest-shape})} & DiffIPC & 22 & 0.60 & 3.2 & 1.1 & 11 & 38 & 2.9 & 12 & 0.49 & 16 & 54 \\
 & DiffGIPC & 0.93 & 0.02 & 0.014 & 0.0017 & 0.00051 & 0.97 & 0.011 & 0.85 & 0.0066 & 0.87 & 1.8 \\
 & Speedup & $23.8\times$ & $29.5\times$ & $232.1\times$ & $653.7\times$ & $21172.2\times$ & $39.3\times$ & $266.4\times$ & $14.7\times$ & $74.2\times$ & $18.3\times$ & $29.4\times$ \\
\specialrule{\lightrulewidth}{0pt}{0pt}
\multirow{3}{*}{Material fitting (Fig.~\ref{fig:material-identification})} & DiffIPC & 0.89 & 0.029 & 0.69 & 0.028 & 0.61 & 2.2 & 0.35 & 0.53 & 0.057 & 0.94 & 3.2 \\
 & DiffGIPC & 0.015 & 0.0021 & 0.0042 & 0.00064 & 0.0032 & 0.025 & 0.0026 & 0.023 & 0.0016 & 0.028 & 0.053 \\
 & Speedup & $60.3\times$ & $14.3\times$ & $161.8\times$ & $43.5\times$ & $190.7\times$ & $90.1\times$ & $135.0\times$ & $22.8\times$ & $36.0\times$ & $34.2\times$ & $60.7\times$ \\
\specialrule{\lightrulewidth}{0pt}{0pt}
\multirow{3}{*}{Friction optimization (Fig.~\ref{fig:friction-identification})} & DiffIPC & 17 & 22 & 21 & 8.1 & 6.8 & 76 & 2.5 & 1.9 & 1.2 & 5.6 & 81 \\
 & DiffGIPC & 3.8 & 1.4 & 1.3 & 0.7 & 0.0093 & 7.2 & 0.037 & 0.36 & 0.039 & 0.44 & 7.7 \\
 & Speedup & $4.6\times$ & $16.3\times$ & $15.7\times$ & $11.7\times$ & $734.8\times$ & $10.5\times$ & $66.3\times$ & $5.3\times$ & $31.8\times$ & $12.8\times$ & $10.6\times$ \\
\specialrule{\lightrulewidth}{0pt}{0pt}
\bottomrule\end{tabularx}
\end{table*}

\begin{figure}[!b]

  \centering
  \begin{subfigure}[t]{0.48\linewidth}
    \centering
    \includegraphics[width=0.9\linewidth,trim=15bp 372bp 15bp 315bp,clip]{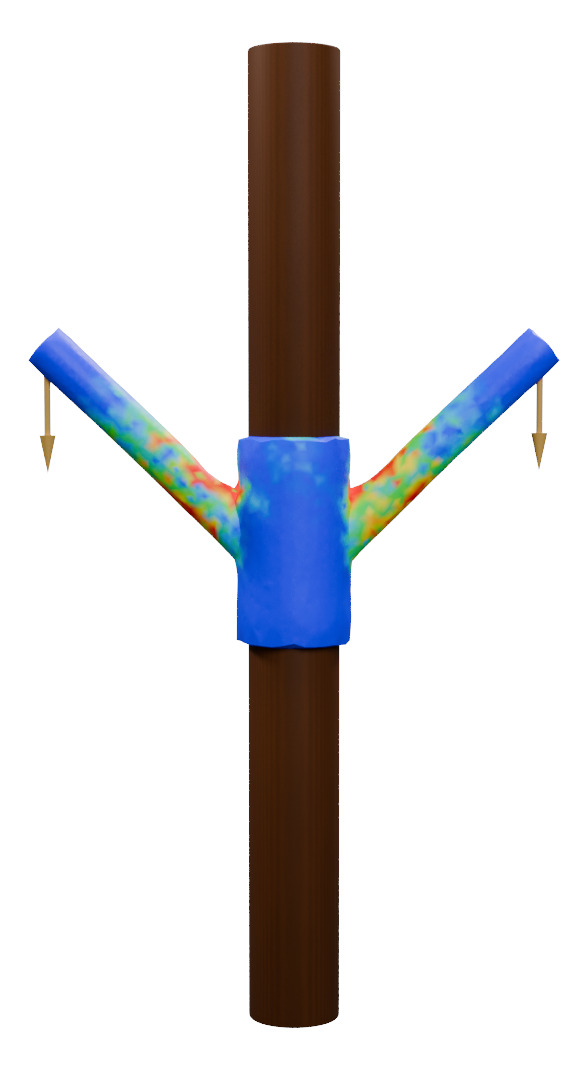}
    \caption{Iteration 0}
  \end{subfigure}\hfill
  \begin{subfigure}[t]{0.48\linewidth}
    \centering
    \includegraphics[width=0.9\linewidth,trim=15bp 372bp 15bp 315bp,clip]{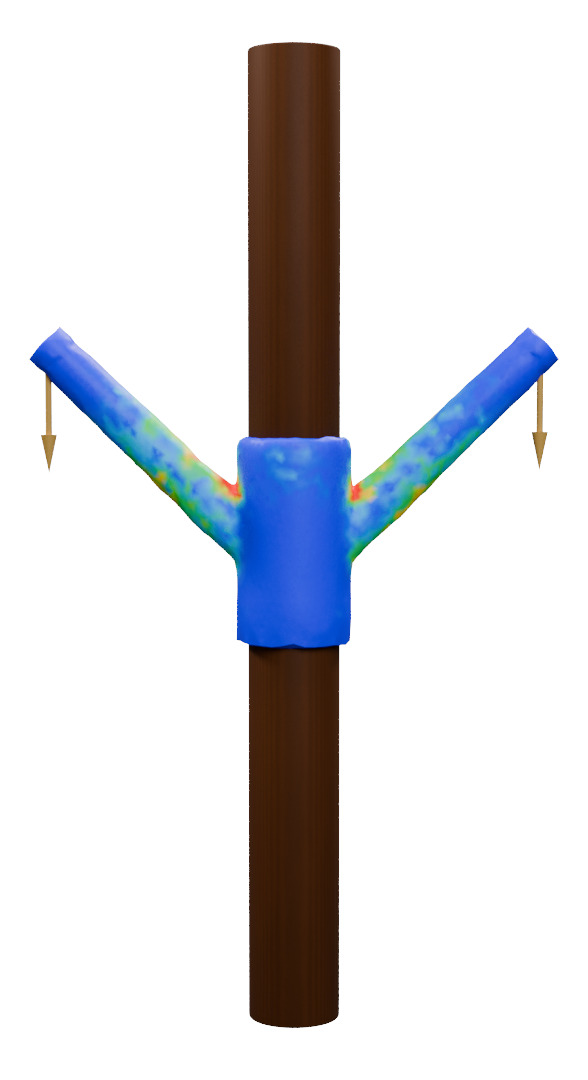}
    \caption{Iteration 20}
  \end{subfigure}\par\smallskip
  \begin{subfigure}[t]{0.48\linewidth}
    \centering
    \includegraphics[width=0.9\linewidth,trim=15bp 372bp 15bp 315bp,clip]{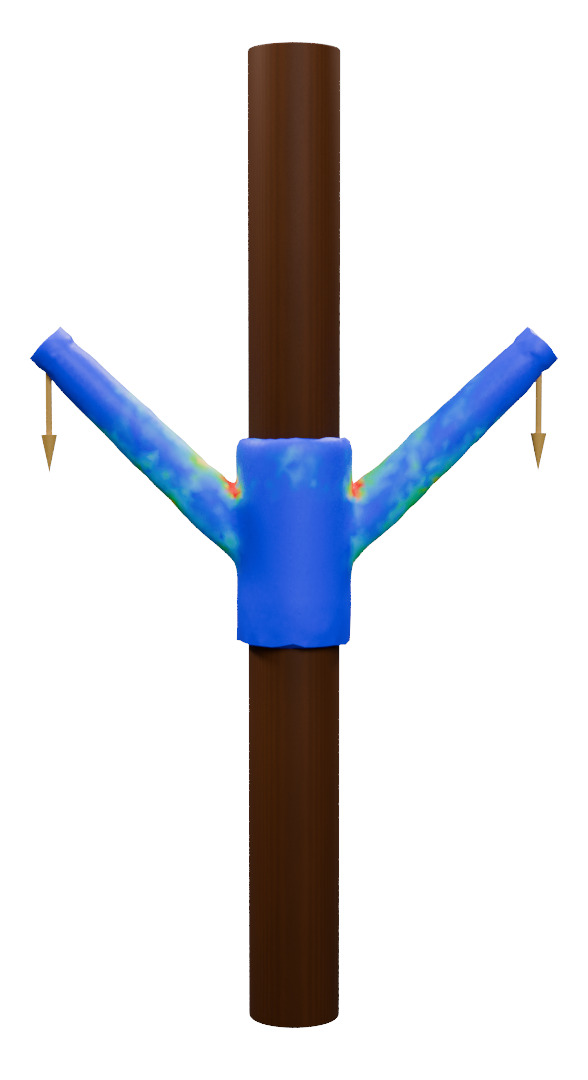}
    \caption{Iteration 50}
  \end{subfigure}\hfill
  \begin{subfigure}[t]{0.48\linewidth}
    \centering
    \includegraphics[width=0.9\linewidth,trim=15bp 372bp 15bp 315bp,clip]{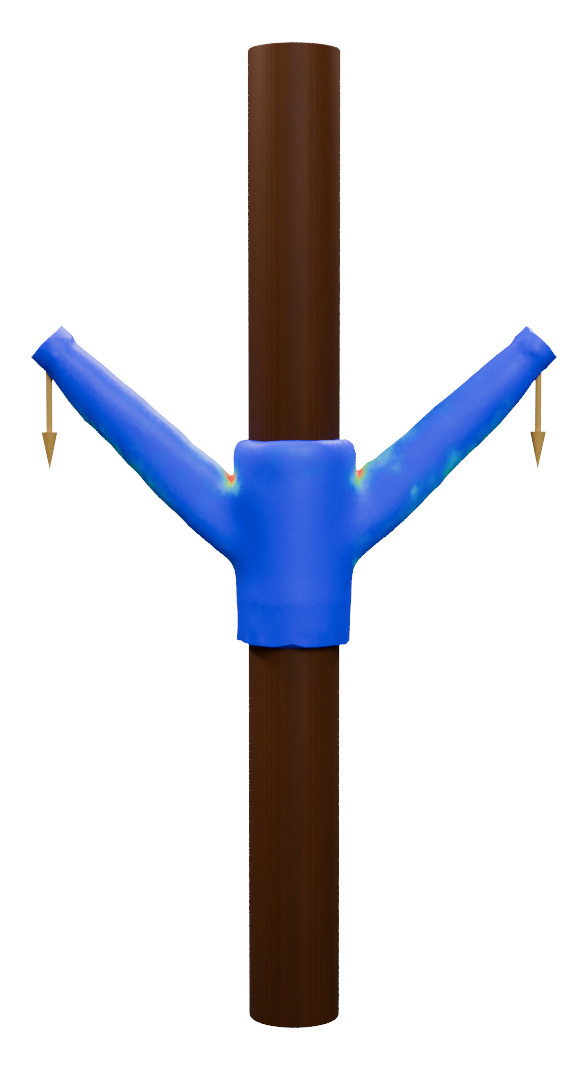}
    \caption{Iteration 200}
  \end{subfigure}
  \caption{{Rest-shape} optimization of a loaded FEM support. Under fixed
  constraints and downward loading, the objective minimizes squared
  first-Piola stress. The panels show the deformed support at frame 20
  before optimization and after 20, 50, and 200 iterations.}
  \Description{Four front-view stress renderings compare the initial support
  and the designs after 20, 50, and 200 updates.}
  \label{fig:fem-rest-shape}
\end{figure}

\paragraph{Loaded support design}

Selected rest vertices of a FEM support are optimized under downward loading, with its supports fixed. At the end of each 20-frame simulation, the squared-stress term is
$L_{\mathrm{stress}}=10^{-8}\sum_{e\in{\mathcal T_{\mathrm d}}}
V_e\|{{\mathbf P}_e}\|_{\mathrm F}^{2}$.
Here $\mathcal T_{\mathrm d}$ is the evaluated tetrahedron set,
{$V_e$ is the rest volume, and $\mathbf P_e$
is the first Piola stress evaluated at $T=20$.}
{
Rest-volume weighting makes larger stressed regions contribute proportionally
more. A surface-smoothing penalty discourages local irregularities in the
designed rest geometry. The optimized loss is
$L_{\mathrm{support}}=L_{\mathrm{stress}}+L_{\mathrm{sm}}$. The smoothing set
$\mathcal V_{\mathrm{sm}}$ contains surface vertices selected in rest coordinates.
For each $i\in\mathcal V_{\mathrm{sm}}$, let $\mathcal N_{\mathrm{sm}}(i)$
be its surface-edge neighbors that also belong to $\mathcal V_{\mathrm{sm}}$.
Using the current rest positions $\bar{\vx}_i$, the penalty is
\begin{equation}
L_{\mathrm{sm}}=
\sum_{i\in\mathcal V_{\mathrm{sm}}}
\left\|
\frac{\sum_{j\in\mathcal N_{\mathrm{sm}}(i)}
      (\bar{\vx}_i-\bar{\vx}_j)}
     {\sum_{j\in\mathcal N_{\mathrm{sm}}(i)}
      \|\bar{\vx}_i-\bar{\vx}_j\|}
\right\|^2.
\label{eq:loaded-support-smoothing}
\end{equation}
Vertices without neighbors in this set contribute zero. This term measures
each selected vertex's deviation from its neighbors' mean position,
normalized by the mean incident edge length.}
Figure~\ref{fig:fem-rest-shape} shows how the redesigned geometry redistributes stress under the same loading and color scale. The loss histories in Figure~\ref{fig:loss-curves-shared} reach similar final objective levels for both methods. Their intermediate jumps occur when remeshing changes the stress discretization. Gradients are evaluated on each current fixed mesh, with feasible rest-shape updates selected by backtracking. The task demonstrates rest-shape design through the combined changes in geometry, element volume, and elastic response.

\begin{figure*}[tp]

  \centering
  \begin{subfigure}[t]{0.24\textwidth}
    \centering
    \includegraphics[width=0.9\linewidth,trim=120bp 170bp 120bp 170bp,clip]{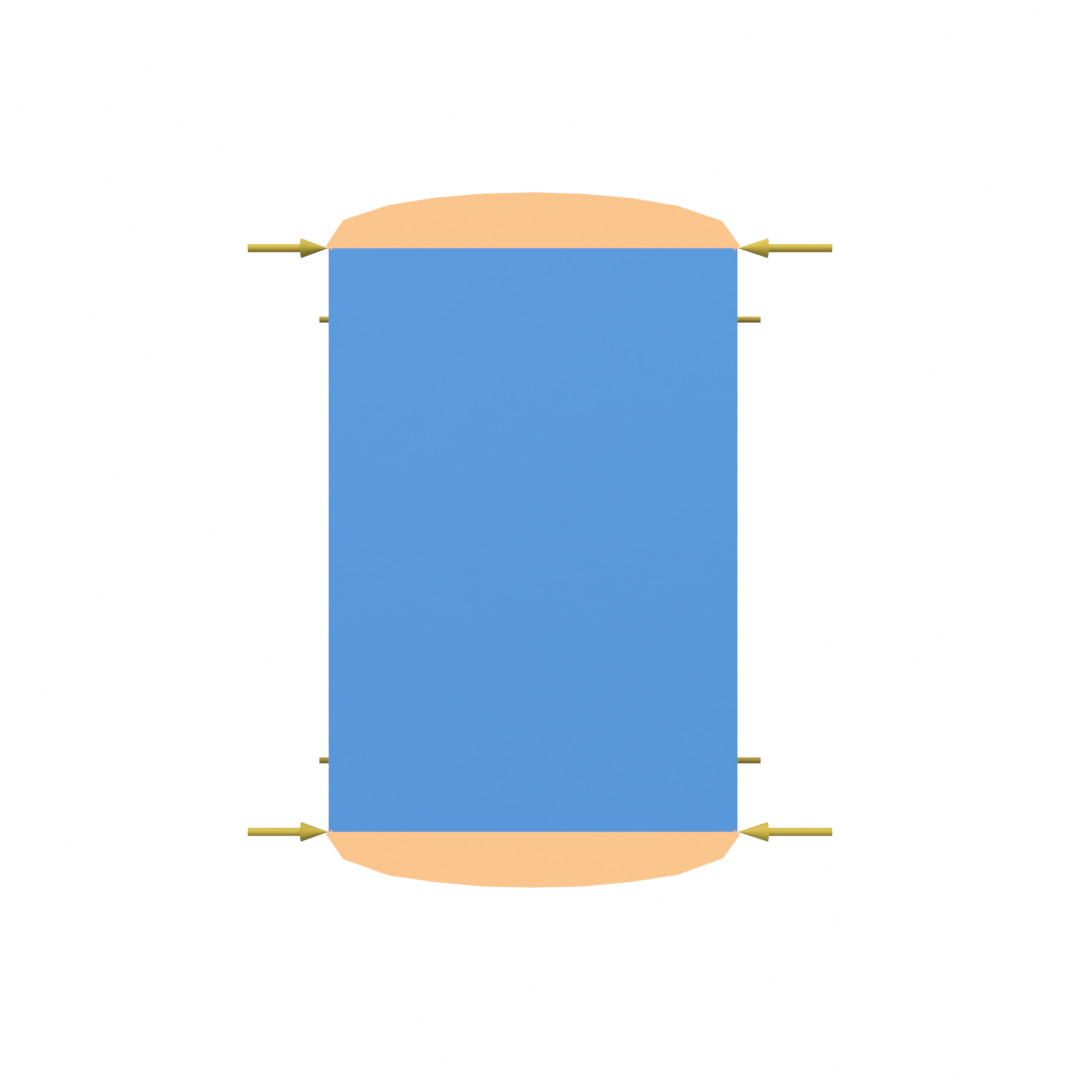}
    \par\vspace{1pt}\nointerlineskip
    \includegraphics[width=0.9\linewidth,trim=120bp 170bp 120bp 240bp,clip]{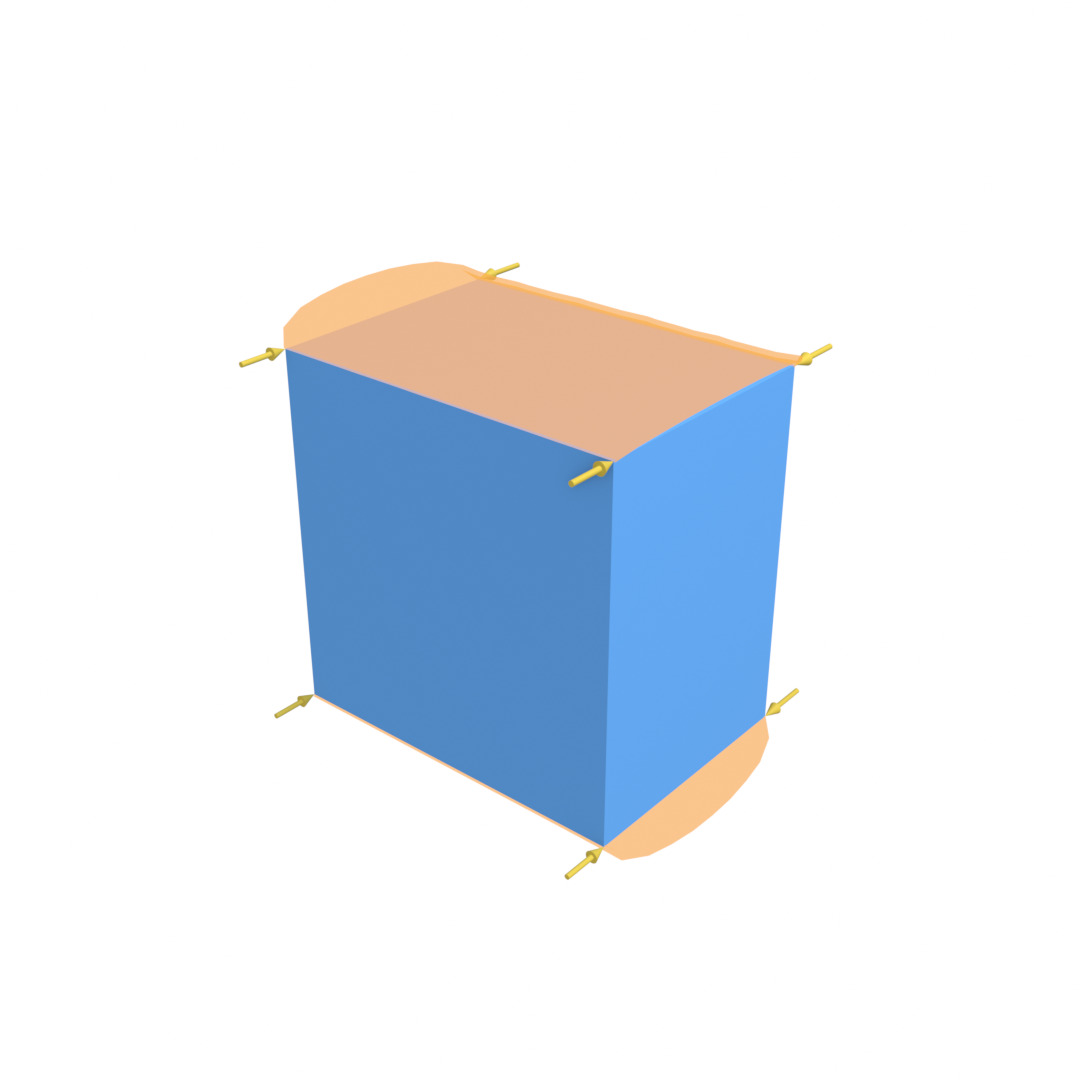}
    \caption{Iteration 0}
  \end{subfigure}\hfill
  \begin{subfigure}[t]{0.24\textwidth}
    \centering
    \includegraphics[width=0.9\linewidth,trim=120bp 170bp 120bp 170bp,clip]{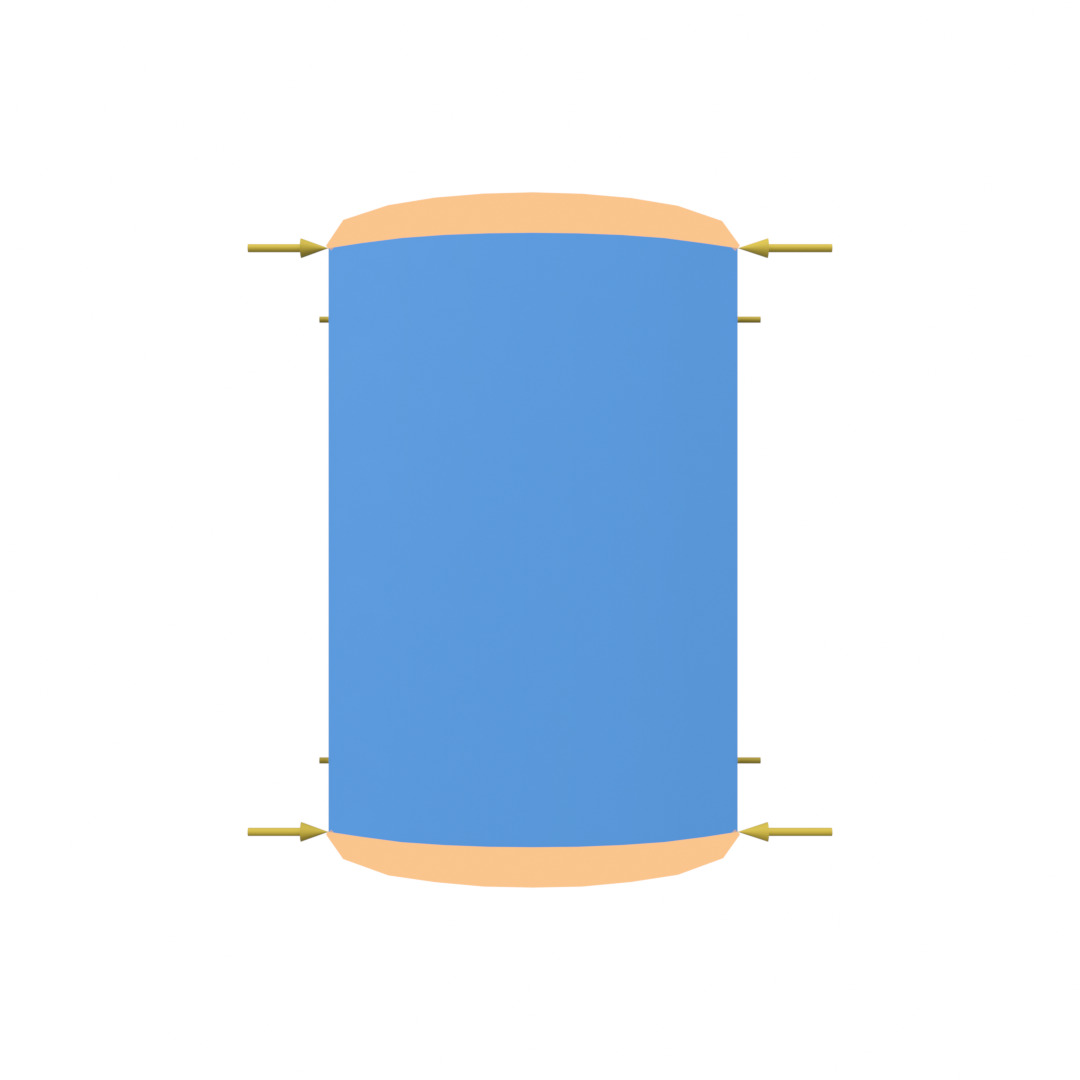}
    \par\vspace{1pt}\nointerlineskip
    \includegraphics[width=0.9\linewidth,trim=120bp 170bp 120bp 240bp,clip]{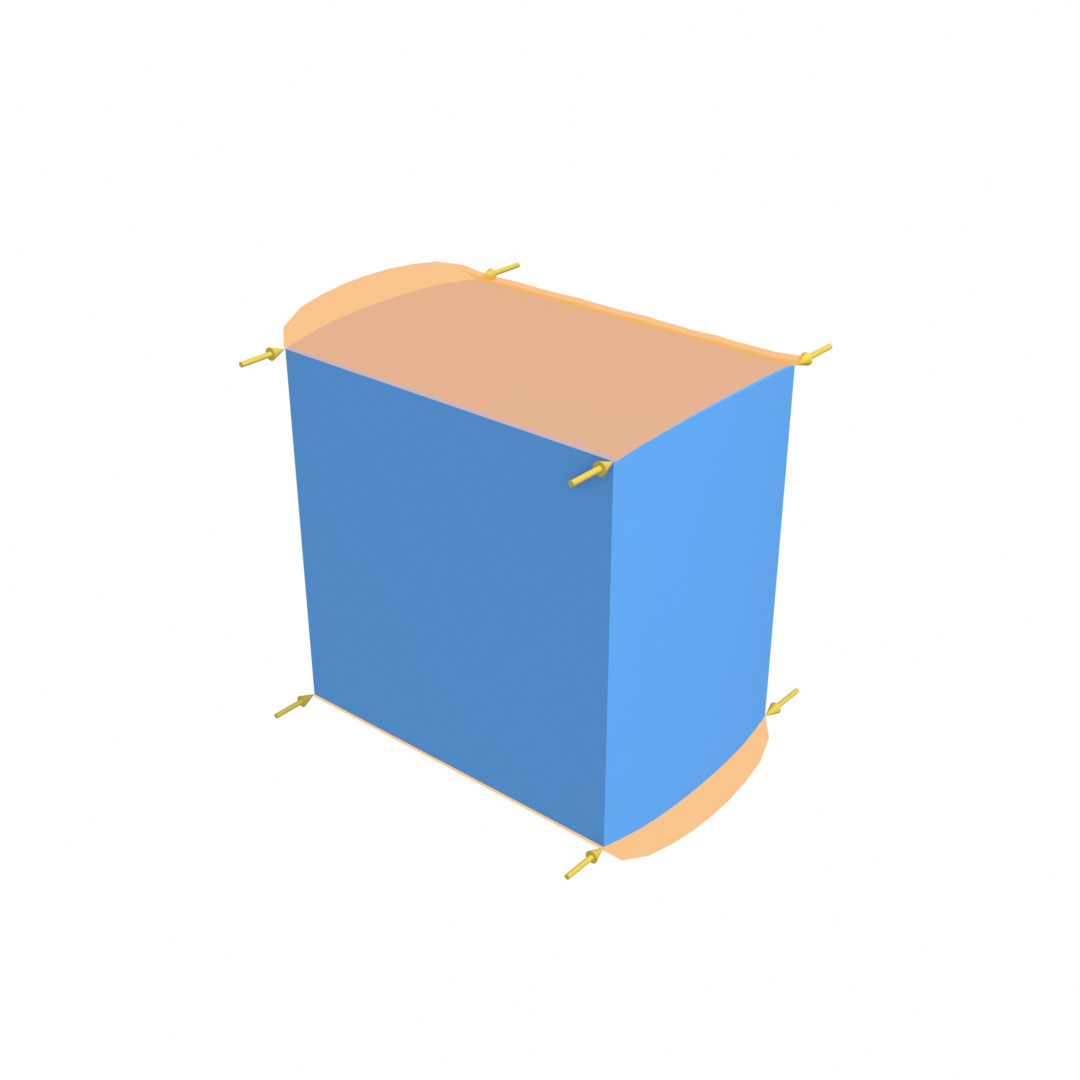}
    \caption{Iteration 2}
  \end{subfigure}\hfill
  \begin{subfigure}[t]{0.24\textwidth}
    \centering
    \includegraphics[width=0.9\linewidth,trim=120bp 170bp 120bp 170bp,clip]{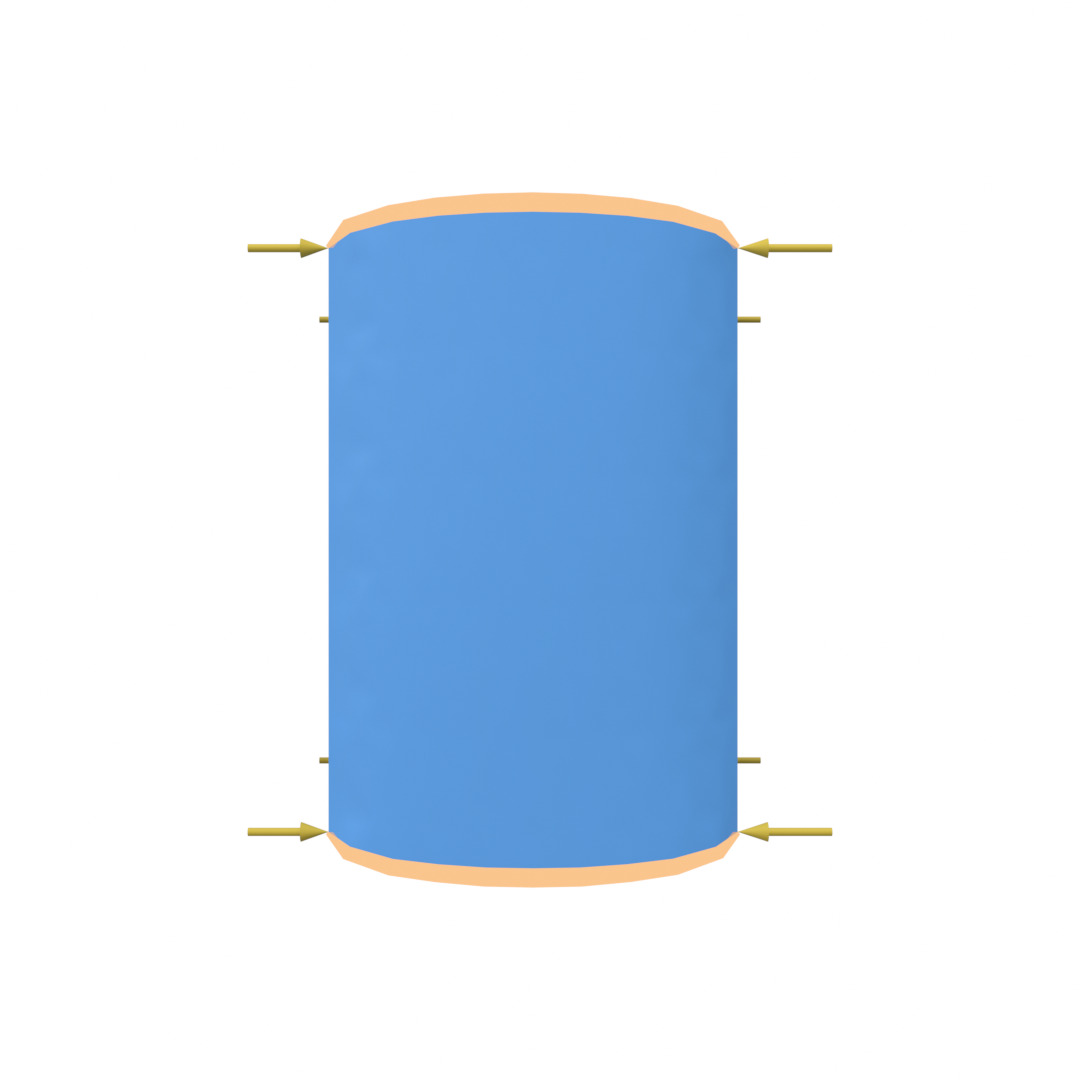}
    \par\vspace{1pt}\nointerlineskip
    \includegraphics[width=0.9\linewidth,trim=120bp 170bp 120bp 240bp,clip]{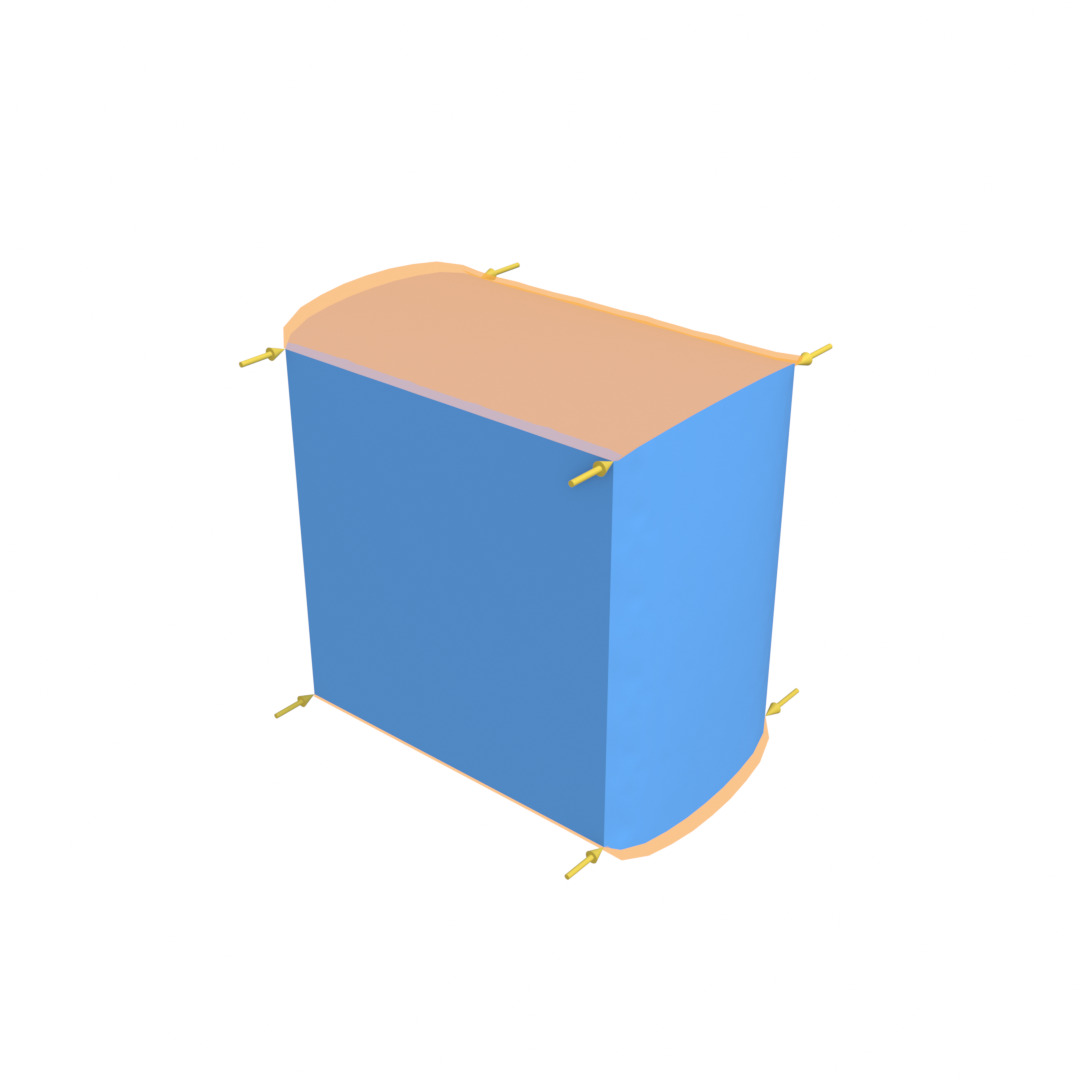}
    \caption{Iteration 5}
  \end{subfigure}\hfill
  \begin{subfigure}[t]{0.24\textwidth}
    \centering
    \includegraphics[width=0.9\linewidth,trim=120bp 170bp 120bp 170bp,clip]{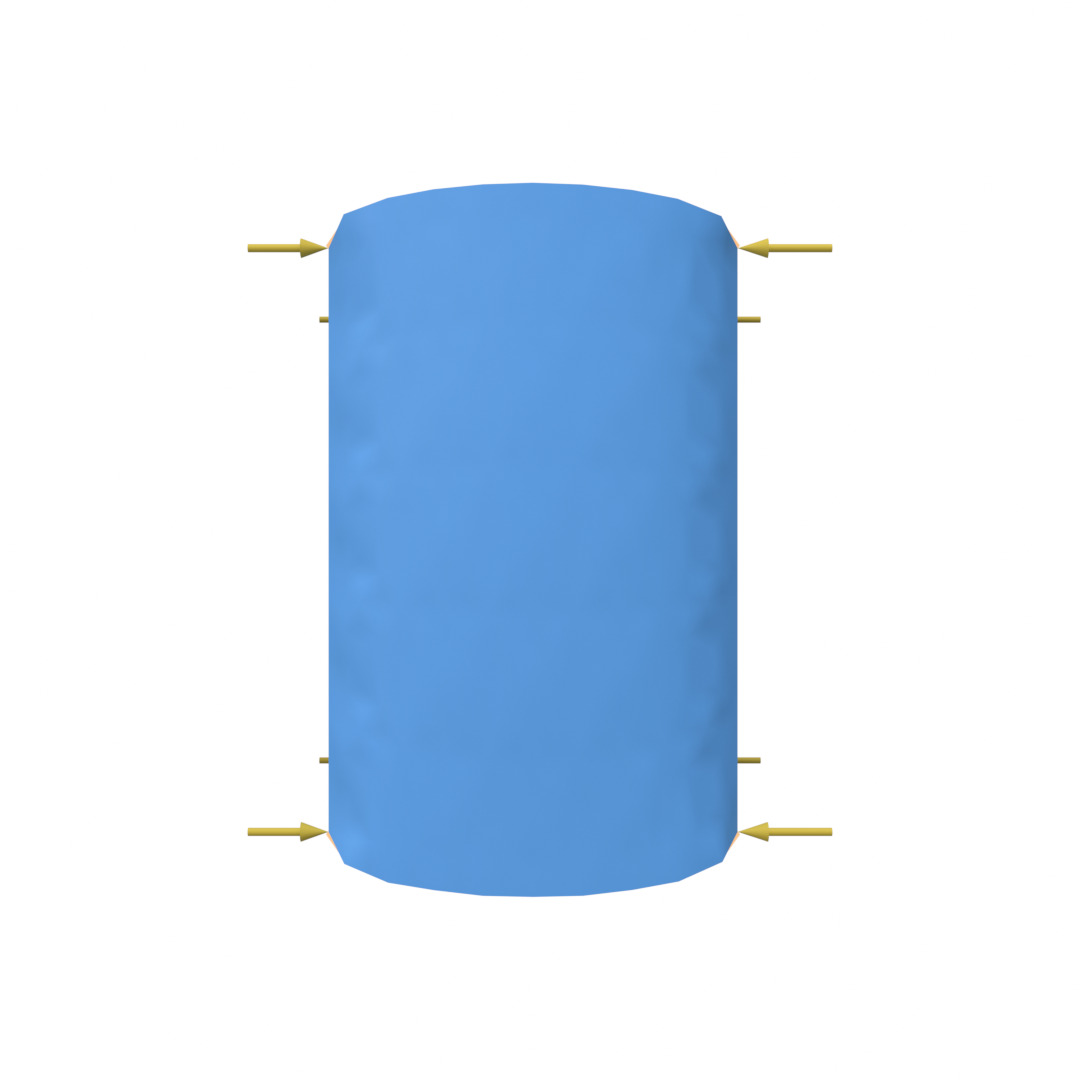}
    \par\vspace{1pt}\nointerlineskip
    \includegraphics[width=0.9\linewidth,trim=120bp 170bp 120bp 240bp,clip]{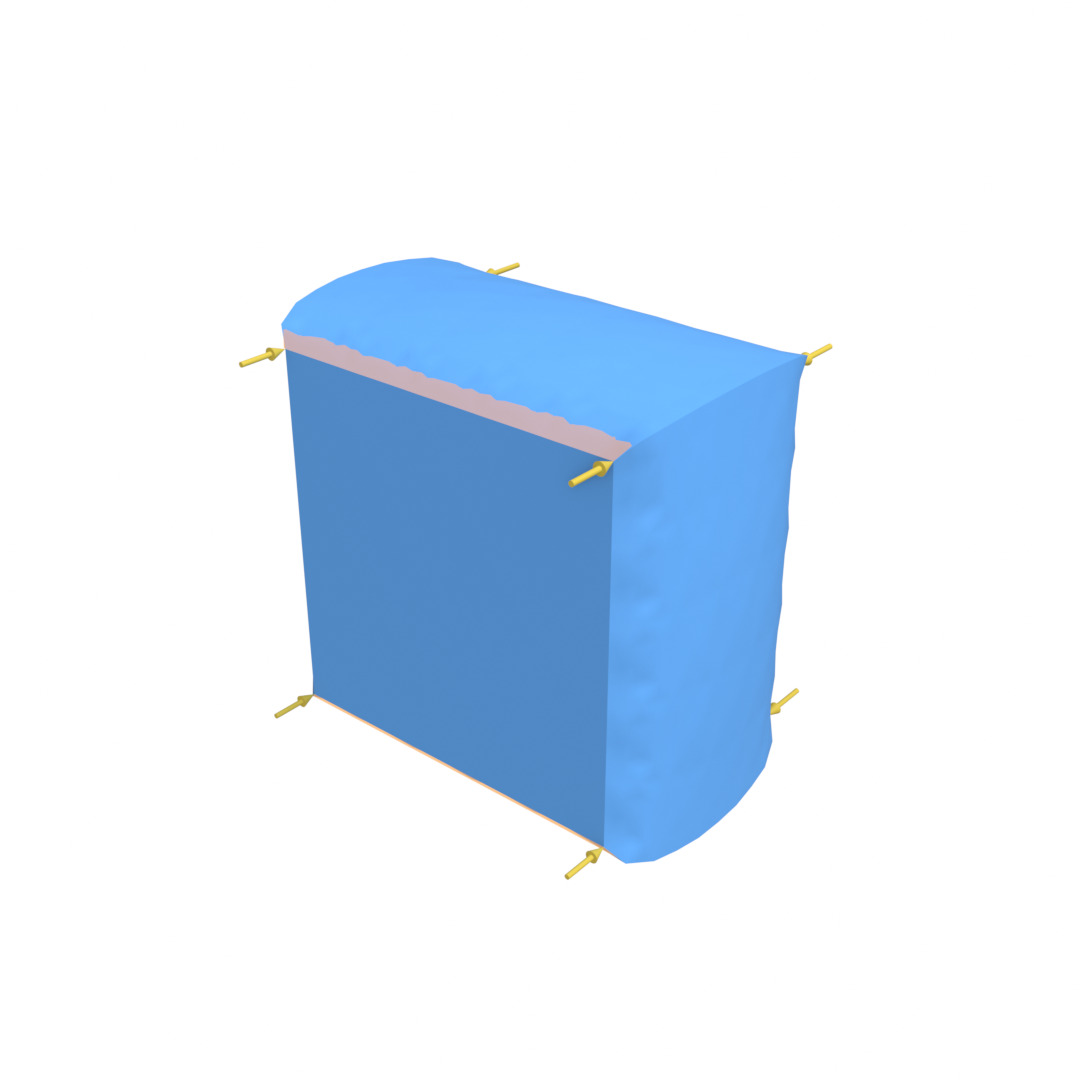}
    \caption{Iteration 20}
  \end{subfigure}
  \caption{Fitting the material response of a constrained cube to marker
  positions. Each column pairs side (top) and oblique (bottom) views at
  frame 10, before optimization and after 2, 5, and 20 iterations.
  Opaque surfaces denote the simulation, translucent surfaces the target,
  and arrows the Dirichlet constraints.}
  \Description{A two-row, four-column image sequence for material identification.
  Side views form the top row and oblique views the bottom row. Each column
  pairs the two views of the same optimization iteration, ordered 0, 2, 5,
  and 20 from left to right. Solid surfaces show the simulated deformation,
  translucent surfaces show the target, and arrows show the constraint directions.}
  \label{fig:material-identification}
\end{figure*}

\paragraph{Material identification}
{A homogeneous FEM cube is compressed by prescribed
displacements on its two opposing faces. Its material response is fitted
to a fixed reference deformation using surface-vertex positions. Two shared
logarithmic variables, $z_{\mathrm{vol}}=\log\lambda_{\mathrm{Lam\acute{e}}}$ and
$ z_{\mathrm{stretch}}=\log\mu_{\mathrm{Lam\acute{e}}}$, parameterize the positive Lam\'e
inputs $\lambda_{\mathrm{Lam\acute{e}}}$ and $\mu_{\mathrm{Lam\acute{e}}}$. They determine the
deformation coefficients of the stable Neo-Hookean energy as
\begin{equation}
 \lambda=e^{z_{\mathrm{vol}}}
       +e^{ z_{\mathrm{stretch}}},\qquad
 \mu=e^{ z_{\mathrm{stretch}}}.
 \label{seq:material-task-map}
\end{equation}
The same coefficient pair is assigned to every tetrahedron. The loss is
$L_{\mathrm{mat}}=5\times10^4\sum_{i\in\mathcal M}
\|\vx_{T,i}-\vx_i^*\|^2$.
Here $\mathcal M$ contains prescribed vertex markers on the cube's face in the rest configuration. The marker indices remain
unchanged during optimization, so $\vx_{T,i}$ and $\vx_i^*$ refer to the
simulated and reference positions of the same vertex, not to a nearest-point
match. The factor $5\times10^4$ sets the numerical scale of the squared-position loss,
which is summed rather than averaged over the markers.
The timing configuration uses $T=10$ simulation steps with $h=0.1$\,s,
and the loss is evaluated only at the final step. Both Lam\'e input parameters
are initialized to $4\times10^5$.
Both methods reach similar marker-position errors at the plotted scale.
Figure~\ref{fig:material-identification} shows the separately rendered
frame-10 configurations approaching the reference deformation. This task
demonstrates fitting a shared material response from surface-position
errors through the positive coefficient map.}

\begin{figure}[!tbp]

  \centering
  \begin{subfigure}[t]{0.48\linewidth}
    \centering
    \includegraphics[width=\linewidth,trim=84bp 37bp 155bp 219bp,clip]{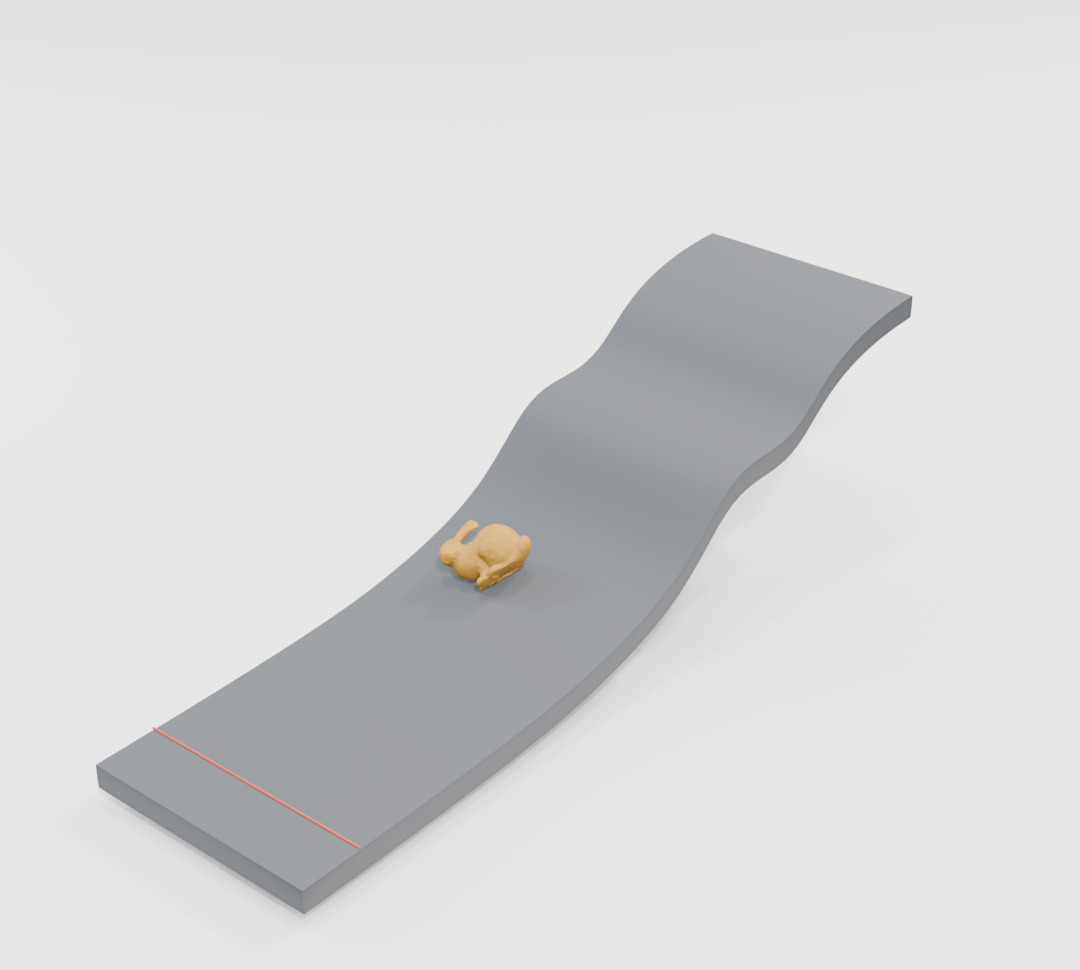}
    \caption{Iteration 0}
  \end{subfigure}\hfill
  \begin{subfigure}[t]{0.48\linewidth}
    \centering
    \includegraphics[width=\linewidth,trim=84bp 37bp 155bp 219bp,clip]{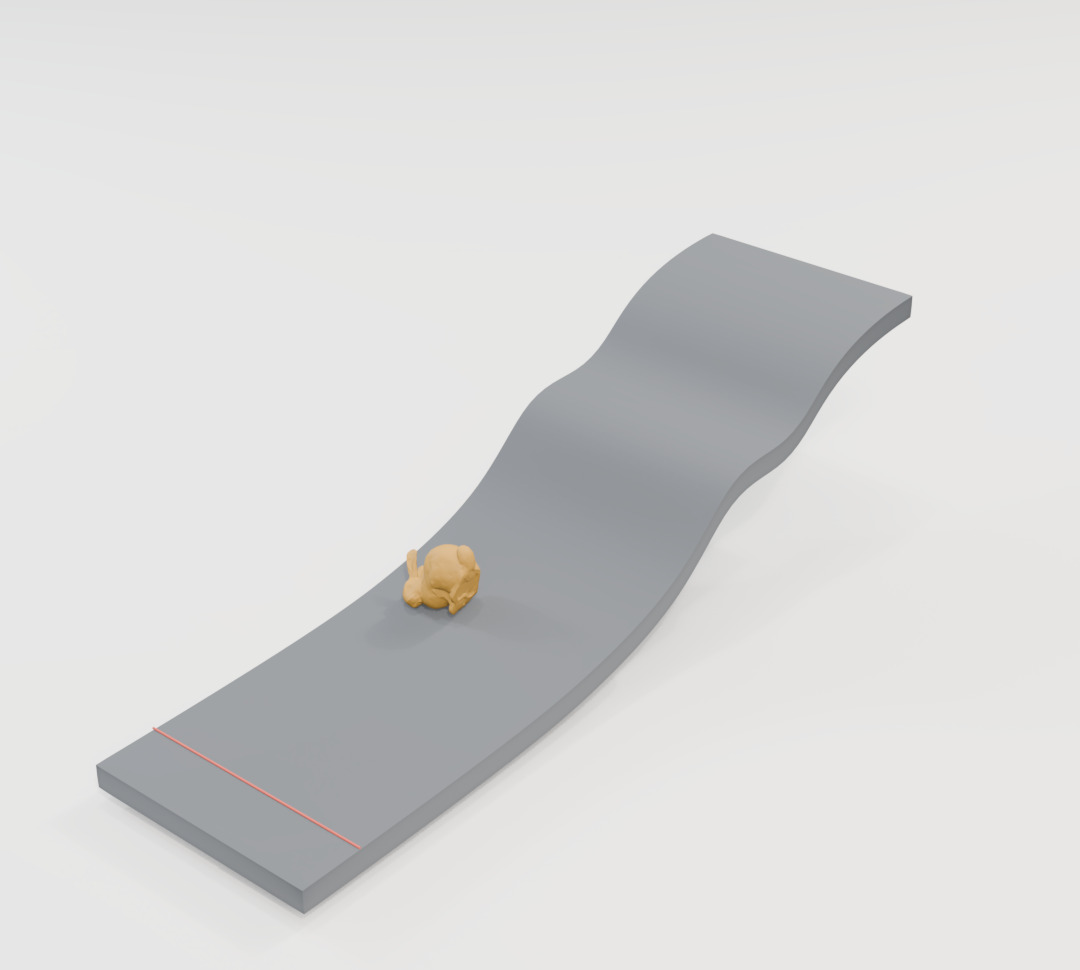}
    \caption{Iteration 1}
  \end{subfigure}\par\smallskip
  \begin{subfigure}[t]{0.48\linewidth}
    \centering
    \includegraphics[width=\linewidth,trim=84bp 37bp 155bp 219bp,clip]{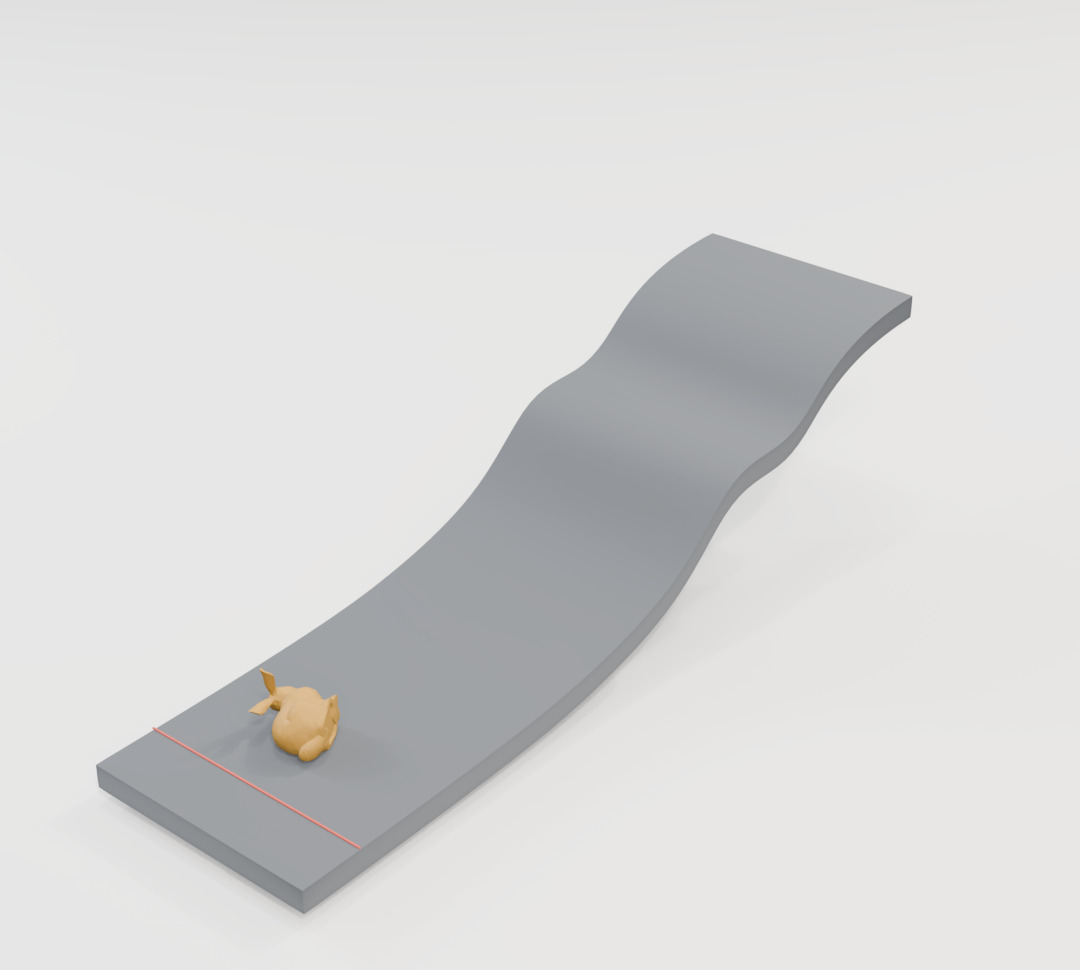}
    \caption{Iteration 3}
  \end{subfigure}\hfill
  \begin{subfigure}[t]{0.48\linewidth}
    \centering
    \includegraphics[width=\linewidth,trim=84bp 37bp 155bp 219bp,clip]{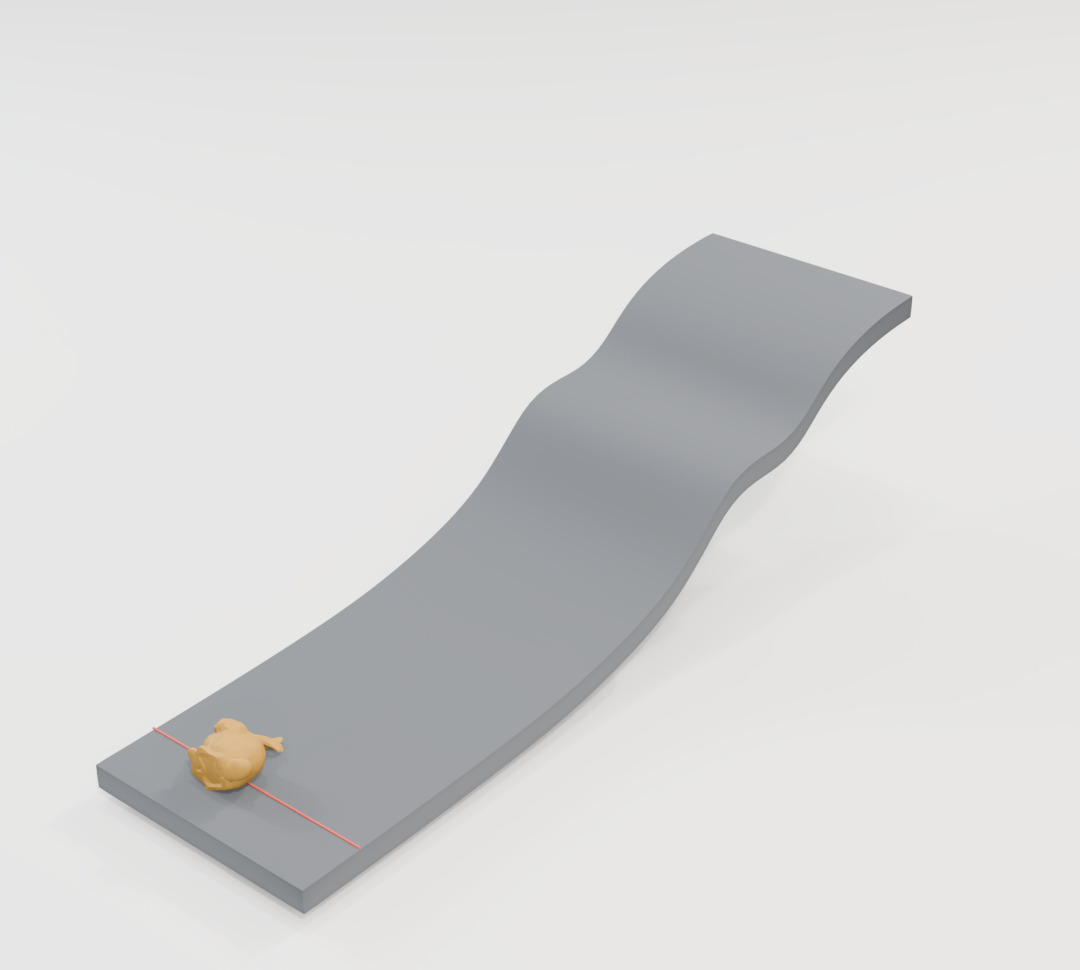}
    \caption{Iteration 5}
  \end{subfigure}
  \caption{Friction optimization on a curved slide. A {bunny-slide friction coefficient} is adjusted so that the bunny
  reaches its target position at frame 100. The panels show the
  unoptimized simulation and results after 1, 3, and 5 iterations.}
  \Description{A deformable bunny slides along a curved surface. The bunny-slide friction coefficient is optimized so that its position at the specified time matches a target position. The panels show final frame 100 for the initial coefficient and after one, three, and five optimization updates.}
  \label{fig:friction-identification}
\end{figure}

\paragraph{Friction optimization}
The bunny-slide friction coefficient is fitted to a target position at frame 100, with $h=0.02$\,s. The loss is
 $L_{\mathrm{fr}}={100}(c_z(\vx_T)-z_{\mathrm{end}})^2$,
where $c_z$ is the bunny's volume-weighted center coordinate,
$z_{\mathrm{end}}$ is the slide endpoint, and $T=100$.
The coefficient changes from $0.5$ to $0.1$
(Table~\ref{tab:sim-settings} in
Appendix~\ref{sec:experimental-details}).
Figure~\ref{fig:friction-identification} shows the corresponding change in sliding distance at the fixed observation time. This task isolates the coefficient-to-motion dependence: geometry and initial motion remain fixed, and the optimized friction changes the cumulative sliding response.

\begin{figure*}[tp]
\centering
{\ignorespaces
\includegraphics[width=\linewidth]{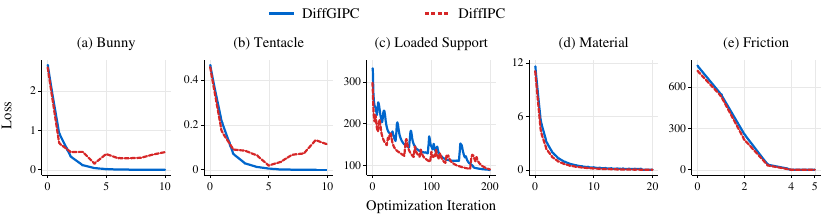}
\par}
\caption{Optimization histories for the shared FEM benchmarks.
 Blue and red denote DiffGIPC and DiffIPC, respectively. The horizontal
 axes count accepted parameter updates from the initial parameters
 (iteration 0). Loss scales are task-specific, and Loaded support is
 remeshed between updates.}
\Description{Five side-by-side plots show loss against optimization iteration
for bunny collision, tentacle matching, loaded-support shape design,
material fitting, and friction optimization. Blue curves denote DiffGIPC, and red curves denote DiffIPC.}
\label{fig:loss-curves-shared}
\end{figure*}

Together, the shared benchmarks test gradients through collision-transmitted motion, observations over time, loaded shape changes, material fitting and friction response. 
Table~\ref{tab:diffipc-phase-comparison} measures the cost of that optimization capability with full-space FEM in both systems. The complete backward pass is {$12.3$--$34.1\times$} faster, including Hessian assembly, the adjoint solve, and gradient accumulation. Combined forward and backward speedups range from {$10.6\times$} for friction fitting to {$60.9\times$} for material fitting, while forward speedups are {$10.5$--$88.1\times$}. These gains demonstrate efficient trajectory-gradient evaluation before introducing affine motion reduction. Section~\ref{sec:representation-study} examines that additional benefit, and Section~\ref{sec:optimization-timing} relates evaluation cost to complete optimization workloads.

\subsection{\texorpdfstring{{Rest-shape design and stable placement}}{Rest-shape design and stable placement}}
\label{sec:affine-design-results}

The design tasks separate the object being changed from the quantities used to judge the result. Cloth fitting maps a size parameter to a deforming garment, hanger design optimizes an affine surface through its effect on supported cloth, and stone placement adjusts initial positions while preserving shape. Together they demonstrate how rest geometry and initial-state parameters act through the same coupled trajectory gradient.

\begin{figure*}[tp]

  \centering
  \begin{subfigure}[t]{0.192\textwidth}
    \centering
    \includegraphics[width=0.9\linewidth,trim=235bp 300bp 235bp 200bp,clip]{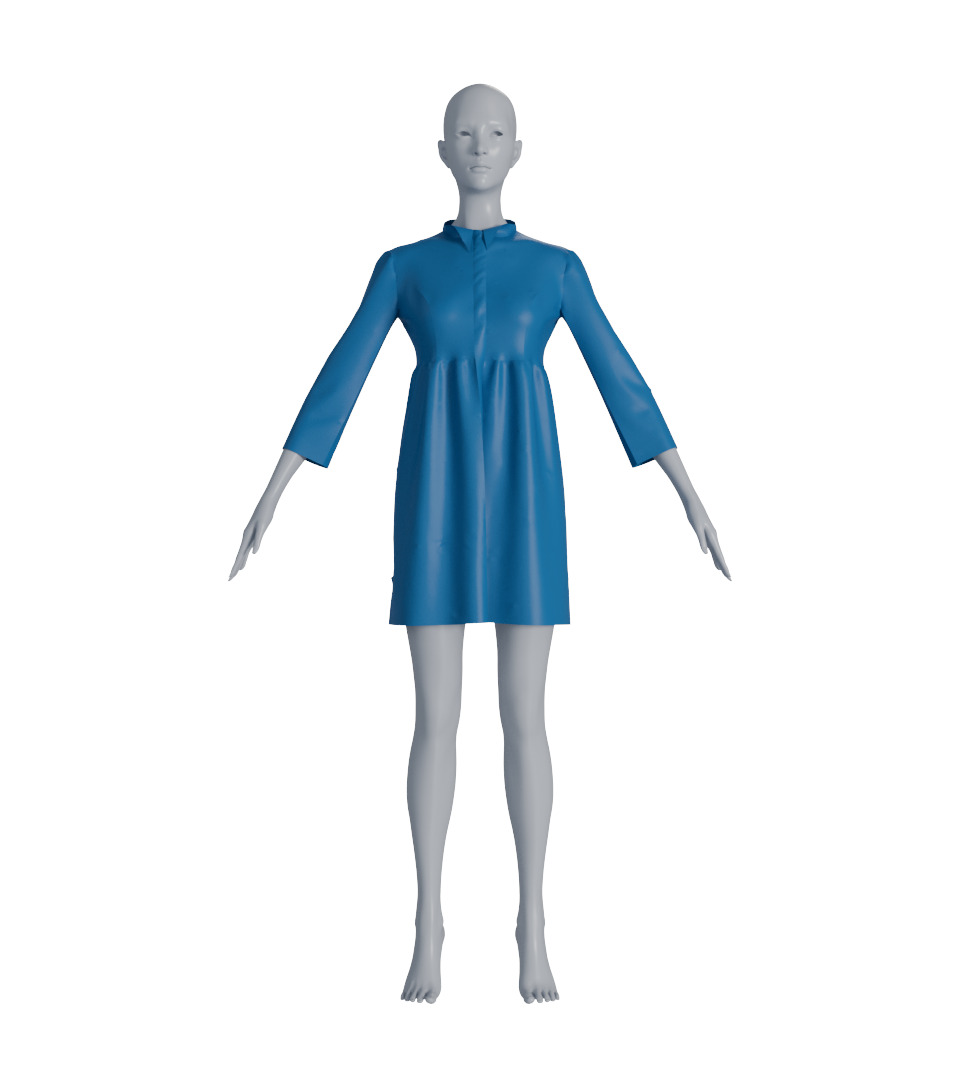}
    \caption{Iteration 0}
  \end{subfigure}\hfill
  \begin{subfigure}[t]{0.192\textwidth}
    \centering
    \includegraphics[width=0.9\linewidth,trim=235bp 300bp 235bp 200bp,clip]{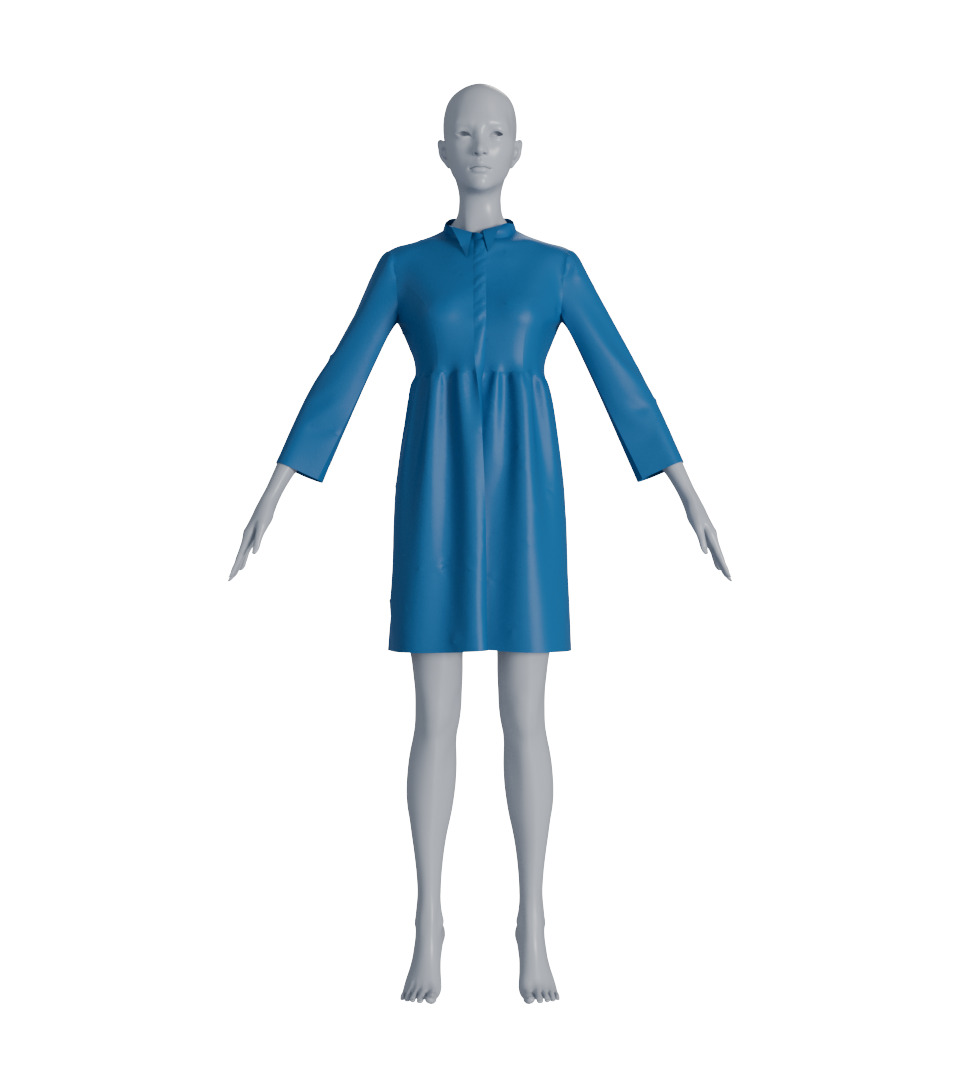}
    \caption{Iteration 1}
  \end{subfigure}\hfill
  \begin{subfigure}[t]{0.192\textwidth}
    \centering
    \includegraphics[width=0.9\linewidth,trim=235bp 300bp 235bp 200bp,clip]{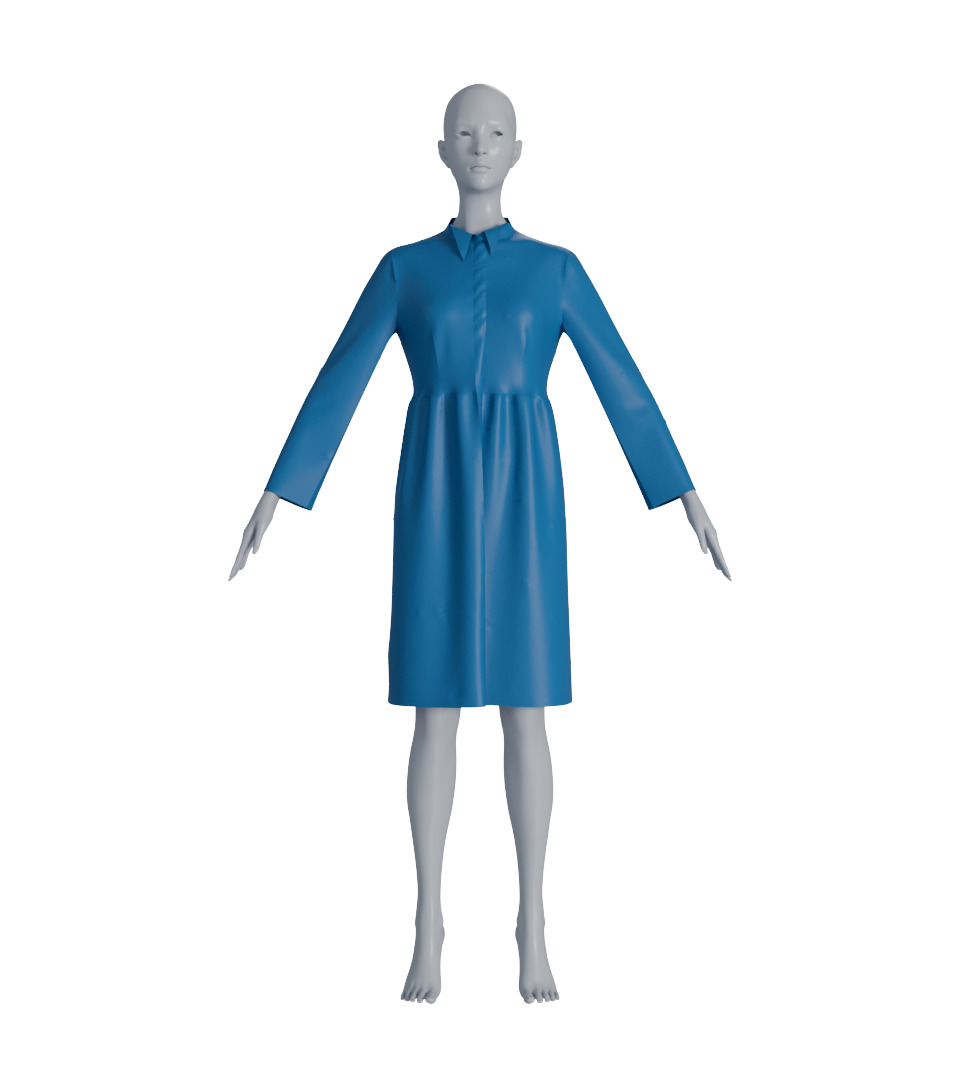}
    \caption{Iteration 3}
  \end{subfigure}\hfill
  \begin{subfigure}[t]{0.192\textwidth}
    \centering
    \includegraphics[width=0.9\linewidth,trim=235bp 300bp 235bp 200bp,clip]{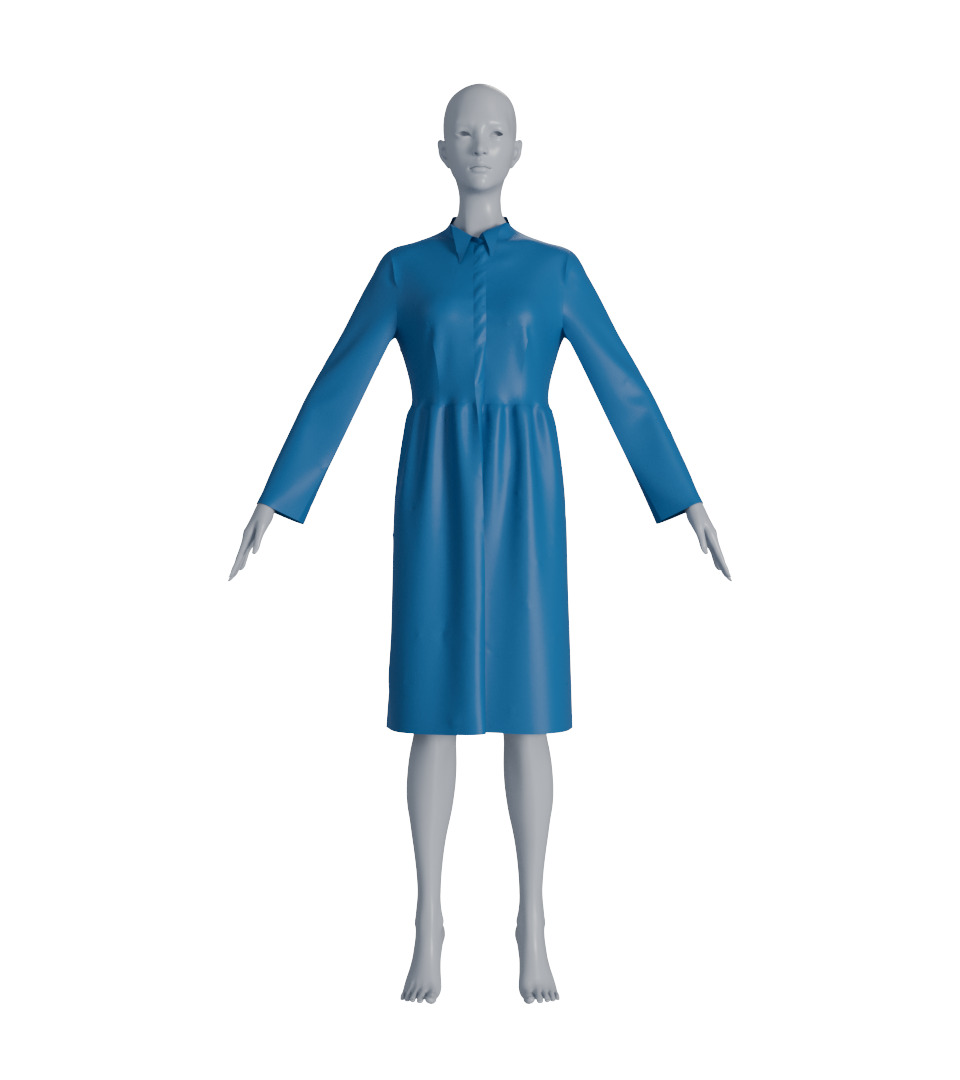}
    \caption{Iteration 5}
  \end{subfigure}\hfill
  \begin{subfigure}[t]{0.192\textwidth}
    \centering
    \includegraphics[width=0.9\linewidth,trim=235bp 300bp 235bp 200bp,clip]{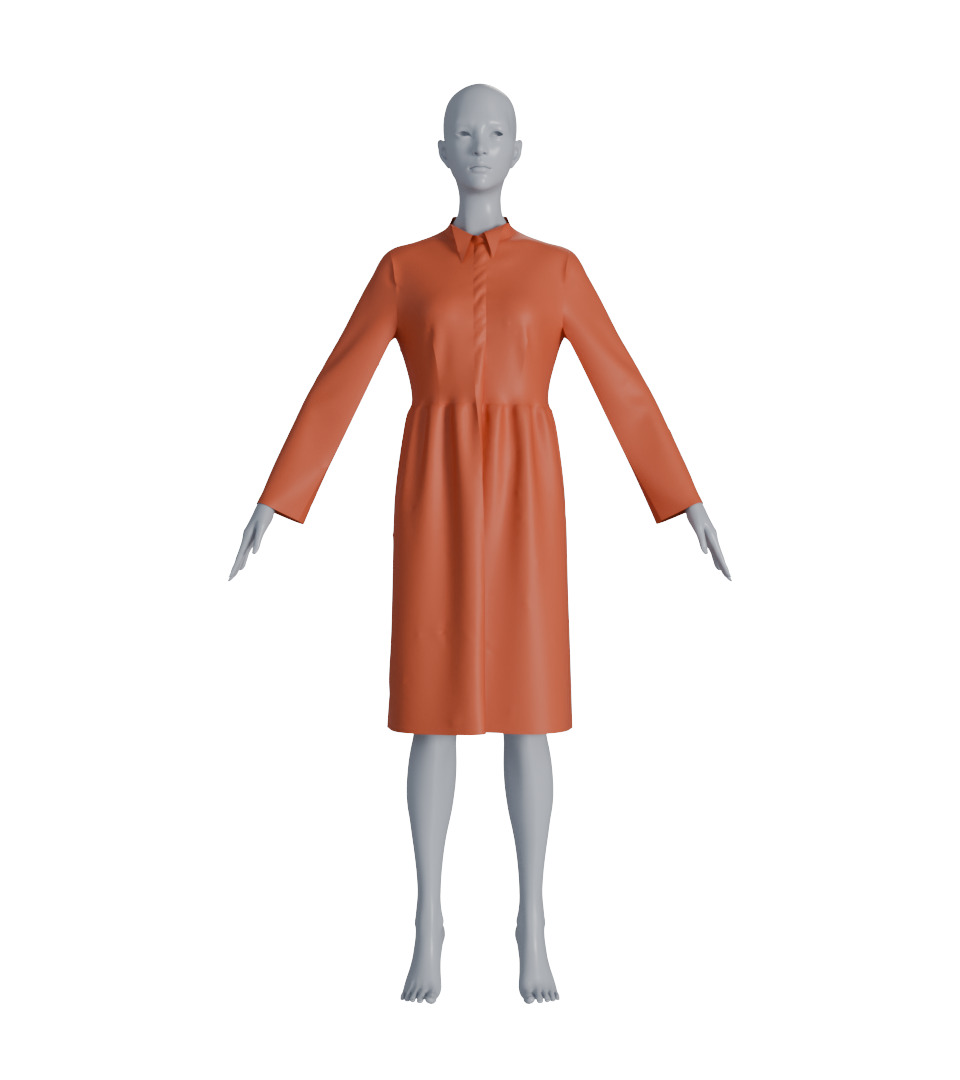}
    \caption{Target}
  \end{subfigure}
  \caption{Cloth rest-scale fitting on a fixed mannequin. The blue panels
  {show the initial drape and results after 1, 3, and 5 updates.
  The orange panel shows the target.}}
  \Description{Five garment-focused views with one common crop defined from the target. The four blue panels show the initial, intermediate, and optimized drapes. The orange panel shows the target. Sleeves and hems remain visible so changes in garment size can be compared at a common scale. Orange identifies the target, not stress.}
  \label{fig:dress10}
\end{figure*}

\paragraph{Cloth rest-scale fitting}

A single optimizer variable scales the garment's rest shape, with a fixed mannequin. This map changes many rest coordinates while preserving full-space cloth deformation. At the observation endpoint $T=48$, the target drape defines $L_{\mathrm{cloth}}=\sum_{i\in\mathcal V_{\mathrm{cloth}}}\|\vx_{T,i}-\vx_i^*\|^2$, where $\mathcal V_{\mathrm{cloth}}$ contains all garment vertices and $\vx_i^*$ is the fixed target position of the corresponding material vertex. After 5 updates, the optimized drape approaches the target in Figure~\ref{fig:dress10}. The result demonstrates how a low-dimensional design parameter controls detailed simulated geometry through rest-shape derivatives.

\paragraph{Garment hanger shoulder design}

{Each shoulder profile is controlled by a
quadratic B\'ezier curve with two fixed endpoints and one movable
middle control point. The design variables $s_x$ and $s_y$ adjust
the middle control point's horizontal and vertical positions.
Starting from $(s_x,s_y)=(1,1)$, decreasing $s_x$ moves the point
toward the hanger's centerline, and decreasing $s_y$ lowers it.
The reflected curve uses the same pair, so both shoulders change
symmetrically without moving their endpoints. The affine hanger retains twelve motion DoFs and
supports a garment with friction coefficient $0.05$.

At the terminal frame, $R_s^{\mathrm{tip}}$ and
$R_s^{\mathrm{strap}}$ denote the horizontal positions of
the hanger's shoulder tip and the corresponding strap's outer edge,
respectively, on the left or right side $s\in\{\mathrm L,\mathrm R\}$.
On each side, horizontal coordinates increase from the hanger's
centerline toward the shoulder tip. Their difference therefore
measures how far the hanger extends beyond the strap. The quantity $d_s$ is the
minimum three-dimensional distance between the selected strap and
corresponding upper-shoulder vertices. For the fixed target margin
$\delta=0.205$\,m, the support margin and loss are
\[
\begin{aligned}
m_s &= R_s^{\mathrm{tip}}-R_s^{\mathrm{strap}}
       -\max(d_s-0.01,0),\\
L_{\mathrm{hanger}} &= 10^4
  \sum_{s\in\{\mathrm L,\mathrm R\}}\max(\delta-m_s,0)^2.
\end{aligned}
\]
The gap penalty reduces the margin when the strap separates from
the shoulder by more than $0.01$\,m. Both surfaces contribute to the
loss, whose shape gradient includes direct geometric and
coupled-motion contributions (Section~\ref{sec:rest-gradient}).

Each optimization rollout ends at the first margin violation
$m_s<\delta$ after the corresponding strap approaches within
$0.01$\,m of its shoulder, or at frame 600.
The optimized control-point scales $(s_x,s_y)=(0.8285,0.8000)$ move
the middle control points inward and downward. The resulting hanger
retains both straps in two independent 600-frame checks
(Figure~\ref{fig:abd-hanger-rest}). After each strap contacts the
hanger, its support margin must remain nonnegative ($m_s\geq0$)
until frame 600. The larger target margin $\delta$ is used to guide
optimization. 
Figure~\ref{fig:loss-curves-remaining} shows optimization progress, and
Section~\ref{sec:representation-study} compares a surface-matched
FEM hanger.}

\paragraph{Seven-stone placement}

We optimize the horizontal initial positions of seven affine bodies with fixed shapes. Their detailed surfaces determine the contacts, while the motion system contains only 84 DoFs. {Let $b_{\mathrm{base}}$ identify the bottom reference
stone. At frame $t$, $d_{t,b}$ is the Euclidean distance between the
horizontal center positions of stone $b$ and that reference
stone. With allowed distance $\delta$, the placement objective is
$L_{\mathrm{stone}}=\sum_{t\in\mathcal F}
\sum_{b\ne b_{\mathrm{base}}}\max(d_{t,b}-\delta,0)^2$,
where $\mathcal F$ is the set of observation frames.
The base is excluded from the inner sum, and only distances exceeding
$\delta$ contribute.}
{Optimization starts from horizontally offset
stone placements. The first 50 updates use $\delta=0.01$ to encourage
the simulated centers of mass to align horizontally with that of the
base stone. The remaining updates use $\delta=0.15$ to allow a broader
set of contact-supported configurations. Both stages take
$\mathcal F=\{t^{\star}\}$, where $t^{\star}$ is the first frame
where $d_{t,b}>\delta$ for any non-base stone.}
{If $d_{t,b}\leq\delta$ for every non-base stone
at every simulated frame, the loss is zero. The optimizer adjusts
the stones' initial horizontal positions. The loss is computed from
the horizontal distances between their simulated centers of mass
and that of the base stone.}
Figure~\ref{fig:stone-placement} shows updates 10, 50, and 78, {with separate 1,000-frame forward simulations testing whether the resulting stacks remain stable.} The task demonstrates contact-aware placement through affine initial-state gradients: detailed surface interactions guide the horizontal positions without changing the stones' shapes.

\subsection{Articulated control}
\label{sec:affine-gradient-evidence}

Articulated tasks extend the coupled gradients from geometry and placement to actuation. Grasp-angle control tests joint-target tracking under contact, and cup catching optimizes joint commands for the motion of an interacting object. {Dual-arm grasping optimizes one vertical offset added to both grippers' target positions to control their normal contact force on the cloth.} These objectives span motion and force while sharing the coupled FEM-affine backward computation.

\begin{figure}[!t]
  \centering
  \setlength{\tabcolsep}{0pt}
  \begin{tabular}{@{}cc@{}}
    \small Initial & \small Optimized\\[2pt]
    \includegraphics[width=0.48\columnwidth,trim=435bp 10bp 70bp 0bp,clip]{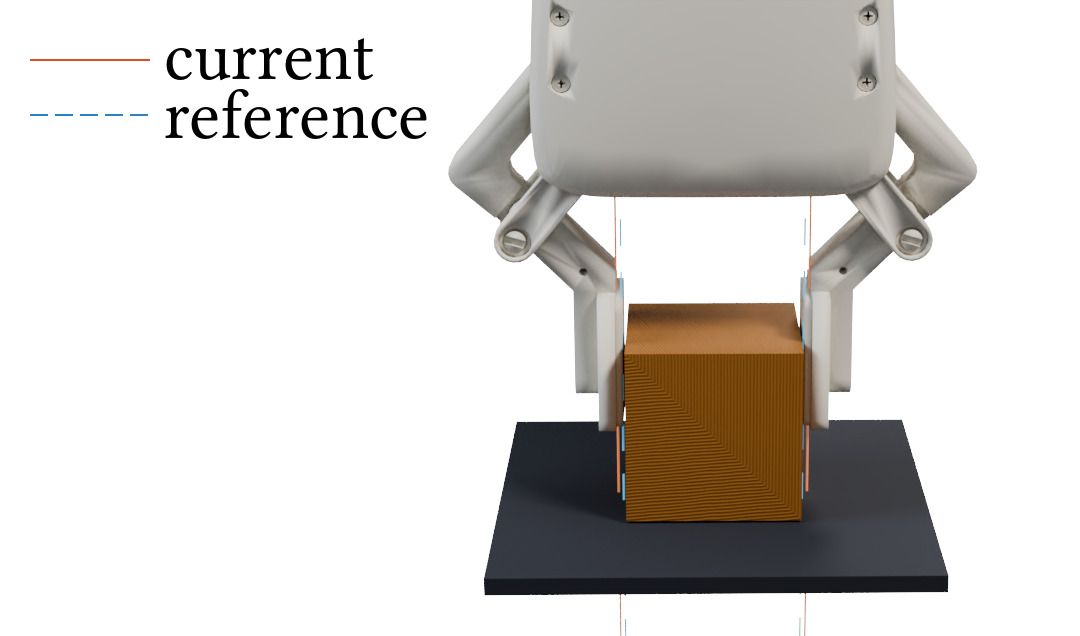} &
    \includegraphics[width=0.48\columnwidth,trim=435bp 10bp 70bp 0bp,clip]{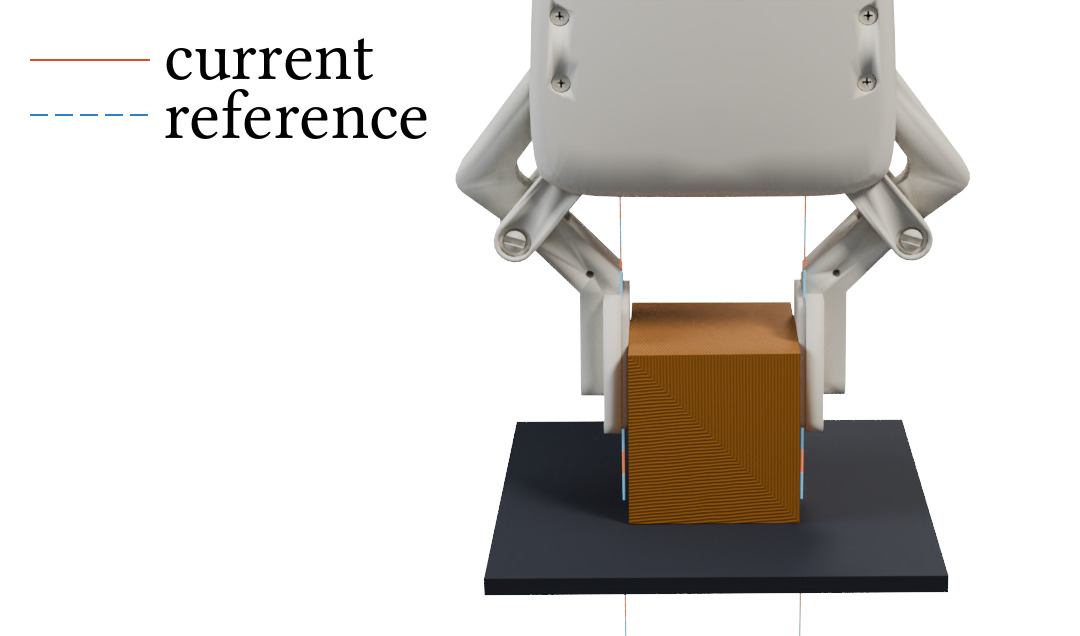}\\[-4pt]
    \multicolumn{2}{c}{\small
      \textcolor[RGB]{244,93,57}{\rule[0.4ex]{1.4em}{0.8pt}}\,current\qquad
      \textcolor[RGB]{38,132,193}{%
        \rule[0.4ex]{0.35em}{0.8pt}\hspace{0.12em}%
        \rule[0.4ex]{0.35em}{0.8pt}\hspace{0.12em}%
        \rule[0.4ex]{0.35em}{0.8pt}}\,reference}
  \end{tabular}
  \captionsetup{skip=3pt}
  \caption{Grasp-angle control. Initial (left) and optimized (right)
  grasps at frame 60. Joint-target optimization moves the grasp toward
  the $0.45$-rad target. {Coral guides follow the fingers. Dashed blue
  guides indicate visual reference directions.}}
  \Description{Initial grasp beside optimized grasp, with finger and
  reference guides.}
  \label{fig:xarm-target-closeup}
\end{figure}

\paragraph{Grasp-angle control}
An articulated affine robot approaches and grasps a FEM object above a
fixed support. {The closure objective measures
two revolute joints in each finger linkage.} The
120-frame horizon includes approach, closure, and return motion.
{Let $\mathcal J_{\mathrm f}$ be {this set of four
gripper joints} and $\alpha_{t,j}$ their simulated revolute angles.
At the closure observation $t=60$, the objective is
$L_{\mathrm{grasp}}=\tfrac12\sum_{j\in\mathcal J_{\mathrm f}}
(\alpha_{60,j}-a^*)^2$, with fixed desired angle $a^*=0.45$ rad.
{The optimized input is the common target
angle. The four angles $\alpha_{60,j}$ are measured from the simulated
response to that target and need not be identical under contact.}}
{The optimized joint-target angle is $0.447843$ rad.} The objective acts on the four simulated angles, so the gradient accounts for the robot's response to the command within the coupled scene. Figure~\ref{fig:xarm-target-closeup} {compares the grasps. Its dashed guides} are visual references and do not enter the loss.

\paragraph{Cup catching}

\begin{figure}[t]
  \centering
  \makebox[0.31\linewidth][c]{\small Static}\hfill
  \makebox[0.31\linewidth][c]{\small Unoptimized}\hfill
  \makebox[0.31\linewidth][c]{\small Optimized}\par\smallskip
  \begin{subfigure}[t]{\linewidth}
    \centering
    \includegraphics[width=0.31\linewidth]{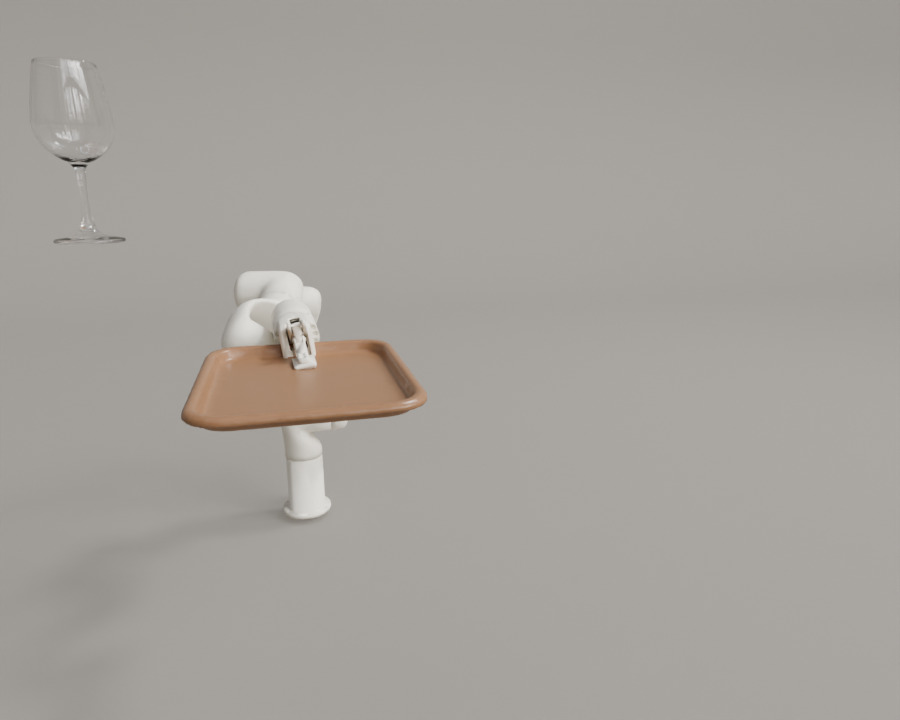}\hfill
    \includegraphics[width=0.31\linewidth]{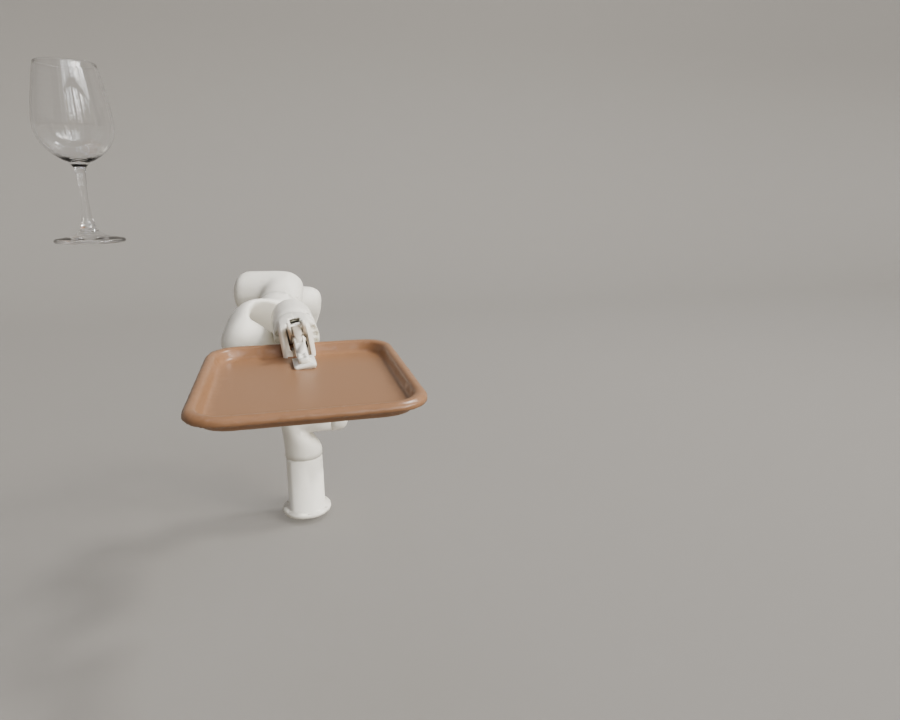}\hfill
    \includegraphics[width=0.31\linewidth]{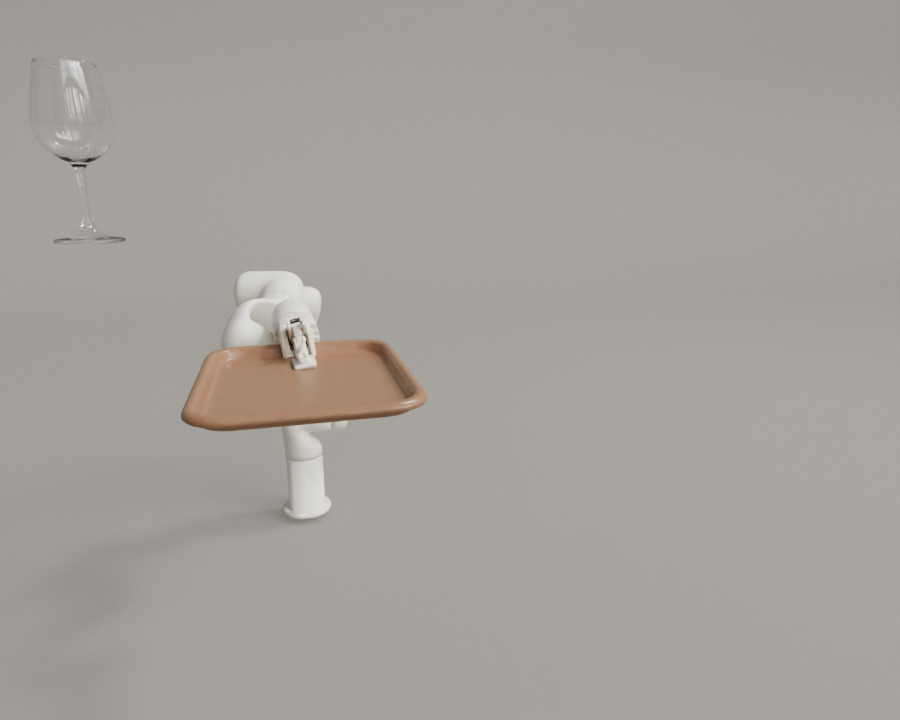}
    \caption{Frame 94}
  \end{subfigure}\par\smallskip
  \begin{subfigure}[t]{\linewidth}
    \centering
    \includegraphics[width=0.31\linewidth]{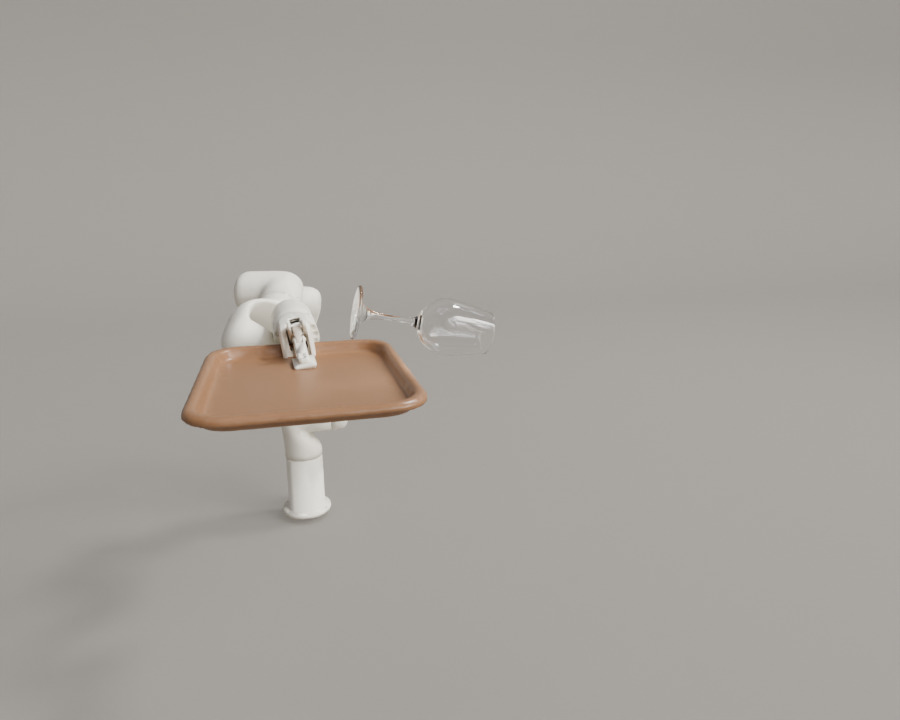}\hfill
    \includegraphics[width=0.31\linewidth]{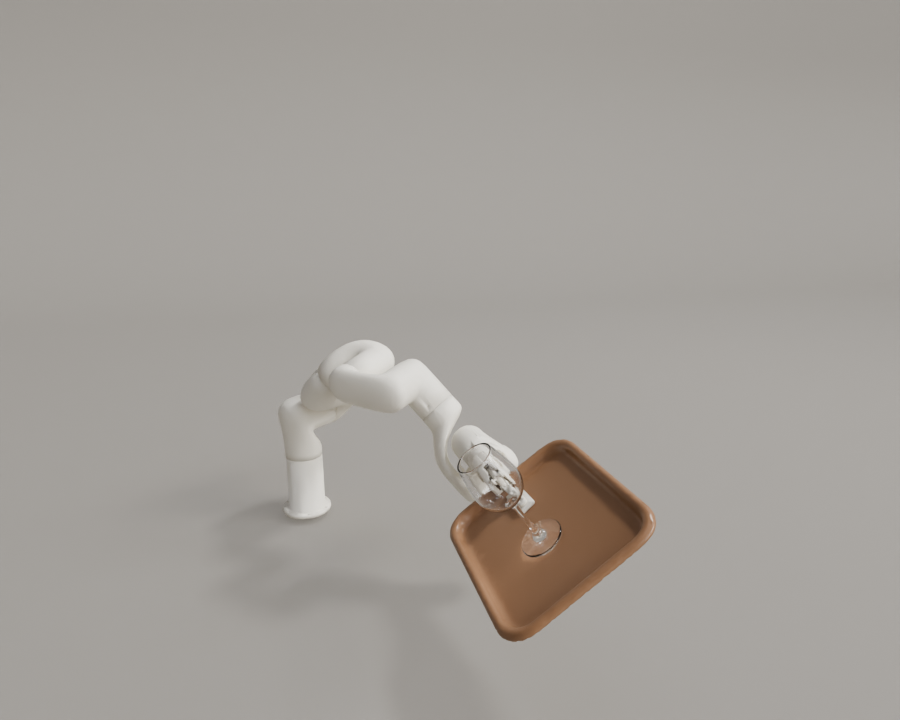}\hfill
    \includegraphics[width=0.31\linewidth]{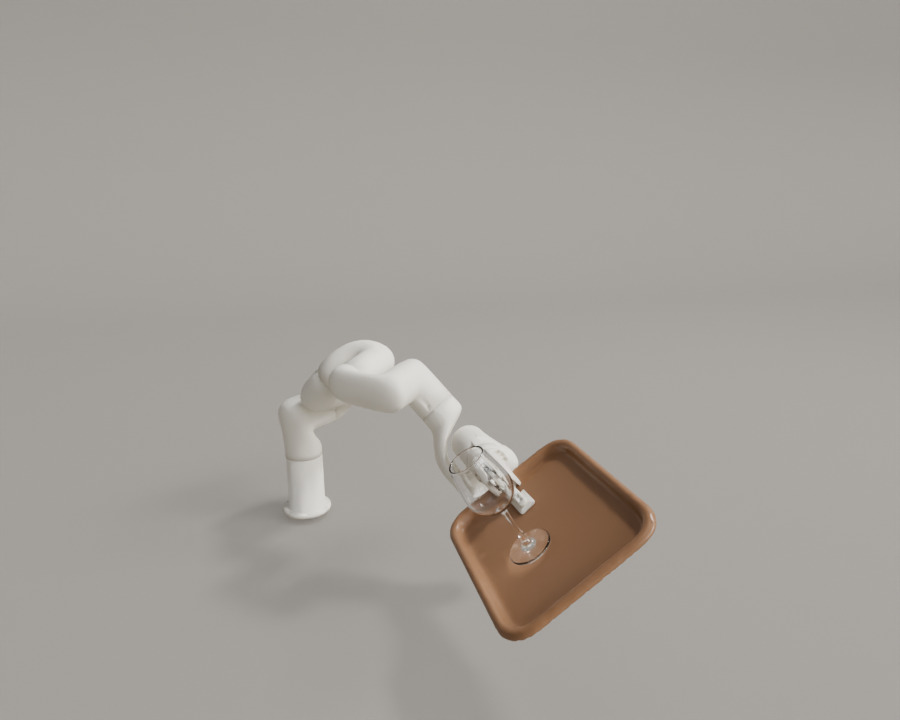}
    \caption{Frame 111}
  \end{subfigure}\par\smallskip
  \begin{subfigure}[t]{\linewidth}
    \centering
    \includegraphics[width=0.31\linewidth]{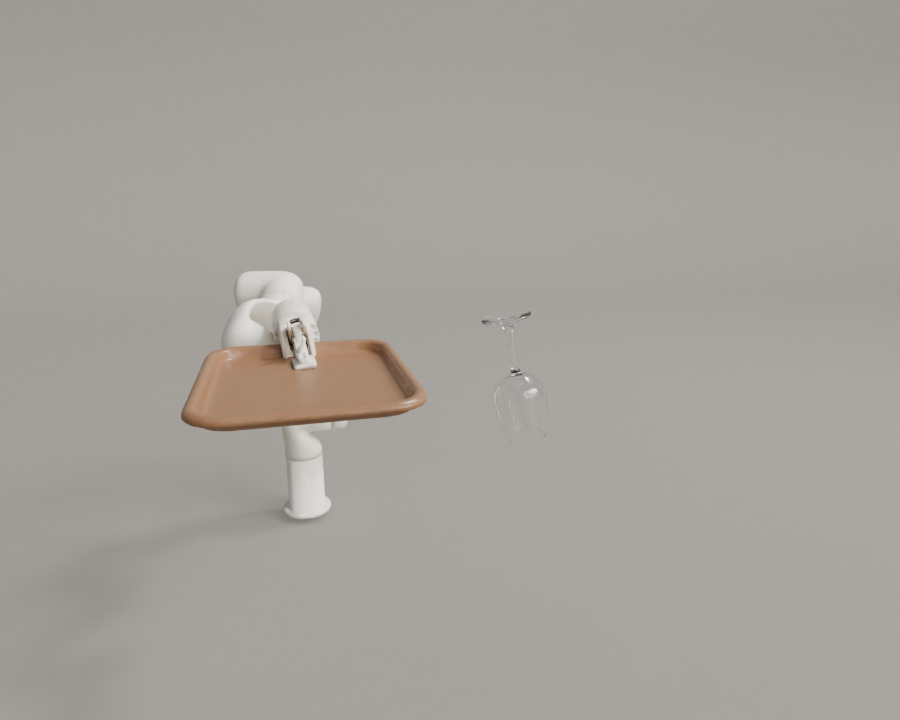}\hfill
    \includegraphics[width=0.31\linewidth]{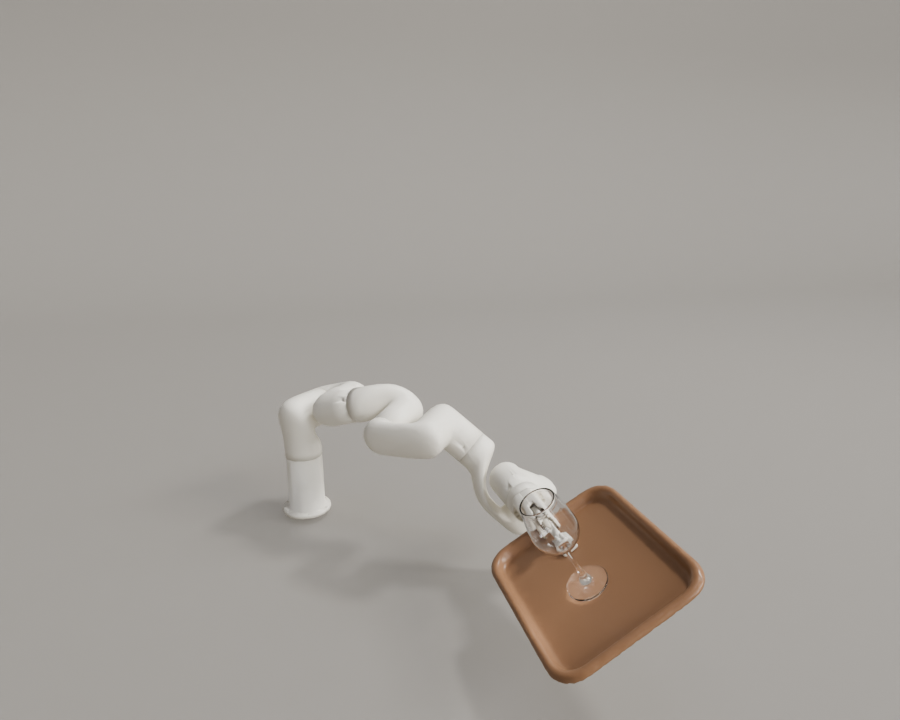}\hfill
    \includegraphics[width=0.31\linewidth]{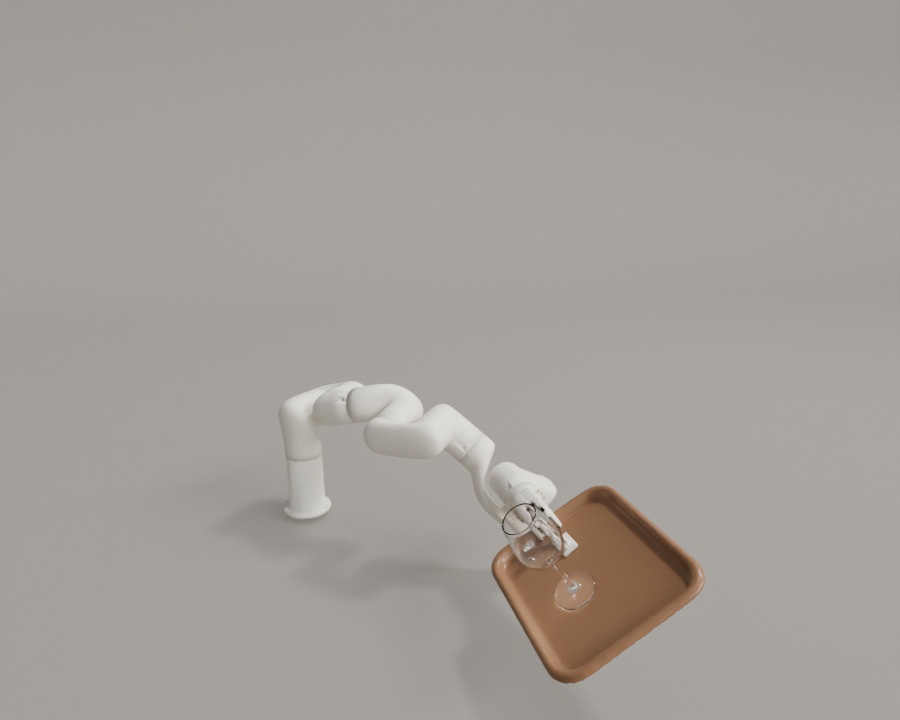}
    \caption{Frame 120}
  \end{subfigure}\par\smallskip
  \begin{subfigure}[t]{\linewidth}
    \centering
    \includegraphics[width=0.31\linewidth]{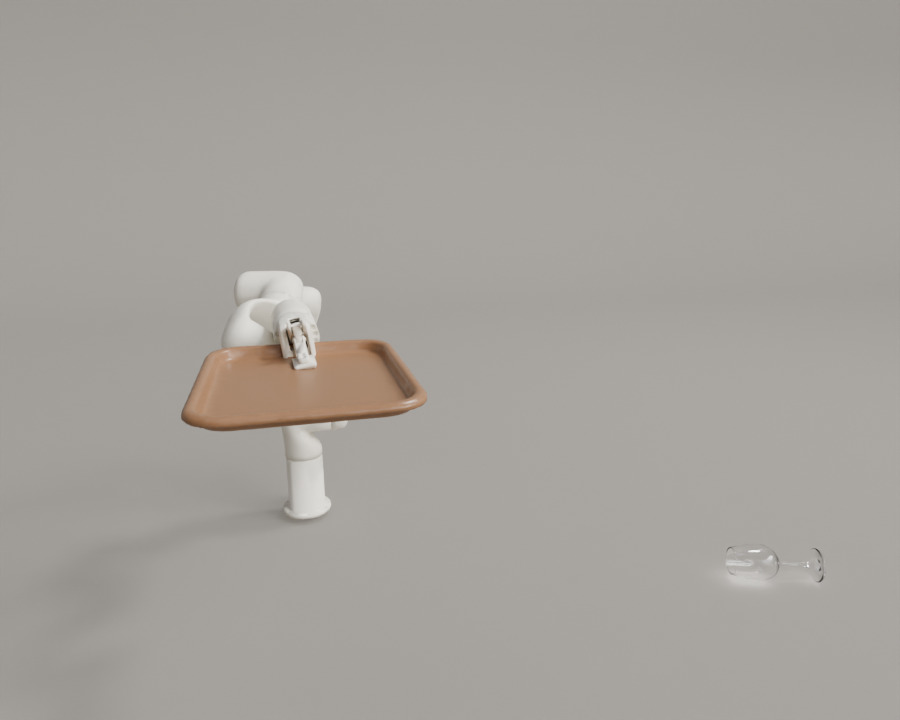}\hfill
    \includegraphics[width=0.31\linewidth]{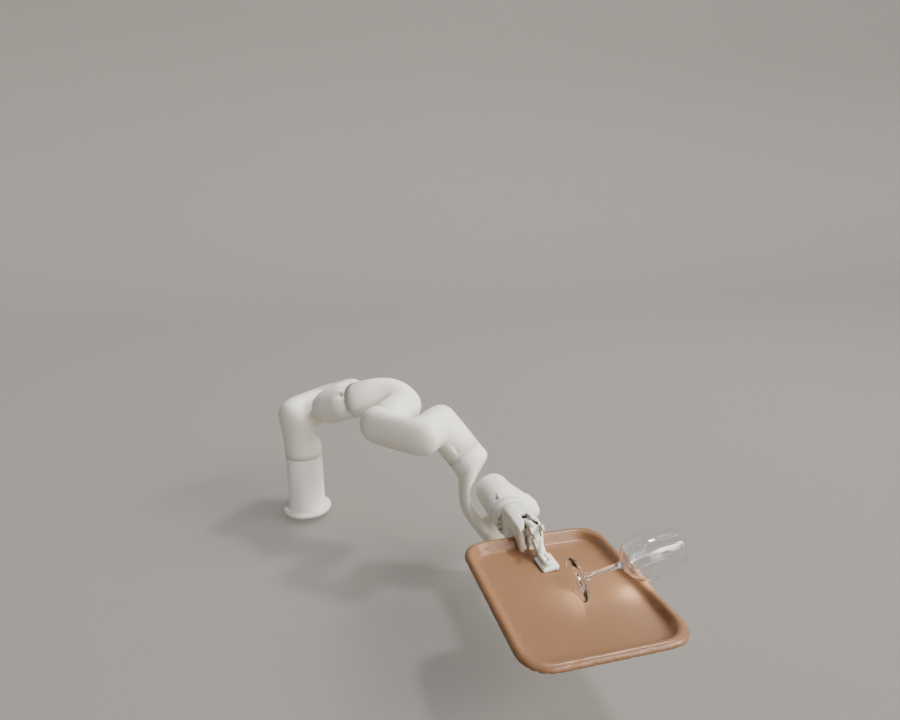}\hfill
    \includegraphics[width=0.31\linewidth]{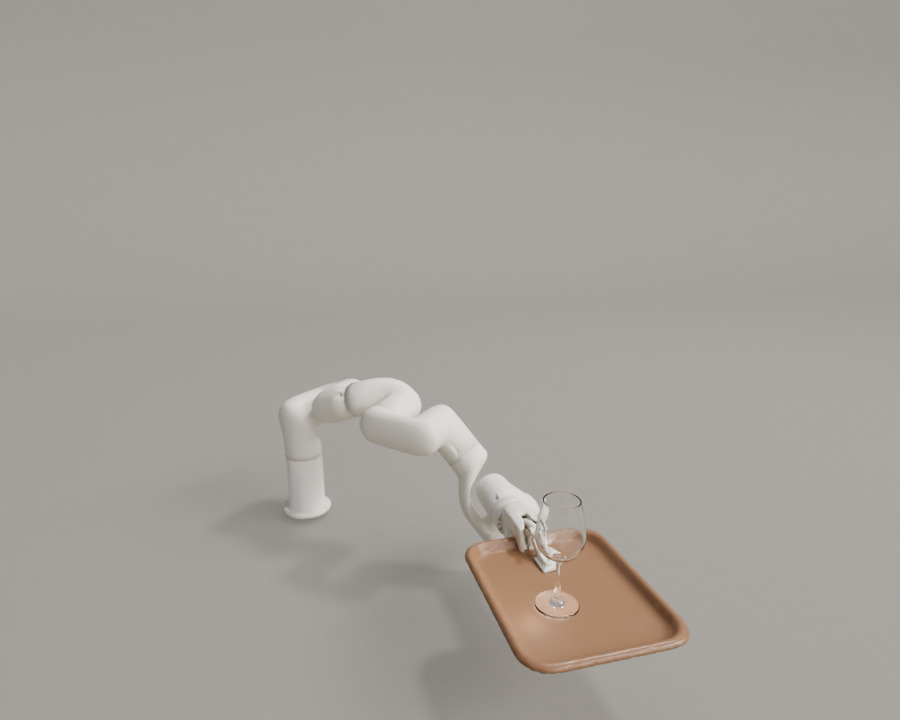}
    \caption{Frame 239}
  \end{subfigure}
  \caption{Articulated cup catching through plate-trajectory optimization.
  {With the {arm held at rest} (left), the cup falls from the plate.
  After {optimization }(right), the plate catches and retains
  it.} The middle column shows the initial IK trajectory, which also fails to retain the cup.}
  \Description{Four rows show the same saved frames (94, 111, 120, and 239). Rest-arm controls are on the left, the failing initial IK trajectory is in the middle, and optimized controls are on the right.}
  \label{fig:xarm-incoming-cup}
\end{figure}

Joint targets move a plate {held firmly by} an xArm gripper.
{The task requires the plate motion to account for the incoming cup's response through contact. A uniformly decelerating free-plate trajectory, constructed from the cup's initial velocity, supplies an initial joint-target motion through sequential URDF inverse kinematics (IK). This initializes the commands geometrically. Differentiable simulation then optimizes their effect on the coupled cup-and-plate dynamics, as compared with the stationary and initial-trajectory cases in Figure~\ref{fig:xarm-incoming-cup}.}

{
The simulation uses 240 frames at \mbox{$h=0.005$\,s} with friction
coefficient $1.0$. The optimized command matrix is
$\mathbf U_{\mathrm{cmd}}=(\mathbf u_1,\ldots,\mathbf u_{100})^{\T}
\in\R^{100\times7}$, where $\mathbf u_k\in\R^7$ contains the relative
target angles of the seven arm joints at control sample $k$. The
command at simulation frame $t$ is
\[
\mathbf a_t^{\mathrm{cmd}}=
\begin{cases}
\mathbf 0, & 1\le t\le98,\\
\mathbf u_{t-98}, & 99\le t\le198,\\
\mathbf u_{100}, & 199\le t\le240.
\end{cases}
\]
The final 42 frames hold $\mathbf u_{100}$ fixed, and their gradients
accumulate into this target.

Five state losses average errors over frames 99--240, starting when
the plate commands begin. For body $b\in\{\mathrm{plate},\mathrm{cup}\}$,
$\mathbf p_t^b$ and $\mathbf A_t^b$ denote its simulated center position
and affine orientation matrix, and
$\mathbf v_t^b=(\mathbf p_t^b-\mathbf p_{t-1}^b)/h$ is its center
velocity. Starred plate quantities denote the prescribed reference.
Position, velocity, and orientation tracking keep the plate near this
path with the intended motion and orientation:
\[
\begin{aligned}
L_p&=\frac{1}{142s_p^2}\sum_{t=99}^{240}
 \|\mathbf p_t^{\mathrm{plate}}-\mathbf p_t^{\mathrm{plate},*}\|^2,\\
L_v&=\frac{1}{142s_v^2}\sum_{t=99}^{240}
 \|\mathbf v_t^{\mathrm{plate}}-\mathbf v_t^{\mathrm{plate},*}\|^2,\\
L_A&=\frac{1}{142s_A^2}\sum_{t=99}^{240}
 \|\mathbf A_t^{\mathrm{plate}}-\mathbf A_t^{\mathrm{plate},*}\|_{\mathrm F}^2.
\end{aligned}
\]
The normalization scales are $s_p=0.02$,
$s_v=\sqrt{8}$, and $s_A={\sqrt{2}}/{20}$.

Cup uprightness discourages tipping without constraining yaw, and a
relative-velocity loss discourages sliding between the cup and plate.
With world up $\mathbf e_{\mathrm{up}}=(-1,0,0)^{\T}$, the cup up vector is
\[
\mathbf z_t^{\mathrm{cup}}=
\frac{\mathbf A_t^{\mathrm{cup}}(\mathbf A_0^{\mathrm{cup}})^{-1}
      \mathbf e_{\mathrm{up}}}
     {\|\mathbf A_t^{\mathrm{cup}}(\mathbf A_0^{\mathrm{cup}})^{-1}
      \mathbf e_{\mathrm{up}}\|}.
\]
The two losses are
\[
\begin{aligned}
L_{\mathrm{up}}&=\frac{1}{142}\sum_{t=99}^{240}
 \|\mathbf z_t^{\mathrm{cup}}-\mathbf e_{\mathrm{up}}\|^2,\\
L_{\mathrm{vel}}&=\frac{1}{142s_v^2}\sum_{t=99}^{240}
 \|\mathbf v_t^{\mathrm{cup}}-\mathbf v_t^{\mathrm{plate}}\|^2.
\end{aligned}
\]
Command smoothness is encouraged by
$L_{\Delta}=\frac{1}{99}\sum_{k=2}^{100}
\|\mathbf u_k-\mathbf u_{k-1}\|^2$, which penalizes changes between
consecutive independent targets. The total objective is
\[
\begin{aligned}
L_{\mathrm{cup}}={}&w_pL_p+w_vL_v+w_AL_A\\
 &+w_{\mathrm{up}}L_{\mathrm{up}}+w_{\mathrm{vel}}L_{\mathrm{vel}}
  +w_{\Delta}L_{\Delta},
\end{aligned}
\]
with fixed weights $w_p=100$, $w_v=20$, $w_A=20$,
$w_{\mathrm{up}}=20$, $w_{\mathrm{vel}}=50$, and $w_{\Delta}=0.001$.

The cup tilt is
$\alpha_t=\arccos((\mathbf z_t^{\mathrm{cup}})^{\T}\mathbf e_{\mathrm{up}})$.
A catch is successful when, in each of the final ten frames,
$\alpha_t\le\pi/18$\,rad, the minimum cup--plate vertex distance is
at most $2$\,mm, and both center speeds
$\|\mathbf v_t^{\mathrm{cup}}\|$ and
$\|\mathbf v_t^{\mathrm{plate}}\|$ are below $0.05$\,m/s.
Optimization stops once all of these conditions are met.
The optimized plate catches and retains the cup
(Figure~\ref{fig:xarm-incoming-cup}). All 700 joint-target variables
are optimized jointly through the arm dynamics and cup--plate contact.
Their gradients share one coupled adjoint solve per frame
(Sections~\ref{sec:frame-computation} and~\ref{sec:joint-derivatives}).
At a fixed simulation horizon, additional control variables therefore
require local derivative accumulation rather than additional coupled
solves.
}

\begin{figure}[t]
  \centering
  \makebox[0.47\linewidth][c]{\small Before}\hfill
  \makebox[0.47\linewidth][c]{\small After}\par\smallskip
  \begin{subfigure}[t]{\linewidth}
    \centering
    \includegraphics[width=0.47\linewidth]{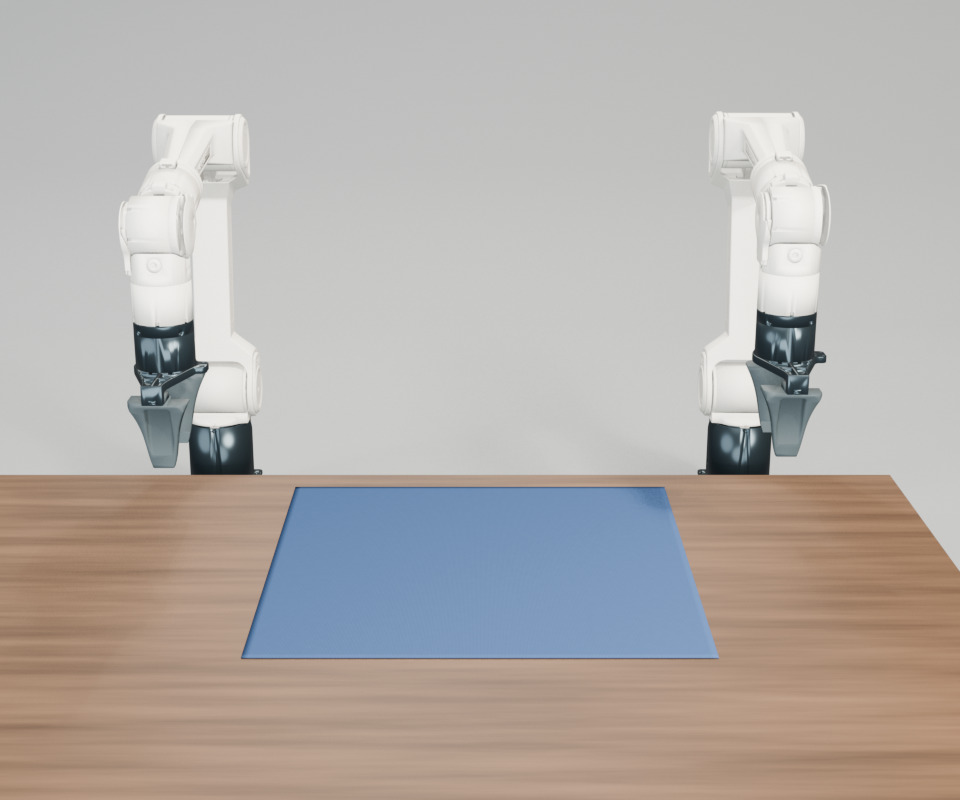}\hfill
    \includegraphics[width=0.47\linewidth]{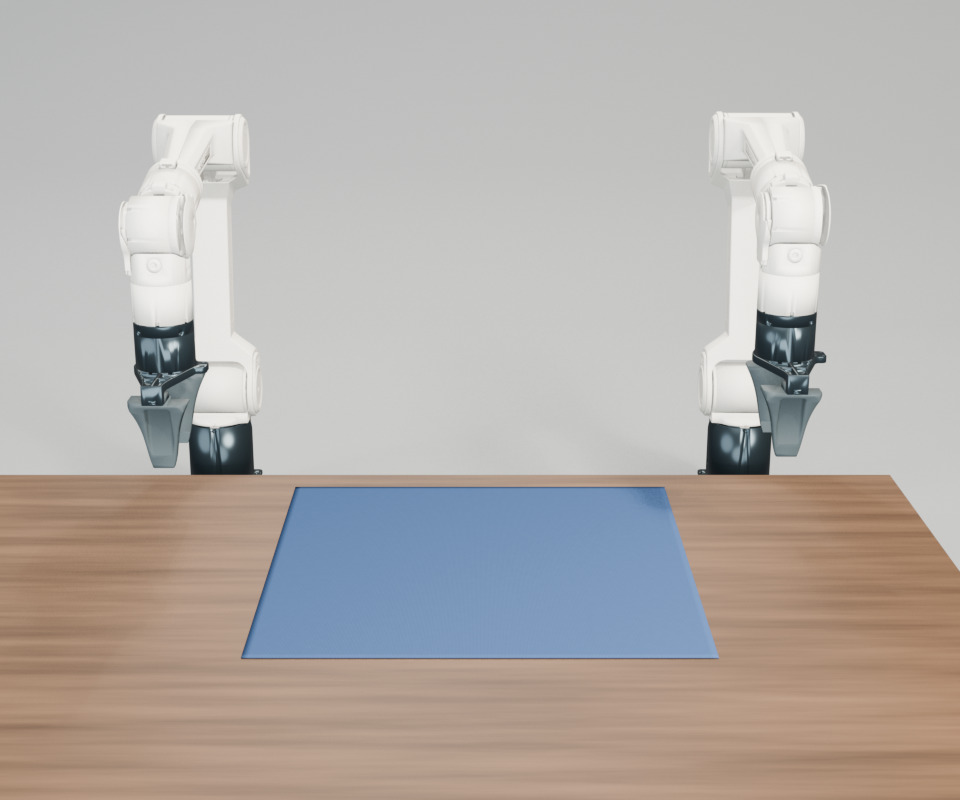}
    \caption{Initial ($0$\,s)}
  \end{subfigure}\par\smallskip
  \begin{subfigure}[t]{\linewidth}
    \centering
    \includegraphics[width=0.47\linewidth]{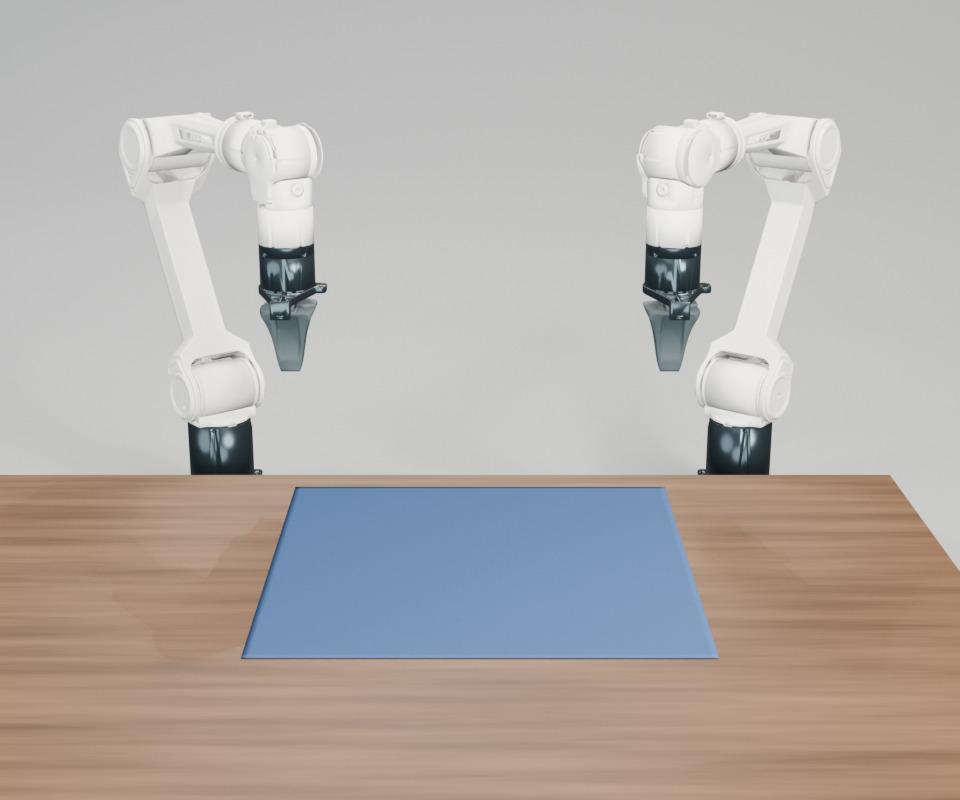}\hfill
    \includegraphics[width=0.47\linewidth]{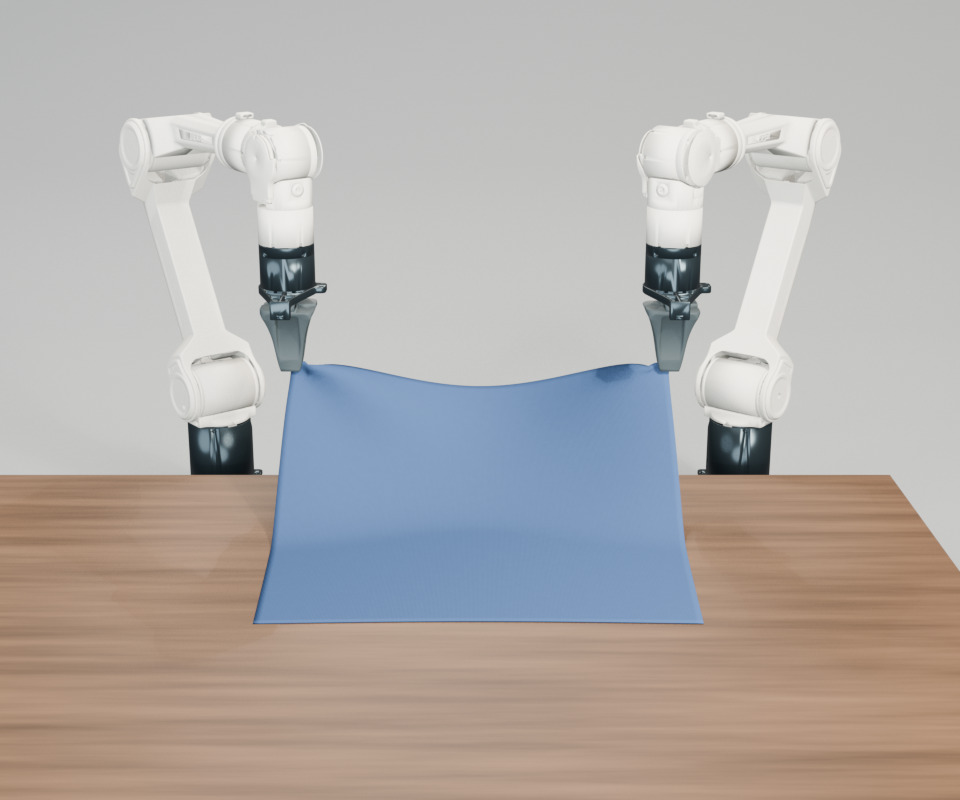}
    \caption{Lifting ($2.00$\,s)}
  \end{subfigure}\par\smallskip
  \begin{subfigure}[t]{\linewidth}
    \centering
    \includegraphics[width=0.47\linewidth]{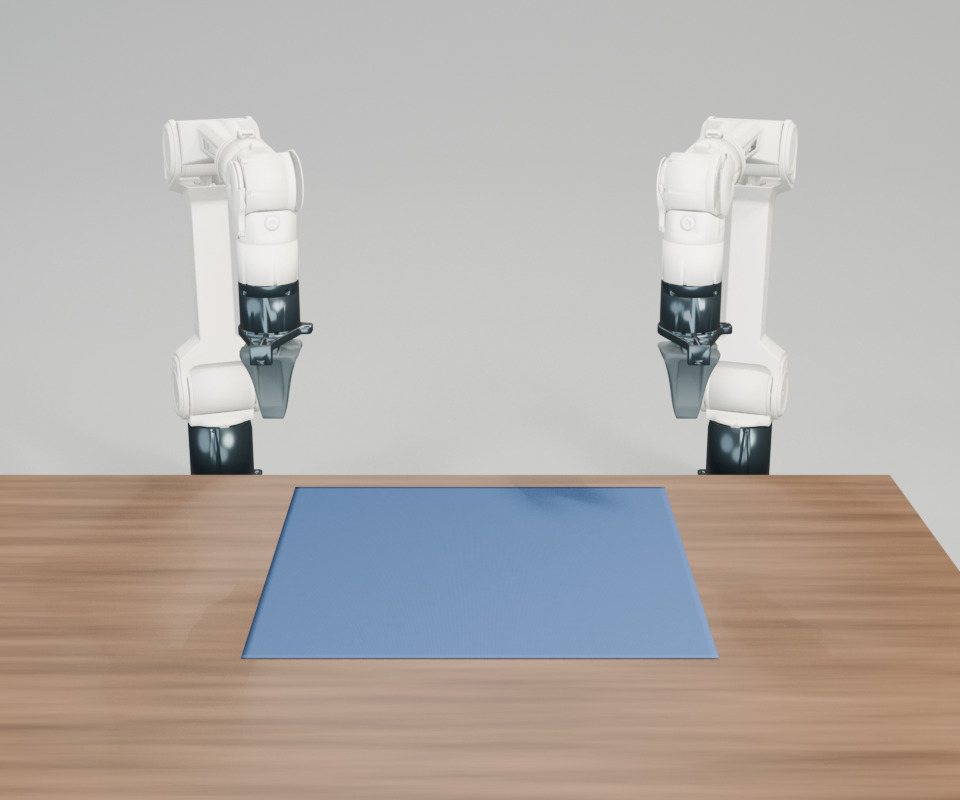}\hfill
    \includegraphics[width=0.47\linewidth]{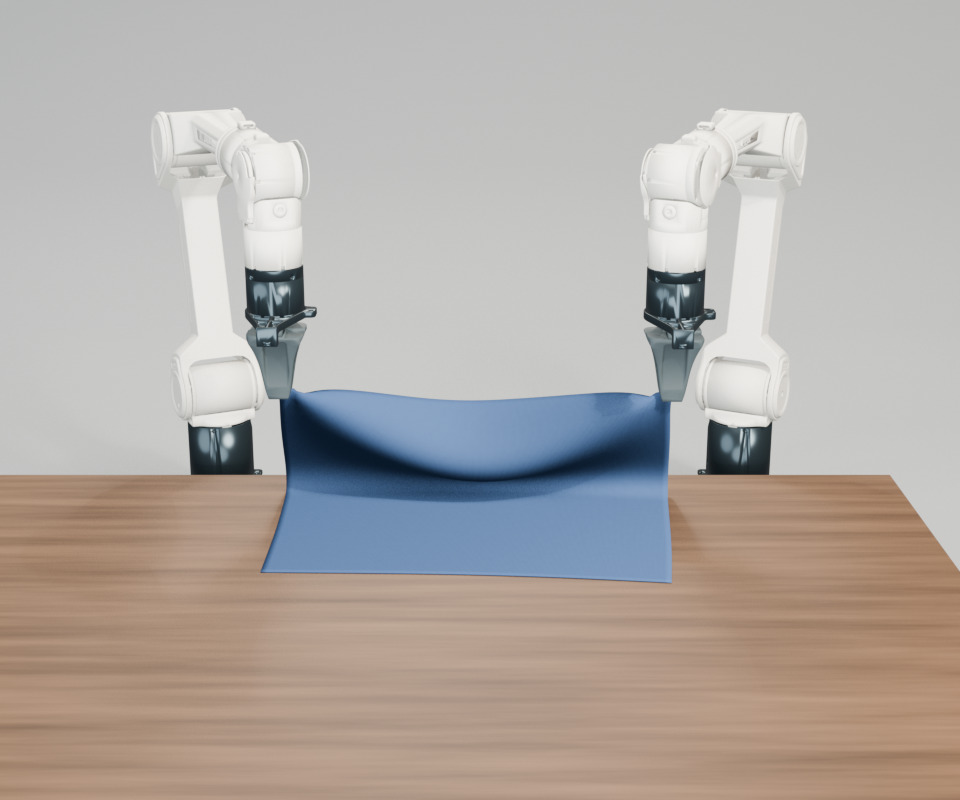}
    \caption{Raised ($2.40$\,s)}
  \end{subfigure}\par\smallskip
  \begin{subfigure}[t]{\linewidth}
    \centering
    \includegraphics[width=0.47\linewidth]{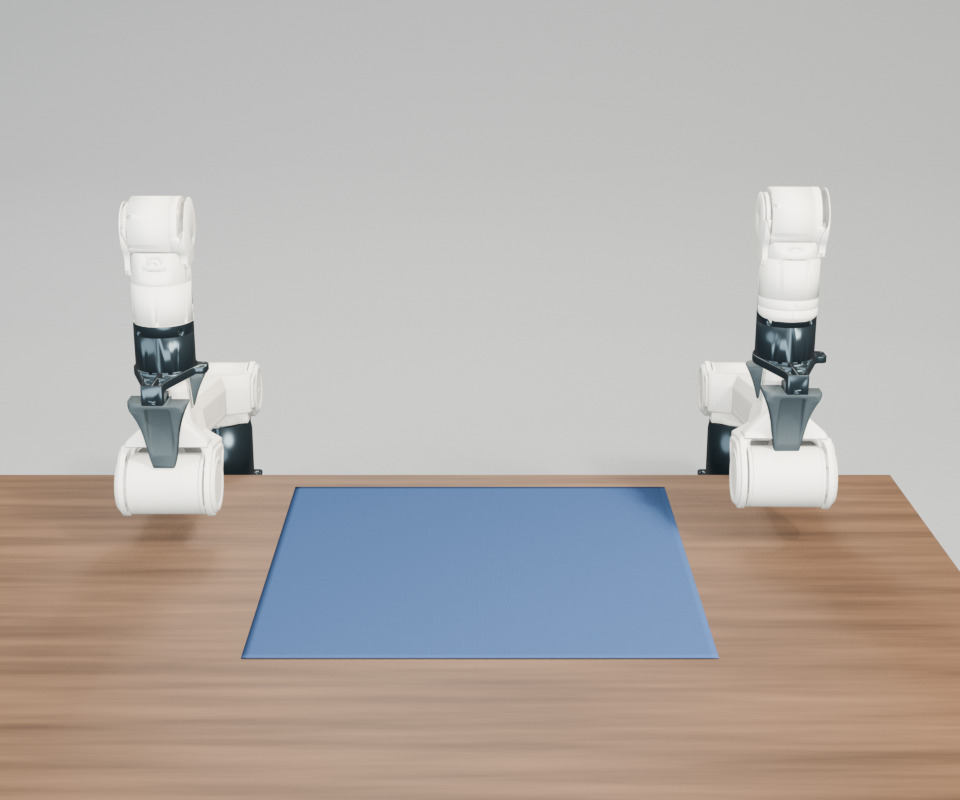}\hfill
    \includegraphics[width=0.47\linewidth]{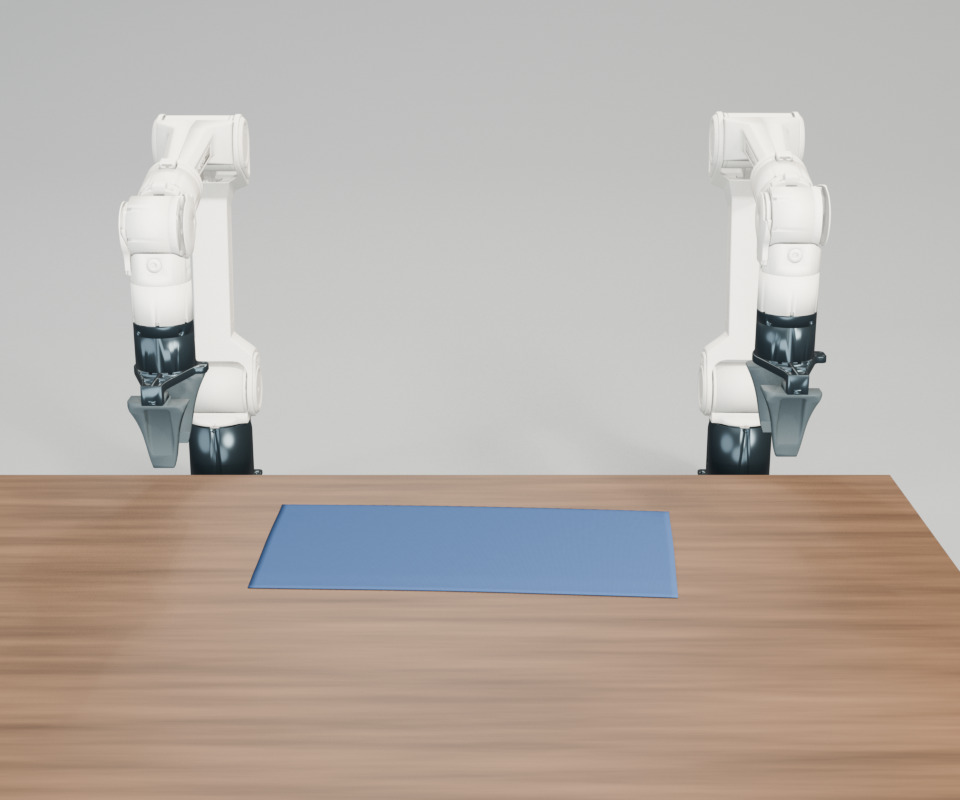}
    \caption{Final ($3.30$\,s)}
  \end{subfigure}
  \caption{{Dual-arm towel grasp by optimizing
  one vertical offset applied to both grippers' target positions.}
  {Before optimization (left), the arms {fail to grasp the towel}.
  After optimization (right), they {grasp the towel and then lift and fold it}.} Rows show the motion
  at $0$, $2.00$, $2.40$, and $3.30$\,s.}
  \Description{Four rows pair the same saved frames of the dual-arm towel grasp. Before optimization is on the left, and after optimization is on the right. The optimized arms grasp, lift, and fold the towel.}
  \label{fig:dual-piper}
\end{figure}

\paragraph{Dual-arm towel grasp}

{The same vertical offset is added to both
grippers' target positions during descent and the subsequent hold.
Optimizing this one value changes how firmly the fingers press the
cloth. The horizontal trajectories and finger-closing commands remain
fixed. Let $F_{120}$ be the total normal contact force between the
cloth and the fingers of both arms at frame 120 during the hold.

The target normal force is estimated from the friction needed to
accelerate the cloth with the closing fingers, using the cloth mass,
prescribed closing acceleration, and cloth--finger friction coefficient.
A fixed factor of three provides additional margin, giving
\mbox{$F^*=1.4573$\,N}. The optimized loss is
$L_{\mathrm{force}}=\tfrac12(F_{120}-F^*)^2$, with $F_{120}$ measured
from the simulated contacts.}
{Starting from a finger-surface height of $1.9$\,mm and contact force $0.0226$\,N, the optimizer produces a grasp that lifts and folds the towel after seven updates (Figure~\ref{fig:dual-piper}).} This task uses a force objective during the press-down hold to set the contact needed for the subsequent cloth motion, linking an affine position command to a deformable-object outcome.

%{
%\revnote{Complete the exact definitions, targets, and observation-frame
%sets for $L_{\mathrm{up}}$, $L_{\mathrm{rel}}$, and
%$L_{\mathrm{arr}}$ in Cup catching, and specify the control-sample
%range for $L_{\Delta}$. Record the numerical task-loss weights.
%For Loaded Support, specify the normalization of $L_{\mathrm{sm}}$.
%For Garment Hanger, specify the strap-vertex sets and the geometric
%measurement used for $m_s$.}
%}

\begin{figure*}[tp]
  \centering

{\ignorespaces
  \includegraphics[width=\linewidth,trim=0 0 0 12bp,clip]{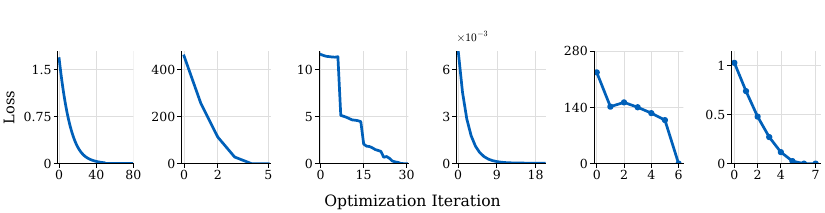}
  % Match the vector plot's 16/396 left gutter at the ACM text width.
  \par\noindent\hspace*{\dimexpr\textwidth*4/99\relax}%
  \begin{subfigure}[t]{\dimexpr\textwidth*95/594\relax}
    \centering
    \caption{Fig.~\ref{fig:stone-placement}}
  \end{subfigure}%
  \begin{subfigure}[t]{\dimexpr\textwidth*95/594\relax}
    \centering
    \caption{Fig.~\ref{fig:dress10}}
  \end{subfigure}%
  \begin{subfigure}[t]{\dimexpr\textwidth*95/594\relax}
    \centering
    \caption{Fig.~\ref{fig:abd-hanger-rest}}
  \end{subfigure}%
  \begin{subfigure}[t]{\dimexpr\textwidth*95/594\relax}
    \centering
    \caption{Fig.~\ref{fig:xarm-target-closeup}}
  \end{subfigure}%
  \begin{subfigure}[t]{\dimexpr\textwidth*95/594\relax}
    \centering
    \caption{Fig.~\ref{fig:xarm-incoming-cup}}
  \end{subfigure}%
  \begin{subfigure}[t]{\dimexpr\textwidth*95/594\relax}
    \centering
    \caption{Fig.~\ref{fig:dual-piper}}
  \end{subfigure}\par
  \caption{{Optimization histories for the additional
  tasks. Stone placement changes its centroid tolerance after 50
  updates. Garment hanger, cup catching, and dual-arm
  grasp use constant offsets that set their final plotted values to
  zero.}}
  \Description{Six loss plots for stone placement, cloth scale,
  garment hanger shape design, grasp angle, cup catching, and dual-arm grasp.
  }
  \label{fig:loss-curves-remaining}
\par}
\end{figure*}

These outcomes connect parameter choice to physical effect: rest geometry establishes support, initial placement forms a stable assembly, and actuation controls contacting objects. All use the same coupled backward equations, with task-specific parameter maps and losses. Figure~\ref{fig:loss-curves-remaining} complements the physical sequences with the corresponding optimization histories.

\subsection{\texorpdfstring{{Gradient validation and representation studies}}{Gradient validation and representation studies}}
\label{sec:representation-study}

\paragraph{Gradient validation}
Table~\ref{tab:initial-fd} reports finite-difference
(FD) checks at the initial optimization parameters for all DiffGIPC
examples and the shared DiffIPC benchmarks. For the checked parameter
vector $\mathbf z$, let ${\mathbf g_{\mathrm{ana}}}$ be its analytic loss
gradient. The direction $\mathbf d$ is the normalized gradient, or
its sign for a scalar parameter. With perturbation size
$\epsilon_{\mathrm{FD}}$, the centered difference is
$D_{\mathrm{FD}}L=[L(\mathbf z+\epsilon_{\mathrm{FD}}\mathbf d)
-L(\mathbf z-\epsilon_{\mathrm{FD}}\mathbf d)]/(2\epsilon_{\mathrm{FD}})$.
The reported relative error is
$|{\mathbf g_{\mathrm{ana}}}^{\T}\mathbf d-D_{\mathrm{FD}}L|/
\max(|{\mathbf g_{\mathrm{ana}}}^{\T}\mathbf d|,|D_{\mathrm{FD}}L|)$.
The shared-case checks use the first 10 frames. The other checks use
the stated task-specific windows.

% Manuscript-local Table 2: preserve shared table data for the other versions.
\begin{table}[!htbp]\centering\small

\caption{Finite-difference checks at the initial
optimization parameters.}
\label{tab:initial-fd}
\setlength{\tabcolsep}{3pt}
\renewcommand{\arraystretch}{1.15}
\setlength{\aboverulesep}{0pt}
\setlength{\belowrulesep}{0pt}
\setlength{\cmidrulesep}{0pt}
\begin{tabularx}{\columnwidth}{@{}c|c|c|>{\centering\arraybackslash}X@{}}\toprule
\multicolumn{1}{c|}{Example} & \multicolumn{1}{c|}{System} & \multicolumn{1}{c|}{Dim} & \multicolumn{1}{c}{{Relative FD error}} \\
\midrule
Stone placement (Fig.~\ref{fig:stone-placement}) & DiffGIPC & 14 & $7.53\times10^{-11}$ \\
\specialrule{\lightrulewidth}{0pt}{0pt}
Garment hanger (Fig.~\ref{fig:abd-hanger-rest}) & DiffGIPC & 2 & $5.45\times10^{-3}$ \\
\specialrule{\lightrulewidth}{0pt}{0pt}
\multirow{2}{*}{Bunny collision (Fig.~\ref{fig:bunny-initial-velocity})} & DiffGIPC & 3 & $1.21\times10^{-6}$ \\
\cmidrule{2-4}
 & DiffIPC & 3 & $1.01\times10^{-4}$ \\
\specialrule{\lightrulewidth}{0pt}{0pt}
\multirow{2}{*}{Tentacle matching (Fig.~\ref{fig:tentacle-trajectory})} & DiffGIPC & 3 & $3.33\times10^{-5}$ \\
\cmidrule{2-4}
 & DiffIPC & 3 & $3.83\times10^{-11}$ \\
\specialrule{\lightrulewidth}{0pt}{0pt}
\multirow{2}{*}{Loaded support (Fig.~\ref{fig:fem-rest-shape})} & DiffGIPC & 9939 & $2.56\times10^{-6}$ \\
\cmidrule{2-4}
 & DiffIPC & 9939 & $4.02\times10^{-4}$ \\
\specialrule{\lightrulewidth}{0pt}{0pt}
\multirow{2}{*}{Material fitting (Fig.~\ref{fig:material-identification})} & DiffGIPC & 2 & $8.37\times10^{-7}$ \\
\cmidrule{2-4}
 & DiffIPC & 2 & $6.71\times10^{-7}$ \\
\specialrule{\lightrulewidth}{0pt}{0pt}
\multirow{2}{*}{Friction optimization (Fig.~\ref{fig:friction-identification})} & DiffGIPC & 1 & $4.38\times10^{-5}$ \\
\cmidrule{2-4}
 & DiffIPC & 1 & $6.27\times10^{-3}$ \\
\specialrule{\lightrulewidth}{0pt}{0pt}
Cloth rest scale (Fig.~\ref{fig:dress10}) & DiffGIPC & 1 & {$9.45\times10^{-3}$} \\
\specialrule{\lightrulewidth}{0pt}{0pt}
Grasp angle (Fig.~\ref{fig:xarm-target-closeup}) & DiffGIPC & 1 & $9.04\times10^{-7}$ \\
\specialrule{\lightrulewidth}{0pt}{0pt}
Cup catching (Fig.~\ref{fig:xarm-incoming-cup}) & DiffGIPC & 700 & $4.48\times10^{-4}$ \\
\specialrule{\lightrulewidth}{0pt}{0pt}
Dual-arm grasp (Fig.~\ref{fig:dual-piper}) & DiffGIPC & 1 & $1.70\times10^{-3}$ \\
\bottomrule\end{tabularx}
\end{table}

% Preview resolution/variant study with TOG figure references.
\begin{table*}[!t]\centering\small
\newtablemark
\caption{Mean forward and backward evaluation times (seconds)
for the bunny-collision resolution study and Garment hanger rest-shape design.
Forward + Backward gives the sum of the two
per-evaluation means.}
\label{tab:fig1-resolution-phase-comparison}
\setlength{\tabcolsep}{3pt}
\renewcommand{\arraystretch}{1.15}
\setlength{\aboverulesep}{0pt}
\setlength{\belowrulesep}{0pt}
\setlength{\cmidrulesep}{0pt}
\begin{tabularx}{\textwidth}{@{}c|c|r|r|r|r|r|>{\centering\arraybackslash}X|r|r|r|>{\centering\arraybackslash}X|>{\centering\arraybackslash}X@{}}\toprule
\multirow{2}{*}{Resolution / case} & \multirow{2}{*}{System} & \multicolumn{6}{c|}{Forward} & \multicolumn{4}{c|}{Backward} & \multicolumn{1}{c}{\multirow{2}{*}{\shortstack{Forward +\\Backward}}} \\
\cmidrule{3-8}\cmidrule{9-12}
 &  & \multicolumn{1}{c|}{Linear} & \multicolumn{1}{c|}{CCD} & \multicolumn{1}{c|}{Hessian} & \multicolumn{1}{c|}{Search} & \multicolumn{1}{c|}{Other} & \multicolumn{1}{c|}{Total} & \multicolumn{1}{c|}{Hessian} & \multicolumn{1}{c|}{Linear} & \multicolumn{1}{c|}{Other} & \multicolumn{1}{c|}{Total} & \multicolumn{1}{c}{} \\
\midrule
\multicolumn{13}{@{}c}{\textbf{Bunny collision (Fig.~\ref{fig:bunny-initial-velocity}): resolution study}} \\
\specialrule{\lightrulewidth}{0pt}{0pt}
\multirow{4}{*}{933 vertices} & DiffGIPC ABD+ABD & 0.025 & 0.039 & 0.041 & 0.021 & 0.0023 & 0.13 & 0.0081 & 0.0042 & 0.0032 & 0.016 & 0.14 \\
 & DiffGIPC ABD+FEM & 0.58 & 0.049 & 0.058 & 0.03 & 0.0025 & 0.72 & 0.011 & 0.22 & 0.0085 & 0.24 & 0.96 \\
 & DiffGIPC FEM+FEM & 0.65 & 0.044 & 0.037 & 0.017 & 0.002 & 0.75 & 0.0069 & 0.22 & 0.0057 & 0.24 & 0.98 \\
 & DiffIPC FEM+FEM & 4.5 & 10 & 4.6 & 1.1 & 1.9 & 22 & 1.4 & 1.1 & 0.44 & 2.9 & 25 \\
\specialrule{\lightrulewidth}{0pt}{0pt}
\multirow{4}{*}{7,991 vertices} & DiffGIPC ABD+ABD & 0.032 & 0.054 & 0.11 & 0.029 & 0.0023 & 0.22 & 0.018 & 0.0043 & 0.0036 & 0.026 & 0.25 \\
 & DiffGIPC ABD+FEM & 1.4 & 0.075 & 0.18 & 0.045 & 0.0033 & 1.7 & 0.028 & 0.32 & 0.025 & 0.37 & 2.1 \\
 & DiffGIPC FEM+FEM & 2.1 & 0.069 & 0.18 & 0.032 & 0.0031 & 2.4 & 0.029 & 0.49 & 0.02 & 0.54 & 3 \\
 & DiffIPC FEM+FEM & 80 & 24 & 54 & 9.7 & 7.6 & 170 & 15 & 17 & 0.98 & 33 & 210 \\
\specialrule{\lightrulewidth}{0pt}{0pt}
\multirow{4}{*}{28,435 vertices} & DiffGIPC ABD+ABD & 0.027 & 0.059 & 0.21 & 0.026 & 0.0026 & 0.32 & 0.039 & 0.0044 & 0.0043 & 0.048 & 0.37 \\
 & DiffGIPC ABD+FEM & 3.5 & 0.091 & 0.53 & 0.059 & 0.0049 & 4.2 & 0.081 & 0.99 & 0.074 & 1.1 & 5.4 \\
 & DiffGIPC FEM+FEM & 6.9 & 0.088 & 0.73 & 0.052 & 0.0043 & 7.8 & 0.11 & 2.3 & 0.057 & 2.5 & 10 \\
 & DiffIPC FEM+FEM & 740 & 58 & 440 & 55 & 32 & 1300 & 100 & 110 & 4.6 & 220 & 1500 \\
\midrule
\multicolumn{13}{@{}c}{\textbf{Garment hanger (Fig.~\ref{fig:abd-hanger-rest}): rest-shape optimization}} \\
\specialrule{\lightrulewidth}{0pt}{0pt}
\multirow{2}{*}{Bezier pair} & DiffGIPC ABD & 3.7 & 0.88 & 1.5 & 0.8 & 0.0088 & 6.9 & 0.033 & 0.21 & 0.022 & 0.27 & 7.1 \\
 & DiffGIPC FEM & 11 & 1.1 & 1.7 & 0.73 & 0.0076 & 15 & 0.029 & 0.95 & 0.02 & 1 & 16 \\
\bottomrule\end{tabularx}
\end{table*}

\paragraph{Bunny resolution study}

Table~\ref{tab:fig1-resolution-phase-comparison}
compares three body representations as bunny resolution increases.
FEM+FEM uses two deformable bodies, ABD+FEM uses one affine and one FEM
body, and ABD+ABD uses two affine bodies. 
As resolution grows from 933 to 28,435 vertices, the
combined forward and backward time for DiffGIPC FEM+FEM increases
from {0.98} to 10\,s, whereas the ABD+ABD time increases from
{0.14} to {0.37}\,s.
The increases are about {$10.0\times$} and {$2.6\times$}, respectively.

At the highest resolution, ABD+ABD has a forward-plus-backward speedup
of {about $27\times$} over DiffGIPC FEM+FEM.
{The time spent in backward linear solves falls} from {2.3} to 0.0044\,s
({about $520\times$}),
while {the time for a complete backward evaluation falls} from
2.5 to {0.048}\,s
({about $52\times$}).
{With twelve motion DoFs per body, ABD reduces
the adjoint-system size without simplifying the surface mesh used for
contact and shape derivatives.}
For FEM+FEM, the speedup of DiffGIPC over DiffIPC grows from {$25.4\times$} to {$150.0\times$} across resolutions. Combining GPU execution with an affine motion representation gives a {$4133.3\times$} speedup for DiffGIPC ABD+ABD over DiffIPC FEM+FEM at the highest resolution. %The within-DiffGIPC comparison measures the additional benefit of the representation choice: ABD+ABD is roughly {$27\times$} faster than FEM+FEM in this study.

\paragraph{Garment hanger: ABD versus FEM}

This comparison keeps the surface discretization fixed at 4,476
vertices and 8,952 triangles while changing the motion representation.
The FEM replacement adds 902 interior vertices and 18,256 tetrahedra,
with a harmonic extension of the two B\'ezier shape fields to the interior.
The mean forward-plus-backward cost is {7.1}\,s for ABD and {$16$}\,s for FEM,
{giving a speedup of {$2.2\times$}. The backward costs are {0.27} and $1.0$\,s.}
The corresponding optimized {control-point scales
$(s_x,s_y)$} are $(0.8285,0.8000)$ and
$(0.8301,0.8000)$.
At fixed surface discretization, the two models allocate computation differently. FEM resolves non-affine deformation through an interior tetrahedral mesh, whereas ABD retains a compact motion state and differentiates mass directly from the surface. {The timings compare forward and backward evaluations of the ABD and FEM models for the same cloth-support design task.} %Appendix~\ref{sec:experimental-details} specifies the optimization and verification horizons.

% Queue the wide timing tables before the subsection introduction crosses a column.
% 6-phase graph-on workloads.
\begin{table*}[!t]\centering\small

\caption{Complete optimization workloads and mean phase times (seconds)
of the eleven examples.
$N_{\mathrm{fwd}}$ counts forward evaluations, including
line-search trials and verification runs. $N_{\mathrm{bwd}}$ counts
complete gradient evaluations used for optimization.
Linear comprises PCG and Others, and Hessian combines local and global assembly.}
\label{tab:eleven-phase-times}
\setlength{\tabcolsep}{2.4pt}
\renewcommand{\arraystretch}{1.15}
\setlength{\aboverulesep}{0pt}
\setlength{\belowrulesep}{0pt}
\setlength{\cmidrulesep}{0pt}
\begin{tabularx}{\textwidth}{@{}c|c|c|r|r|r|r|r|r|>{\centering\arraybackslash}X|r|r|r|r|>{\centering\arraybackslash}X|>{\centering\arraybackslash}X@{}}\toprule
\multirow{3}{*}{Example} & \multirow{3}{*}{$N_{\mathrm{fwd}}$} & \multirow{3}{*}{$N_{\mathrm{bwd}}$} & \multicolumn{7}{c|}{Forward mean} & \multicolumn{5}{c|}{Backward mean} & \multicolumn{1}{c}{\multirow{3}{*}{$T_{\mathrm{eval}}$}} \\
\cmidrule{4-10}\cmidrule{11-15}
 & & & \multicolumn{2}{c|}{Linear} & \multirow{2}{*}{CCD} & \multirow{2}{*}{Hessian} & \multirow{2}{*}{Search} & \multirow{2}{*}{Other} & \multirow{2}{*}{Total} & \multirow{2}{*}{Hessian} & \multicolumn{2}{c|}{Linear} & \multirow{2}{*}{Other} & \multirow{2}{*}{Total} & \\
\cmidrule{4-5}\cmidrule{12-13}
 & & & PCG & Others & & & & & & & PCG & Others & & & \\
\midrule
Stone placement (Fig.~\ref{fig:stone-placement}) & 80 & 78 & 0.097 & 0.0093 & 0.079 & 0.12 & 0.024 & 0.0088 & 0.34 & 0.051 & 0.067 & 0.003 & 0.023 & 0.14 & 39 \\
\specialrule{\lightrulewidth}{0pt}{0pt}
Bunny collision (Fig.~\ref{fig:bunny-initial-velocity}) & 15 & 12 & 0.62 & 0.027 & 0.044 & 0.037 & 0.017 & 0.002 & 0.75 & 0.0069 & 0.22 & 0.005 & 0.0057 & 0.24 & 14 \\
\specialrule{\lightrulewidth}{0pt}{0pt}
Tentacle matching (Fig.~\ref{fig:tentacle-trajectory}) & 11 & 10 & 6.8 & 1.3 & 2.4 & 2.2 & 1.4 & 0.011 & 14 & 0.04 & 0.99 & 0.024 & 0.024 & 1.1 & 170 \\
\specialrule{\lightrulewidth}{0pt}{0pt}
Garment hanger (Fig.~\ref{fig:abd-hanger-rest}) & 32 & 30 & 2.8 & 0.87 & 0.88 & 1.5 & 0.8 & 0.0088 & 6.9 & 0.033 & 0.2 & 0.01 & 0.022 & 0.27 & 230 \\
\specialrule{\lightrulewidth}{0pt}{0pt}
Loaded support (Fig.~\ref{fig:fem-rest-shape}) & 231 & 200 & 0.92 & 0.012 & 0.02 & 0.014 & 0.0017 & 0.00051 & 0.97 & 0.011 & 0.84 & 0.0065 & 0.0066 & 0.87 & 400 \\
\specialrule{\lightrulewidth}{0pt}{0pt}
Material fitting (Fig.~\ref{fig:material-identification}) & 21 & 20 & 0.011 & 0.0035 & 0.0021 & 0.0042 & 0.00064 & 0.0032 & 0.025 & 0.0026 & 0.021 & 0.002 & 0.0016 & 0.028 & 1.1 \\
\specialrule{\lightrulewidth}{0pt}{0pt}
Friction optimization (Fig.~\ref{fig:friction-identification}) & 6 & 5 & 3.2 & 0.6 & 1.4 & 1.3 & 0.7 & 0.0093 & 7.2 & 0.037 & 0.35 & 0.014 & 0.039 & 0.44 & 45 \\
\specialrule{\lightrulewidth}{0pt}{0pt}
Cloth rest scale (Fig.~\ref{fig:dress10}) & 7 & 6 & 13 & 0.38 & 0.54 & 0.68 & 0.31 & 0.0046 & 15 & 0.0029 & 0.48 & 0.00087 & 0.001 & 0.48 & 110 \\
\specialrule{\lightrulewidth}{0pt}{0pt}
Grasp angle (Fig.~\ref{fig:xarm-target-closeup}) & 44 & 20 & 0.34 & 0.044 & 0.074 & 0.27 & 0.12 & 0.0035 & 0.85 & 0.048 & 0.058 & 0.0073 & 0.064 & 0.18 & 41 \\
\specialrule{\lightrulewidth}{0pt}{0pt}
Cup catching (Fig.~\ref{fig:xarm-incoming-cup}) & 7 & 6 & 2.3 & 0.17 & 2.1 & 2.2 & 1.1 & 0.017 & 7.9 & 0.2 & 0.68 & 0.009 & 0.13 & 1 & 62 \\
\specialrule{\lightrulewidth}{0pt}{0pt}
Dual-arm grasp (Fig.~\ref{fig:dual-piper}) & 9 & 7 & 3.4 & 0.24 & 0.84 & 1.6 & 0.4 & 0.012 & 6.5 & 0.2 & 0.9 & 0.024 & 0.58 & 1.7 & 70 \\
\specialrule{\lightrulewidth}{0pt}{0pt}
\bottomrule\end{tabularx}
\end{table*}

% Paired graph timings: two rows per example, with speedups inline.
\begin{table*}[!t]\centering\small
\newtablemark
\caption{CUDA Graph enabled (On) versus disabled (Off).
Times are mean seconds per evaluation; parentheses in the Off rows give
speedups from enabling the graph.
Linear comprises PCG and Others; the separate Other is Total minus Linear. The rightmost Total combines forward and
backward time.}
\label{tab:graph-on-off-selected-comparison}
\setlength{\tabcolsep}{2.4pt}
\renewcommand{\arraystretch}{1.1}
% Keep vertical separators continuous through the horizontal rules.
\setlength{\aboverulesep}{0pt}
\setlength{\belowrulesep}{0pt}
\setlength{\cmidrulesep}{0pt}
% Stretch timing columns rather than gaps so header rules meet group borders.
\begin{tabularx}{\textwidth}{@{}c|c|>{\raggedleft\arraybackslash}X@{\hspace{2pt}}l|r|r|>{\raggedleft\arraybackslash}X@{\hspace{2pt}}l|>{\raggedleft\arraybackslash}X@{\hspace{2pt}}l|r|r|>{\raggedleft\arraybackslash}X@{\hspace{2pt}}l|>{\raggedleft\arraybackslash}X@{\hspace{2pt}}l@{}}\toprule
\multirow{3}{*}{Example} & \multirow{3}{*}{Graph} & \multicolumn{6}{c|}{Forward (s)} & \multicolumn{6}{c|}{Backward (s)} & \multicolumn{2}{c}{\multirow{3}{*}{Total (s)}} \\
\cmidrule{3-8}\cmidrule{9-14}
 & & \multicolumn{3}{c|}{Linear} & \multirow{2}{*}{Other} & \multicolumn{2}{c|}{\multirow{2}{*}{Total}} & \multicolumn{3}{c|}{Linear} & \multirow{2}{*}{Other} & \multicolumn{2}{c|}{\multirow{2}{*}{Total}} & \multicolumn{2}{c}{} \\
\cmidrule{3-5}\cmidrule{9-11}
 & & \multicolumn{2}{c|}{PCG} & Others & & \multicolumn{2}{c|}{} & \multicolumn{2}{c|}{PCG} & Others & & \multicolumn{2}{c|}{} & \multicolumn{2}{c}{} \\
\midrule
\multirow{2}{*}{Stone placement (Fig.~\ref{fig:stone-placement})} & On & 0.097 &  & 0.0093 & 0.24 & 0.34 &  & 0.067 &  & 0.003 & 0.074 & 0.14 &  & 0.49 &  \\
 & Off & 0.18 & $(1.8\times)$ & 0.0094 & 0.24 & 0.42 & $(1.2\times)$ & 0.12 & $(1.8\times)$ & 0.003 & 0.073 & 0.2 & $(1.4\times)$ & 0.62 & $(1.3\times)$ \\
\specialrule{\lightrulewidth}{0pt}{0pt}
\multirow{2}{*}{Bunny collision (Fig.~\ref{fig:bunny-initial-velocity})} & On & 0.62 &  & 0.027 & 0.1 & 0.75 &  & 0.22 &  & 0.005 & 0.013 & 0.24 &  & 0.98 &  \\
 & Off & 0.97 & $(1.6\times)$ & 0.026 & 0.1 & 1.1 & $(1.5\times)$ & 0.35 & $(1.6\times)$ & 0.0051 & 0.011 & 0.36 & $(1.5\times)$ & 1.5 & $(1.5\times)$ \\
\specialrule{\lightrulewidth}{0pt}{0pt}
\multirow{2}{*}{Tentacle matching (Fig.~\ref{fig:tentacle-trajectory})} & On & 6.8 &  & 1.3 & 6 & 14 &  & 0.99 &  & 0.024 & 0.063 & 1.1 &  & 15 &  \\
 & Off & 10 & $(1.5\times)$ & 1.3 & 6.3 & 18 & $(1.3\times)$ & 1.5 & $(1.5\times)$ & 0.024 & 0.067 & 1.6 & $(1.4\times)$ & 20 & $(1.3\times)$ \\
\specialrule{\lightrulewidth}{0pt}{0pt}
\multirow{2}{*}{Garment hanger (Fig.~\ref{fig:abd-hanger-rest})} & On & 2.8 &  & 0.87 & 3.2 & 6.9 &  & 0.2 &  & 0.01 & 0.055 & 0.27 &  & 7.1 &  \\
 & Off & 4.1 & $(1.5\times)$ & 0.8 & 3.2 & 8.1 & $(1.2\times)$ & 0.27 & $(1.4\times)$ & 0.01 & 0.049 & 0.33 & $(1.2\times)$ & 8.4 & $(1.2\times)$ \\
\specialrule{\lightrulewidth}{0pt}{0pt}
\multirow{2}{*}{Loaded support (Fig.~\ref{fig:fem-rest-shape})} & On & 0.92 &  & 0.012 & 0.036 & 0.97 &  & 0.84 &  & 0.0065 & 0.017 & 0.87 &  & 1.8 &  \\
 & Off & 1.2 & $(1.4\times)$ & 0.012 & 0.032 & 1.3 & $(1.3\times)$ & 1.2 & $(1.4\times)$ & 0.0064 & 0.019 & 1.2 & $(1.4\times)$ & 2.5 & $(1.4\times)$ \\
\specialrule{\lightrulewidth}{0pt}{0pt}
\multirow{2}{*}{Material fitting (Fig.~\ref{fig:material-identification})} & On & 0.011 &  & 0.0035 & 0.01 & 0.025 &  & 0.021 &  & 0.002 & 0.0042 & 0.028 &  & 0.053 &  \\
 & Off & 0.017 & $(1.5\times)$ & 0.0034 & 0.011 & 0.032 & $(1.3\times)$ & 0.033 & $(1.5\times)$ & 0.0018 & 0.0049 & 0.039 & $(1.4\times)$ & 0.071 & $(1.4\times)$ \\
\specialrule{\lightrulewidth}{0pt}{0pt}
\multirow{2}{*}{Friction optimization (Fig.~\ref{fig:friction-identification})} & On & 3.2 &  & 0.6 & 3.4 & 7.2 &  & 0.35 &  & 0.014 & 0.076 & 0.44 &  & 7.7 &  \\
 & Off & 5 & $(1.6\times)$ & 0.63 & 3.7 & 9.3 & $(1.3\times)$ & 0.55 & $(1.6\times)$ & 0.015 & 0.082 & 0.64 & $(1.5\times)$ & 9.9 & $(1.3\times)$ \\
\specialrule{\lightrulewidth}{0pt}{0pt}
\multirow{2}{*}{Cloth rest scale (Fig.~\ref{fig:dress10})} & On & 13 &  & 0.38 & 1.5 & 15 &  & 0.48 &  & 0.00087 & 0.004 & 0.48 &  & 16 &  \\
 & Off & 18 & $(1.4\times)$ & 0.4 & 1.6 & 20 & $(1.3\times)$ & 0.64 & $(1.3\times)$ & 0.00093 & 0.0038 & 0.64 & $(1.3\times)$ & 21 & $(1.3\times)$ \\
\specialrule{\lightrulewidth}{0pt}{0pt}
\multirow{2}{*}{Grasp angle (Fig.~\ref{fig:xarm-target-closeup})} & On & 0.34 &  & 0.044 & 0.47 & 0.85 &  & 0.058 &  & 0.0073 & 0.11 & 0.18 &  & 1 &  \\
 & Off & 0.51 & $(1.5\times)$ & 0.045 & 0.48 & 1 & $(1.2\times)$ & 0.084 & $(1.4\times)$ & 0.0074 & 0.11 & 0.21 & $(1.2\times)$ & 1.2 & $(1.2\times)$ \\
\specialrule{\lightrulewidth}{0pt}{0pt}
\multirow{2}{*}{Cup catching (Fig.~\ref{fig:xarm-incoming-cup})} & On & 2.3 &  & 0.17 & 5.4 & 7.9 &  & 0.68 &  & 0.009 & 0.33 & 1 &  & 8.9 &  \\
 & Off & 3.7 & $(1.6\times)$ & 0.13 & 5.5 & 9.3 & $(1.2\times)$ & 1 & $(1.5\times)$ & 0.0099 & 0.33 & 1.3 & $(1.3\times)$ & 11 & $(1.2\times)$ \\
\specialrule{\lightrulewidth}{0pt}{0pt}
\multirow{2}{*}{Dual-arm grasp (Fig.~\ref{fig:dual-piper})} & On & 3.4 &  & 0.24 & 2.9 & 6.5 &  & 0.9 &  & 0.024 & 0.78 & 1.7 &  & 8.2 &  \\
 & Off & 5 & $(1.5\times)$ & 0.21 & 2.8 & 7.9 & $(1.2\times)$ & 1.3 & $(1.5\times)$ & 0.024 & 0.79 & 2.1 & $(1.2\times)$ & 10 & $(1.2\times)$ \\
\bottomrule\end{tabularx}
\end{table*}

\subsection{\texorpdfstring{{GPU execution and optimization cost}}{GPU execution and optimization cost}}
\label{sec:optimization-timing}

{This section measures the speedup from CUDA Graphs
and the time spent on forward simulation and backward differentiation
during optimization.} Conditional CUDA Graphs reduce coordination during PCG, while parallel assembly and derivative accumulation process the local work surrounding each solve.
{Table~\ref{tab:eleven-phase-times} gives a
detailed breakdown of the time spent on forward simulation and backward
differentiation, together with the evaluation counts and accumulated
evaluation time for each optimization.
Table~\ref{tab:graph-on-off-selected-comparison} compares execution
with CUDA Graphs enabled and disabled, reporting PCG and total
evaluation times for forward simulation and backward differentiation,
together with the resulting speedups.}

{CUDA Graph execution provides
$1.4$--$1.8\times$ speedups in forward PCG and
$1.3$--$1.8\times$ in backward PCG over conventional kernel launches
(Table~\ref{tab:graph-on-off-selected-comparison}). The PCG recurrence
and convergence criterion are unchanged. These gains demonstrate the
benefit of reducing kernel-launch overhead and CPU--GPU synchronization
within the iterative solve (Section~\ref{sec:parallel-solve}).

PCG is the largest contributor to backward time in most tasks
(Table~\ref{tab:eleven-phase-times}), making the adjoint solver a major
source of savings in gradient evaluation. The acceleration remains
visible in the complete computation: combined forward and backward
evaluations achieve $1.2$--$1.5\times$ speedups across all reported
tasks. These timings include collision detection, Hessian assembly,
line search, and derivative accumulation, demonstrating a reduction in
the cost of simulation and differentiation beyond the PCG loop itself.

CUDA Graphs also accelerate the additional forward simulations used to
evaluate candidate parameters and verify the optimized motion.
Grasp-angle optimization, for example,
uses $44$ forward evaluations and $20$ backward evaluations
(Table~\ref{tab:eleven-phase-times}). The same graph-based forward
solver serves both the trajectories used for gradient computation and
the additional objective checks and verification runs. Together with
the representation study, these results establish complementary
benefits: compact coordinates reduce the numerical workload, and
graph-based execution reduces the recurring overhead of solving these
systems throughout optimization.}

\section{\texorpdfstring{Conclusion and Discussion}{Conclusion and Discussion}}
\label{sec:conclusion-discussion}

DiffGIPC combines mixed-coordinate trajectory adjoints with GPU execution for
contact-rich inverse problems. Full-space FEM and twelve-DoF affine bodies
remain coupled through contact, friction, and joints, while affine shape and
control derivatives expose these interactions to optimization. Parallel
accumulation, CUDA Graph-based PCG, and a PyTorch interface provide the GPU
backward and task-level pipeline.
Across the shared full-space FEM benchmarks, complete backward evaluation is
{$12.3$--$34.1\times$} faster than DiffIPC, with {$10.6$--$60.9\times$}
combined forward-and-backward speedups. Affine coordinates provide further
savings when near-rigid motion is appropriate, although surface resolution
still determines contact and assembly work.

Differentiation uses the contact pairs, closest-feature choices, and control operations recorded during forward simulation. A pair of objects that remains outside the barrier's activation range contributes no direct contact gradient, which can limit the guidance available when the initial trajectory does not establish the desired interaction.
{Rest-shape} derivatives assume fixed connectivity within each differentiated trajectory. Remeshing can occur between optimization steps, but its decisions and vertex transfers are not differentiated. 
Improving the underlying contact model and the gradients used for optimization offers a direction for future work. Geometric Contact Potential~\cite{huang2025geometriccontact} could reduce spurious contact forces and sensitivity to surface discretization. Long-range contact potentials~\cite{ye2026longrange} could provide gradient information before objects enter the local barrier range, helping optimization establish new contacts. Randomized smoothing with bundled gradients~\cite{suh2022bundled} could mitigate rapid changes in contact gradients. Adapting these approaches to coupled FEM-affine dynamics and evaluating their effects on simulation accuracy, optimization convergence, and computational cost remain future work.
% Keep all main-text floats before the references, without stopping at each section.
\FloatBarrier
\bibliographystyle{ACM-Reference-Format}
\bibliography{references}
% The appendix follows the references in reading order, without forcing an empty column.
\appendix

\section{Derivatives for the Coupled FEM--Affine Adjoint}
\label{sec:method-details}

{The frame-wise adjoint requires local derivatives
for assembling the coupled Hessian and accumulating parameter and
earlier-state gradients
(Sections~\ref{sec:adjoint}--\ref{sec:parallel-accumulation}).
This appendix derives these quantities for friction, elasticity,
inertia, and joints. The derivations account for current and previous
contact geometry, the dependence of mass and elastic response on
rest shape, and the effects of joint geometry and control parameters.}

\subsection{Friction derivatives}
\label{sec:friction-details}

\subsubsection{Smoothed friction energy}
{The friction energy uses current-step sliding and
contact quantities computed from previous-frame geometry \cite{Li2020IPC}.
For contact $c$, the previous positions $\vx_{t-1}^c$ determine the
closest-point interpolation weights, contact plane, and normal force.} Let
$\vx_j^-=\vx_{t-1,i_j}$ and
$\Delta\vx_j=\vx_{t,i_j}-\vx_j^-$ denote the previous position and
displacement of contact vertex $j$.

The matrix $\mathbf{W}\in\R^{3\times3n_c}$ converts stacked vertex
displacements into the relative displacement of the two closest points:
\[
 \mathbf{W}=
 [w_1^{\mathrm{fr}}\mathbf{I}_3\ \cdots\
  w_{n_c}^{\mathrm{fr}}\mathbf{I}_3],
 \qquad
 \mathbf{W}\,(\vx_t^c-\vx_{t-1}^c)
 =\sum_{j=1}^{n_c}w_j^{\mathrm{fr}}\Delta\vx_j.
\]
Adding the same displacement to all contact vertices leaves their relative displacement unchanged because the sum of the weights is zero. With
$\boldsymbol w^{\mathrm{fr}}=(w_1^{\mathrm{fr}},\ldots,w_{n_c}^{\mathrm{fr}})$,
the weights for the four mesh-contact types of IPC are
\[
 \boldsymbol w^{\mathrm{fr}}=
 \begin{cases}
 (1,-1),&\text{point--point},\\
 (1,\eta-1,-\eta),&\text{point--edge},\\
 (1,-\beta_1,-\beta_2,-\beta_3),&\text{point--triangle},\\
 (1-s_A,s_A,s_B-1,-s_B),&\text{edge--edge}.
 \end{cases}
\]
The point is numbered first for point--edge and point--triangle contact. For
point--edge contact, the other side has endpoints $2,3$, and its closest
point is $\vx_2^-+\eta\mathbf e$, where
$\mathbf e=\vx_3^--\vx_2^-$ and
$\eta=\mathbf e^{\T}(\vx_1^--\vx_2^-)/(\mathbf e^{\T}\mathbf e)$.
For point--triangle contact, $\beta_1,\beta_2,\beta_3$ are the barycentric
coordinates of the closest point on triangle $(2,3,4)$:
\[
 \begin{aligned}
 \mathbf{E}_{\triangle}
   &=[\,\vx_3^--\vx_2^-\quad\vx_4^--\vx_2^-\,],\\
 \begin{bmatrix}\beta_2\\\beta_3\end{bmatrix}
   &=(\mathbf{E}_{\triangle}^{\T}\mathbf{E}_{\triangle})^{-1}
     \mathbf{E}_{\triangle}^{\T}(\vx_1^--\vx_2^-),
 \qquad \beta_1=1-\beta_2-\beta_3.
 \end{aligned}
\]
For edge--edge contact, the two edges have endpoints $(1,2)$ and $(3,4)$, with closest
points $\vx_1^-+s_A\mathbf e_A$ and $\vx_3^-+s_B\mathbf e_B$, where
$\mathbf e_A=\vx_2^--\vx_1^-$ and $\mathbf e_B=\vx_4^--\vx_3^-$. The
coefficients satisfy
\[
 \begin{bmatrix}
  \mathbf e_A^{\T}\mathbf e_A&-\mathbf e_A^{\T}\mathbf e_B\\
  -\mathbf e_A^{\T}\mathbf e_B&\mathbf e_B^{\T}\mathbf e_B
 \end{bmatrix}
 \begin{bmatrix}s_A\\s_B\end{bmatrix}
 =
 \begin{bmatrix}
  \mathbf e_A^{\T}(\vx_3^--\vx_1^-)\\
  -\mathbf e_B^{\T}(\vx_3^--\vx_1^-)
 \end{bmatrix}.
\]
The edge and triangle formulas apply when the closest points lie in
the interiors of the respective features. When a closest point lies
at an edge endpoint or on a triangle boundary, the contact uses the
corresponding point--point or point--edge formula.

The columns of $\mathbf{U}\in\R^{3\times2}$ are two orthogonal unit
directions in the contact plane. For mesh--mesh contact, the unit normal is
$\mathbf n=\mathbf{W}\vx_{t-1}^c/\|\mathbf{W}\vx_{t-1}^c\|$. An explicit
construction using a Cartesian unit vector $\mathbf e_{\mathrm{ref}}$ not
parallel to $\mathbf n$ is
\[
 \begin{aligned}
 \mathbf t_1
 &=\frac{(\mathbf{I}_3-\mathbf n\mathbf n^{\T})\mathbf e_{\mathrm{ref}}}
         {\|(\mathbf{I}_3-\mathbf n\mathbf n^{\T})\mathbf e_{\mathrm{ref}}\|},
 &\mathbf t_2&=\mathbf t_1\times\mathbf n,\\
 \mathbf{U}&=[\,\mathbf t_1\quad\mathbf t_2\,],
 &\mathbf{U}^{\T}\mathbf{U}&=\mathbf{I}_2.
 \end{aligned}
\]
Any orthonormal basis of this plane gives the same isotropic friction energy
and world-space gradient, since
$\mathbf{U}\mathbf{U}^{\T}=\mathbf{I}_3-\mathbf n\mathbf n^{\T}$.
For the fixed horizontal ground, the normal is fixed and only the
moving vertex is included in $\vx_t^c$. In this case,
$\mathbf{W}=\mathbf{I}_3$ and
$\mathbf{U}=[\,\mathbf e_x\quad\mathbf e_z\,]$, with
$\mathbf e_x=(1,0,0)^{\T}$ and $\mathbf e_z=(0,0,1)^{\T}$.

For mesh--mesh contact, $\mathbf{W}$ and $\mathbf{U}$ depend on
$\vx_{t-1}^c$. Both remain fixed when differentiating with respect
to current positions. For the
stacked displacement $\Delta\vx_t^c=\vx_t^c-\vx_{t-1}^c$, the sliding
displacement is
\begin{equation}
 \boldsymbol\xi=\mathbf{U}^{\T}\mathbf{W}\,\Delta\vx_t^c,
 \qquad \sigma=\|\boldsymbol\xi\|^2.
 \label{seq:friction-displacement}
\end{equation}
Here $\boldsymbol\xi\in\R^2$ contains the two in-plane displacement
components, and $\sigma$ is its squared length.

The friction energy is $E_{t,c}^{\mathrm{fr}}={\mu_{\mathrm{fr}}} f_n\varphi(\sigma)$, where
${\mu_{\mathrm{fr}}}$ is the friction coefficient.
{The normal-force magnitude $f_n$ is determined by
the barrier in Equation~\eqref{eq:rank2-barrier}. For
$0<d<\widehat d$, let $\Delta_d=d-\widehat d$ and
$\zeta_d=\log(d/\widehat d)$. The first and second barrier derivatives are}
\begin{align}
 b_{\mathrm{bar}}'(d)&=2\kappa\Delta_d\zeta_d
                         (\zeta_d+\Delta_d/d),\notag\\
 b_{\mathrm{bar}}''(d)&=2\kappa\left[\zeta_d^2
       +4\Delta_d\zeta_d/d+\Delta_d^2(1-\zeta_d)/d^2\right].
 \label{seq:barrier-derivatives}
\end{align}
{At the previous squared contact distance
$d^-$, the normal-force magnitude is
$f_n=-2\sqrt{d^-}\,b_{\mathrm{bar}}'(d^-)$.
It is held fixed during the current-frame solve. Its dependence on
the previous positions is included in the history derivative.}
{The unsmoothed friction energy is proportional to
$\|\boldsymbol{\xi}\|$, which is not differentiable at zero sliding
displacement. IPC uses smoothed friction to avoid discontinuities. With
$\epsilon_{\mathrm{fr}}=h\widehat d_f$, where $\widehat d_f$ is the
friction-smoothing velocity threshold, the smooth function for
mesh--mesh contact is}
\begin{equation}
 \varphi_{\mathrm{self}}(\sigma)=\begin{cases}
 \dfrac{\sigma(-\sqrt\sigma/3+\epsilon_{\mathrm{fr}})}
       {\epsilon_{\mathrm{fr}}^2}+\dfrac{\epsilon_{\mathrm{fr}}}{3},
       &\sigma\leq\epsilon_{\mathrm{fr}}^2,\\
 \sqrt\sigma,&\sigma>\epsilon_{\mathrm{fr}}^2.
 \end{cases}
 \label{seq:self-friction-law}
\end{equation}
For ground friction, the smoothed function is
\begin{equation}
 \varphi_{\mathrm{ground}}(\sigma)=\begin{cases}
 \sigma/(2\epsilon_{\mathrm{fr}}),&\sigma\leq\epsilon_{\mathrm{fr}}^2,\\
 \sqrt\sigma-\epsilon_{\mathrm{fr}}/2,&\sigma>\epsilon_{\mathrm{fr}}^2.
 \end{cases}
 \label{seq:ground-friction-law}
\end{equation}

The required energy gradient is with respect to current vertex positions
$\vx_t^c$. Since $\mathbf{W}$, $\mathbf{U}$, and $f_n$ are fixed for this
derivative, $\partial\boldsymbol\xi/\partial\vx_t^c=\mathbf{U}^{\T}\mathbf{W}$.
The chain rule gives
\begin{equation}
 \begin{aligned}
 \vg_{t,c}^{\mathrm{fr},x}
 &=\nabla_{\vx_t^c}E_{t,c}^{\mathrm{fr}}\\
 &={\mu_{\mathrm{fr}}} f_n
   \left(\frac{\partial\boldsymbol\xi}{\partial\vx_t^c}\right)^{\T}
   \nabla_{\boldsymbol\xi}\varphi(\|\boldsymbol\xi\|^2)\\
 &={\mu_{\mathrm{fr}}} f_n\mathbf{W}^{\T}\mathbf{U}
       \bigl[2\varphi'(\sigma)\boldsymbol\xi\bigr].
 \end{aligned}
 \label{seq:friction-residual}
\end{equation}
The vector $2\varphi'(\sigma)\boldsymbol\xi$ is the
gradient with respect to the two-dimensional sliding displacement.
For mesh--mesh friction below the smoothing threshold,
\begin{equation}
 \nabla_{\boldsymbol\xi}\varphi_{\mathrm{self}}
             (\|\boldsymbol\xi\|^2)
 =\frac{2\epsilon_{\mathrm{fr}}-\|\boldsymbol\xi\|}
        {\epsilon_{\mathrm{fr}}^2}\boldsymbol\xi.
 \label{seq:small-slip-gradient}
\end{equation}
This slip gradient is mapped to vertex positions by the
factor ${\mu_{\mathrm{fr}}} f_n\mathbf{W}^{\T}\mathbf{U}$ in
Equation~\eqref{seq:friction-residual}. Above the threshold, the slip
gradient is $\boldsymbol\xi/\|\boldsymbol\xi\|$ for both laws.
For ground friction below the threshold, it is
$\boldsymbol\xi/\epsilon_{\mathrm{fr}}$.

Within each smooth branch, differentiating again gives the Hessian:
\[
 \begin{aligned}
 \mathbf{H}_{\xi}
 &=2\varphi'(\sigma)\mathbf{I}_2
   +4\varphi''(\sigma)\boldsymbol\xi\boldsymbol\xi^{\T},\\
 \mathbf{H}_{t,c}^{\mathrm{fr},x}
 &={\mu_{\mathrm{fr}}} f_n\mathbf{W}^{\T}\mathbf{U}
   \mathbf{H}_{\xi}\mathbf{U}^{\T}\mathbf{W}.
 \end{aligned}
\]
At zero slip, the gradient is zero and the limiting $\mathbf{H}_{\xi}$ is
$2\mathbf{I}_2/\epsilon_{\mathrm{fr}}$ for mesh--mesh friction or
$\mathbf{I}_2/\epsilon_{\mathrm{fr}}$ for ground friction.
{The dependence of $\mathbf{W}$, $\mathbf{U}$, and $f_n$
on previous-frame positions supplies the temporal coupling in
Equation~\eqref{eq:friction-history-gradient}.}

\subsubsection{Friction-coefficient and history derivatives}
{Point-space friction gradient supports
coefficient fitting and differentiation through time. Its coefficient
derivative measures how frictional resistance changes with
$\mu_{\mathrm{fr}}$. Its history derivative also accounts for changes
in closest-point weights, contact-plane orientation, normal force,
and relative displacement.} Let
$\mathbf g_{\xi}=2\varphi'(\sigma)\boldsymbol\xi$ denote the two-dimensional
slip derivative in Equation~\eqref{seq:friction-residual}.
{At fixed frame configurations,
$\mathbf{W}$, $\mathbf{U}$, and $f_n$ are independent of
$\mu_{\mathrm{fr}}$. The coefficient derivative and its loss-gradient
contribution $\Delta\bar{\mu}_{\mathrm{fr}}^{(t,c)}$ from contact $c$ at
frame $t$ are}
{
\begin{equation}
 \begin{aligned}
 \frac{\partial\vg_{t,c}^{\mathrm{fr},x}}{\partial\mu_{\mathrm{fr}}}
 &=f_n\mathbf{W}^{\T}\mathbf{U}\mathbf g_{\xi},\\
 \Delta\bar{\mu}_{\mathrm{fr}}^{(t,c)}&=-
 \left(\frac{\partial\vg_{t,c}^{\mathrm{fr},x}}{\partial\mu_{\mathrm{fr}}}\right)^{\T}
 \mathbf{K}_{t,c}\vlambda_t.
 \end{aligned}
 \label{seq:friction-coefficient}
\end{equation}
}
{The contributions
$\Delta\bar{\mu}_{\mathrm{fr}}^{(t,c)}$ are summed over all contacts and
frames that share $\mu_{\mathrm{fr}}$. Together with the direct loss
derivative, this sum gives the total coefficient gradient
$\bar{\mu}_{\mathrm{fr}}$ in
Equation~\eqref{eq:dynamic-parameter-gradient}.}

A change in $\vx_{t-1}^c$ changes both the measured displacement and the
geometry defining friction. The corresponding matrix is
\begin{equation}
 \mathbf{C}_{t,c}^{\mathrm{fr},x}
 =\frac{\partial\vg_{t,c}^{\mathrm{fr},x}}
        {\partial\vx_{t-1}^c}
 \in\R^{3n_c\times3n_c}.
 \label{seq:friction-history-matrix}
\end{equation}
With the contact vertex indices and contact type fixed, vary
$\vx_{t-1}^c$ while holding $\vx_t^c$,
${\mu_{\mathrm{fr}}}$, and
$\epsilon_{\mathrm{fr}}$ fixed. Since
$\mathrm{d}(\Delta\vx_t^c)=-\mathrm{d}\vx_{t-1}^c$,
Equation~\eqref{seq:friction-displacement} gives
\[
 \begin{aligned}
 \mathrm{d}\boldsymbol\xi
 &=(\mathrm{d}\mathbf{U})^{\T}\mathbf{W}\Delta\vx_t^c
   +\mathbf{U}^{\T}(\mathrm{d}\mathbf{W})\Delta\vx_t^c\\
 &\quad-\mathbf{U}^{\T}\mathbf{W}\,\mathrm{d}\vx_{t-1}^c.
 \end{aligned}
\]
The first two terms account for changes in the contact-plane basis and
closest-point weights. The last accounts for the previous positions in
the displacement difference. Applying the product rule to
Equation~\eqref{seq:friction-residual} yields
\[
 \begin{aligned}
 \mathbf{C}_{t,c}^{\mathrm{fr},x}\,\mathrm{d}\vx_{t-1}^c
 &={\mu_{\mathrm{fr}}}\,(\mathrm{d}f_n)\mathbf{W}^{\T}\mathbf{U}\mathbf g_{\xi}\\
 &\quad+{\mu_{\mathrm{fr}}} f_n(\mathrm{d}\mathbf{W})^{\T}\mathbf{U}\mathbf g_{\xi}\\
 &\quad+{\mu_{\mathrm{fr}}} f_n\mathbf{W}^{\T}(\mathrm{d}\mathbf{U})\mathbf g_{\xi}\\
 &\quad+{\mu_{\mathrm{fr}}} f_n\mathbf{W}^{\T}\mathbf{U}
               \mathbf{H}_{\xi}\,\mathrm{d}\boldsymbol\xi .
 \end{aligned}
\]
The weight differential is
$\mathrm{d}\mathbf{W}
=[(\mathrm{d}w_1^{\mathrm{fr}})\mathbf{I}_3\ \cdots\
  (\mathrm{d}w_{n_c}^{\mathrm{fr}})\mathbf{I}_3]$.
For point--point contact, the weights are constant, so
$\mathrm{d}\mathbf{W}=0$.
{
For the remaining contact types, the interpolation weights depend
on the previous-frame vertex positions
through the closest-point parameters. These parameters
locate the closest points on the edges or triangle and form the vector
\[
 \boldsymbol\omega_{\mathrm{cp}}=
 \begin{cases}
  [\,\eta\,],&\text{point--edge},\\
  [\,\beta_2\quad\beta_3\,]^{\T},&\text{point--triangle},\\
  [\,s_A\quad s_B\,]^{\T},&\text{edge--edge}.
 \end{cases}
\]
The vector $\boldsymbol\omega_{\mathrm{cp}}$ is determined by the linear system
\[
 \mathbf{D}_{\mathrm{cp}}\boldsymbol\omega_{\mathrm{cp}}
 =\mathbf r_{\mathrm{cp}}.
\]
Here $\mathbf{D}_{\mathrm{cp}}$ is the coefficient matrix and
$\mathbf r_{\mathrm{cp}}$ is the right-hand side, both determined
by the previous-frame vertex positions.

For point--edge contact, the closest point on edge
$(2,3)$ is $\vx_2^-+\eta\mathbf e$, with
$\mathbf e=\vx_3^--\vx_2^-$. At an interior closest point, the
vector $\vx_1^--(\vx_2^-+\eta\mathbf e)$ is perpendicular to the
edge direction $\mathbf e$, giving
\[
 \mathbf{D}_{\mathrm{cp}}=\mathbf e^{\T}\mathbf e,
 \qquad
 \mathbf r_{\mathrm{cp}}=\mathbf e^{\T}(\vx_1^--\vx_2^-).
\]

For point--triangle contact, $\beta_2,\beta_3$ are the two independent
barycentric coordinates of the closest point on triangle $(2,3,4)$,
with $\beta_1=1-\beta_2-\beta_3$. The matrix
\[
 \mathbf{E}_{\triangle}
 =[\,\vx_3^--\vx_2^-\quad\vx_4^--\vx_2^-\,]
 \in\R^{3\times2}
\]
has as its columns the two triangle edges starting at vertex $2$.
The closest point is
$\vx_2^-+\mathbf{E}_{\triangle}\boldsymbol\omega_{\mathrm{cp}}$.
For a closest point inside the triangle, the vector
$\vx_1^--(\vx_2^-+\mathbf{E}_{\triangle}\boldsymbol\omega_{\mathrm{cp}})$
is perpendicular to both triangle edges, giving
\[
 \mathbf{D}_{\mathrm{cp}}=\mathbf{E}_{\triangle}^{\T}\mathbf{E}_{\triangle},
 \qquad
 \mathbf r_{\mathrm{cp}}=\mathbf{E}_{\triangle}^{\T}(\vx_1^--\vx_2^-).
\]

For edge--edge contact, $s_A,s_B$ locate the closest points
along edges $(1,2)$ and $(3,4)$.
With $\mathbf e_A=\vx_2^--\vx_1^-$ and
$\mathbf e_B=\vx_4^--\vx_3^-$, the closest points are
$\vx_1^-+s_A\mathbf e_A$ and $\vx_3^-+s_B\mathbf e_B$.
For closest points in the interiors of both edges, the vector
$\vx_1^--\vx_3^-+s_A\mathbf e_A-s_B\mathbf e_B$ connecting them
is perpendicular to both edge directions, yielding
\[
 \begin{aligned}
 \mathbf{D}_{\mathrm{cp}}
 &=\begin{bmatrix}
   \mathbf e_A^{\T}\mathbf e_A&-\mathbf e_A^{\T}\mathbf e_B\\
   -\mathbf e_A^{\T}\mathbf e_B&\mathbf e_B^{\T}\mathbf e_B
  \end{bmatrix},\\
 \mathbf r_{\mathrm{cp}}
 &=\begin{bmatrix}
   \mathbf e_A^{\T}(\vx_3^--\vx_1^-)\\
   -\mathbf e_B^{\T}(\vx_3^--\vx_1^-)
  \end{bmatrix}.
 \end{aligned}
\]

Differentiating the linear system
$\mathbf{D}_{\mathrm{cp}}\boldsymbol\omega_{\mathrm{cp}}
=\mathbf r_{\mathrm{cp}}$ with respect to the previous-frame
vertex positions gives
\[
 \mathrm{d}\boldsymbol\omega_{\mathrm{cp}}
 =\mathbf{D}_{\mathrm{cp}}^{-1}
  \bigl(\mathrm{d}\mathbf r_{\mathrm{cp}}
       -(\mathrm{d}\mathbf{D}_{\mathrm{cp}})
        \boldsymbol\omega_{\mathrm{cp}}\bigr).
\]
The product rule applied to the explicit expressions above gives
$\mathrm{d}\mathbf{D}_{\mathrm{cp}}$ and $\mathrm{d}\mathbf r_{\mathrm{cp}}$
from the vertex variations. The edge--edge coordinate variations
$\mathrm{d}s_A$ and $\mathrm{d}s_B$ include the effects of changes
in both edge directions and their relative positions. The resulting
weight variations are
\[
 \mathrm{d}\boldsymbol w^{\mathrm{fr}}=
 \begin{cases}
  (0,\mathrm{d}\eta,-\mathrm{d}\eta),&\text{point--edge},\\
  (0,\mathrm{d}\beta_2+\mathrm{d}\beta_3,-\mathrm{d}\beta_2,-\mathrm{d}\beta_3),
    &\text{point--triangle},\\
  (-\mathrm{d}s_A,\mathrm{d}s_A,\mathrm{d}s_B,-\mathrm{d}s_B),
    &\text{edge--edge},
 \end{cases}
\]
where $\mathrm{d}\beta_1=-\mathrm{d}\beta_2-\mathrm{d}\beta_3$
has been substituted in the triangle case. These entries determine
$\mathrm{d}\mathbf W$ through its block definition above.
}

For mesh--mesh contact, let
$\mathbf r_{\mathrm{gap}}=\mathbf{W}\vx_{t-1}^c$ be the vector
between the closest points and
$\mathbf t_{\mathrm{ref}}=(\mathbf{I}_3-\mathbf n\mathbf n^{\T})
\mathbf e_{\mathrm{ref}}$ the reference direction projected into
the contact plane. Holding $\mathbf e_{\mathrm{ref}}$ fixed,
differentiation of the normalized directions gives
\[
 \begin{aligned}
 \mathrm{d}\mathbf r_{\mathrm{gap}}
  &=(\mathrm{d}\mathbf{W})\vx_{t-1}^c
      +\mathbf{W}\,\mathrm{d}\vx_{t-1}^c,\\
 \mathrm{d}\mathbf n
  &=\frac{\mathbf{I}_3-\mathbf n\mathbf n^{\T}}
         {\|\mathbf r_{\mathrm{gap}}\|}
      \mathrm{d}\mathbf r_{\mathrm{gap}},\\
 \mathrm{d}\mathbf t_{\mathrm{ref}}
  &=-\bigl((\mathrm{d}\mathbf n)\mathbf n^{\T}
          +\mathbf n(\mathrm{d}\mathbf n)^{\T}\bigr)
      \mathbf e_{\mathrm{ref}},\\
 \mathrm{d}\mathbf t_1
  &=\frac{\mathbf{I}_3-\mathbf t_1\mathbf t_1^{\T}}
         {\|\mathbf t_{\mathrm{ref}}\|}
      \mathrm{d}\mathbf t_{\mathrm{ref}},\\
 \mathrm{d}\mathbf t_2
  &=(\mathrm{d}\mathbf t_1)\times\mathbf n
      +\mathbf t_1\times\mathrm{d}\mathbf n,\\
 \mathrm{d}\mathbf{U}
  &=[\,\mathrm{d}\mathbf t_1\quad\mathrm{d}\mathbf t_2\,].
 \end{aligned}
\]
These expressions supply the weight and contact-plane derivatives
required by the history derivative.
{Differentiating the normal force using
Equation~\eqref{seq:barrier-derivatives} gives}
\[
 \mathrm{d}f_n=
 \left[-\frac{b_{\mathrm{bar}}'(d^-)}{\sqrt{d^-}}
       -2\sqrt{d^-}\,b_{\mathrm{bar}}''(d^-)\right]\mathrm{d}d^-.
\]
For the fixed ground plane, $\mathrm{d}\mathbf{W}$ and
$\mathrm{d}\mathbf{U}$ vanish. Its history derivative includes changes in
sliding displacement and in the normal force set by the previous
vertex-to-plane distance.

{The mapped derivative
$\mathbf{K}_{t,c}^{\T}\mathbf{C}_{t,c}^{\mathrm{fr},x}\mathbf{K}_{t-1,c}$
places FEM and affine contact points in the same temporal coupling.
Its transposed product contributes to $\mathbf b_{t-1}$ through
Equation~\eqref{eq:friction-history-gradient}, carrying the effect of
later friction back to earlier motion DoFs. Second-frame friction
therefore contributes to initial-state gradients through the first-frame
adjoint. At $t=1$, the previous positions are initial inputs and their
parameter dependence contributes directly to the initial-state gradient.}

\subsubsection{Rest-shape derivatives under friction}
\label{sec:friction-rest-details}
{With solved motion DoFs fixed, an affine rest-shape
change moves contact points at both the current and previous frames.
The direct derivative therefore includes both configurations and the
map from point forces to affine DoFs. These derivatives hold
$\vy_1,\ldots,\vy_T$, the friction coefficient, and the smoothing
threshold fixed. The frame adjoints supply the complementary change
in simulated motion.}

{For a surface vertex $i$ on affine body $b$,
let $\theta_k=\bar x_{i,a}$ be one component of its rest position
$\bar\vx_i$, where $a=1,2,3$ indexes the $x$, $y$, and $z$
components, respectively. The vector $\mathbf e_a$ is the unit
vector along the selected coordinate axis.}
Equation~\eqref{eq:abd-map} gives
$\partial\vx_{s,i}/\partial\theta_k=\mathbf{A}_{s,b}\mathbf e_a$
at both $s=t$ and $s=t-1$. Thus the same rest-coordinate change moves current
and previous contact points through their respective affine transformations.
The point map in Equation~\eqref{seq:point-map} also changes:
\[
 \frac{\partial\mathbf{J}_i}{\partial\theta_k}
 =
 [\,\mathbf{0}_{3\times3}\quad
    \mathbf e_1\mathbf e_a^{\T}\quad
    \mathbf e_2\mathbf e_a^{\T}\quad
    \mathbf e_3\mathbf e_a^{\T}\,].
\]
For a contact containing vertex $i$, this block occupies the rows
of that vertex and the columns of body $b$ in
$\partial\mathbf{K}_{t,c}/\partial\theta_k$. For a shape parameter
controlling several rest coordinates, its derivative follows by applying
the chain rule through the rest-shape parameterization.

The induced change in the point-space friction gradient is
\begin{equation}
 \mathrm{d}\vg_{t,c}^{\mathrm{fr},x}
 =\mathbf{H}_{t,c}^{\mathrm{fr},x}\,\mathrm{d}\vx_t^c
  +\mathbf{C}_{t,c}^{\mathrm{fr},x}\,\mathrm{d}\vx_{t-1}^c.
 \label{eq:friction-rest-point-chain}
\end{equation}
The current-position term uses the Hessian obtained by differentiating
Equation~\eqref{seq:friction-residual}. The previous-position term uses
Equation~\eqref{seq:friction-history-matrix}, including the changes in
weights, basis, and normal force derived above. Combining these geometry
terms with the point-map derivative gives
Equation~\eqref{eq:affine-rest-friction-map}.
{For contact $c$ at frame $t$, the friction
contribution $\Delta\bar\theta_k^{(t,c),\mathrm{fr}}$ to the loss gradient is}
{
\[
 \begin{aligned}
 \Delta\bar\theta_k^{(t,c),\mathrm{fr}}={}&-
 (\vg_{t,c}^{\mathrm{fr},x})^{\T}
 \frac{\partial\mathbf{K}_{t,c}}{\partial\theta_k}\vlambda_t\\
 &-\left(\frac{\partial\vx_t^c}{\partial\theta_k}\right)^{\T}
   \mathbf{H}_{t,c}^{\mathrm{fr},x}\mathbf{K}_{t,c}\vlambda_t\\
 &-\left(\frac{\partial\vx_{t-1}^c}{\partial\theta_k}\right)^{\T}
   (\mathbf{C}_{t,c}^{\mathrm{fr},x})^{\T}
   \mathbf{K}_{t,c}\vlambda_t .
 \end{aligned}
\]
}
{The three terms account for the
force map and the contact geometry at both time levels. Their sum
over the relevant contacts and frames is added to the other physical
contributions and any direct loss derivative to obtain the total
rest-shape gradient $\bar\theta_k$.}

For free FEM vertices, the solved positions are the motion DoFs and
their point maps are identity matrices. The effect of rest shape on
simulated motion is propagated through the frame adjoints. Affine
surfaces also contribute the direct geometry and point-map derivatives
above. If rest shape determines the initial vertex positions, their
direct contribution is included in the first-frame friction derivatives
with respect to those initial positions.

\subsection{Volumetric FEM and cloth parameter derivatives}
\label{sec:fem-derivative-details}
{Element energies supply both the frame Hessian and the
parameter derivatives needed for material identification and
deformable rest-shape design. Volumetric elements use stable Neo-Hookean~\cite{smith2018stable}.
Cloth combines the finite-element Baraff--Witkin \cite{kim2020fembw} with StiffGIPC's cubic extension penalty. Differentiating these energies with
respect to material and rest-shape parameters supplies local gradient
contributions at fixed current positions. The coupled adjoint accounts
for the resulting deformation and its interactions with other objects.}

\subsubsection{Tetrahedral elasticity}
{For a tetrahedron with local vertex indices $0,1,2,3$,
define its rest and current edge matrices as}
\[
 \begin{aligned}
 \mathbf{D}_e^{\mathrm{rest}}
   &=[\,\bar\vx_1-\bar\vx_0\quad
        \bar\vx_2-\bar\vx_0\quad\bar\vx_3-\bar\vx_0\,],\\
 \mathbf{D}_e^{\mathrm{cur}}
   &=[\,\vx_1-\vx_0\quad\vx_2-\vx_0\quad\vx_3-\vx_0\,].
 \end{aligned}
\]
Its deformation gradient is
$\mathbf{F}_e=\mathbf{D}_e^{\mathrm{cur}}(\mathbf{D}_e^{\mathrm{rest}})^{-1}$,
and its rest volume is $V_e=|\det\mathbf{D}_e^{\mathrm{rest}}|/6$. Let
$J_e=\det\mathbf{F}_e$.
{The stable Neo-Hookean \cite{smith2018stable} is}
\[
 \psi_e=
 \frac{{\mu}}{2}(\|\mathbf{F}_e\|_F^2-3)
 +\frac{{\lambda}}{2}
  \left(J_e-1-\frac{{\mu}}{{\lambda}}\right)^2.
\]
The coefficients are per element, with element subscript suppressed. The
incremental-potential contribution is $h^2V_e\psi_e$.
{The determinant derivative is
$\partial J_e/\partial\mathbf{F}_e=\operatorname{cof}(\mathbf{F}_e)$,
where $\operatorname{cof}(\mathbf{F}_e)$ denotes the cofactor matrix.
Differentiating the energy density with respect to $\mathbf{F}_e$
gives the first Piola stress:}
\[
 \mathbf{P}_e
 =\frac{\partial\psi_e}{\partial\mathbf{F}_e}
 ={\mu}\mathbf{F}_e+
   [{\lambda}(J_e-1)-{\mu}]
     \operatorname{cof}(\mathbf{F}_e).
\]
Let $\operatorname{vec}$ stack matrix columns and
$\mathbf{H}_{F,e}=\partial\operatorname{vec}(\mathbf{P}_e)/
 \partial\operatorname{vec}(\mathbf{F}_e)$.
The same geometric Jacobian maps stress, constitutive Hessian, and material
derivatives into vertex DoFs:
\begin{equation}
 \begin{aligned}
 \mathbf{R}_e
   &=\frac{\partial\operatorname{vec}(\mathbf{F}_e)}{\partial\vx_e},
 &\vg_e&=h^2V_e\mathbf{R}_e^{\T}\operatorname{vec}(\mathbf{P}_e),\\
 \mathbf{H}_e
   &=h^2V_e\mathbf{R}_e^{\T}\mathbf{H}_{F,e}\mathbf{R}_e .
 \end{aligned}
 \label{seq:tet-residual}
\end{equation}
Here $\vx_e$ stacks the four current vertex positions and
$\mathbf{R}_e\in\R^{9\times12}$ is constant with respect to them. With
$\mathbf d_{e,a}=(\mathbf{D}_e^{\mathrm{rest}})^{-\T}\mathbf e_a$ for
$a=1,2,3$ and $\mathbf d_{e,0}=-\sum_{a=1}^3\mathbf d_{e,a}$,
$\mathbf{R}_e=[\,\mathbf d_{e,0}\otimes\mathbf{I}_3\ \cdots\
 \mathbf d_{e,3}\otimes\mathbf{I}_3\,]$, where $\otimes$ is the Kronecker
product.

{The constitutive Hessian follows by differentiating the stress:}
\[
 \begin{aligned}
 \mathrm{d}\mathbf{P}_e
 &={\mu}\,\mathrm{d}\mathbf{F}_e
   +{\lambda}(\mathrm{d}J_e)\operatorname{cof}(\mathbf{F}_e)\\
 &\quad+[{\lambda}(J_e-1)-{\mu}]\,
       \mathrm{d}\operatorname{cof}(\mathbf{F}_e),\\
 \mathrm{d}J_e&=\operatorname{cof}(\mathbf{F}_e):\mathrm{d}\mathbf{F}_e .
 \end{aligned}
\]
The colon is the Frobenius inner product. For
$\mathbf{F}_e=[\,\mathbf f_1\ \mathbf f_2\ \mathbf f_3\,]$, the cofactor
columns are $\mathbf f_2\times\mathbf f_3$,
$\mathbf f_3\times\mathbf f_1$, and $\mathbf f_1\times\mathbf f_2$. Their
differentials follow the cross-product rule, e.g.,
$\mathrm{d}(\mathbf f_2\times\mathbf f_3)
 =\mathrm{d}\mathbf f_2\times\mathbf f_3+
   \mathbf f_2\times\mathrm{d}\mathbf f_3$.
These expressions determine every entry of $\mathbf{H}_{F,e}$.

At fixed $\mathbf{F}_e$, the material derivatives are
\begin{equation}
 \begin{aligned}
 \frac{\partial\mathbf{P}_e}{\partial{\mu}}
    &=\mathbf{F}_e-\operatorname{cof}(\mathbf{F}_e),\\
 \frac{\partial\mathbf{P}_e}{\partial{\lambda}}
    &=(J_e-1)\operatorname{cof}(\mathbf{F}_e).
 \end{aligned}
 \label{seq:material-derivatives}
\end{equation}
Their contributions to $\partial\vg_e/\partial\theta_k$ are
$h^2V_e\mathbf{R}_e^{\T}
 \operatorname{vec}(\partial\mathbf{P}_e/\partial\theta_k)$.
{The material parameterization relates Young's modulus $E$ and Poisson's
ratio $\nu$ to the energy coefficients through the Lam\'e input parameters
$\mu_{\mathrm{Lam\acute{e}}}=E/[2(1+\nu)]$ and
$\lambda_{\mathrm{Lam\acute{e}}}=E\nu/[(1+\nu)(1-2\nu)]$, followed by
$\mu=\mu_{\mathrm{Lam\acute{e}}}$ and
$\lambda=\lambda_{\mathrm{Lam\acute{e}}}+\mu_{\mathrm{Lam\acute{e}}}$.}
Gradients with respect to $E$, $\nu$, or the optimizer variables in
Equation~\eqref{seq:material-task-map} follow through these maps.

{Rest-shape design changes both the reference geometry
and the elastic response.}
{Rest-shape differentiation accounts for
changes in the inverse rest-edge matrix
$(\mathbf{D}_e^{\mathrm{rest}})^{-1}$, the element rest volume $V_e$,
and the deformation-gradient Jacobian $\mathbf R_e$.}
{With current positions fixed,}
\begin{equation}
 \begin{aligned}
 \mathrm{d}(\mathbf{D}_e^{\mathrm{rest}})^{-1}
   &=-(\mathbf{D}_e^{\mathrm{rest}})^{-1}
      (\mathrm{d}\mathbf{D}_e^{\mathrm{rest}})
      (\mathbf{D}_e^{\mathrm{rest}})^{-1},\\
 \mathrm{d}\mathbf{F}_e
   &=\mathbf{D}_e^{\mathrm{cur}}\,
      \mathrm{d}(\mathbf{D}_e^{\mathrm{rest}})^{-1},\\
 \mathrm{d}V_e
   &=V_e\operatorname{tr}\!\left[
      (\mathbf{D}_e^{\mathrm{rest}})^{-1}
      \mathrm{d}\mathbf{D}_e^{\mathrm{rest}}\right].
 \end{aligned}
 \label{seq:rest-inverse}
\end{equation}
Differentiating the block expression for $\mathbf{R}_e$ gives
$\mathrm{d}\mathbf{R}_e$. The complete elastic contribution is
\[
 \begin{aligned}
 \mathrm{d}\vg_e=h^2\bigl[
 &(\mathrm{d}V_e)\mathbf{R}_e^{\T}\operatorname{vec}(\mathbf{P}_e)\\
 &+V_e(\mathrm{d}\mathbf{R}_e)^{\T}\operatorname{vec}(\mathbf{P}_e)\\
 &+V_e\mathbf{R}_e^{\T}\operatorname{vec}(\mathrm{d}\mathbf{P}_e)\bigr].
 \end{aligned}
\]
{The three terms account for variations in
the rest volume $V_e$, the deformation-gradient Jacobian
$\mathbf R_e$, and the first Piola stress $\mathbf P_e$, respectively.}
{The adjoint
converts their effect on the frame equations into loss gradients.
For stress-based objectives such as Loaded support
(Figure~\ref{fig:fem-rest-shape}), the loss's explicit dependence on
rest geometry contributes separately through $\boldsymbol\ell_\theta$.}

\subsubsection{Cloth membrane elasticity}
{
For a nondegenerate triangle with local vertices $0,1,2$, define the rest
edges $\bar{\mathbf e}_1=\bar\vx_1-\bar\vx_0$ and
$\bar{\mathbf e}_2=\bar\vx_2-\bar\vx_0$, and the current edge matrix
$\mathbf{D}_e^{\mathrm{cur}}=[\,\vx_1-\vx_0\quad\vx_2-\vx_0\,]
\in\R^{3\times2}$. Aligning the local horizontal axis with the rest edge
$\bar{\mathbf e}_1$ gives
\[
 \mathbf{D}_e^{\mathrm{rest}}=
 \begin{bmatrix}
  \|\bar{\mathbf e}_1\| &
  \dfrac{\bar{\mathbf e}_1^{\T}\bar{\mathbf e}_2}
        {\|\bar{\mathbf e}_1\|}\\[6pt]
  0 &
  \dfrac{\|\bar{\mathbf e}_1\times\bar{\mathbf e}_2\|}
        {\|\bar{\mathbf e}_1\|}
 \end{bmatrix}.
\]
The entries depend on edge
lengths and their included angle, so a rigid rotation of the rest
triangle leaves $\mathbf{D}_e^{\mathrm{rest}}$ unchanged.
At a fixed rest shape, this matrix remains constant
throughout the forward trajectory. The deformation gradient is
$\mathbf{F}_e=\mathbf{D}_e^{\mathrm{cur}}
(\mathbf{D}_e^{\mathrm{rest}})^{-1}
=[\,\mathbf f_1\quad\mathbf f_2\,]$.
Let $r_a=\|\mathbf f_a\|$ and $c_e=\mathbf f_1^{\T}\mathbf f_2$.
}
{The two local coordinate axes specify the
orthogonal warp and weft directions in the stretch and shear terms
of \citet{kim2020fembw}.} Including StiffGIPC's cubic extension
penalty~\cite{stiffgipc2025} gives
\[
 \begin{aligned}
 \psi_e^{\mathrm{mem}}
   &=\kappa_e^{\mathrm{str}}\,[s(r_1)+s(r_2)]
       +\kappa_e^{\mathrm{shr}}\,c_e^2,\\
 s(r)&=(r-1)^2+\eta_{\mathrm{sl},e}\langle r-1\rangle_+^3 .
 \end{aligned}
\]
Here $\langle z\rangle_+=\max(z,0)$ and $\eta_{\mathrm{sl},e}$ is the fixed
cubic-penalty coefficient. The cubic term adds resistance in extension, with
its derivatives included in the same stress and Hessian. The incremental potential is \(h^2 A_e \psi_e^{\mathrm{mem}}\), with \(A_e\) the rest area. The constitutive parameterization is
$\kappa_e^{\mathrm{str}}=E_{\mathrm{str}}\tau_e/(1-\nu_{\mathrm{str}}^2)$
and
$\kappa_e^{\mathrm{shr}}=E_{\mathrm{shr}}/[2(1+\nu_{\mathrm{shr}})]$.

Define $a(r)=s'(r)/r$, where
$s'(r)=2(r-1)+3\eta_{\mathrm{sl},e}\langle r-1\rangle_+^2$. Since
$\partial r_a/\partial\mathbf f_a=\mathbf f_a/r_a$ and
$\partial c_e/\partial\mathbf f_1=\mathbf f_2$, the stress columns are
\[
 \begin{aligned}
 \mathbf{P}_e(:,1)
 &=\kappa_e^{\mathrm{str}}a(r_1)\mathbf f_1
   +2\kappa_e^{\mathrm{shr}}c_e\mathbf f_2,\\
 \mathbf{P}_e(:,2)
 &=\kappa_e^{\mathrm{str}}a(r_2)\mathbf f_2
   +2\kappa_e^{\mathrm{shr}}c_e\mathbf f_1 .
 \end{aligned}
\]
For $a_a=a(r_a)$ and
$a'_a=s''(r_a)/r_a-s'(r_a)/r_a^2$, with
$s''(r)=2+6\eta_{\mathrm{sl},e}\langle r-1\rangle_+$, the four $3\times3$
blocks of $\mathbf{H}_{F,e}$ are
\[
 \begin{aligned}
 (\mathbf{H}_{F,e})_{11}
 &=\kappa_e^{\mathrm{str}}
    \left(a_1\mathbf{I}_3+\frac{a'_1}{r_1}\mathbf f_1\mathbf f_1^{\T}\right)
   +2\kappa_e^{\mathrm{shr}}\mathbf f_2\mathbf f_2^{\T},\\
 (\mathbf{H}_{F,e})_{22}
 &=\kappa_e^{\mathrm{str}}
    \left(a_2\mathbf{I}_3+\frac{a'_2}{r_2}\mathbf f_2\mathbf f_2^{\T}\right)
   +2\kappa_e^{\mathrm{shr}}\mathbf f_1\mathbf f_1^{\T},\\
 (\mathbf{H}_{F,e})_{12}
 &=2\kappa_e^{\mathrm{shr}}
      (\mathbf f_2\mathbf f_1^{\T}+c_e\mathbf{I}_3),\\
 (\mathbf{H}_{F,e})_{21}&=(\mathbf{H}_{F,e})_{12}^{\T}.
 \end{aligned}
\]
The parameter derivatives follow from the linear dependence of the stress on
the two stiffnesses:
\[
 \begin{aligned}
 \frac{\partial\mathbf{P}_e}{\partial\kappa_e^{\mathrm{str}}}
    &=[\,a_1\mathbf f_1\quad a_2\mathbf f_2\,],\\
 \frac{\partial\mathbf{P}_e}{\partial\kappa_e^{\mathrm{shr}}}
    &=2c_e[\,\mathbf f_2\quad\mathbf f_1\,].
 \end{aligned}
\]
{The assembly in Equation~\eqref{seq:tet-residual}
maps these constitutive derivatives to cloth vertex DoFs, using
$\mathbf{R}_e\in\R^{6\times9}$ and replacing $V_e$ by $A_e$.}

{
Rest-shape derivatives follow directly from the edge-based reference
coordinates. Let $L_{e,1}=\|\bar{\mathbf e}_1\|$ be the first edge
length, and let $(u_{e,2},v_{e,2})$ be vertex $2$'s local coordinates,
given by the second column of $\mathbf{D}_e^{\mathrm{rest}}$.
With rest area $A_e=\|\bar{\mathbf e}_1\times\bar{\mathbf e}_2\|/2$
and unit normal
$\bar{\mathbf n}=(\bar{\mathbf e}_1\times\bar{\mathbf e}_2)/(2A_e)$,
the geometric differentials are
\[
 \begin{aligned}
 \mathrm{d}L_{e,1}
  &=\bar{\mathbf e}_1^{\T}\mathrm{d}\bar{\mathbf e}_1/L_{e,1},\\
 \mathrm{d}A_e
  &=\tfrac12\bar{\mathbf n}^{\T}
   (\mathrm{d}\bar{\mathbf e}_1\times\bar{\mathbf e}_2
    +\bar{\mathbf e}_1\times\mathrm{d}\bar{\mathbf e}_2),\\
 \mathrm{d}u_{e,2}
  &=\frac{(\mathrm{d}\bar{\mathbf e}_1)^{\T}\bar{\mathbf e}_2
          +\bar{\mathbf e}_1^{\T}\mathrm{d}\bar{\mathbf e}_2}{L_{e,1}}
    -\frac{u_{e,2}}{L_{e,1}}\mathrm{d}L_{e,1},\\
 \mathrm{d}v_{e,2}
  &=\frac{2\,\mathrm{d}A_e}{L_{e,1}}
    -\frac{v_{e,2}}{L_{e,1}}\mathrm{d}L_{e,1}.
 \end{aligned}
\]
Consequently,
\[
 \mathrm{d}\mathbf{D}_e^{\mathrm{rest}}=
 \begin{bmatrix}
  \mathrm{d}L_{e,1} & \mathrm{d}u_{e,2}\\
  0 & \mathrm{d}v_{e,2}
 \end{bmatrix}.
\]
The inverse-matrix identity in Equation~\eqref{seq:rest-inverse}
gives $\mathrm{d}(\mathbf{D}_e^{\mathrm{rest}})^{-1}$.
At fixed current positions,
$\mathrm{d}\mathbf{F}_e=\mathbf{D}_e^{\mathrm{cur}}
\mathrm{d}(\mathbf{D}_e^{\mathrm{rest}})^{-1}$ changes the constitutive
stress, and the same inverse-map derivative changes $\mathbf{R}_e$.
$\mathrm{d}A_e$
supplies the integration-weight variation.
}

\subsubsection{Cloth bending derivatives}
{The quadratic bending model of
\citet{quadraticbending} uses a geometric matrix determined by the
adjacent triangle areas and cotangent weights. This matrix supplies
the position gradient, Hessian, and bending-stiffness derivative.
Its rest-shape derivative includes changes in both the area
normalization and the cotangent weights.} An interior
cloth edge defines a hinge with rest vertices
$(\bar\vx_0,\bar\vx_1,\bar\vx_2,\bar\vx_3)$, where $0,1$ are endpoints and
$2,3$ are opposite vertices. Let
$\bar{\mathbf e}_0=\bar\vx_1-\bar\vx_0$,
$\bar{\mathbf e}_1=\bar\vx_2-\bar\vx_0$, and
$\bar{\mathbf e}_2=\bar\vx_3-\bar\vx_0$.
{The adjacent rest areas are}
$A_1=\|\bar{\mathbf e}_0\times\bar{\mathbf e}_1\|/2$ and
$A_2=\|\bar{\mathbf e}_0\times\bar{\mathbf e}_2\|/2$. Using
$\operatorname{cot}(\mathbf a,\mathbf b)
=\mathbf a^{\T}\mathbf b/\|\mathbf a\times\mathbf b\|$, define
\[
 \begin{aligned}
 c_{01}&=\operatorname{cot}(\bar{\mathbf e}_0,\bar{\mathbf e}_1),
 &c_{02}&=\operatorname{cot}(\bar{\mathbf e}_0,\bar{\mathbf e}_2),\\
 c_{03}&=\operatorname{cot}(-\bar{\mathbf e}_0,\bar{\mathbf e}_1-\bar{\mathbf e}_0),
 &c_{04}&=\operatorname{cot}(-\bar{\mathbf e}_0,\bar{\mathbf e}_2-\bar{\mathbf e}_0),\\
 \mathbf k_e&=
 \begin{bmatrix}c_{03}+c_{04}\\c_{01}+c_{02}\\-c_{01}-c_{03}\\-c_{02}-c_{04}\end{bmatrix},
 &s_e&={\frac{3}{A_1+A_2}}.
 \end{aligned}
\]
The geometric matrix is
$\mathbf{Q}_e^{\mathrm{bend}}=s_e(\mathbf k_e\mathbf k_e^{\T})\otimes\mathbf{I}_3$.
With $\vx_e\in\R^{12}$ stacking current hinge vertices, the energy is
\begin{equation}
 E_e^{\mathrm{bend}}
 =\tfrac12\kappa_e^{\mathrm{bend}}\vx_e^{\T}
        \mathbf{Q}_e^{\mathrm{bend}}\vx_e .
 \label{seq:bending}
\end{equation}
{Its contribution to the incremental
potential is $h^2E_e^{\mathrm{bend}}$.}
{The hinge coefficient is the average of its adjacent
triangles' bending stiffnesses
$E_{\mathrm{bend}}\tau^3/[12(1-\nu_{\mathrm{bend}}^2)]$.
The matrix $\mathbf{Q}_e^{\mathrm{bend}}$ is symmetric and fixed during
differentiation with respect to current positions, giving}
\[
 \begin{aligned}
 \vg_e^{\mathrm{bend}}
   &=h^2\kappa_e^{\mathrm{bend}}\mathbf{Q}_e^{\mathrm{bend}}\vx_e,
 &\mathbf{H}_e^{\mathrm{bend}}
   &=h^2\kappa_e^{\mathrm{bend}}\mathbf{Q}_e^{\mathrm{bend}},\\
 \frac{\partial\vg_e^{\mathrm{bend}}}{\partial\kappa_e^{\mathrm{bend}}}
   &=h^2\mathbf{Q}_e^{\mathrm{bend}}\vx_e .
 \end{aligned}
\]
At fixed stiffness and current positions, the rest derivative is
$\mathrm{d}\vg_e^{\mathrm{bend}}
=h^2\kappa_e^{\mathrm{bend}}
 (\mathrm{d}\mathbf{Q}_e^{\mathrm{bend}})\vx_e$, where
\[
 \begin{aligned}
 \mathrm{d}\mathbf{Q}_e^{\mathrm{bend}}
  &=\bigl[(\mathrm{d}s_e)\mathbf k_e\mathbf k_e^{\T}
    +s_e(\mathrm{d}\mathbf k_e)\mathbf k_e^{\T}\\
  &\qquad+s_e\mathbf k_e(\mathrm{d}\mathbf k_e)^{\T}\bigr]
      \otimes\mathbf{I}_3,\\
 \mathrm{d}s_e
  &={-s_e
      \frac{\mathrm{d}A_1+\mathrm{d}A_2}{A_1+A_2}}.
 \end{aligned}
\]
{The area derivatives use the triangle
formula above.} Each cotangent
derivative is explicit:
\[
 \begin{aligned}
 \mathrm{d}\operatorname{cot}(\mathbf a,\mathbf b)
 &=\frac{(\mathrm{d}\mathbf a)^{\T}\mathbf b+
          \mathbf a^{\T}\mathrm{d}\mathbf b}{\|\mathbf a\times\mathbf b\|}\\
 &\quad-\frac{(\mathbf a^{\T}\mathbf b)(\mathbf a\times\mathbf b)^{\T}
  (\mathrm{d}\mathbf a\times\mathbf b+\mathbf a\times\mathrm{d}\mathbf b)}
  {\|\mathbf a\times\mathbf b\|^3}.
 \end{aligned}
\]
{These cotangent derivatives determine
$\mathrm{d}\mathbf k_e$ and complete the hinge's rest-shape derivative.}

{Rest-shape gradients also account for the mass
redistribution induced by geometry changes. A tetrahedron contributes
$\rho_eV_e/4$ to each incident vertex, and a cloth triangle contributes
$\rho_eA_e\tau_e/3$. Differentiating these weights gives
$\mathrm{d}\mathbf{M}$ and the inertial contribution}
\[
 \mathrm{d}\vg_t^{\mathrm{in}}
 =(\mathrm{d}\mathbf{M})(\vy_t-\widehat\vy_t)
   -\mathbf{M}\,\mathrm{d}\widehat\vy_t
\]
{at fixed solved states. The second term accounts for
parameter-dependent positions or velocities in the predictor.
Elastic, bending, and inertial terms provide local contributions to
$\mathbf{G}_t^{\T}\vlambda_t$ in
Equation~\eqref{eq:parameter-gradient}. Their products share the same
coupled adjoint and can be accumulated without forming the global
parameter-derivative matrix, as described in
Section~\ref{sec:parallel-accumulation}.}

\subsection{Joint geometry and control derivatives}
\label{sec:joint-actuation-details}
{Joint targets and forces influence the coupled scene
through the same frame equations as collision and friction.}
{Geometric constraints maintain the connection
between the two bodies $p$ and $c$.}
{The control energy sets a target for the joint's
permitted motion or applies a joint force. Differentiating both
energies in affine coordinates preserves their cross-body Hessian
blocks, allowing the adjoint to connect a joint command to the response
of interacting FEM and affine objects. The local expressions use
$\mathbf q=[\vq_p^{\T},\vq_c^{\T}]^{\T}\in\R^{24}$ and
$\nabla=\nabla_{\mathbf q}$. Joint anchors, reference directions, and
stiffnesses are held fixed when differentiating control parameters.
%The effective driving stiffness $k$ is the product of the prescribeddriving strength, the joint-type scale, and the initial total mass of the two bodies.
}

\subsubsection{Local maps and geometric constraints}
A material-space anchor $\bar\vx_b$ has world position
$\mathbf{J}_b\vq_b$, where $\mathbf{J}_b=\mathbf{J}(\bar\vx_b)$ is the point
map from Equation~\eqref{seq:point-map}. A material-space direction
$\bar{\mathbf a}$ has no translation component. Its map is
\begin{equation}
 \mathbf{D}(\bar{\mathbf a})=
 [\,\mathbf{0}_{3\times3}\quad\mathbf e_1\bar{\mathbf a}^{\T}\quad
       \mathbf e_2\bar{\mathbf a}^{\T}\quad
       \mathbf e_3\bar{\mathbf a}^{\T}\,],
 \label{eq:joint-direction-map}
\end{equation}
so $\mathbf{D}(\bar{\mathbf a})\vq_b=\mathbf{A}_b\bar{\mathbf a}$. The joint
energy uses these affine-transformed directions directly. For each body,
denote the transformed joint axis and two transverse reference directions by
$\mathbf a_b$, $\mathbf n_b$, and $\mathbf m_b$. Let
$\boldsymbol\Delta=\mathbf{J}_c\vq_c-\mathbf{J}_p\vq_p$.

A revolute joint matches two distinct points on its hinge line. For endpoint
$i\in\{0,1\}$, define
$\boldsymbol\Delta_i
=\mathbf{J}_{c,i}\vq_c-\mathbf{J}_{p,i}\vq_p$,
using the two bodies' material-space copies of that endpoint. A prismatic
joint constrains the anchor separation to lie along the joint axes and
matches transverse directions. The geometric residuals are
\[
 \begin{aligned}
 \mathbf c_g^{\mathrm{rev}}
   &=\begin{bmatrix}\boldsymbol\Delta_0\\\boldsymbol\Delta_1\end{bmatrix},
 &\mathbf c_g^{\mathrm{pri}}
   &=\begin{bmatrix}
       \boldsymbol\Delta\times\mathbf a_p\\
       -\boldsymbol\Delta\times\mathbf a_c\\
       \mathbf n_p-\mathbf n_c\\
       \mathbf m_p-\mathbf m_c
      \end{bmatrix}.
 \end{aligned}
\]
A fixed joint uses anchor coincidence and matching transverse directions:
$\mathbf c_g^{\mathrm{fix}}
=[-\boldsymbol\Delta^{\T},
  (\mathbf n_p-\mathbf n_c)^{\T},
  (\mathbf m_p-\mathbf m_c)^{\T}]^{\T}$.
For any of these residuals, the constraint energy is
$\Phi_g=\tfrac12 k_g\|\mathbf c_g\|^2$. With
$\mathbf J_g=\partial\mathbf c_g/\partial\mathbf q$, the chain and product
rules give
\begin{equation}
 \begin{aligned}
 \nabla\Phi_g&=k_g\mathbf J_g^{\T}\mathbf c_g,\\
 \nabla^2\Phi_g
 &=k_g\left(\mathbf J_g^{\T}\mathbf J_g+
       \sum_m c_{g,m}\nabla^2c_{g,m}\right).
 \end{aligned}
 \label{eq:joint-geometry-full-hessian}
\end{equation}
{The affine coordinate maps make point- and
direction-matching residuals linear in $\mathbf q$, with zero second
derivatives. Prismatic cross-product residuals are bilinear and supply
additional cross-body Hessian terms.}
The two cross-product terms in $\mathbf c_g^{\mathrm{pri}}$,
$\boldsymbol\Delta\times\mathbf a_p$ and
$-\boldsymbol\Delta\times\mathbf a_c$, contribute second derivatives
to the constraint-energy Hessian. To derive these terms, let
$\mathbf c_g^\times$ denote either cross-product residual:
\[
 \mathbf c_g^\times
 =(\mathbf B_1\mathbf q)\times(\mathbf B_2\mathbf q).
\]
With the joint reference geometry fixed,
$\mathbf B_1,\mathbf B_2\in\R^{3\times24}$ are constant
with respect to $\mathbf q$. For
the first prismatic block they are
{$\mathbf B_1=[-\mathbf J_p\ \mathbf J_c]$ and
$\mathbf B_2=[\mathbf D(\bar{\mathbf a}_p)\ \mathbf0]$.
For the second block, they are
$\mathbf B_1=[\mathbf J_p\ -\mathbf J_c]$ and
$\mathbf B_2=[\mathbf0\ \mathbf D(\bar{\mathbf a}_c)]$.}
Writing $[\mathbf v]_\times\mathbf w=\mathbf v\times\mathbf w$, the block
Jacobian is
\[
 \mathbf J_\times
 =-[\mathbf B_2\mathbf q]_\times\mathbf B_1
   +[\mathbf B_1\mathbf q]_\times\mathbf B_2 .
\]
For component $m$, let $\mathbf E_m=-[\mathbf e_m]_\times$, so
$\mathbf u^{\T}\mathbf E_m\mathbf v=(\mathbf u\times\mathbf v)_m$. A second
differentiation yields
\begin{equation}
 \nabla^2c_{g,m}^{\times}
 =\mathbf B_1^{\T}\mathbf E_m\mathbf B_2
  +\mathbf B_2^{\T}\mathbf E_m^{\T}\mathbf B_1.
 \label{eq:joint-cross-product-hessian}
\end{equation}
These constant component Hessians supply the residual-weighted term in
Equation~\eqref{eq:joint-geometry-full-hessian}.

\subsubsection{Prismatic displacement}
{The prismatic coordinate depends on the
vector joining the two joint anchors and the world-space
joint-axis vector of body $c$. Differentiating both vectors accounts for
the motion of both connected bodies in its gradient and Hessian.} Let
$\mathbf D_c=\mathbf D(\bar{\mathbf a}_c)$ and
$\mathbf a=\mathbf D_c\vq_c$. Differentiating
$\xi=\boldsymbol\Delta^{\T}\mathbf a$ gives
$\mathrm d\xi=(\mathrm d\boldsymbol\Delta)^{\T}\mathbf a+
 \boldsymbol\Delta^{\T}\mathrm d\mathbf a$, hence
\begin{equation}
 \xi=\boldsymbol\Delta^{\T}\mathbf a,\qquad
 \nabla\xi=
 \begin{bmatrix}
  -\mathbf J_p^{\T}\mathbf a\\
  \mathbf J_c^{\T}\mathbf a+\mathbf D_c^{\T}\boldsymbol\Delta
 \end{bmatrix}.
 \label{eq:prismatic-coordinate}
\end{equation}
{The gradient with respect to $\vq_c$
includes both anchor motion and axis variation. Differentiating
$\nabla\xi$ with respect to $\mathbf q$ gives}
\begin{equation}
 \nabla^2\xi=
 \begin{bmatrix}
  \mathbf0_{12\times12}&-\mathbf J_p^{\T}\mathbf D_c\\
  -\mathbf D_c^{\T}\mathbf J_p&
    \mathbf J_c^{\T}\mathbf D_c+\mathbf D_c^{\T}\mathbf J_c
 \end{bmatrix}.
 \label{eq:prismatic-coordinate-hessian}
\end{equation}
{The bilinear coordinate gives a constant Hessian at
fixed joint geometry. Its off-diagonal blocks retain the dependence
on both connected bodies.}

\subsubsection{Revolute orientation}
{Revolute actuation is expressed directly in the two
bodies' affine coordinates. Bilinear orientation quantities give
explicit gradients and constant cross-body Hessian blocks, retaining
the coupled angular response in compact motion coordinates.} Let
$\mathbf n_b=\mathbf D(\bar{\mathbf n}_b)\vq_b$ and
$\mathbf m_b=\mathbf D(\bar{\mathbf m}_b)\vq_b$ be the two transformed
reference directions on body $b\in\{p,c\}$. The orientation quantities used
by the control energy are
\begin{equation}
 \begin{aligned}
 s_{\mathrm r}&=\tfrac12(\mathbf n_c^{\T}\mathbf m_p-
                              \mathbf m_c^{\T}\mathbf n_p),\\
 c_{\mathrm r}&=\tfrac12(\mathbf n_p^{\T}\mathbf n_c+
                              \mathbf m_p^{\T}\mathbf m_c).
 \end{aligned}
 \label{eq:revolute-raw-signals}
\end{equation}
{The two quantities measure relative turning and
alignment. For rigid rotations with orthonormal reference frames and
aligned joint axes, they equal the sine and cosine of the relative
angle. The target energy uses their affine expressions directly,
so the derivatives retain changes in both the lengths and orientations
of the transformed reference vectors.}

{To express their bilinear form, let}
$\mathbf D_{n,b}=\mathbf D(\bar{\mathbf n}_b)$ and
$\mathbf D_{m,b}=\mathbf D(\bar{\mathbf m}_b)$, {and define}
\begin{equation}
 \begin{aligned}
 \mathbf Q_s&=\tfrac12(\mathbf D_{m,p}^{\T}\mathbf D_{n,c}
                         -\mathbf D_{n,p}^{\T}\mathbf D_{m,c}),\\
 \mathbf Q_c&=\tfrac12(\mathbf D_{n,p}^{\T}\mathbf D_{n,c}
                         +\mathbf D_{m,p}^{\T}\mathbf D_{m,c}).
 \end{aligned}
 \label{eq:revolute-signal-cross-hessians}
\end{equation}
Then $s_{\mathrm r}=\vq_p^{\T}\mathbf Q_s\vq_c$ and
$c_{\mathrm r}=\vq_p^{\T}\mathbf Q_c\vq_c$, giving
\begin{equation}
 \nabla s_{\mathrm r}=
 \begin{bmatrix}\mathbf Q_s\vq_c\\\mathbf Q_s^{\T}\vq_p\end{bmatrix},
 \qquad
 \nabla c_{\mathrm r}=
 \begin{bmatrix}\mathbf Q_c\vq_c\\\mathbf Q_c^{\T}\vq_p\end{bmatrix},
 \label{eq:revolute-signal-gradients}
\end{equation}
and
\begin{equation}
 \nabla^2s_{\mathrm r}=
 \begin{bmatrix}\mathbf0&\mathbf Q_s\\\mathbf Q_s^{\T}&\mathbf0\end{bmatrix},
 \qquad
 \nabla^2c_{\mathrm r}=
 \begin{bmatrix}\mathbf0&\mathbf Q_c\\\mathbf Q_c^{\T}&\mathbf0\end{bmatrix}.
 \label{eq:revolute-signal-hessians}
\end{equation}
The joint angle is
$\alpha=\operatorname{atan2}(s_{\mathrm r},c_{\mathrm r})$.
The target quantities $S_{\mathrm r},C_{\mathrm r}$ are defined in
Equation~\eqref{eq:revolute-target-signals}. For
$s_{\mathrm r}^2+c_{\mathrm r}^2>0$ on the selected angle branch,
differentiation gives
\begin{equation}
 \begin{aligned}
 \nabla\alpha
 &=\frac{c_{\mathrm r}\nabla s_{\mathrm r}
          -s_{\mathrm r}\nabla c_{\mathrm r}}
         {s_{\mathrm r}^2+c_{\mathrm r}^2},\\
 \nabla S_{\mathrm r}
 &=\cos a\,\nabla s_{\mathrm r}-\sin a\,\nabla c_{\mathrm r},\\
 \nabla C_{\mathrm r}
 &=\cos a\,\nabla c_{\mathrm r}+\sin a\,\nabla s_{\mathrm r}.
 \end{aligned}
 \label{eq:joint-revolute-coordinate-gradients}
\end{equation}
At fixed applied target $a$, the second derivatives of the target quantities
use the same linear combinations:
\begin{equation}
 \begin{aligned}
 \nabla^2S_{\mathrm r}
 &=\cos a\,\nabla^2s_{\mathrm r}-\sin a\,\nabla^2c_{\mathrm r},\\
 \nabla^2C_{\mathrm r}
 &=\cos a\,\nabla^2c_{\mathrm r}+\sin a\,\nabla^2s_{\mathrm r}.
 \end{aligned}
 \label{eq:revolute-target-hessians}
\end{equation}
For the angle, set $d_\alpha=s_{\mathrm r}^2+c_{\mathrm r}^2$,
$\mathbf u_s=\nabla s_{\mathrm r}$, and
$\mathbf u_c=\nabla c_{\mathrm r}$. Differentiating the numerator and
denominator of $\nabla\alpha$ yields
\begin{equation}
 \begin{aligned}
 \nabla^2\alpha
 &=\frac{c_{\mathrm r}\nabla^2s_{\mathrm r}
          -s_{\mathrm r}\nabla^2c_{\mathrm r}}{d_\alpha}\\
 &\quad+\frac{2s_{\mathrm r}c_{\mathrm r}}{d_\alpha^2}
       (\mathbf u_c\mathbf u_c^{\T}-\mathbf u_s\mathbf u_s^{\T})\\
 &\quad+\frac{s_{\mathrm r}^2-c_{\mathrm r}^2}{d_\alpha^2}
       (\mathbf u_s\mathbf u_c^{\T}+\mathbf u_c\mathbf u_s^{\T}).
 \end{aligned}
 \label{eq:revolute-angle-hessian}
\end{equation}

\begin{table*}[!t]
\centering
\small
\captionsetup{skip=3pt}
\caption{{Simulation settings for the design,
identification, and control tasks in Section~\ref{sec:evaluation}.
The columns report time step $h$,
relative collision activation distance $d_{\mathrm{rel}}$, barrier
coefficient $\kappa$, friction-smoothing velocity threshold
$\widehat d_f$, motion DoFs, relative Newton tolerance $\epsilon_N$,
friction coefficient, Young's modulus $E$ for volumetric FEM or cloth
stretching, and affine stiffness $\kappa_{\mathrm A}$.
For Dual-arm grasp, $0.3$ is the arm--cloth friction
coefficient and $0.2$ is the cloth--table coefficient.}}
\label{tab:sim-settings}
\setlength{\tabcolsep}{2.2pt}
\renewcommand{\arraystretch}{1.1}
\setlength{\aboverulesep}{0pt}
\setlength{\belowrulesep}{0pt}
\setlength{\cmidrulesep}{0pt}
\begin{tabularx}{\textwidth}{@{}c|*{9}{>{\centering\arraybackslash}X|}>{\centering\arraybackslash}X@{}}
\toprule
\multicolumn{1}{c|}{Example} & \multicolumn{1}{c|}{$h$} & \multicolumn{1}{c|}{Horizon} & \multicolumn{1}{c|}{$d_{\mathrm{rel}}$} & \multicolumn{1}{c|}{$\kappa$} & \multicolumn{1}{c|}{$\widehat d_f$} & \multicolumn{1}{c|}{\#DOF} & \multicolumn{1}{c|}{Newton $\epsilon_N$} & \multicolumn{1}{c|}{Friction} & \multicolumn{1}{c|}{$E$} & \multicolumn{1}{c}{$\kappa_{\mathrm A}$} \\
\midrule
Stone placement (Fig.~\ref{fig:stone-placement})
  & $0.03$ & $1000$
  & $10^{-3}$ & $10^{5}$ & $10^{-3}$
  & $84$
  & $10^{-3}$
  & $0.3$
  & -- & $10^7$ \\
\specialrule{\lightrulewidth}{0pt}{0pt}
Bunny collision (Fig.~\ref{fig:bunny-initial-velocity})
  & $0.05$ & $40$
  & $10^{-3}$ & $10^{10}$ & --
  & $5598$
  & $3\times10^{-4}$
  & $0$
  & $4{\times}10^7$ & -- \\
\specialrule{\lightrulewidth}{0pt}{0pt}
Tentacle matching (Fig.~\ref{fig:tentacle-trajectory})
  & $0.005$ & $100$
  & $10^{-3}$ & $10^{7}$ & $10^{-3}$
  & $20688$
  & $10^{-2}$
  & $1$
  & $2{\times}10^4$ & -- \\
\specialrule{\lightrulewidth}{0pt}{0pt}
Garment hanger (Fig.~\ref{fig:abd-hanger-rest})
  & $1/120$ & $600$
  & $10^{-3}$ & $10^{5}$ & $10^{-3}$
  & $51357$
  & $10^{-2}$
  & $0.05$
  & $5{\times}10^4$ & $10^8$ \\
\specialrule{\lightrulewidth}{0pt}{0pt}
Loaded support (Fig.~\ref{fig:fem-rest-shape})
  & $0.02$ & $20$
  & $3{\times}10^{-4}$ & $10^{6}$ & --
  & $22545$
  & $10^{-3}$
  & $0$
  & $10^8$ & -- \\
\specialrule{\lightrulewidth}{0pt}{0pt}
Material fitting (Fig.~\ref{fig:material-identification})
  & $0.1$ & $10$
  & $10^{-3}$ & $10^{7}$ & --
  & $12567$
  & $10^{-2}$
  & $0$
  & $10^6$ & -- \\
\specialrule{\lightrulewidth}{0pt}{0pt}
Friction optimization (Fig.~\ref{fig:friction-identification})
  & $0.02$ & $100$
  & $10^{-3}$ & $10^{6}$ & $10^{-3}$
  & $2799$
  & $10^{-5}$
  & $0.5{\to}0.1$
  & $10^5$ & -- \\
\specialrule{\lightrulewidth}{0pt}{0pt}
Cloth rest scale (Fig.~\ref{fig:dress10})
  & $1/24$ & $48$
  & $10^{-3}$ & $10^{5}$ & --
  & $102732$
  & $10^{-3}$
  & $0$
  & $10^5$ & $10^8$ \\
\specialrule{\lightrulewidth}{0pt}{0pt}
Grasp angle (Fig.~\ref{fig:xarm-target-closeup})
  & $0.02$ & $120$
  & $10^{-3}$ & $10^{5}$ & $10^{-3}$
  & $195$
  & $5{\times}10^{-3}$
  & $0.1$
  & $5{\times}10^5$ & $10^7$ \\
\specialrule{\lightrulewidth}{0pt}{0pt}
Cup catching (Fig.~\ref{fig:xarm-incoming-cup})
  & $0.005$ & $240$
  & $10^{-3}$ & $10^{5}$ & $10^{-3}$
  & $192$
  & $10^{-3}$
  & $1.0$
  & -- & $10^7$ \\
\specialrule{\lightrulewidth}{0pt}{0pt}
Dual-arm grasp (Fig.~\ref{fig:dual-piper})
  & $0.01$ & $330$
  & $10^{-3}$ & $10^{5}$ & $10^{-3}$
  & $12504$
  & $10^{-3}$
  & $0.3/0.2$
  & $3\times10^4$ & $10^9$ \\
\bottomrule
\end{tabularx}
\end{table*}

\begin{table}[!t]
\caption{{Physical parameter families exposed to
task-specific design variables through the interface in
Section~\ref{sec:interface-backend}. Each family contributes local
derivatives to the shared coupled adjoint.}}
\centering\small
\setlength{\tabcolsep}{2.4pt}
\captionsetup{skip=3pt}
\setlength{\aboverulesep}{0pt}
\setlength{\belowrulesep}{0pt}
\renewcommand{\arraystretch}{0.9}
\renewcommand{\tabularxcolumn}[1]{m{#1}}
\begin{tabularx}{\linewidth}{@{}>{\centering\arraybackslash}m{0.21\linewidth}|>{\raggedright\arraybackslash}m{0.33\linewidth}|Y@{}}
\toprule
\multicolumn{1}{c|}{Physical group} & \multicolumn{1}{c|}{Differentiated quantities} & \multicolumn{1}{c}{Role} \\
\midrule
Tetrahedral material & {$\mu$, $\lambda$} & Stable Neo-Hookean material response \\
\midrule[0.3pt]
Initial FEM motion & Selected initial vertex velocities & Initial dynamic state \\
\midrule[0.3pt]
Deformable rest shape & Selected rest coordinates & Volumetric and cloth rest geometry \\
\midrule[0.3pt]
Cloth material & Stretch, shear, and bending coefficients & Cloth constitutive response \\
\midrule[0.3pt]
Friction & Mesh self and ground friction coefficients & Frictional contact \\
\midrule[0.3pt]
Initial affine motion & $\vq_{0,b}$, $\dot\vq_{0,b}$ & Initial affine state \\
\midrule[0.3pt]
Affine rest shape & Selected surface-rest coordinates & Mass, inertia, shape energy, collision, and friction \\
\midrule[0.3pt]
Joint controls & Joint targets and forces & Revolute or prismatic actuation on selected joints and frames \\
\midrule[0.3pt]
Affine position targets & {Target positions} & {Position control} \\
\bottomrule
\end{tabularx}

\label{stab:registered-parameters}
\end{table}

\subsubsection{Control derivatives and coupled assembly}
{The coordinate derivatives determine the Hessians of the joint energies
in Section~\ref{sec:joint-derivatives} with respect to the motion DoFs.
Differentiating the energy gradients with respect to joint targets
and joint forces gives the control derivatives. For a prismatic joint, the
scalar multiplying $\nabla\xi$ is $k(\xi-a)-f$.
Its derivatives with respect to $a$ and $f$ are $-k$ and $-1$,
giving Equation~\eqref{eq:prismatic-control-derivatives}.}
For a revolute target,
$\partial_a S_{\mathrm r}=-C_{\mathrm r}$ and
$\partial_a C_{\mathrm r}=S_{\mathrm r}$. Differentiating their state
gradients also gives
$\partial_a\nabla S_{\mathrm r}=-\nabla C_{\mathrm r}$ and
$\partial_a\nabla C_{\mathrm r}=\nabla S_{\mathrm r}$. Substitution into the
product rule yields Equation~\eqref{eq:revolute-control-derivatives}. The
joint-force terms use coordinate gradients $-\nabla\xi$ and $-\nabla\alpha$.
At fixed $a$, $f$, and $k$, differentiating the state gradients gives the
full control-energy Hessians:
\begin{equation}
 \begin{aligned}
 \mathbf H^{\mathrm{pri}}
 &=k\nabla\xi\,\nabla\xi^{\T}
   +[k(\xi-a)-f]\nabla^2\xi,\\
 \mathbf H^{\mathrm{rev}}
 &=k(\nabla S_{\mathrm r}\nabla S_{\mathrm r}^{\T}
       +S_{\mathrm r}\nabla^2S_{\mathrm r})\\
 &\quad+kw_c(\nabla C_{\mathrm r}\nabla C_{\mathrm r}^{\T}
       -(1-C_{\mathrm r})\nabla^2C_{\mathrm r})\\
 &\quad-f\nabla^2\alpha .
 \end{aligned}
 \label{eq:joint-solve-blocks}
\end{equation}
Here $w_c$ is the fixed cosine weight from
Equation~\eqref{eq:revolute-joint-gradient}.

{Geometric and control Hessians use the same
$[\vq_p^{\T},\vq_c^{\T}]^{\T}$ ordering. Their off-diagonal blocks
couple the joint-connected bodies within the frame matrix, alongside
the contact blocks coupling them to the rest of the scene.
The local control derivative is multiplied by the corresponding
free-body entries of the global adjoint in
Equation~\eqref{eq:joint-control-gradient}. A shared control accumulates
contributions from every joint and frame to which it is applied.
All selected controls therefore use the same frame adjoints rather
than separate coupled solves for individual control variables.}

% Experimental settings and differentiated-parameter tables.

\section{Experimental Settings and Parameters}
\label{sec:experimental-details}

{Tables~\ref{tab:sim-settings} and
\ref{stab:registered-parameters} summarize the simulation settings
and differentiated parameter families, respectively, for the
experiments in Section~\ref{sec:evaluation}.}
{The task losses define the
observation frames and optimization stopping events. For tasks with
separate optimization and verification phases, the differentiated
trajectory provides the gradient at the specified observation or
event, and the full listed horizon assesses the resulting motion.
Forward and backward evaluation counts follow the definitions in
Section~\ref{sec:evaluation}.}

For scene diagonal $L_{\mathrm{box}}$, the collision activation gap is
$d_{\mathrm{rel}}L_{\mathrm{box}}$, whose square is $\widehat d$ in the
barrier energy. The friction-smoothing velocity $\widehat d_f$ gives the
displacement scale $\epsilon_{\mathrm{fr}}=h\widehat d_f$. The coefficients
$\kappa$ and $\kappa_{\mathrm A}$ control the collision barrier and affine
shape energy, respectively. With $\delta\vx$ denoting the Newton search
direction in world coordinates, the stopping criterion is
\begin{equation}
 \|\delta\vx\|_\infty<\epsilon_N hL_{\mathrm{box}}.
 \label{eq:benchmark-newton-test}
\end{equation}
{The shared FEM benchmarks use {the same}
convergence tolerances to align numerical accuracy in the timing
comparisons. The forward PCG relative residual tolerance is $10^{-4}$.}

{The parameter families in
Table~\ref{stab:registered-parameters} connect task-specific design
variables to material response, geometry, initial motion, and
actuation through the interface in Section~\ref{sec:interface-backend}.
Different parameterizations and objectives share the same coupled
backward computation. Each parameter family is differentiated in its
associated motion representation. The FEM rest-shape configuration
uses free FEM vertex positions. Affine rest-shape derivatives include
the effects of coupled contact and active friction.}

\clearpage

\end{document}